\documentclass{ws-book975x65}
\pdfoutput=1
\usepackage{ws-book-thm}           % comment this line when `amsthm` is used
\usepackage{ws-book-har}           % citation - Author-Date system
\usepackage[pdfpagelabels=false]{hyperref}  % \label, \ref and \cite are recommended
\usepackage{pdflscape}
\title{Science cases for a visible interferometer}              % set your book title here for even page running title

\newcommand{\apjl}{ApJL}
\newcommand{\apjs}{ApJS}

\newcommand{\aap}{A\&A}

\makeindex

\begin{document}

\begin{figure}
\includegraphics[width = 0.9\textwidth]{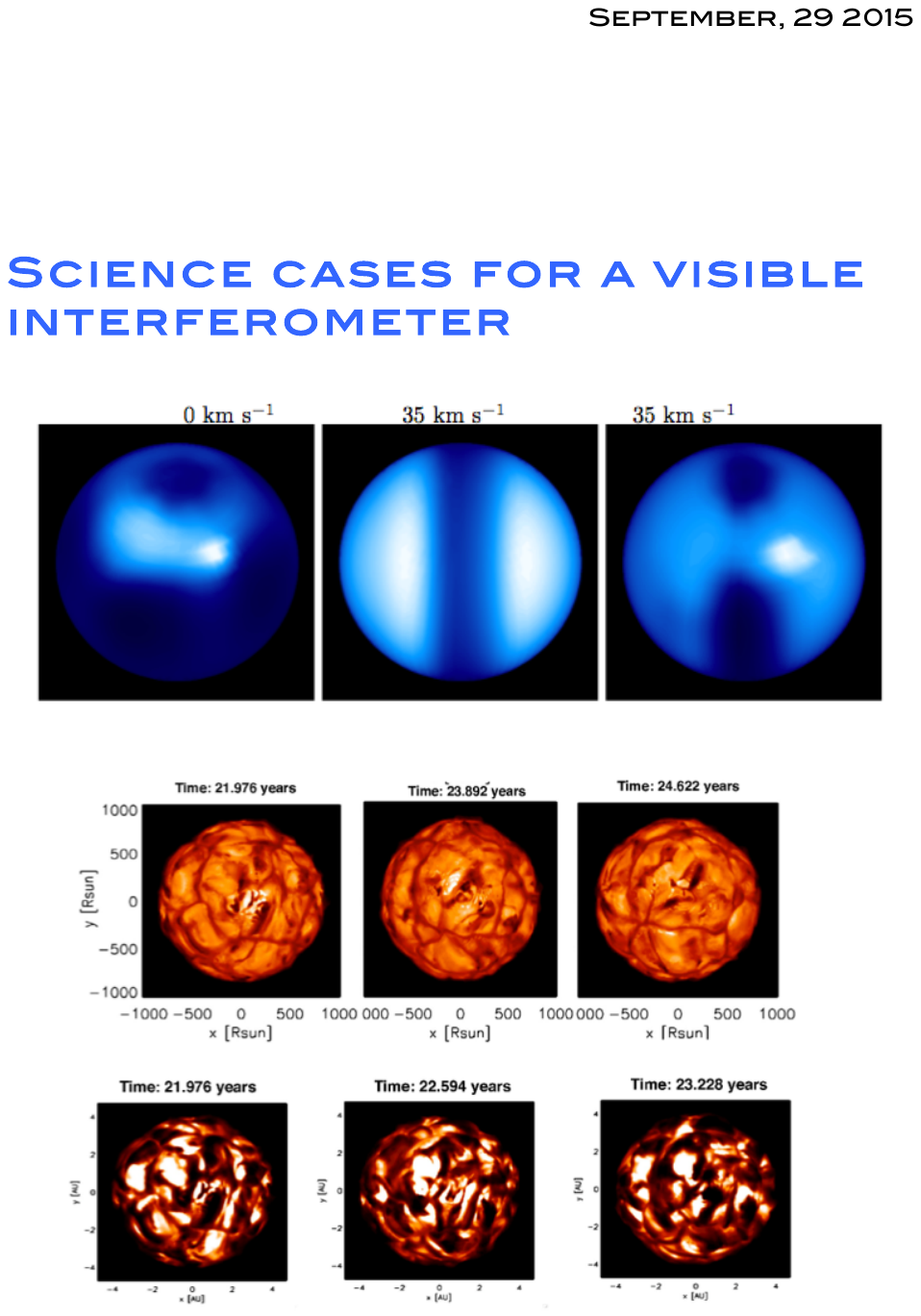}
\end{figure}

\titlepages                        % pls. do not remove this line

\begin{dedication}
\large This book is dedicated to the memory of our colleague Olivier Chesneau who passed away at the age of 41.
%\large (optional)
\end{dedication}

%\blankpage                         % blank page with no running heads

\begin{preface}
High spatial resolution is the key for the understanding of various 
astrophysical phenomena. But even with the future E-ELT, single dish
instrument are limited to a spatial resolution of about 4 mas in the visible 
whereas, for the closest objects within our Galaxy, most of the stellar 
photosphere remain smaller than 1 mas.\\

Part of these limitations was the success of long baseline interferometry with the AMBER
\citep{petrov2007} instrument on the VLTI, operating in the near infrared (K band) of the MIDI 
instrument \citep{leinert2003} in the thermal infrared (N band). One of the key point of the 
AMBER beam combiner was its capability of combining high spatial resolution
observations with spectrally resolved measurements thanks to its spectrograph
\citep{petrov1988} and the fact that it can combine up to 3 telescopes.
MIDI was limited to 2 telescopes baselines and a poor spectral resolution
allowing for broad band observations but with a higher sensitivity.
Thanks to its 3 beams combiner, AMBER was also able to measure closure
phases which was very useful to evidence various asymmetries in many
objects. \\

In 2010, VEGA  \citep{mourard11} was able to combine simultaneously
up to 4 telescopes while the fringes were externally stabilized with MIRC  \citep{monnier08}
in medium spectral resolution, i.e. 5000. More recently, PIONIER \citep{lebouquin2011} was 
also able to combine routinely up to 4 telescopes with a limited spectral resolution but a 
very good sensitivity which enables image reconstructions of faint binaries or extended sources, 
such as circumstellar disks around pre-main sequence stars.\\

In the following year, GRAVITY \citep{eisenhauer2011}, a four way beam combination second 
generation instrument for the VLTI will be installed in Paranal. GRAVITY will be able to measure 
astrometric distances between objects located within the 2$^{''}$ field-of-view of the VLTI. With the sensitivity 
of the 8 meters telescopes (UTs) and the $\sim$10 $\mu$as astrometric precision, it will allow to measure 
orbital motions within the galactic center with unprecedented precision. More or less at the same
period, MATISSE \citep{lopez2014},  a mid-infrared spectro-interferometer (from 3 to 13 $\mu$m) combining the beams of up to 4 UTs or Auxiliary 
1.8m Telescopes (ATs) of the VLTInterferometer, will extend the astrophysical potential of the VLTI 
by overcoming the ambiguities existing in the interpretation of simple visibility measurements from 
MIDI and will enable direct image reconstruction. MATISSE key science programs 
cover the formation and evolution of planetary systems, the birth of massive stars as well as the 
observation of the high-contrast environment of hot and evolved stars.\\

In the visible, after decades of spectrally resolved interferometry with respectively I2T, GI2T and 
VEGA on the CHARA array \citep{mourard09}  we have decide to push for a new instrument in the 
visible which will allow direct image reconstruction, following the PIONIER success as well as 
high spectral resolution, from the VEGA concept. In this direction a prototype, named
FRIEND (Fibered and spectrally Resolved Interferometric Experiment - New Design) as been
designed and was already tested onto the sky \citep{berio2014}. Having a visible instrument
on the VLTI seems to be promising since by reducing by a factor 4 the observing wavelength 
(going from the K band to the visible) is equivalent, regarding the spatial resolution, to an increase 
by a factor 4 of the baseline length, i.e. reaching a nearly kilometric baseline. This future instrument
can also be installed on the CHARA array, an US facility with 6 telescopes of 1m which will
be equipped with adaptive optics.\\

With  the idea to develop strong science cases for a future visible interferometer, we have organized 
a science group around the following main topics: Pre-main sequence stars, Main sequence stars, 
Fundamental parameters, Asteroseismology, Pulsating stars, Evolved stars, Massive Stars, 
Active Galactic Nuclei (AGNs) and Imaging technics. A meeting was organized
in January 15 \& 16, 2015 in Nice with the support of the Action Specific in Haute R\'esolution Angulaire (ASHRA),
the Programme National en Physique Stellaire (PNPS), the Lagrange Laboratory and the Observatoire de la C\^ote d'Azur, 
in order to present and discuss possible science cases for a future visible interferometers. This book presents 
these science cases developed with the help of the following people (last name, first name):\\

- Allard France
- Benisty Myriam
- Bigot Lionel
- Blind Nicolas
- Boffin Henri
- Borges Fernandes Marcelo
- Carciofi Alex
- Chiavassa Andrea
- Creevey Orlagh 
- Cruzalebes Pierre
- de Wit Willem-Jan
- Domiciano de Souza Armando
- Elvis martin
- Fabas Nicolas
- Faes Daniel
- Gallenne Alexandre
- Guerrero Pena Carlos
- Hillen Michel
- Hoenig Sebastian
- Irland Michael
- Kervella Pierre
- Kishimoto Makoto
- Kostogryz Nadia
- Kraus Stefan
- Labeyrie Antoine
- Le Bouquin Jean-Baptiste 
- Lebre Agn\`es
- Ligi Roxanne
- Marconi Alessandro
- Marsh Thomas
- Meilland Anthony
- Millour Florentin
- Monnier John
- Mourard Denis
- Nardetto Nicolas
- Ohnaka Keiichi
- Paladini Claudia
- Perraut Karine
- Perrin Guy
- Petit Pascal
- Petrov Romain
- Rakshit Suvendu
- Schaefer Gail
- Schneider Jean
- Shulyak Denis
- Simon Michal
- Soulez Ferreol
- Stee Philippe
- Steeghs Danny
- Tallon-Bosc Isabelle
- Tallon Michel
- ten Brummelaar Theo
- Thiebaut Eric
- Thevenin Frederic
- Van Winckel Hans
- Wittkowski Markus
- Zorec Juan

I warmly thanks all the contributors for their participation and the redaction of the following chapters.

\begin{flushright}
{\it Ph. Stee}
\end{flushright}

\end{preface}
                 %\begin{preface}...\end{preface}

%\input foreword.tex               %\begin{foreword}...\end{foreword}

%\input acknowledgments.tex        %\begin{acknowledgments}...\end{acknowledgments}

\tableofcontents

\setcounter{page}{1}

\chapter[Executive summary]{Executive summary\label{ch0}}

In this executive summary I have tried to extract the most important science cases from the various topics discussed during the Nice meeting. For more details please read the corresponding contribution in the following chapters.\\

{\it Fundamental properties of main sequence and sub-giant stars:}

A major topic is the  diameters estimation of planet-hosting stars for characterization of planetary systems and in particular to determine the planet's radius. This requires up to 2\% precision on the visibility measurements. Determining the effective temperatures of metal-poor stars for setting the temperature scale is also fundamental. The interferometric temperatures provide reference points for fixing the effective temperature scale.  This thus allows a more precise and accurate determination of the chemical abundances of more distant stars for studying the chemical evolution of our Galaxy. Radii and effective temperatures can be used for calibrating the 1D mixing-length parameter in stellar evolution models and determining ages of stars. Interferometric radii are complementary to asteroseismic data for solar-like stars, for determining precise masses (3\%) and ages (10\%). Magnetic inflation of cool, rapidly rotating stars can also be studied. Closure phases and imaging of stellar granulation and detection of planetary transits are possible in the visible as well as limb-darkening and surface spots and surface inhomogeneities determination in stars across the HR diagram. For that purpose, the following requirements are the most important in terms of fundamental parameters of main-sequence and sub-giant stars: Increase of the angular resolution and sensitivity and for higher precision, more telescope time is needed.\\
\\

{\it Pre-Main Sequence Stars:}

The young stellar objects (YSOs) present a strong science case for interferometry in the visible. Interferometric imaging and spectroscopy will provide unique and complementary data for understanding star and planet formation. The techniques will probe the innermost regions of protoplanetary disks and will enable diameter measurements of stars still contracting to the main sequence. The science case is very challenging mainly because of the brightness of targets. Determining the fundamental parameters of pre-main sequence stars is certainly a key subject. For instance measuring the angular diameter of Nearby Young Moving Groups (NYMGs) provide their physical diameter and hence an independent measurement of their age. Revealing complex structures around young stellar objects such as spirals, gaps, and holes at distances of few tens of astronomical units, probing the innermost regions are key processes for the star-disk-protoplanet(s) interactions. Accretion-ejection tracers is also possible by studying the HeI 10830 nm line as well as recombination and forbidden lines.
Companions hunting by high-contrast imaging is a key issue for pre-main sequence stars formation studies and scattered light studies tin the visible can reveal asymmetric structures that might be linked to planet formation.\\
\\

{\it Breaking the frontier to the cosmic distance scale using visible interferometry of cepheids and eclipsing binaries}

The general concept is to strengthen the physics of Cepheids. High-spectral resolution visible interferometry can be a very robust tool to bring new constrain on the dynamical structure of Cepheids' atmosphere, but could also provide a novel method of distance determination. The study of the projection factor derived from Cepheids in binaries can be done by the detection and characterization of more Cepheids in binary systems. Such detections are crucial since they bring new constrain on the mass of Cepheids and could help in resolving the Cepheid mass discrepancy and they allow a geometrical determination of the projection factor. The limb-darkening of Cepheids as a constrain on the projection facto is also a key topic. A precision of 5\% on the visibility in the second lobe would give the average value of the  limb-darkening of the star at about 1\% of precision. But even better, 1\% on the visibility in the second lobe would provide the time-dependency of the limb-darkening of the star with a 5$\sigma$ detection. The precision of the projection factor and the physics of the photosphere of Cepheids would be greatly improved with such unique measurement. Characterizing the circumstellar environment of Cepheids in the visible is mandatory  since the circumstellar envelopes we  have discovered around several Galactic Cepheids create a positive bias on their apparent luminosity. The robust Baade-Wesselink method combined with Gaia can put a constrain on the period-projection factor relation. For instance a 2\% precision on the distance (or similarly 2\% on the projection factor) for 75 Cepheids with CHARA and 40 with VLTI would lead to a 2\% precision on the individual projection factors bringing exciting result on the period-projection factor relation. High spectral resolution interferometry of Cepheids as an indicator of distances is very promising. Spectro-interferometry within metallic or even H$\alpha$ lines of Cepheids will become possible with future visible interferometers. Spectro-interferometry could be used to apply a new method of distance determination (with a 15$\sigma$ detection or 5\% in precision on the individual angular diameter for only one measurement. Moreover, the surface-brightness color-relation (SBCR) of early and late-type stars allows the determination of distances in the local group. With a new generation of visible long-baseline interferometry it could be possible to measure orbital separations of a large sample of Galactic eclipsing binaries. Thus,
we could compare several types of distances:
\begin{itemize}
\item The distance derived from the interferometric measurement of the separation (combined with spectroscopy)
\item The distance derived from the classical approach: photometry, spectroscopy are used to derive the individual radii and are combined with the angular diameter obtained from SBCR relation
\item If the individual angular diameter of the system are resolved by interferometry, once combined with individual radii (classical approach), it provides another estimate of the distance
\item The distance from trigonometric parallaxes with Gaia.
\end{itemize}  

These different approaches will greatly help in resolving some possible systematics in the method. \\

Finally, a next generation interferometer in the visible, with high precision measurements, 6 telescopes with baselines of 300m, but also high spectral resolution would be extremely important to reach the 1\% precision and accuracy on the distances in our Galaxy and beyond, but also to understand the physics of Cepheids and eclipsing binaries. Such improvements are crucial to reduce the precision and systematics on the Hubble constant.\\
\\

{\it Massive Stars}

Massive multiple stars studies is actually a key topic in massive stars formation theories. Interferometry in the visible is able to resolve binaries with separations ranging between 1 and 40 mas (at 0.6 $\mu$m). Establishing the distribution of binary separations and of mass ratios is essential to constrain the influence of mass loss and mass transfer on the evolution of high-mass binaries and to ascertain the origin and properties of their remnants. The nature of the circumstellar environment (CE) of active massive stars is central to understand many issues related to these objects. As a natural tracer of mass-loss, the study of this material is crucial to improve massive star evolution models. Moreover, the effect of rotation on the non-spherical distribution of mass-loss for these luminous objects is an active subject of debate. This is especially for classical Be stars which are known to be the fastest-rotating non-degenerated stars. To make progress in the understanding of these objects and the role of rotation in mass-loss processes, one needs to accurately determine the structure of their CE. By providing simultaneously high spatial and spectral resolution, interferometry in the visible is particularly well suited for the study of these objects. Measuring differential rotation on the surface of Bn star is a new and challenging subject. The measurement of the ratio of the centrifugal and the gravitational force at the stellar equator is crucial if we would like to know how close are Be stars to the critical rotation. Unfortunately, Be stars can have photospheric spectral lines marred by emission/absorptions due to their circumstellar disc. Bn stars, which are nearly as fast rotators as Be stars, don't have spectra perturbed by circumstellar matter, so that the study of their apparent geometry can be carried out more properly and reliably. Moreover, due to the rapid rotation, the surface geometry of Bn and Be stars is highly deformed. The surface angular velocity can be dependent on the stellar latitude. Then, not only the centrifugal force acting on the stellar surface, but also the effective temperature distribution (Von Zeipel effect) should depend on the particular surface rotation law. In order to constrain this differential rotation we definitely need an interferometric instrument in the visible since it the only way to combine high spatial resolution (for our purpose we need to reach $\sim$ 0.2 mas) and spectral resolution as high as 100000 in order to reach 0.1 $\AA$ in spectral lines. The direct detection of Non Radial Pulsations of massive stars is foreseen with a visible instrument. We propose to detect non-radial pulsations by differential interferometry using the dynamic spectra of photocenter shift variability characterized by bumps traveling from blue to red within the spectral lines. The theoretical estimation of expected signal-to-noise ratios in differential speckle interferometry demonstrated the practical applicability of the technique to a wide number of sources. We finally remind that up to now, interferometric studies of active hot and massive stars have produced the largest contributions of referred papers using the VLTI/AMBER and CHARA/VEGA instruments. This is certainly due to the fact that they are bright (many sources have m$_{v}$ $<$ 7) with diameters well suited for current interferometric baselines, i.e. central stars with diameters $\sim$ 0.5 mas and circumstellar disk of a few mas. These properties will certainly also benefit from a future visible interferometric instrument.\\
\\

{\it Evolved stars, Planetary Nebulae, $\delta$~Scuti, and RR~Lyrae  as seen by a visible interferometer}

Determining the dust distribution around Asymptotic Giant Branch (AGB) Stars is certainly a key topic. The atmosphere of an AGB star is cool enough to allow formation of various dust species. Dust plays a crucial role in the stellar wind, although there are some cases where the dynamics is not completely understood. The gain of going to the visible to study the dust distribution around AGBs is that the scattering from dust contribution becomes more important. The challenge is the fact that dusty stars become fainter. 
The size of the photosphere in the visible will be in average $<10$~mas,but the dust component is expected to be more extended, especially for the nearby objects. Mapping magnetic fields and velocity fields over the surface of stars will also be possible. The spatially resolved spectra of regions with strong local
magnetic fields (e.g., starspots) would show the Zeeman splitting, which enables us to map the magnetic fields over the surface of stars. Spatially resolving the chromosphere and its dynamics is very interesting. We propose to image the chromosphere and derive its physical properties by interferometry of the H$\alpha$ and Ca II lines. High resolution imaging would help to understand the heating mechanism of the chromosphere, its role in the mass loss, and why it can coexist with the molecular component.  Another promising subject is to probe the MOLsphere with TiO lines thanks to a visible instrument. The MOLsphere is quasi-static, dense, warm molecular layers extending to several stellar radii and exists not only in Mira stars but also in normal red giants and red supergiants.  Spatially resolving the stars in the TiO lines helps to better understand the structure of the outer atmosphere.  Given that TiO may serve as seed nuclei for dust formation, it is important to know the physical properties of TiO layers for understanding the dust formation. The geometrical characterization of shock waves in Mira stars is foreseen in the visible. Radially pulsating AGB stars (a.k.a. Mira stars) are characterized by strong emission in the Balmer lines linked to the propagation of a strong hypersonic radiative shock waves. Spectro-interferometry in the visible would help establish the shape and time evolution of these shock waves, all of this in comparison to
the shape and evolution of the photosphere. The study of AGB Stars is very promising. AGBs are very luminous and have extended atmosphere, therefore they become naturally the first objects to be observed with infrared interferometry. When moving to the visible though one has to be careful. In fact many evolved AGBs will be faint because of dust obscuration. Typical magnitudes for the most evolved Mira variables will vary between 8 and 13 mag in the V-band. One has also to take into account the fact that Mira variable will change their brightness of up to 9 magnitude within one year. Imaging programs will have to take this into account, and a proper number of telescopes (6+) or a fast system to change configuration (in the current VLTI scheme: 3 configurations within 2 weeks maximum) are needed. Massive evolved stars with masses between roughly 10 and 25~M$_{\odot}$ spend some time as red supergiant (RSG) making them the largest stars in the universe. The understanding of the physics of their convective envelope is crucial for these stars that contribute extensively to the dust and chemical enrichment of the galaxies. Moreover, the mass-loss significantly affects the evolution of massive stars, and it is a key to understanding the progenitors of core-collapse supernovae. The effects of convection and non-radial waves can be represented by numerical multi-dimensional time-dependent radiation hydrodynamics (RHD) simulations with realistic input physics carried out with CO$^5$BOLD code. These simulations have been used extensively employed to predict and interpret interferometric observations. Optically-bright post-AGB stars are commonly found in binary systems with separations of typically 1~AU and that are surrounded by rather compact, stable circumbinary disks. These disks are very similar to the protoplanetary disks (PPDs) around young stars. However, the formation history of post-AGB disks is clearly different from PPDs, as is the further evolution of the illuminating star. A visible interferometric instrument is highly needed to extend the constraints on the parameter space of the models 
beyond what the ALMA and the second-generation instruments on the VLTI are able to provide. Visible interferometry can provide unique additional insights and constraints.
There are, in particular, two fundamental questions that can be addressed by such an instrument: Locating the continuum scattered light around post-AGB binaries and 
tracing the inner gas streams in post-AGB binaries: accretion disk and/or jet ? Finally, as for Cepheids, the projection factor of $\delta$ Scuti and RR Lyrae stars can be constrained. The objective is to derive the expected period-projection factor relation of RR Lyrae and High-Amplitude $\delta$ Scuti stars (HADS) by applying the inverse BW method. HADS are generally supposed to pulsate radially. However, weak non-radial pulsation could be detected using long baseline visible interferometry. In that case, the Baade-Wesselink method, if revised, can constitute an interesting tool to distinguish radial and non-radial modes of pulsation. Last but not least, another very interesting aspect is to study pulsating stars in binaries in order to measure their orbital solution with interferometry and derive their mass, which put important constrains on the evolutionary models. It concerns not only  $\delta$ Scuti stars, but also RR Lyrae, $\beta$ Cepheids, $\gamma$ Dor and RoAp stars.  \\ \\

{\it Interacting binaries}\\

The presence of a companion will be betrayed by its signature in the interferometric signal. If the difference in magnitude between the two objects is less than 6, i.e. the contrast in flux is about 200 or less, then current interferometers such as PIONIER on ESO's Very Large Telescope can detect the companion, using visibilities and closure phases provided there is an error of 0.5 degrees or less on the latter (Le Bouquin et al. 2011). Once the companion has been detected, it is then possible to follow the relative orbital motion of the two components, providing a visual orbit, which coupled with radial velocities measurements lead to the measurement of the masses of the two stars. Working at high spectral resolution  allows one to probe the smallest interacting systems by spectro-astrometry. While the giant is in generally dominating the continuum emission in the visible domain, the faint companion is situated in the energetic accretion zone presenting strong emission line like H$\alpha$. The photocentre of the binary, centred on the giant companion in the continuum, shifts towards the accretor in the line, which could be measured by help of differential phases. A simple calculation and the experience gained on VEGA show that reaching precision of order 10 $\mu$as or better is well within reach, and is therefore able to probe the most compact semi-detached systems. Mass is the most crucial input in stellar internal structure modeling. It predominantly influences the luminosity of a star and, therefore, its lifetime. Unfortunately, the mass of a star can generally only be determined when the star belongs to a binary system. Therefore, modeling stars with extremely accurate masses (better than 1 \%), in different ranges of masses, would allow to firmly anchor the models of the more loosely constrained single stars.
Interferometry in the visible is quite suitable for that, as already demonstrated by PIONIER: it  observes the relative orbits instead of the photocentric ones, and this compensates the fact that the accuracy of the instrument is a bit worse than that of Gaia. For systems which are double-lined spectroscopic binaries, that is, for which we already have the mass ratio, as well as the orbital period and eccentricity, using interferometry to make it a visual binary, allows us to determine the remaining parameters, i.e. the inclination and the individual masses. This thus allows obtaining precise stellar masses. Symbiotic stars show a composite spectrum, composed of the absorption features of a cool giant, in addition to strong hydrogen and helium emission lines, linked to the presence of a hot star and a nebula. It is now well established that such a ``symbiosis'' is linked to the fact that these stars are active binary systems, with orbital periods between a hundred days and several years. In such systems, the red giant is losing mass that is partly transferred to the accreting companion -- either a main sequence (MS) star or a white dwarf (WD). One of the main questions related to symbiotic stars is how the mass transfer takes place: by stellar wind? Through Roche-lobe overflow (RLOF)? Or through some intermediate process?  Answering this question requires  being able to compare the radii of the stars to their Roche lobe radius (which depends on the separation and the mass ratio). However, determining the radius of the red giant in symbiotic systems is not straightforward, and there has always been some controversy surrounding this. Optical interferometry is currently the only available technique that can achieve this. It allows determining the size and the distortion of the giant star, and in some cases, the orbital parameters of the system, without any a priori on their characteristics.

{\it Imaging, technics and the FRIEND prototype.}

For the CHARA array an important limitation for imaging purposes is the limitation of short baselines.  Adding one telescope would be an interesting idea to boost its imaging capabilities. The VLTI has larger telescopes but no visible beam combiner up to now. The VLTI is not yet ready for visible light combination, and a few technical tweaks need to be implemented to enable UBV bands (RIJ do go through already). The MACAO dichroic would need to be changed and we are wondering if one would be able to select between different dichroic/semi-reflecting plates. The guiding strategy should also be modified. Regarding the polychromatic data there is already some existing tools thanks to the JMMC to prepare interferometric observations and make some model fitting such as ASPRO, LITPro and searchCal. On the other side, a new visible instrument will provide differential visibilities and differential phases, as do VEGA, AMBER and MIDI today but current tools (LITpro, MIRA, BSMEM) used squared visibilities and closure phases only. Some developments have occurred to take chromatic information into account. The POLCA project has successfully developed tools for handling polychromatic data. In image reconstruction, new algorithms are now able to reconstruct full 3D polychromatic maps from all available interferometric data. Project POLCA has also allowed improvements in LITpro model fitting software, by including for instance VEGA differential visibilities taking nutrimental chromatic artifacts into account. Spectral resolution is also important for helping regularizations in image reconstruction, for instance when there is a large continuum with various spectral lines. PIONIER instrument was missing the spectral information which would have been better for data interpretation.  VEGA has spectroscopic capabilities but is limited due to saturation effects on the detector. Varying spectral resolution as a function of baseline length is a powerfull method in some hyperspectral remote sensing studies since high frequency information position is not going to change much depending on high frequency baseline. The spectrum of the object is also quite important for image reconstruction. For the number of telescopes and the baselines length, a possibility would be many telescopes with few delay lines that can be switched quickly wheres another option would be a continuously following of one baseline along time with a high frequency of recording. A minimum is certainly 4 telescopes and different configurations to make imaging. Experience at CHARA with MIRC shows that 6 telescopes is a strict minimum for snapshot imaging. In the same direction we recall that the Plateau de Bure interferometer of IRAM was only able to make direct images when the telescope network was extended to 6 antennas  within 10 years and that actually the NOEMA project is extended the array to 12 antennas. Regarding the baselines length, short baseline are really mandatory for imaging purposes in order to fix features in the field of view. Longest baselines are not always usable due to SNR issues, and short baselines are needed to get the low frequencies. This pushes the need of additional telescopes. In the next 2 or 3 years, the two major interferometric arrays, VLTI and CHARA, will equip their telescopes of 1.8m and 1m respectively with Adaptive Optics (AO hereafter) systems. This improvement will permit to apply with a reasonable efficiency in the visible domain, the principle of spatial filtering with single mode fibers demonstrated in the near-infrared. It will clearly open new astrophysical fields by taking benefit of an improved sensitivity and state-of-the-art precision and accuracy on interferometric observables. To prepare this future possibility, we started the development of a demonstrator called FRIEND (Fibered and spectrally Resolved Interferometric Experiment - New Design). FRIEND combines the beams coming from 3 telescopes after injection in single mode optical fibers and provides some spectral capabilities for characterization purposes as well as photometric channels. It operates in the R spectral band (from $600nm$ to $750nm$) and uses the world's fastest and more sensitive analogic detector OCAM2. Tests on sky at the focus of the CHARA interferometer have been successfully done in December 2014. \\
\\

{\it AGNs}\\
Active Galactic Nuclei (AGN) are extremely bright sources powered by the accretion of material on a central super massive black hole (SMBH). They emit more than 1/5 of the electromagnetic power in the universe and a majority of galaxies might host a central BH triggering some level of nuclei activity. AGNs can be considered as important contributors and markers for the global history of mass accretion and galaxy evolution in the Universe. If well understood, they could be used as standard candles for the evaluation of cosmological distances at redshifts $z>3$. Quasars make the current reference grid for the calibration of GAIA astrometry but, at the tens of $\mu$as accuracy of GAIA, the structure of these sources could have a significant impact on the definition of their GAIA photocenter.
We have an unified model of AGNs (\citet{Antonucci1993}, \citet{Urry1995}) that is increasingly considered as over simplistic but still offers a simple and useful structure to discuss typical sizes and magnitudes and therefore to contribute to a first evaluation of the potential of visible long baseline interferometry. This model features a very compact accretion disk (AD) around the central SMBH, a Broad Line Region (BLR) composed of high velocity gas clouds producing broad emission lines when this central region is not obscured by the clumpy dust torus located after the dust sublimation radius. This dust torus (DT) collimates the light from the central source that can ionize some narrow line region (NLR) lower velocity gas clouds placed in the non-obscured double cone and producing narrow emission lines. When an AGN is close to equator-on, the dust torus shields the BLR, that can be detected only in polarized light reprocessed by the more far away NLR clouds. Some AGNs emit high velocity jets.

Observations in the visible would bring a decisive improvement, if the interferometer can be made sensitive enough in that spectral band. First we will gain in resolution by a factor 4 compared to near-infrared observations. Second we will be allowed to use Balmer lines (for low z sources) that are 3 to 20 times stronger than the Paschen and Bracket lines available in the near IR. The combination of these two effects would yield a significant gain in the accuracy of quasar parallax distance measurements, as well as in the other parameters measures such as the masses. It will also strongly enhance the possibility to image gas in direct relation with the dust torus.

Another major advantage of visible observations is to allow a direct combination with RM observations that have so far been made almost exclusively in the visible, without the need to use models to correct the relative scales in the visible and in the IR, nor having to launch very long term IR RM campaigns.
The condition for visible observation of BLRs with the VLTI to give a significant contribution is to be able to observe with a sufficient limiting magnitude in V. A very preliminary results of a feasibility study of BLR observation in visible using VLTI from \citet{Rakshit2015}, extrapolated from his work for the K band. It shows that:

\begin{itemize}

\item At V=14, we would obtain visibility, differential visibility and differential phase for all the GRAVITY targets (without FT), but with an improved distance accuracy by typically a factor 10. Such a magnitude could be achieved with the ATs if they have adaptive optics system providing a Strehl ratio of 0.5 in the (red) visible. In a few cases we will obtain images of the gas in or above the dust torus.

\item At V=15, we will be able to observe more than 120 quasars. For all K band VLTI targets we should have substantial distance and mass accuracy improvements. The exact gains need to be assessed but the very first estimate is that we would have more than 50 targets with Quasars parallax more accurate than 10\%. Such a magnitude can be achieved on the UTs if they are equipped with an AO system providing a Strehl ratio of 10\% in the red visible.

\end{itemize}

\chapter[Fundamental properties of main sequence stars from interferometric measurements]{Fundamental properties of main sequence stars from interferometric measurements\label{ch1}}

\newcommand{\teff}{$T_{\rm eff}$}
\newcommand{\feh}{[Fe/H]}
\newcommand{\mh}{[M/H]}
\newcommand{\logg}{$\log g$}
\newcommand{\mlsep}{$\langle \Delta \nu \rangle$}
\newcommand{\numax}{$\nu_{\rm max}$}
\newcommand{\pmm}{$\pm$}
\newcommand{\msol}{M$_{\odot}$}
\newcommand{\rsol}{R$_{\odot}$}
\newcommand{\lsol}{L$_{\odot}$}
\newcommand{\red}{\textcolor{red}}
\newcommand{\blue}{\textcolor{blue}}
\newcommand{\rad}{$R$}
\newcommand{\zx}{$Z_{\rm i}/X_{\rm i}$}
\newcommand{\mhz}{$\mu$Hz}
\newcommand{\gmb}{Gmb\,1830}
\newcommand{\hd}{\object{HD\,140283}}
\newcommand{\kms}{km s$^{-1}$}
\newcommand{\fbol}{$F_{\rm bol}$}
\newcommand{\es}{erg cm$^{2}$ s$^{-1}$}
\newcommand{\chisqr}{$\chi^2_{\rm R}$}
\newcommand{\chisq}{$\chi^2$}
\newcommand{\av}{A$_{\rm V}$}
\newcommand{\thet}{$\theta$}
\newcommand{\vmic}{$v_{\rm micro}$}
\newcommand{\vmac}{$v_{\rm macro}$}
\newcommand{\vsini}{$v\sin i$}
\newcommand{\pii}{$\pi$}
\newcommand{\hdd}{HD\,140283}
\def\uma{$\varepsilon$~UMa}

 \author{O.~L.~Creevey$^{1,2}$,
 L.~Bigot$^{2}$,
 A.~Chiavassa$^{2}$,
   P.~Petit$^{3}$,
F.~Allard$^{4}$}
 \author{R.~Ligi$^{2}$,
  N.~Nardetto$^{2}$,
  D.~Shulyak$^{5}$,
M.~Wittkowski$^{6}$
 }

\begin{center} 
$^{1}$~Insitut d'Astrophysique Spatiale, UMR 8617, Universit\'e de Paris-Sud, 91405, Orsay, France\\
$^{2}$~Laboratoire Lagrange, UMR 7293, CNRS, Observatoire de la C\^ote d'Azur, Universit\'e de Nice Sophia-Antipolis,
Nice, France\\
$^{3}$~Institut de Recherche en Astrophysique et PlanŽtologie, 14 Avenue Edouard, Belin, F-31400 Toulouse, France\\
$^{4}$~Centre de Recherche Astrophysique de LyonÑEcole Normale Sup\'erieure de Lyon, 46 All\'ee dÕItalie, 69364 Lyon Cedex 07, France\\
$^{5}$~Institute of Astrophysics, Georg-August University, Friedrich-Hund-Platz 1, D-37077 G\"ottingen, Germany\\
$^{6}$~ESO, Karl-Schwarzschild-St. 2, 85748, Garching bei M\"unchen, Germany\\
\end{center} 

\section{Introduction}

Visible interferometry has many applications to main sequence stars:
studying activity, determining fundamental parameters, and galactic
archaeology.
and complementary observations for asteroseismology.
In this chapter we discuss several science cases in terms of 
current and future interferometers.  In particular we emphasize the 
need for pushing the sensitivity limits to reach fainter targets and thus
better calibrators and higher signal-to-noise target observations which
in turn bring more confidence to observations such as squared visibilities
higher in the visibility curve, thus giving access to diameters as low
as 0.1 mas.
In this chapter we cover the topics of visible interferometry for 
fundamental parameters, main sequence and sub-giant stellar evolution and
asteroseismology.

To begin with we may ask why we care about main sequence stars of which we 
apparently know a lot about.  
Remember that main sequence stars cover a huge percentage of the 
observable Galaxy, 
and they are also the main hosts for searching for Earth-like planets. 
While we have a good understanding
of these stars and thus we can delve into details, there are also
some very basic properties that we have a poor knowledge of.  The poor 
knowledge of such fundamental parameters has implications for 
understanding the star.
Here fundamental parameters implies stellar radii, effective temperatures, 
masses and ages.
Imagine we have an excellent grasp on all of these parameters, then we
can calibrate age-rotation relations (gyrochronology) and thus date
whole populations of stars and planet-hosts to constrain their evolution, we can
determine very accuracte and precise chemical abundances (because we know the
effective temperature and surface gravity of a star) and in fact we can then
use this precise knowledge to make improvements to stellar models 
(evolution and atmosphere) because we have direct indisputable measurements for 
which, for example, diffusion theories need to heed or non LTE corrections can
be calibrated.  Here we list a non-exhaustive list of some of the 
main science drivers 
for determining fundamental parameters and studying main sequence stars
 and in the following sections we describe
in more detail how a selection of these topics can benefit {\it only} from 
long-baseline optical/visible interferometry, and we address the topics
in order of technical requirements.

The following sections are organised as follows, where we note that all require high angular resolution ($>0.1$ mas) hence the need to observe in the visible domain.\\

{\bf Low/Medium spectral resolution:}

\begin{itemize}
\item[{\bf A}] Sect.~\ref{fundsec:diameters} No time constraints (sizes, diameters, V2 measurements)
\item[{\bf B}] Sect.~\ref{fundsec:conv-planet} Time sensitive and time series (convection, planet-detection,
closure phases and imaging measurements)
\end{itemize}

{\bf Medium/High spectral resolution:}

\begin{itemize}
\item[{\bf C}] Sect.~\ref{fundsec:spots-ldark} Spots and limb-darkening and
needing V2 and closure phases
\end{itemize}

%\section{Low/Medium spectral resolution}
\section{\hspace{0.2cm}{\bf A}\hspace{0.2cm} Diameters of stars from non time-constrained V2 observations\label{fundsec:diameters}}

Angular diameters from interferometric {visibility measurements} provide an important input for many different science cases.  
From a measured angular diameter, and using complementary information, we determine with
very little model dependence the radius and effective temperature of a star.
In many science cases we require $<2$\% precision on the angular diameter, 
with typical ``single-shot'' observations of between 4 -- 6 hours, 
over 2 or 3 nights to ensure
reliability, with 2 or more telescopes.  Depending on the nature of the scientific program,
this 2\% precision can be relaxed.
The advantage of visible wavelengths over infra-red is notably the ability to reach a higher angular resolution, e.g. compare 0.2 mas with VEGA \citep{Mourard2009, Ligi2013} on CHARA to 0.9 mas with Classic on CHARA.
The CHARA array \citep{McAlister2012} holds the longest baselines in the world (up to $330$~m), and thus presently provides the ability for the highest angular resolution.  It hosts two visible interferometers VEGA  and PAVO \citep{Ireland2008}, and thus benefit from the best angular resolution in the world ($\sim 0.3$~mas), despite a limiting magnitude in the V band of around $\sim 8$. 
The NPOI's current longest baseline is 79 m, but in the near future this will reach 432m \citep{Baines2014}.

\begin{figure}
\includegraphics[width = 0.7\textwidth]{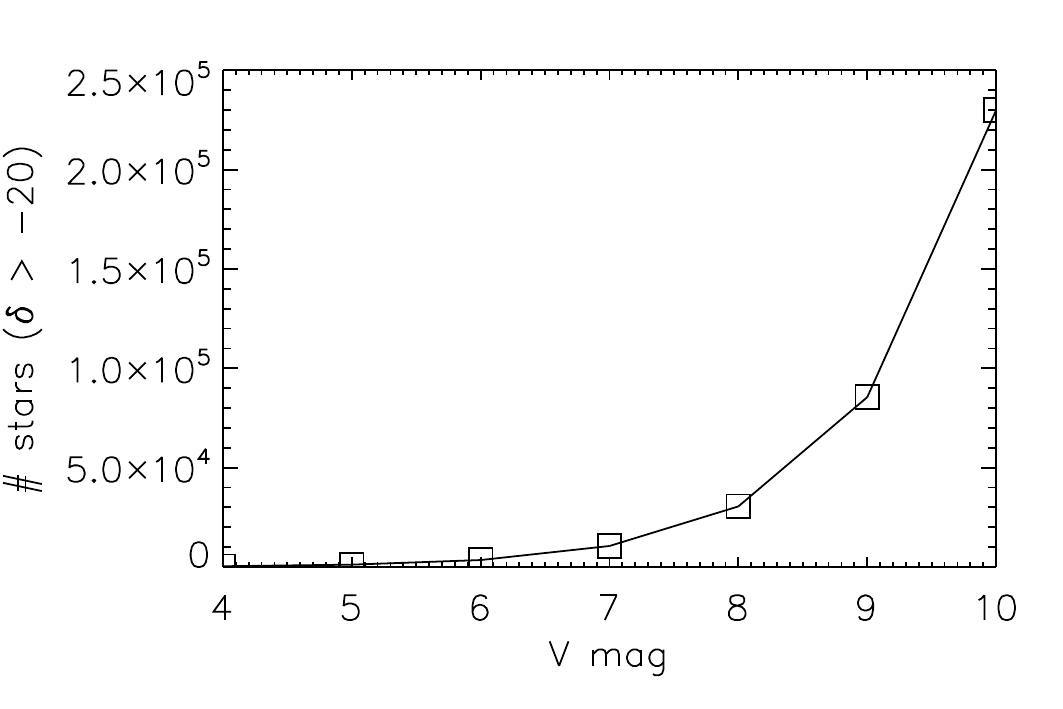}
\caption{The number of stars with declinations above --20$^{\circ}$ (observable from CHARA) as a function of magnitude\label{fundfig:numstars}}
\end{figure}

\begin{figure}
\includegraphics[width = 0.7\textwidth]{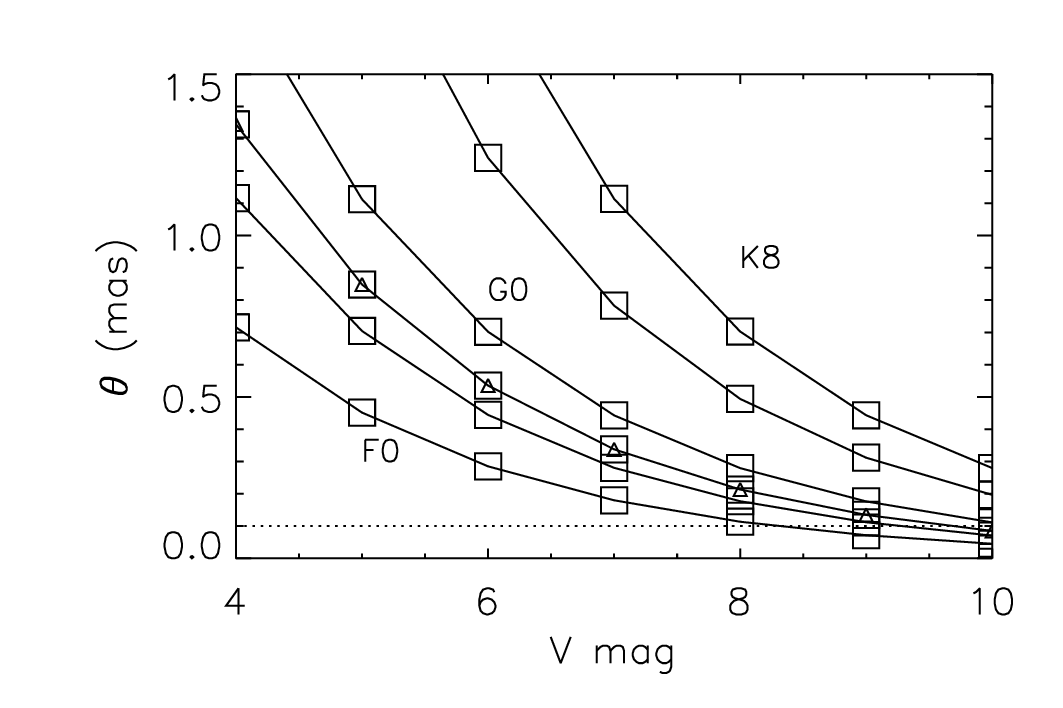}
\caption{The predicted angular diameters of stars of different spectral types as a function of magnitude number of stars grows exponentially as the magnitude limit is extended.  The dotted line is the hard lower limit in angular diameter.\label{fundfig:mag}}
\end{figure}

A new prototype of VEGA, called FRIEND \citep{berio14}, is currently being tested on CHARA and observes in a similar visible wavelength region, but can reach higher sensitivity.
This will result in higher precision measurements for brighter stars along with the
ability to probe a few magnitudes deeper $m_{\rm V} < 10$.   These few extra magnitudes contribute significantly to number of interesting targets to be measured (figure \ref{fundfig:numstars}).  Of course not all of these stars will be resolved, but if we can measure precise
visibilities even high on the visibility curve we can attain a 0.1 mas limit (precision to be tested).  In Fig.~\ref{fundfig:mag} we show the predicted angular diameter of ZAMS stars as 
a function of apparent magnitude and spectral type. As can be seen the ability to probe
higher magnitudes is of great interest considering upcoming missions, such as the TESS mission, which will target bright G and K stars.  In the following section we will discuss 
a series of scientific drivers.


\subsection{Planetary Systems: stellar diameters for characterisation of planetary systems and determining planet radii}

\begin{figure}
\includegraphics[scale=0.5]{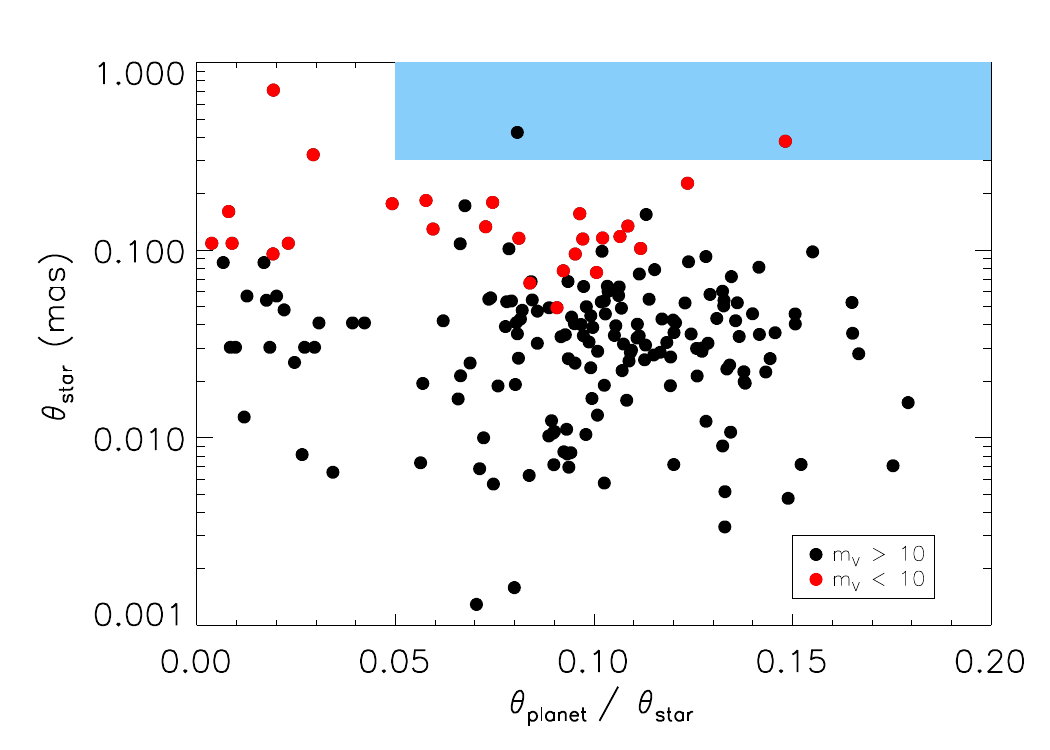}
\includegraphics[scale=0.5]{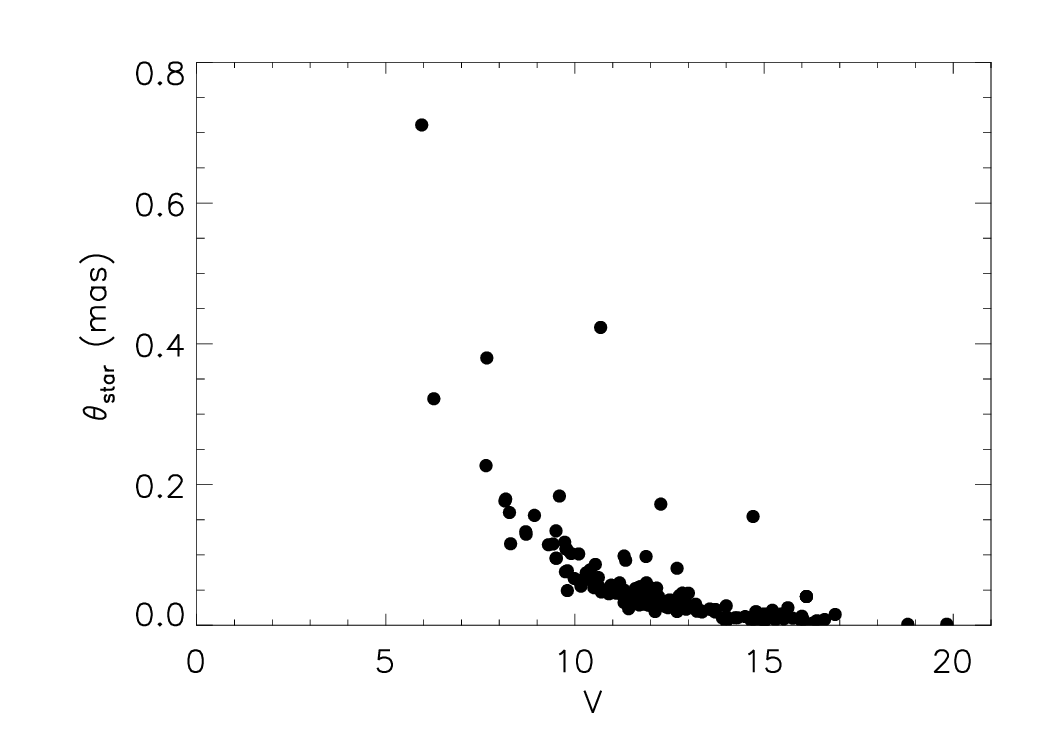}
\caption{(Top) Transiting exoplanets host stars with available distance and radius that allow us to  derive their angular diameter. They are plotted according to the ratio $\theta_{\rm p}/ \theta_{\star}$, with $\theta_{\rm p}$ and $\theta_{\star}$ the planetary and stellar angular diameters. Stars with $m_{\rm V} < 10$ are plotted in red and those with $m_{\rm V} > 10$ in black. The blue box represents the detecting ability of VEGA \citep{Ligi2015}.
(Bottom) Angular diameters of transiting exoplanet host stars as a function of $V$ magnitude.
\label{fundfig:ExoplanetsKnown}}
\end{figure}

Using transit photometry, the ratio of the planetary-to-star radius $R_{\rm p}/R_{\star}$ can 
be determined with very high precision.  The precision on the individual planetary radius $R_{\rm p}$ thus 
depends only on the precision on the stellar radius $R_{\star}$.
When Gaia parallaxes become available (2017), the limiting factor for determining a precise measurement of the
stellar radius (for single stars) is the precision on the angular diameter for which 
we require up to 2\% precision.  
Currently there are a very limited number of exo-planetary host stars discovered by the transit method, 
that are measureable using interferometry, and these are shown in Fig.~\ref{fundfig:ExoplanetsKnown}.  The blue box indicates 
the region where the current visible VEGA interferometer can reach.
The new FRIEND instrument mentioned above which will probe fainter magnitudes, thus allowing higher sensitivity and better calibrators,  will allow us to drop the lower limit of the blue box to about 0.1 mas.  In this case, the star will not be fully resolved, but with the better precision, accurate measurements high in the visibility curve will enable these angular diameters to be measured.  
In this figure we also indicate the cut-off limit in magnitude of FRIEND.  One can see that 
this then opens the possibility of adding about 15 stars to the current 1 observable (red dot shown in the blue box) \citep{Ligi2015}.

In Figure ~\ref{fundfig:ExoplanetsKnown} (bottom) we show the number of known transiting exoplanet host stars as a function of magnitude.  Already we can see that a small increase in number is possible down to a magnitude of about $V\sim10$, while within a couple of years the number of known exoplanet hosts (transiting or RV) will increase exponentially (TESS/CHEOPS/PLATO).

Apart from characterising directly the planetary radius, the stellar diameter also provides a very important constraint for determining the mass of the star, in particular if the stellar density can be derived from light curve fitting.  Constraining the mass, along with the stellar diameter, indicates the evolutionary stage of the stellar system.

\subsection{Effective temperatures of Population I and II stars and setting the temperature scale. }

Deriving effective temperatures of stars using {\it classical} methods such as spectroscopic analyses often encounters difficulties due to missing physics in the atmosphere models, in particular for those metal-poor or more metal-rich stars.  Therefore, using such methods often leads to large differences in the derived temperatures, surface gravity and as a consequence the metallicity and chemical abundances and are very dependent on the method and the models used.   A largely model-independent way of determining their effective temperatures is by using interferometric diameters and coupling this with the bolometric flux of the star.  Combining the diameter with the parallax yields the radius, and even with a very modest estimate of the mass of the star, \logg\ can be determined with precisions of less than 0.05 dex \citep{creevey12,creevey14}.

The effective temperature itself is not in fact a real physical stellar quantity.  It is defined as the temperature that the star would have if it radiated like a blackbody, such that the integral of the total radiated flux is equivalent to the integral of the blackbody.  Typically with spectroscopy and photometry observations do not span the full range of where flux is radiated, but just a very small section, such as a few tens or hundreds of nm.  The most correct way to 
measure \teff\ is thus to measure its total radiation at the surface of the star in all directions and compare this directly to the integral of a blackbody.  However, we can not do this.  We can, however, measure the amount of flux we receive on the top of the Earth's atmosphere per unit of radiating surface.  If we additionally know the true radius of the star and its distance (to convert the received flux to absolute flux) we can calculate the total radiation.
This is equivalent to measuring the received bolometric flux of the star and combining this with its (apparent, seen from the Earth) angular radius.  
With interferometry we measure the true definition of the effective temperature.

\subsubsection{Effective temperature scale}
Knowing the true absolute effective temperature scale has vast implications 
in all of astrophysics \citep{casagrande11}.  While we can't measure it
directly for all stars due to the interferometric constraints, we can measure
it for bright stars and use these along with spectroscopic and photometric 
measurements to define a zero-point.
The infra-red flux method is an extremely powerful way of estimating the 
effective temperatures of stars and can be applied to all stars with 
infra-red photometry, in principle to stars of any magnitude.
Casagrande et al. has developed such a method and has shown that this method
provides excellent agreement with interferometric measurements (see Fig.~2 of 
\cite{casagrande14}) and thus provides a method to  
set the temperature scale and 
then provide reliable estimates for determining metallicities and 
alpha-abundances of stars for studying galactic astrophysics.
\citet{casagrande10} derive colour-temperature-metallicity relations given in the 
form of polynomial expansions, and for the Johnson V and 2MASS Ks dereddened 
magnitudes
this equation can be written as:
%\begin{equation}
\begin{eqnarray}
\theta_{\rm eff} &=& 0.5057 + 0.2600 (V-K_s) -0.0146 (V-K_S)^2   \nonumber \\
& & - 0.0131 (V-K_s) [{\rm Fe/H}] \nonumber \\
& & +  0.0288 [{\rm Fe/H}] + 0.0016 [{\rm Fe/H}]^2
\label{fundeqn:casagrande10}
\end{eqnarray}
%\end{equation}
where $\theta_{\rm eff}$ = 5040/$T_{\rm eff}$ applicable in the ($V-K_s$) range 
of 0.78 -- 3.15 and [Fe/H] between --5.0 and 0.4.
A small error is introduced in the absence of [Fe/H] and for the case of the 
Sun, considering $\pm$ 0.5 dex error this corresponds to $\pm$30 K on the 
solar effective temperature.

\begin{figure}
\center{\includegraphics[width=0.7\textwidth]{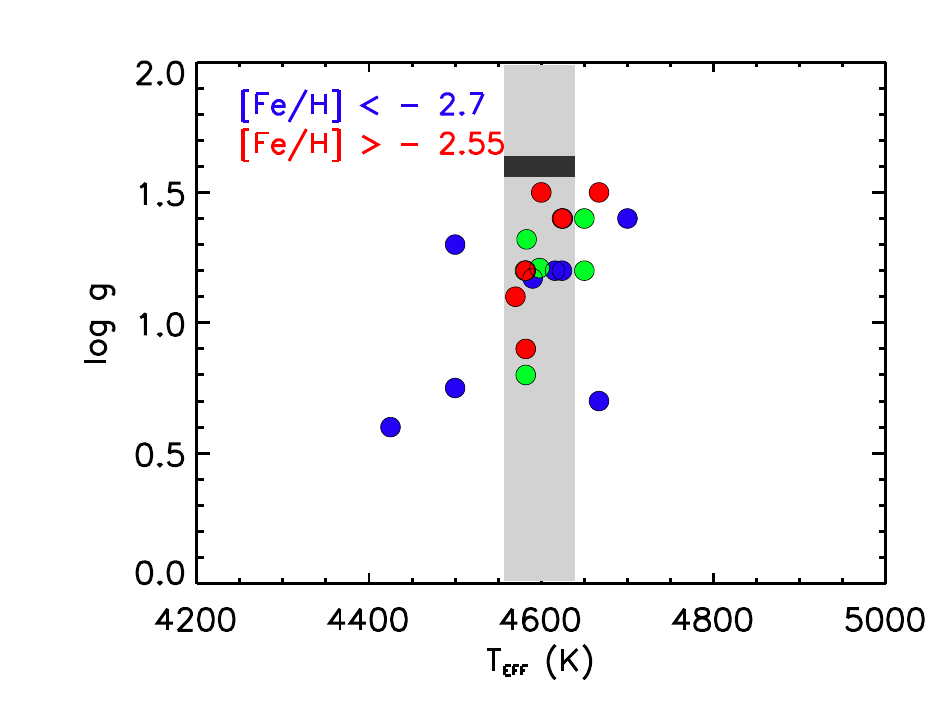}}
\caption{Spectroscopic determinations of the \teff, \logg\ and \feh\ 
of the metal-poor giant HD\,122563 \citep{soubiran10}.  
The interferometric \teff$\pm 1\sigma$ is illustrated by the grey shaded region
\citep{creevey12}.\label{figfund:hd122563}}
\end{figure}

\subsubsection{Calibrating large scale survey data and their methods}
Such interferometric measurements are used also to validate spectroscopic
models and methods.  In particular if \teff\ and \logg\ are known a priori,
studies targeting improvements of stellar atmospheres can be performed.  
Spectroscopic analyses of these stars then provides, theoretically, more 
accurate and precise metallicities and abundances.  
For large scale missions, such as Gaia and the Gaia-ESO survey, use of 
interferometry plays a foremost role for validating the methods that will 
be applied to analyse hundreds of thousands of fainter targets and for 
ensuring the calibration of the data.  
Heiter et al. (submitted) published a list of 34 benchmark stars for which
\teff\ and \logg\ from the literature were critically analysed and 
compared with interferometric measurements.  They conclude that indeed
largest discrepancies are found for the coolest and most metal-poor stars.
Figure \ref{figfund:hd122563} shows a comparison between the 
effective temperature determined by \citet{creevey12} and spectroscopic values
determined from the literature \citet{soubiran10}.  It illustrates the 
consequences on the metallicity and \logg\  determination if all three
parameters need to be fitted in spectroscopic analyses.  Understanding 
the biases in these determinations is important for eliminating 
systematic errors arising from either the method or the (simplified) models used.
\citet{jofre14} published iron abundances for these same stars by fixing
the effective temperatures and \logg.  Some of these are reproduced in 
Table~\ref{fundtab:jofre} listed in order of increasing iron abundance.
The largest sources of errors (thus contributing to the uncertainties) 
arise from non LTE corrections for the poorest metal-poot stars.

\begin{table}
\begin{center}
\caption{Iron abundances of nearby stars for which \teff\ has been fixed from
interferometric observations.}\label{fundtab:jofre}
\begin{tabular}{lll}
\hline\hline
Star & [Fe/H] \\ 
\hline\hline
HD 122563 & --2.64\\
HD 140283 & --2.36\\
HD 84937 & --2.03\\
$\mu$ Cas & --0.81\\
$\epsilon$ For & --0.60\\
$\tau$ Cet & --0.49\\
18 Sco & +0.03\\
Sun & +0.03\\
$\delta$ Eri & +0.06\\
\hline\hline
\end{tabular}
\end{center}
\end{table}

\subsubsection{Metal-poor stars}

\begin{figure}
\includegraphics[width=0.7\textwidth]{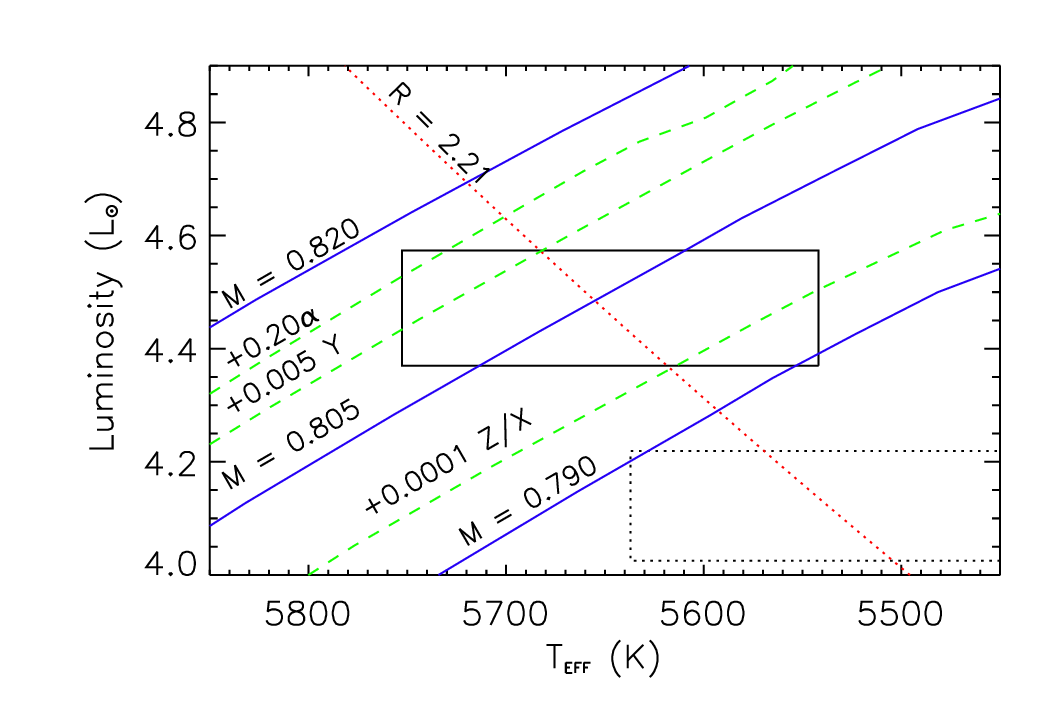}
\caption{Observational constraints in the HR diagram for HD\,140283 (box continuous lines)
and several stellar models that pass through the error box.  
The stellar models vary by mass (blue) and other adjustable quantities (green).
The measured radius is shown by the red line. 
\label{fundfig:agehd140283}}
\end{figure}

Of particular interest are the measurements of fundamental parameters
for metal-poor stars.  One reason, as stated above, is to indeed aim to 
improve our understanding of the physical processes in stellar atmospheres.  
But the ability to reduce the error bar on \teff, \logg\ and [Fe/H] has 
very positive consequences for determining the ages of these stars, some
of the oldest stars in our Galaxy.  Additionally such HR diagram models
often fail to match the observational constraints and thus corrections to 
the evolution models need to be made.
The ages of some metal-poor stars was studied by \citet{creevey12,creevey14}
using tailored stellar models, while such determinations using 
grids of existing models was also done by authors such 
as \citet{vandenberg14} and \citet{casagrande11}.
Tailoring stellar models allows one to understand the impact of all 
variables on the determination of the stellar age, as shown in 
Fig.~\ref{fundfig:agehd140283}.
The mean ages of the stars studied by \citet{creevey12,creevey14} 
is 12.2 Gyr consistent with theory that
predicts Halo stars to appear at roughly 1 billion years after the Big Bang.

Today there are about six metal-poor stars that are 
measureable but with better precision allowing more confidence in a 0.1 mas star
and increasing the magnitude range, this sample can be extended to about 27 stars.

%\subsection{\it Interferometric radii as complementary observations to asteroseismic data for solar-like stars, for determining precise masses (3\%) and ages (10\%).} 
\subsection{Interferometry and Asteroseismology}

Interferometric and asteroseismic  approaches are complementary in two 
main ways: 
the first one concerns the validation of seismic scaling relations, while
the second concerns using the diameter as a complementary constraint for 
stellar modelling along with the seismic diagnostics.
The validation of scaling laws opens a window for galactic archaeology 
\citep{miglio13} and determinations of planetary radii with transit photometry.
A complementary observation of radius in stellar modelling yields high
precision masses and ages, along with constraining other stellar parameters 
such as the mixing-length parameter and the initial helium abundance.
It also constrains the interior structure by constraining its absolute dimensions.

 Asteroseismic data has huge potential for determining stellar properties of stars such as mass, radius, and age.  The frequencies that are detected are sensitive to the density structure of the star and thus provide very precise information about the quantity $M/R^3$.  It has been shown, for example, that $\log g$ can be determined to a precision of $< 0.03$ dex for main sequence stars, with accuracies as high as 0.01 dex (e.g. \citealt{creevey13}).

\begin{figure}
\center{\includegraphics[width = 0.7\textwidth]{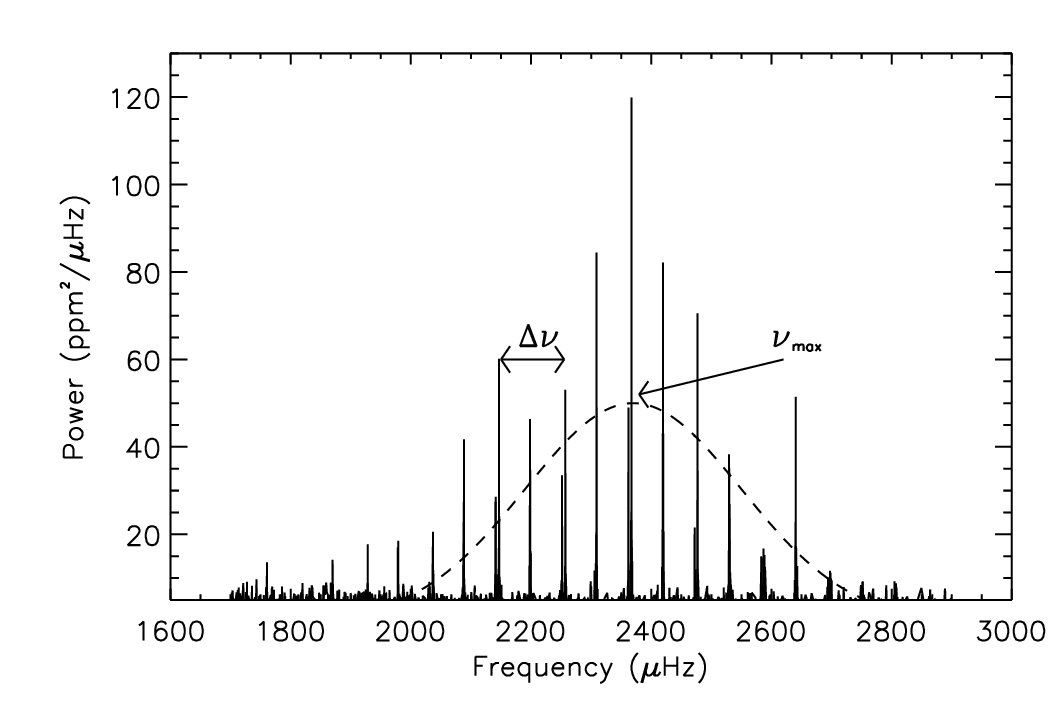}}
\caption{A typical oscillation spectrum of a solar-like star displaying
many oscillation frequencies. The global seismic quantites 
$\langle \Delta \nu \rangle$ and $\nu_{\rm max}$ are indicated.\label{fundfig:kicps}}
\end{figure}

Measurements of oscillation modes have been made in thousands of targets.  For
a star like our Sun a large range of frequencies may be observed as shown in 
Fig.~\ref{fundfig:kicps}.  
In this figure the individual frequencies can be seen,
but it can also be noted that the amplitudes of the frequencies vary
with frequency.  The quantity $\nu_{\rm max}$ is the approximate frequency
where the largest amplitudes are seen.  
It can also be seen that a regular spacing exists between the frequencies.
Twice the value of this spacing (because modes with degree 0 and 1 are separated by about half of this value) is called the 
{\it large frequency spacing} or 
$\langle \Delta \nu \rangle$.
These two global values are of particular interest because along with
an estimate of \teff, the radius and mass can be estimated independently of
stellar models by 
applying the so-called seismic scaling relations.  
\begin{equation}
\left (\frac{R}{R_{\odot}} \right )= 
\left ( \frac{\nu_{\rm max}}{\nu_{\rm max\odot}} \right )
\left (\frac{ \langle \Delta \nu \rangle }{ \langle \Delta \nu \rangle_{\odot} } \right )^{-2}
\left ( \frac{T_{\rm eff}}{T_{\rm eff\odot}} \right )^{1/2}
\label{eqn:radseismic}
\end{equation}
\begin{equation}
\left (\frac{M}{M_{\odot}} \right )= 
\left ( \frac{\nu_{\rm max}}{\nu_{\rm max\odot}} \right )^3
\left (\frac{ \langle \Delta \nu \rangle }{ \langle \Delta \nu \rangle_{\odot} } \right )^{-4}
\left ( \frac{T_{\rm eff}}{T_{\rm eff\odot}} \right )^{3/2}
\label{eqn:masseismic}
\end{equation}
If the radius is measured using interferometry, then Eq.~\ref{eqn:radseismic} can be validated using this external measurement. \citet{huber12} have shown 
that the relation holds well for solar-metallicity main sequence stars, and
\citet{white13} applies the relations to determine model-independent 
masses of several targets.  
Going  a step further \citet{metcalfe14} include some interferometric 
diameters as constraints in asteroseismic analyses which result in 
better determined masses and ages.

\begin{figure}
\includegraphics[width=0.7\textwidth]{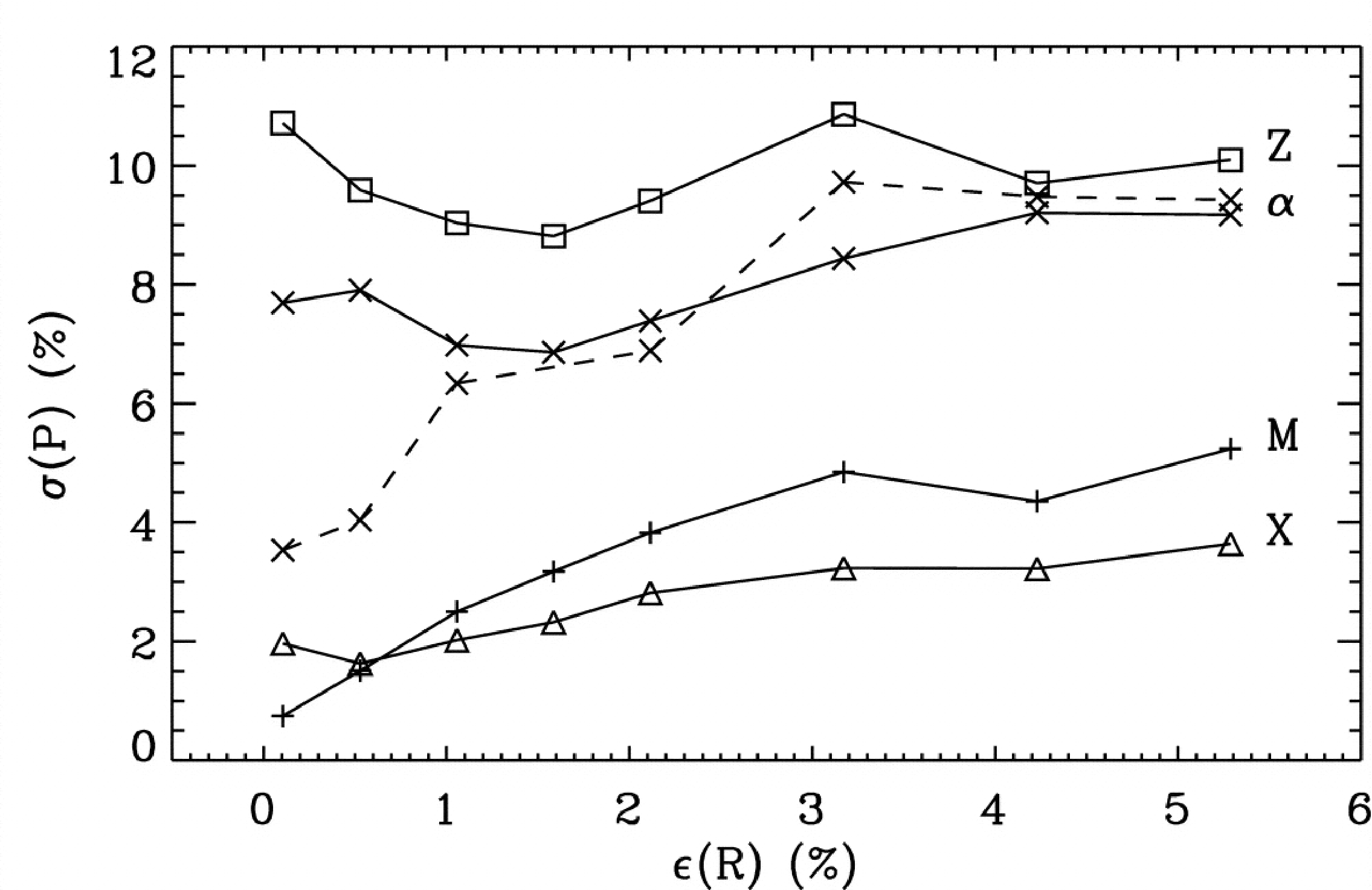}
\caption{Determination of stellar parameters from simulations as a function
of precision in interferometric radius. M, Z, X, and $\alpha$ denote mass, 
initial
metallicity and hydrogen mass fraction, and mixing-length parameter.
All simulations include seismic diagnostics.  The dashed line for $\alpha$
also include the separations between degree modes 0 and 2, adapted from 
\citet{cre07}
\label{fundfig:cre07}}
\end{figure}

How well the mass can be determined depends on the precision in 
the measured diameter. 
In some case, the  mass can be determined with  a precision of as 
low as 1.5\% (e.g. Creevey et al. 2007, Figure~\ref{fundfig:cre07}).   
Along with this high precision mass comes a well-determined age and initial helium abundance.  These parameters are fundamental for studying stellar populations and the history of our Galaxy, along with high-precision characterisation of planetary systems and their members.

Today observational restrictions limit the number of bright asteroseismic targets from ground-based campaigns (about 40 in total for $V<5$), while space-photometry is providing hundreds of asteroseismic data on main sequence and sub-giant stars but for magnitudes in the range of $V>9$.  The fainter targets are not accessible with interferometric instruments, while the brigher targets having been mostly observed from the southern hemisphere have angular diameters that are too small for southern interferometric diameters.  
The Stellar Observations Network Group\footnote{\url{song.au.dk}} (SONG) is a network of telescopes (currently 2) dedicated to bright star asteroseismology.  All of these targets will be accesible to interferometric instruments.  
The TESS\footnote{\url{http://tess.gsfc.nasa.gov/}} and PLATO\footnote{\url{http://www.oact.inaf.it/plato/PPLC/Home.html}} missions, both aiming to find Earth-like planetary systems by searching around bright targets are the most promising missions for a better overlap between asteroseismic and interferometric targets.  In both cases, the targets are generally $4<V<11$ main sequence stars.  For TESS, global seismic quantities (see below) will be attainable for the targets, while for PLATO High precision asteroseismic data will be available for 1000s of stars up to magnitude 8.
At the same time continued efforts to conduct asteroseismic campaigns from ground-based telescopes is needed and will improve the overlap.  

%\end{enumerate}

\subsection{Surface brightness relations, binaries and other applications}
In this final subsection we give a brief summary of other applications of 
interferometric diameters.

{\bf Binaries}\\
In this section we have not discussed interferometric orbits and the huge
potential for determining fundamental parameters in particular the mass.
An interferometric orbit (several data points) provides the inclination and separation of the components (scaled to absolute units when the parallax is known) along with an orbital period.  Combining this with radial velocity data which also derives the orbital period and the mass ratios and applying Kepler's Law, allows us to deduce the individual mass components.

{\bf Surface brightness relations}\\
Relating measured angular diameters with easily measured magnitudes and colours provides a method for estimating angular sizes of stars \citep{ker04} that are too distant (or small) to be measured directly.  Gaia will deliver parallaxes for stars up to a magnitude of 20 and combining these complementary data yields an independent determination of stellar radii, a parameter notably important for determining planetary radii in transiting planetary systems.

%{\bf Diffusion in stellar interiors and atmospheres}\\
{\bf Mixing-length parameter}\\
1D stellar evolution models are characterised by the generally unobservable properties of mass, age and initial chemical composition.  Such models are usually used to determine masses and ages for single stars and/or populations of stars for many different astrophysical reasons.  Few observations and error bars implies that the solutions are generally degenerate.  To add to the problem there is another parameter of stellar models that is used to describe the efficiency of convection.  It is called the {\it mixing-length parameter} and is unfortunately an unconstrained parameter that we usually fix as to be equal to a solar-calibrated one for the same physics.  Observations have shown this not to be correct, however, there are little observational methods to help understand how this parameter varies with stellar evolution stage, mass and temperature.
Generally we are concerned only with the outer stellar convective envelope and so the adjustable parameter in the models effects only the {\it surface-most} properties such as radius and effective temperature, but not luminosity.  Angular diameters therefore have a huge potential for constraining this parameter in cases where other constraints are also available.  Such examples are metal-poor stars, where we already can constrain the mass and age of the star, or binary systems, or clusters where mass, age  and/or chemical composition are known {\it a priori}, or through asteroseismic analysis where the combination of radius and
seismic data allows one to narrow down the plausible range.

\section{\hspace{0.2cm}{\bf B}\hspace{0.2cm} Planet-detection and convection from time sensitive and time series observations using closure phases and imaging
\label{fundsec:conv-planet}}

%\red{ROXANNE: do you want to add in figures here or text?}
To detect and characterise stellar granulation and attempt to detect planetary transits we 
require time-sensitive closure phases and imaging.
 Stellar activity and, in particular, convection-related surface structures, potentially cause bias in the characterisation of stellar parameters and, eventually, in planet detection. 
\cite{2014A&A...567A.115C} showed that the stellar surface asymmetries in the brightness distribution mostly affect closure phases, because of either convection-related structures or a faint companion \citep{2014A&A...567A.115C,2012A&A...540A...5C,2010A&A...524A..93C}. The levels of asymmetry and inhomogeneity of stellar disk images reach high values with stronger effects from frequencies corresponding to the 2nd/3rd visibility lobe on, depending on the stellar parameters: asymmetries are larger for low surface gravity K giant stars and smaller for sub-giant and solar type stars. \cite{2014A&A...567A.115C} presented two possible benchmark solar type and sub-giant targets: Beta Com (1.1 mas) and Procyon (5.4 mas); the choice of the authors is due to the angular diameter to illustrate observations with good UV coverage, but many more can be also observed. 
%(\redoxanne: shall I list the stars that you mentioned? or is it hussshh?}).
 
Interferometry can also help to disentangle the signal of a transiting planet. Closure phases and imaging  are required to do this.  The observations are
time sensitive: before and after planetary transit and for a duration of many hours to a day. \cite{2014A&A...567A.115C} also showed that there is also a starspot signal, due to magnetic activity, on closure phases that can be of the same order as the transiting planet signal. Nevertheless, it should be possible to differentiate between them because the time-scale of a planet crossing the stellar disk is much smaller than the typical rotational modulation of the star. It is important to note that when probing high spatial frequencies, the signal to noise ratio of the measurements would be very low due to low fringe visibilities, greatly deteriorating the closure phase precision and affecting the instrument capability. Moreover, this would influence the capability and sensitivity of detecting the signatures of granulation and disentangling the planetary signal. The future visible interferometer must pay attention to overcome these limitations.

\begin{figure}
  	\begin{center}
  	 \includegraphics[width=0.6\textwidth]{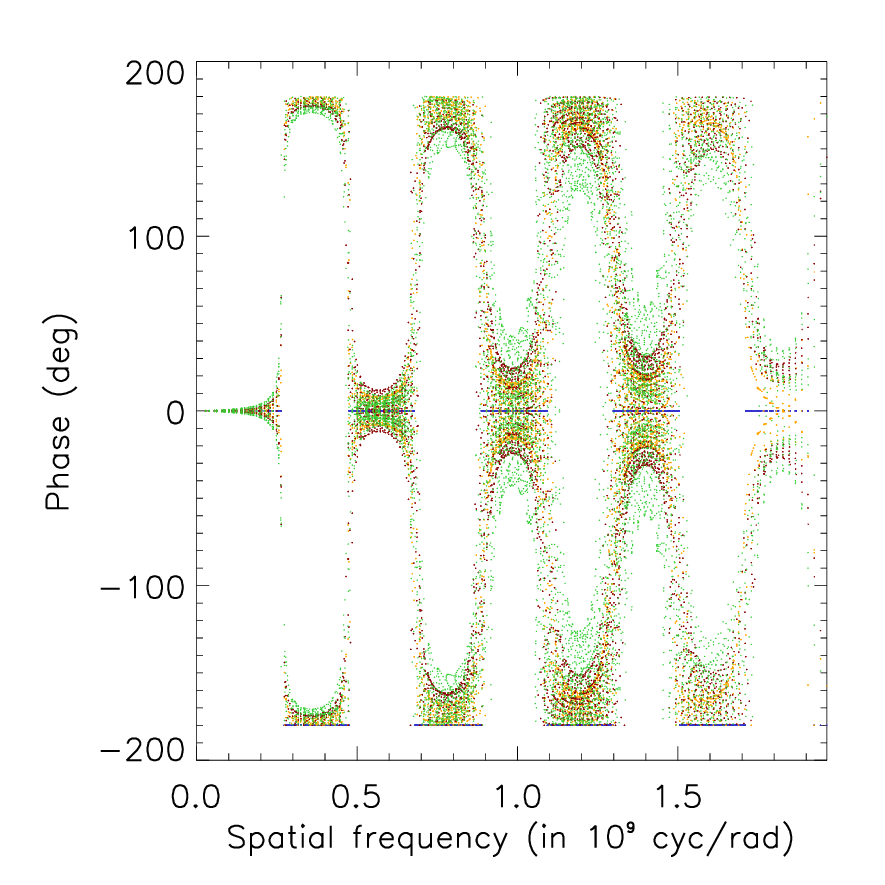}
  	\end{center}
  	\caption{Phases of a single star (blue), a star with a spot (orange) at (0.2, 0.2)~mas, a star with a transiting exoplanet (red) at (0.2, 0.0)~mas on the stellar disk and a star with both (green). The spot's diameter is $\theta_s = 0.10$ mas, the star's diameter is 1~mas and the exoplanet's diameter is $\theta_p = 0.10$~mas \citep{Ligi2014}.}
  	\label{figfund:ligi14}
\end{figure}

To differentiate a planetary signal from a spot signal, instrumental improvements are also needed.
{To directly detect a $0.10$~mas exoplanet crossing a 1 mas star, represented by a limb-darkened disk, at visible wavelength with interferometry, an accuracy better than $\sim 0.5\%$ from the first null is required in squared visibilities, at best observing conditions. A precision better than $\sim 1^{\circ}$ on phases is necessary in the first lobe, or better than $\sim 6^{\circ}$ in the second lobe (see Figure~\ref{figfund:ligi14}). At present, no instrument can reach these accuracies. To detect a $0.05$~mas exoplanet, the accuracy needed on squared visibilities and phases are $\sim 0.1\%$ and $\sim 1^{\circ}$ from the first null. Magnetic spot signals can easily mimic these exoplanet signals, and distinguishing between them is essential for an exoplanet characterization, but requires measurements generally beyond the third lobe of visibility \citep{Ligi2015}.
As stated before, temporal variations should be taken into account for these
measurements since the timescale of stellar spots is different from the transity time, and such accurate measurements would require a very good S/N.}

\section{\hspace{0.2cm}{\bf C}\hspace{0.2cm} Spots and limb-darkening using medium-to-high resolution V2 and imaging
\label{fundsec:spots-ldark}}

Surface inhomogeneities are found in stars throughout the H-R diagram.
These inhomogeneities were first observed
in the Sun in form of cool and therefore dark localized regions and they have been known as spots ever since.
We will refer to spots as to any long-lived inhomogeneity in a stellar photosphere
that could induce strong variations in some observed parameters (e.g., surface brightnest, magnetic field strength, etc.).
Of course, stellar spots and magnetic activity can be constrainted with interferometric visibilities and closure phases
provided that instruments can reach required angular resolution.

In stars with outer convective envelopes, the dynamo action triggers 
magnetic activity which then produces temperature spots on the surface of these stars,
just like we observe in the Sun. 
Therefore, studying starspots is essential if we want to understand 
the generation of magnetic fields and activity cycles in stars other than the Sun.
Combined with alternative observables, like, say, photometric light curves 
and  variations of surface magnetic fields, interferometry can put strong constrains
on spot contrast (and therefore their physics) and morphology (e.g., number of spots)
which is essintial for tuning modern dynamo theories \citep{2015MNRAS.449.3471S,2015A&A...573A..68Y,2013A&A...549L...5G}.

Temperature spots are not the only spots that are found in stars.
There is a particular class of main-sequence stars called chemically peculiar (CP) stars that
have no (or very weak) convection in their atmospheres but still have surface spots. However,
unlike in cool stars, spots in CP stars are not of temperature but abundance origin. 
They appear when atoms and ions of certain elements tend to accumulate 
at different regions on the stellar surface driven by diffusion processes \citep{1970ApJ...160..641M}.
Studying these abundance spots brings important constrains on the evolution of chemical elements, 
physics of diffusion processes, and evolution of fossil magnetic fields 
(the later are frequently found in CP stars and may rich tens of kG~--~strongest fields
among non-degenerate stars \citep{1960ApJ...132..521B,2010MNRAS.402.1883E}). 
The abundance spots have been detected in a number
of CP stars of different spectral types \citep{2012A&A...537A.151N,2010A&A...509A..71L,2010A&A...513A..13K,1999A&A...348..924K},
and even spot evolution was detected for the first time in one presumably non-magnetic CP star \citep{2007NatPh...3..526K}.

Optical and infrared interferometry is a powerful observational technique capable of reaching a very high angular resolution.
Potentially, interferometry allows one to derive not only the sizes of stellar objects, but, under the conditions
of sufficient spatial coverage, even to reconstruct the details on the stellar surfaces.
With the development of the long baseline interferometry it will be possible to resolve and analyze stellar spots
by means of interferometric observable such as absolute visibility and closure/differential phase 
\citep{2002AN....323..241W,2003SPIE.4838..587J,2004A&A...422..193R,2014MNRAS.443.1629S}. 
Orientation of the stellar rotation axis and geometry of local magnetic fields can also be constrained.
Thus, interferometry is a powerful technique to study starspots that can provide a complementary
information about stellar surface structures.

In addition it should be remembered that a method of Doppler Imaging (DI) that is commonly and very successfully
used to recover the surface structures of stars is limited to stars rotating fast enough so that rotation dominates
the broadening of spectroscopic lines. However, there are stars that rotate very slowly or have small projected rotational velocities, 
but still show significant spectral variability indicating existence of spots in their atmospheres. 
For those stars no DI is possible, and interferometry thus appears to be a promising technique to study their 
surface morphology.

The application of interferometry to  main-sequence (MS) stars, unfortunately, is still limited to only brightest objects.
Modern facilities are capable of resolving the closest and/or largest MS stars and measure their diameters already on a regular
basis \citep[e.g.][]{2013ApJ...771...40B,2013MNRAS.434.1321M}. 
Recently, interferometry has been successfully applied to CP stars resulting in a first estimate of the radii
of a few of them: $\alpha$~Cir \citep[HD~128898,][]{2008MNRAS.386.2039B}, $\beta$~CrB  \citep[HD~137909,][]{2010A&A...512A..55B},
$\gamma$~Equ \citep[HD~201601,][]{2011A&A...526A..89P}, and $10$~Aql \citep[HD~176232,][]{2013A&A...559A..21P}.
Detailed studies of surface morphology of MS stars with spots, however, remain challenging.

The essential requirements for the present and planned interferometric facilities
to study spotty structures on MS stars are large (hundred of meters) baselines,
high sensitivity ($V^2<10^{-2}$ and below), and from medium to high spectral resolution
($R\geqslant10^3$).

For instance, in their case study of the one of the brightest CP star 
\uma\ (which also has the largets angular diameter among CP stars known by date; $\theta=1.54$~mas)
\citet{2014MNRAS.443.1629S} showed that an instrument like VEGA or its successor
based on the principle of the demonstrator FRIEND
should be able to detect the effect of spots and stellar rotation from spectroscopic features, provided
that the instrument is able to measure squared visibility  down to $\approx10^{-3}$, and/or closure phase in visual. 
An instrument with the spectral resolution around $R=6\,000$ like, e.g., VLTI instruments 
AMBER (present) or GRAVITY (future) would be able to measure rotation and spots,
but baselines longer than currenly provided $140$~m were needed.

\begin{figure}
\centerline{$0$~\kms \hspace{1.8cm} $35$~\kms \hspace{1.8cm} $35$~\kms}
\centerline{
\includegraphics[width=0.33\hsize]{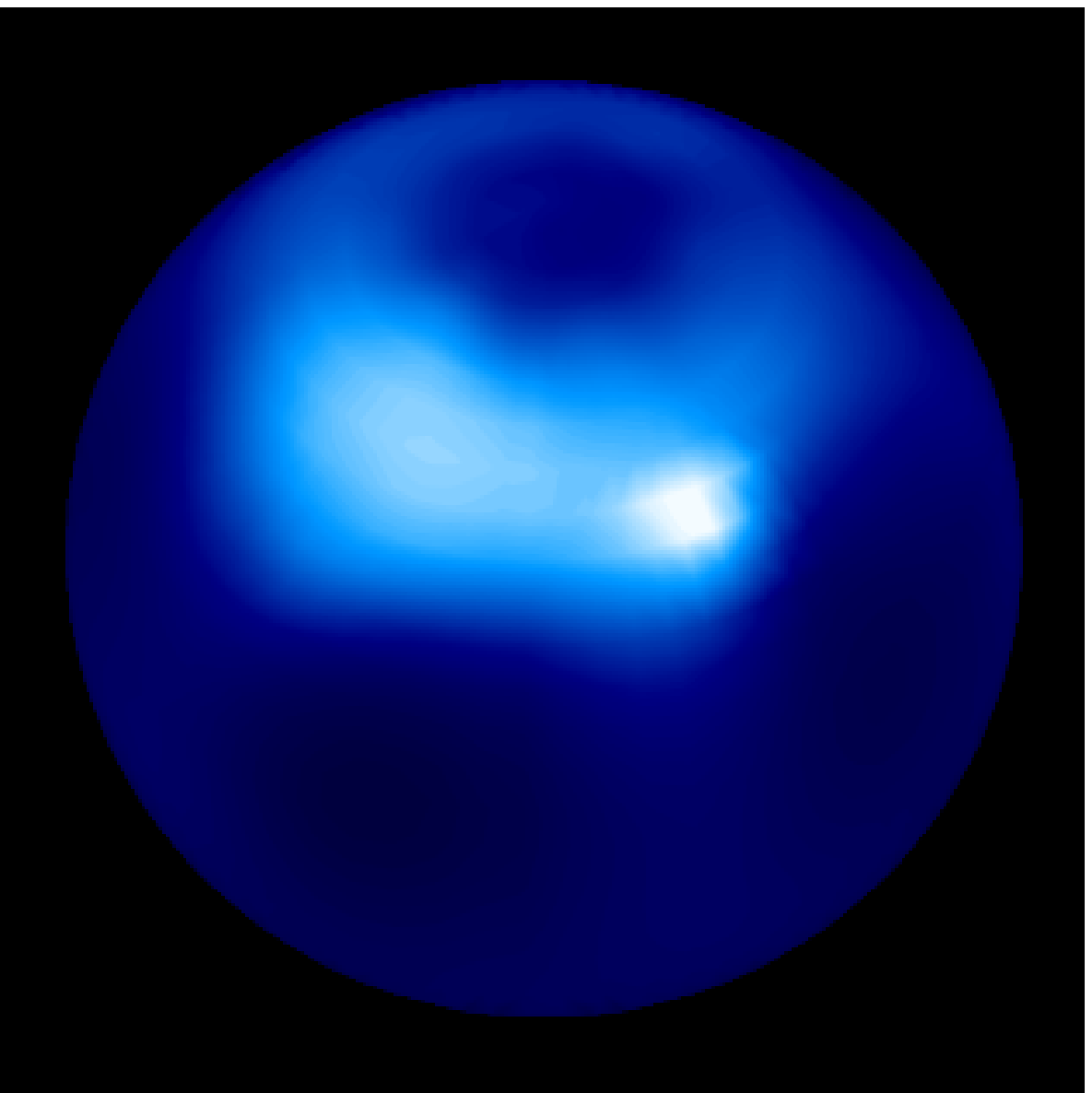}
\includegraphics[width=0.33\hsize]{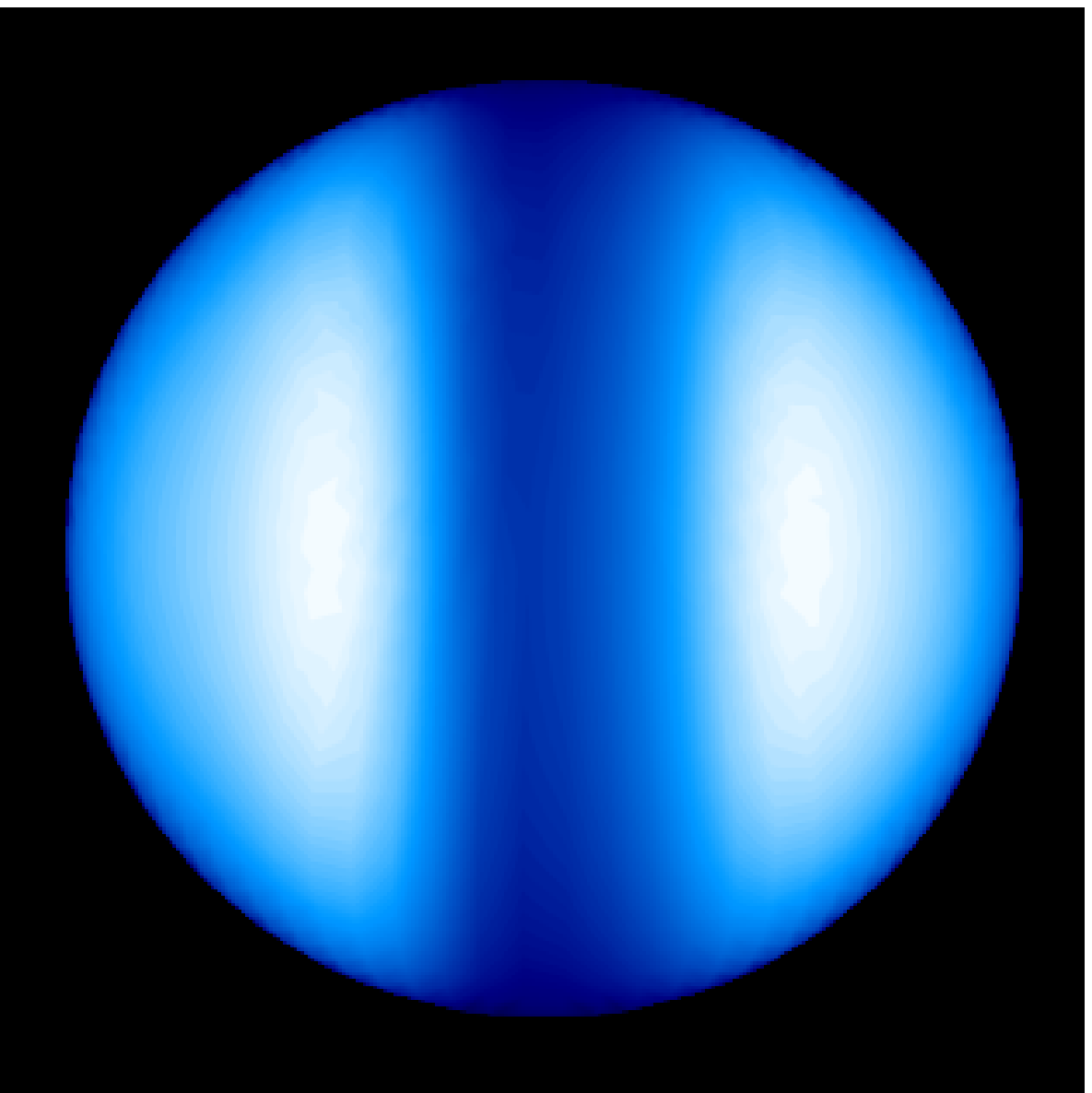}
\includegraphics[width=0.33\hsize]{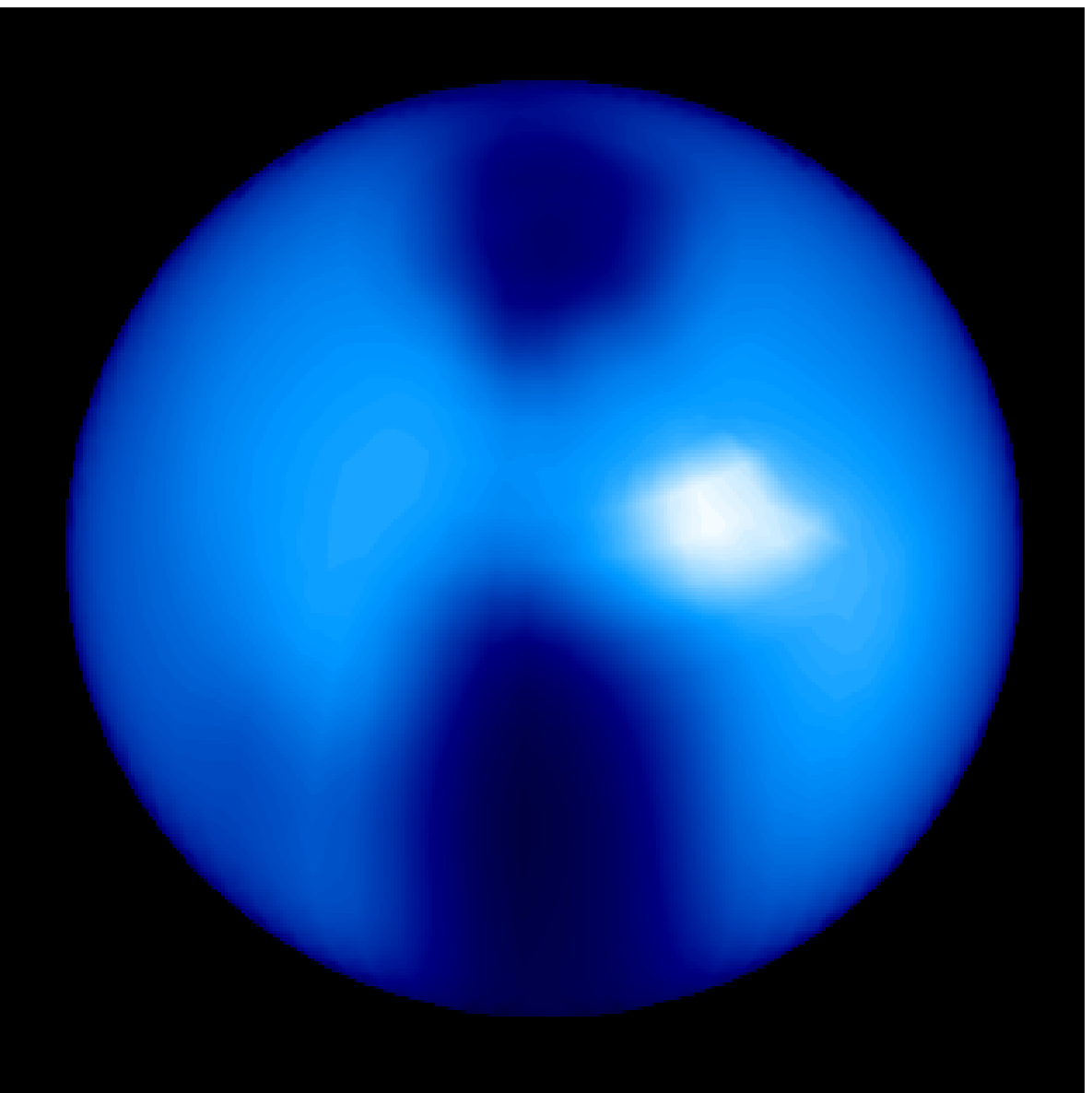}
}
\par\bigskip
\centerline{
\includegraphics[width=\hsize]{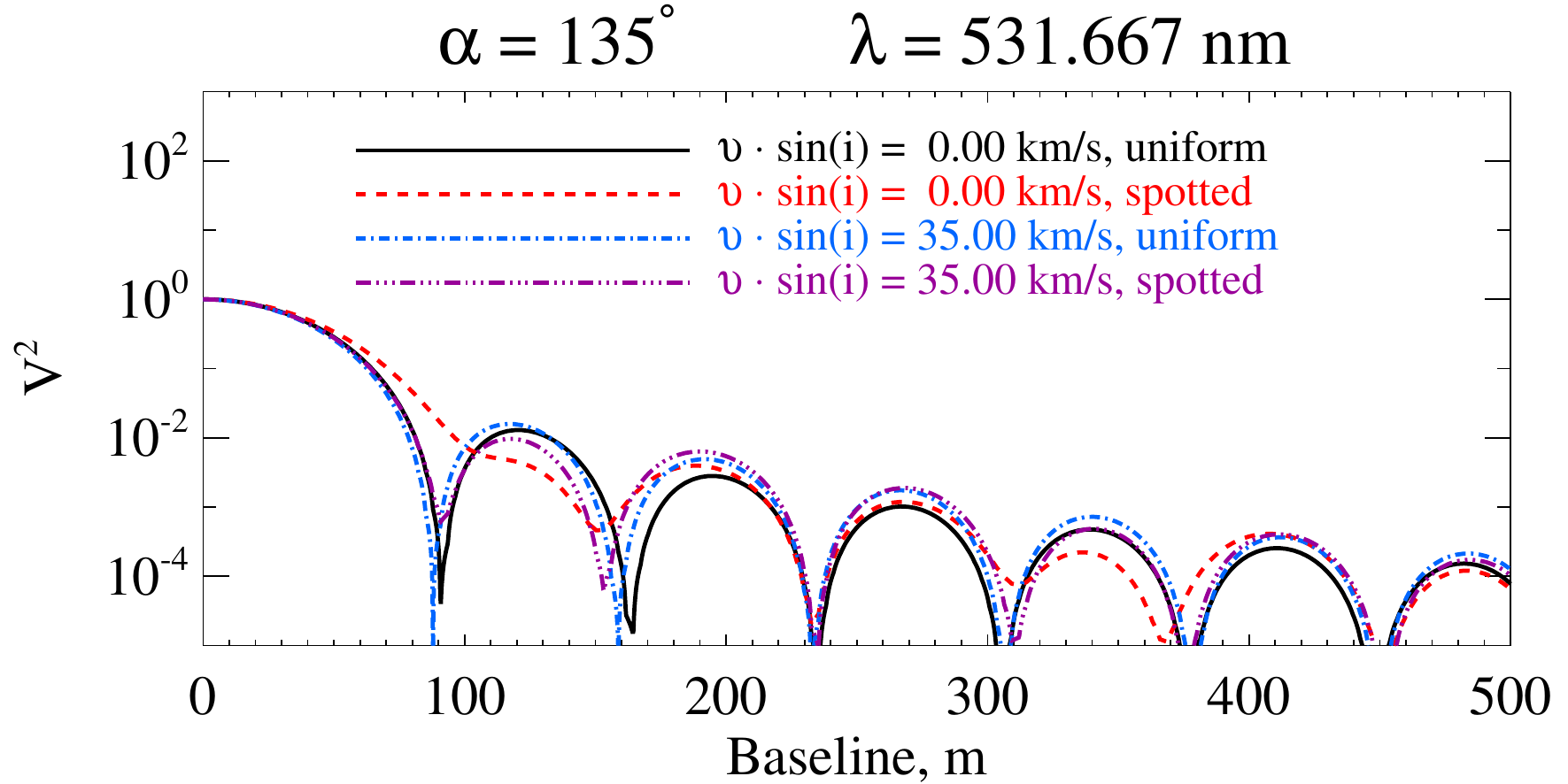}
}
\centerline{
\includegraphics[width=\hsize]{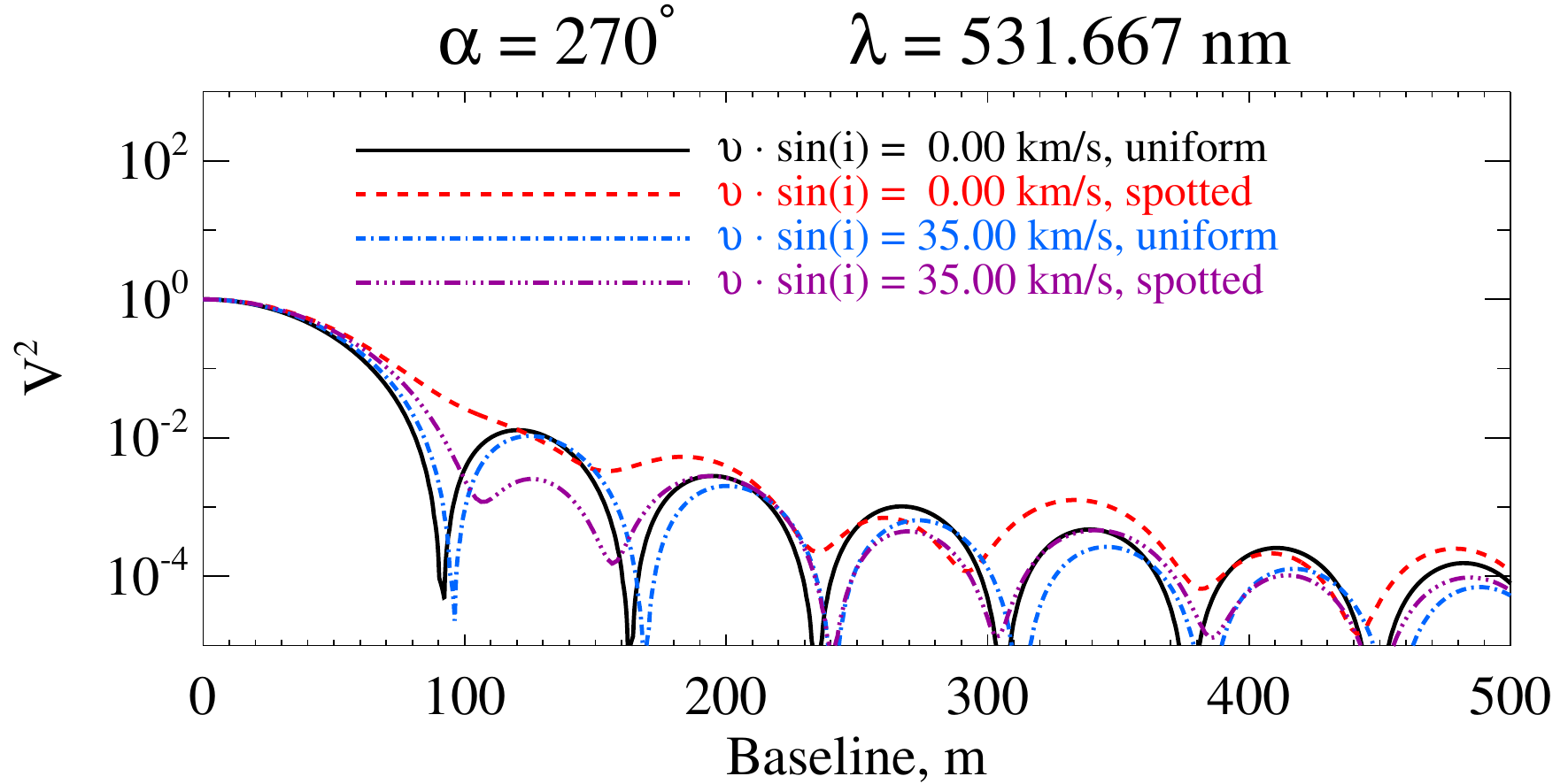}
}
\caption{Stellar intensity images (top panel) for non-rotating spotted star (\vsini$=0$~\kms), rotating homogeneous star, 
and rotating spotted star (\vsini$=35$~\kms),
and squared visibility at the center of strong {Fe}{i}~$531.6$~nm line calculated for the two position
angle of $135^\circ$ (middle panel) and $270^\circ$ (bottom panel) respectively.}
\label{fig:vis-mono-v-1}
\end{figure}

As an example, Fig.~\ref{fig:vis-mono-v-1} illustrates predicted spot and rotation signals in $V^2$ for \uma\ calculated
for a set of baselines at the center of strong {Fe}{i}~$531.6$~nm line. One can clearly see the
bright Fe spot and the dark vertical stripes induced by stellar rotation. Both these effects can be unambiguosly
detected in $V^2$ if the observations are taken at different baseline orientations.

\begin{figure}
\includegraphics[width=\hsize]{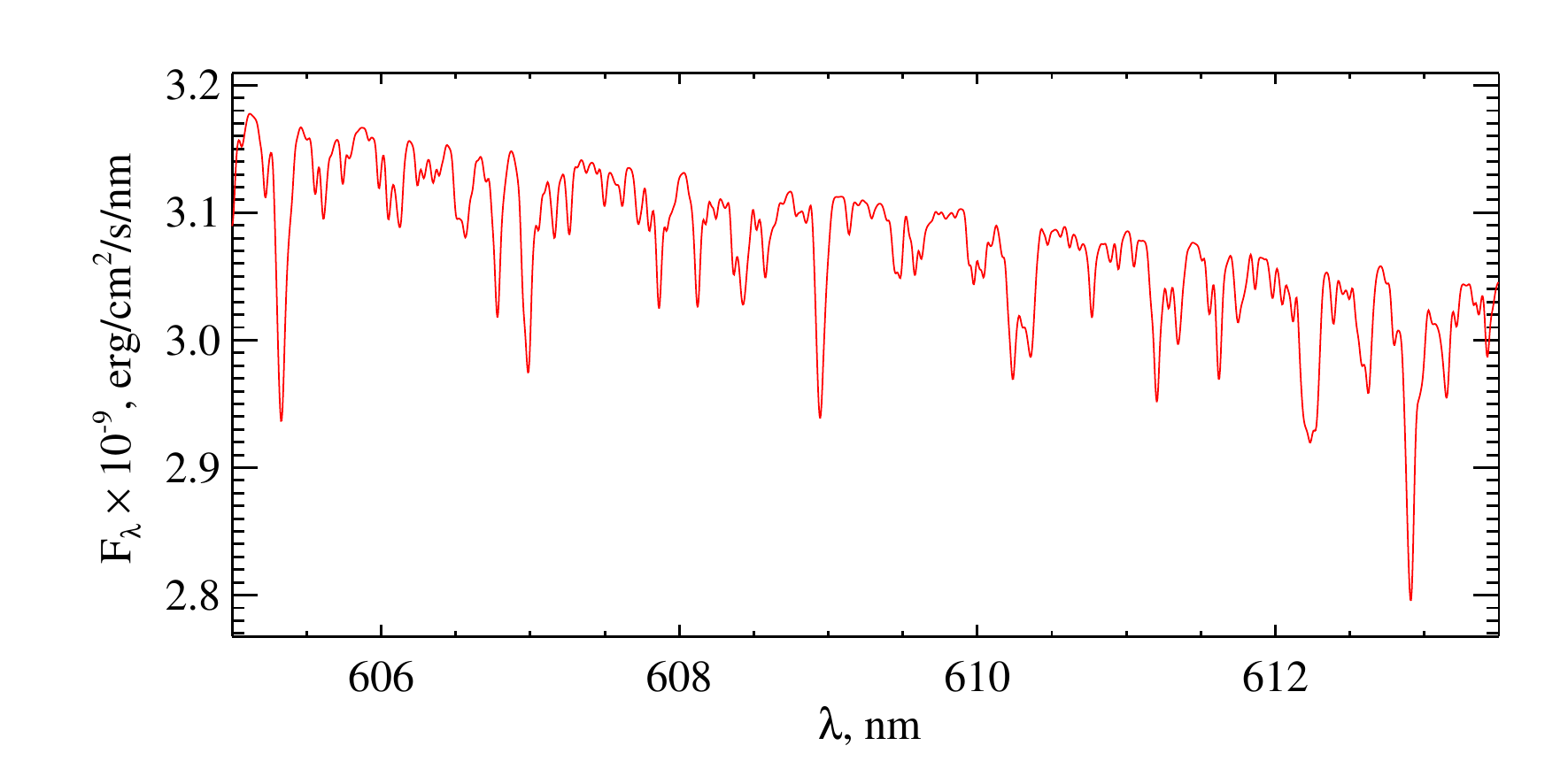}
\centerline{Homogeneous}\vspace{-0.2cm}
\includegraphics[width=\hsize]{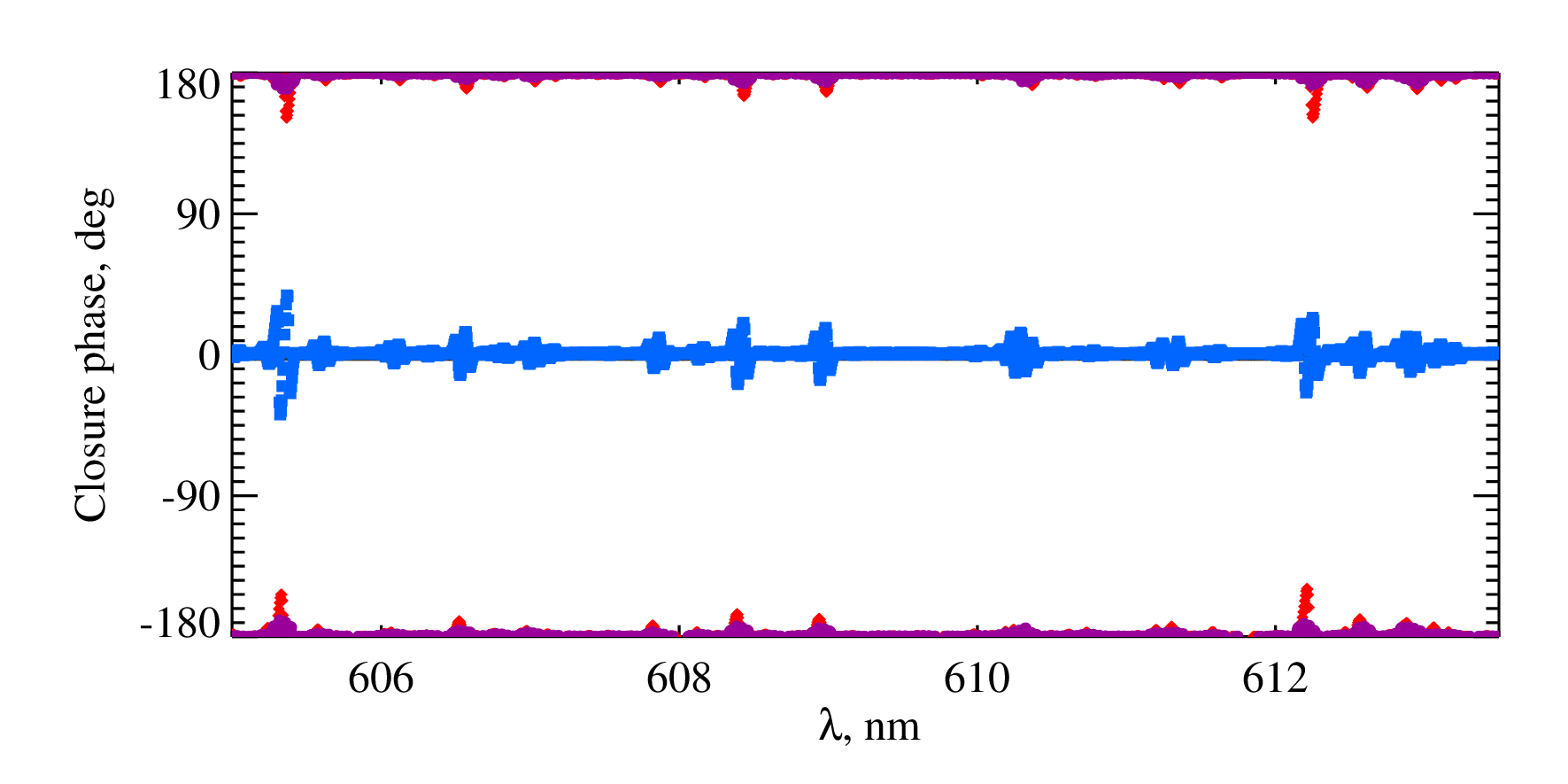}
\centerline{Spotted}\vspace{-0.2cm}
\includegraphics[width=\hsize]{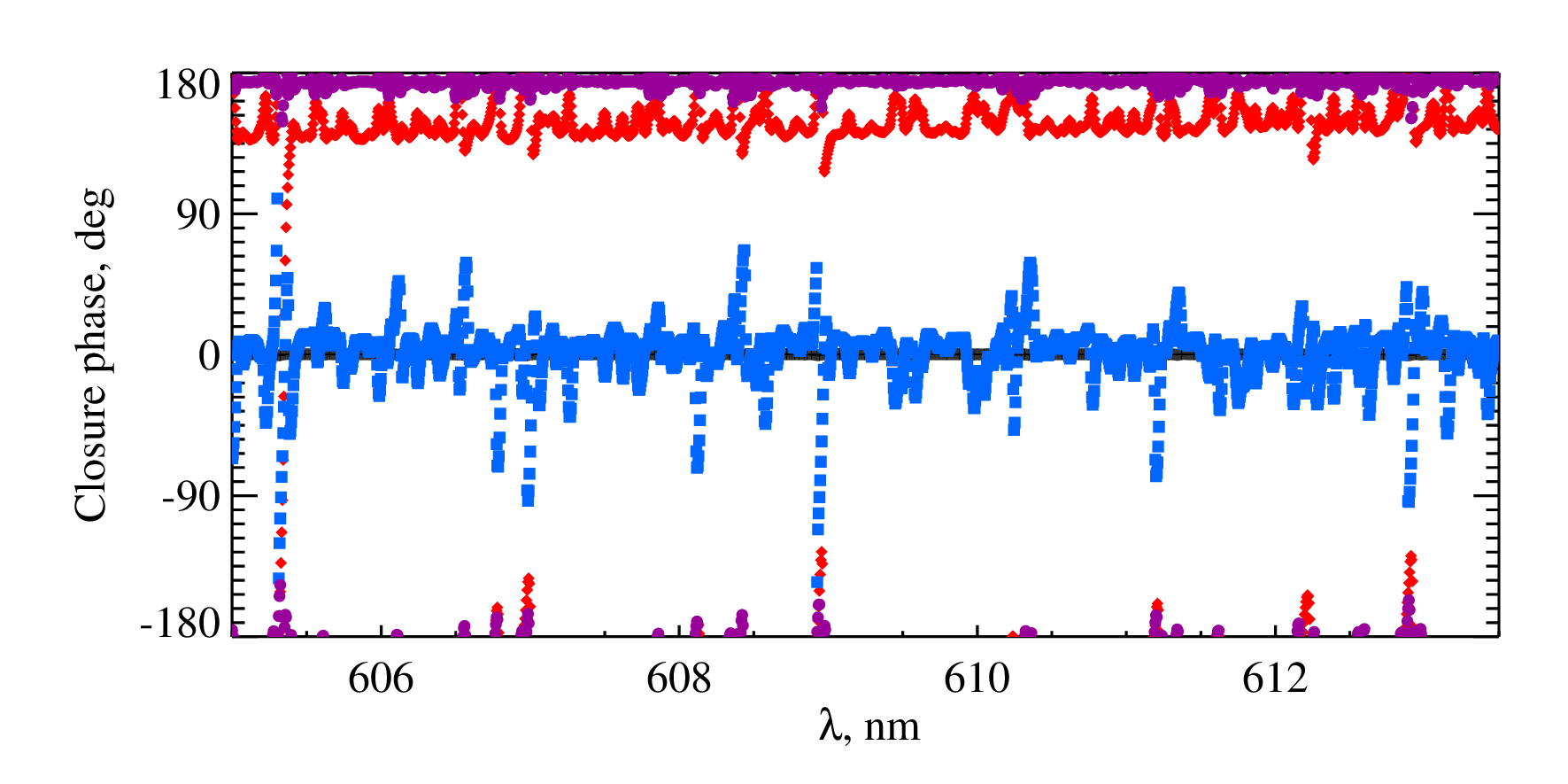}
\caption{Closure phases as a function of wavelength.
First row~--~spectrum predicted by spotted model; second and third rows~--~closure phase predicted by homogeneous
and spotted models, respectively. In all plots $R=30\,000$, \vsini$=35$~\kms. 
Closure phases were computed for the following configurations:
($0^\circ,40$m)$+$($270^\circ,40$m)$+$($135^\circ,57$m)~--~black crosses; 
($0^\circ,100$m)$+$($270^\circ,100$m)$+$($135^\circ,141$m)~--~red diamonds;
($0^\circ,180$m)$+$($270^\circ,180$m)$+$($135^\circ,255$m)~--~blue squares;
($0^\circ,320$m)$+$($270^\circ,320$m)$+$($135^\circ,453$m)~--~violet circles. See online version for colored symbols.}
\label{fig:cp-lambda-v-r30000}
\end{figure}

If closure phases are provided, they can bring very rich information on spots and rotation, especially when
observing at high spectral resolution.
For instance, for the same case of \uma\ there are many telescope configurations 
for which the spotted star looks different
compared to a homogeneous one. This is illustrated on Fig.~\ref{fig:cp-lambda-v-r30000}, where it can be seen
that a rotating star with
uniform surface produces closure phases that are symmetric
relative to the core of spectral lines, whereas spots induce more rich and complex closure phase patterns.
More details can be found in \citet{2014MNRAS.443.1629S}.

Analysis of known CP stars show that most of them have angular diameters below $1$~mas. 
Therefore, both long baselines of hundreds of meters
and detectors sensitive to values of $V^2<10^{-2}$ are required. 
About ten of CP stars can already be subject for spot detection
using existing interferometric facilities.
Note that none of the CP stars can be observed with VLTI
and the reason for this is a short baseline range provided by VLTI compared to other existing interferometers.
Alternatively, there are two objects that could be observed already with available maximum baseline of $140$~m
but detectors operating in visual were needed.

\section{Science case summary}
In this section we summarize the science driver information in terms of scientific
objectives in tabular form for quick reference. 

\begin{enumerate}
\item Diameters of planet-hosting stars
\item Effective temperatures of metal-poor stars
\item Calibrating 1D mixing-length in stellar models
\item Stellar interiors, masses, ages
\item Magnetic inflation of cool, rapidly rotating stars
\item Stellar granulation and planetary-transit detection
\item Spots and magnetic activity
\item Limb-darkening and surface spots
\end{enumerate}

\begin{landscape}
\begin{table*}
\begin{center}
\caption{Summary of observational and instrumental requirements for each science case. {\label{tab:table1}}}
%\begin{tabular}{l|llll|llllllllllllll}
\small
\begin{tabular}{|p{1.0cm}|*{40}{@{\hskip.8mm}c@{\hskip.5mm}l|llll|llllllllllll}}
\hline\hline
Science & \multicolumn{4}{l}{Observables} & \multicolumn{7}{l}{Technical Requirements}\\
Case & V$^2$ & V$^2_{\lambda,i}$ & $\Delta \phi$ & Imaging & Spatial & Spectral & Temporal & \# Telescopes & Time per & Precision  \\
& & & & & (mas) & & (hrs) & & source & (\%) & \\
\hline
1& X & X &--- &--- & $<0.1$ & L-M &--- & $>2$&4-6 & $<2\%$\\
2& X & X &--- &--- & $<0.1$ & L-M &--- & $>2$&4-6 &$<2\%$\\
3& X & X &--- &--- & $<0.1$ & L-M &--- & $>2$&4-6 &$<2\%$\\
4& X & X &--- &--- & $<0.1$ & L-M &--- &$>2$ &4-6 &$<2\%$\\
5& X & --- & X & X&       &    \\
6& ---  &--- & X & X&       & L  &B/A Tr&$>2$ (6) & $4-24$\\
7&   & X & X &---&$<0.1$ & H & & $>2$\\
8& X & X & X &---&$>0.5$ & M & & $>3$ (6) &\\ 
\hline\hline
\end{tabular}
\normalsize
\end{center}
X = Required\\
--- = Not required\\
L, M, H = Low-, Medium-, High-spectral resolution\\
B/A Tr = Before/After planetary transit
\end{table*}
\end{landscape}

\section{Future Requirements}
We have identified the following requirements as the most important in terms of fundamental parameters of main-sequence and sub-giant stars:
\begin{itemize}
\item Increase of angular resolution
\item For higher precision, more telescope time needed
\item Increase sensitivity
\end{itemize}

\chapter[Interferometry in the visible - Pre-Main Sequence stars]{Interferometry in the visible - Pre-Main Sequence stars\label{ch2}}

\addtolength{\textwidth}{2cm}
\addtolength{\textheight}{2cm}
\addtolength{\oddsidemargin}{-1cm}
\addtolength{\topmargin}{-1cm}

\author{K. Perraut$^{1}$, M. Benisty$^{1}$, S. Kraus,$^{2}$ G. Schaefer$^{3}$, M. Simon$^{1}$}

\begin{center} 
$^{1}$~Institut d'Astrophysique et de Plan\'etologie de Grenoble, CNRS-UJF UMR 5571, 414 rue de la Piscine, 38400, St Martin d'H\`eres, France\\
$^{2}$University of Michigan, Astronomy Department, 941 Dennison Bldg, Ann Arbor, MI 48109-1090, USA\\
$^{3}$~The CHARA Array of Georgia State University, Mount Wilson Observatory, Mount Wilson, California 91023, USA
\end{center} 

\section{Introduction}

\begin{figure}[h]
  \centering
  \includegraphics[angle=0,scale=0.5]{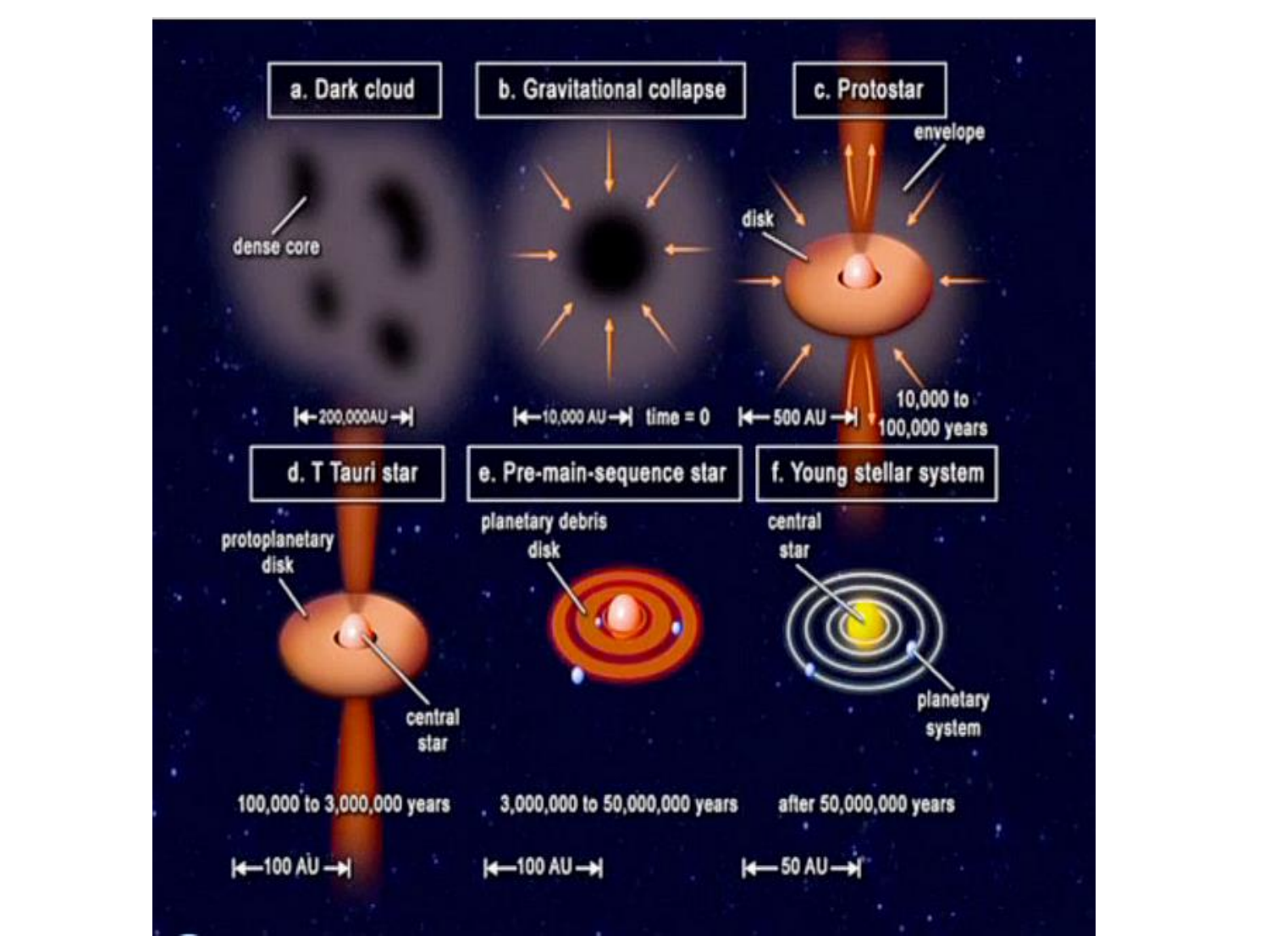}
  \caption{The different phases of star formation.}
  \label{fig:intro}
\end{figure}

Most star formation proceeds along the same lines with: an initial collapse of an interstellar molecular cloud caused by gravity, the cloud fragmentation into clumps, the fragmentation of clumps in protostars, and finally the star formation. The initial rotation of the collapsing cloud leads to the distribution of gas along the equatorial plane in a circumstellar disk. Crucial to the formation of a protostar and active for most of the pre-main sequence (PMS) phase a large scale outflow is ejected to conserve the angular momentum: this is one of the most spectacular manifestation of star formation. The generation of jets from young stars involves a complex interplay, still poorly understood, between gravity, turbulence, and magnetic forces that may have strong implications on conditions for planet formation (Figure~\ref{fig:intro}).

\section{Determining their fundamental parameters}

\subsection{Scientific rationale}

The Hertzsprung-Russel Diagram (HRD) is the single most useful
diagnostic tool in observational stellar astronomy. Astronomers
estimate the mass and age of a young star by its location on the HRD
relative to theoretical calculations of collapse to the main
sequence. It is well known that for stars less massive than $\sim 1.0 $
M$_\odot$ there is considerable scatter in these estimates (e.g., Simon
2001; Hillenbrand \& White 2004).  As a consequence, the mass spectrum
of the stars produced in a star-forming region and the region's
star- and planet-forming histories are imprecisely known. The  main
reasons for the scatter are differences among the several theoretical
calculations of pre-main sequence (PMS) evolution owing to different
treatments of convection, the equation of state,  and the photospheric spectrum.

\parindent =1.0cm
A powerful way to distinguish among these calculations and test the
input physics is to compare predicted masses to masses obtained by
dynamical methods; binary stars are well-known as the ideal targets.
At the distances to nearby star-forming regions ($\sim$ 120--140 pc),
long-baseline optical/infrared interferometry is required to resolve
double-lined spectroscopic binaries (SB2s; e.g., Boden et al.\ 2005;
Simon et al.\ 2013; Le Bouquin et al.\ 2014). Recent measurements have renewed
the interest of theorists in improving the models (e.g., Tognelli et al.\ 2011;
Feiden \& Chaboyer 2012).  Binaries that can be studied both as visual
binaries (VB) and SB2s are particularly precious because the orbital
analysis yields their component masses {\it and} distance.

As young stars collapse to the main sequence their diameters decrease.
Members of the Nearby Young Moving Groups (NYMGs) are near enough,
bright enough, and young enough that their diameters are measurable
interferometrically  (e.g.,  Simon \& Schaefer 2011;  McCarthy \&
White 2012; Schaefer \& White 2015). Since the
theoretical evolutionary tracks of stars more massive than $\sim 1$~M$_\odot$
are reliable,  and {\it Hipparcos} distances are already available
for such stars in the NYMG, their measured angular diameters provide
their physical diameter and hence an independent measurement of their age.

\subsection{Candidates}

\begin{figure}[t]
  \centering
  \includegraphics[angle=0,scale=0.4]{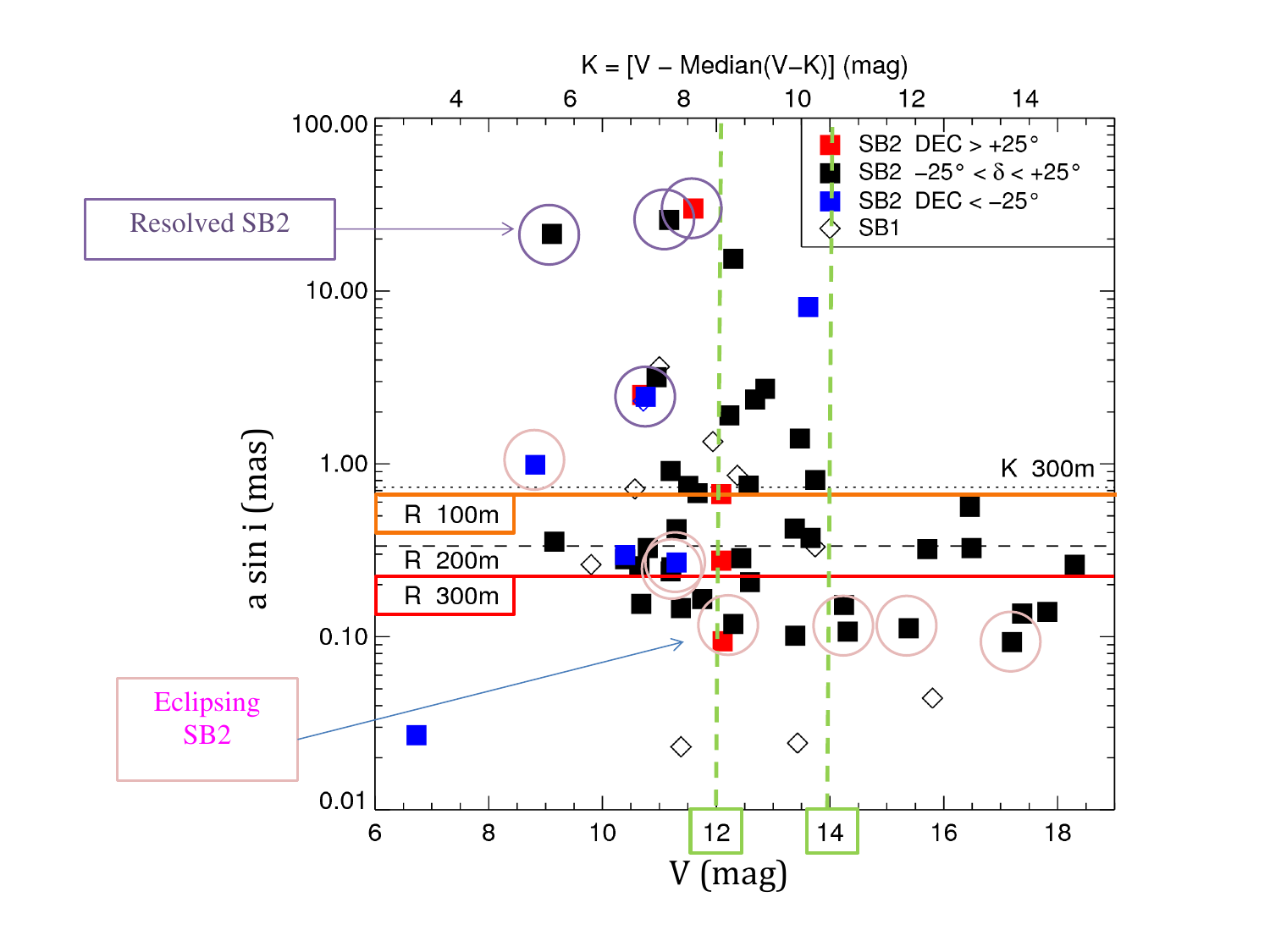}
  \caption{Separations vs. limiting magnitudes for spectroscopic binaries of star-forming regions (Taurus-Auriga, Ophiuchus, Sco-Cen, Chamaeleon, Orion, NGC~2264). Spectral types range from F to M.}
  \label{fig:binaries}
\end{figure}

Extending the long-baseline interferometry technique to the visible,
 say to 0.6 or 0.8~$\mu$m offers the promise of resolutions three to four
times finer than in the infrared at 2.2 $\mu$m. For the PMS stars in the
Taurus and Ophiuchus star-forming regions, astrometric orbital measurements
will be able to reach orbits with smaller angular sizes than possible by
purely infrared studies.  These have the advantages that their orbital
periods are shorter and the studies advance faster. In particular,
more angularly resolvable SB2s will become accessible. Figure~\ref{fig:binaries} shows that
nearly twice as many VB+SB2 systems will be resolvable at $0.6~\mu$m
as at $2.2~\mu$m. Reaching a large sample of these systems requires a limiting
magnitude of $V \sim 14$ mag.

\parindent=1.0cm

While most members of the NYMGs are in the southern hemisphere, the
$\beta$~ Pic (age $\sim$ 8--10 Myr) and AB Dor (age $\sim$ 40--100 Myr)
NYMG also have members in the north and are accessible to both
the CHARA and VLTI arrays.  Table~\ref{tab:NYMG} shows that in the $\beta$ Pic NYMG
alone, about 15 stars are accessible for diameter measurement with the
VLTI.  Successful diameter measurements would increase the measured
stars four to five-fold and enable for the first time the study of the
possible age spread of their formation.

\begin{table}[h]
\begin{center}
\label{tab:NYMG}
\begin{tabular}{ccc}
  \hline
  V$_{mag}$ range & Number of stars & Spectral types \\
  \hline
  3-4 & 1 & A \\
  4-5 & 1 & A \\
  5-6 & 3 & 2 A - 1 F \\
  6-7 & 5 & 5 F \\
  7-8 & 6 & 4 F - 1 G - 1 K \\
  8-9 & 3 &  Mostly G and K \\
  9-10 & 4 & Mostly G and K \\
  10-11 & 13 & Mostly G and K \\
  11-12 & 11 & Mostly M \\
  12-13 & 6 & Mostly M \\
  13-14 & 2 & Mostly M \\
  14-15 & 1 & Mostly M \\
  \hline
\end{tabular}
  \caption{Distribution by visible light magnitude bin of known members of the $\beta$ Pic NYMG in 2008
(Torres et al. 2008).}
\end{center}
\end{table}
\clearpage

\section{Studying their complex environment}

\subsection{Scientific rationale}

\begin{figure}[h]
  \centering
  \includegraphics[angle=0,scale=0.35]{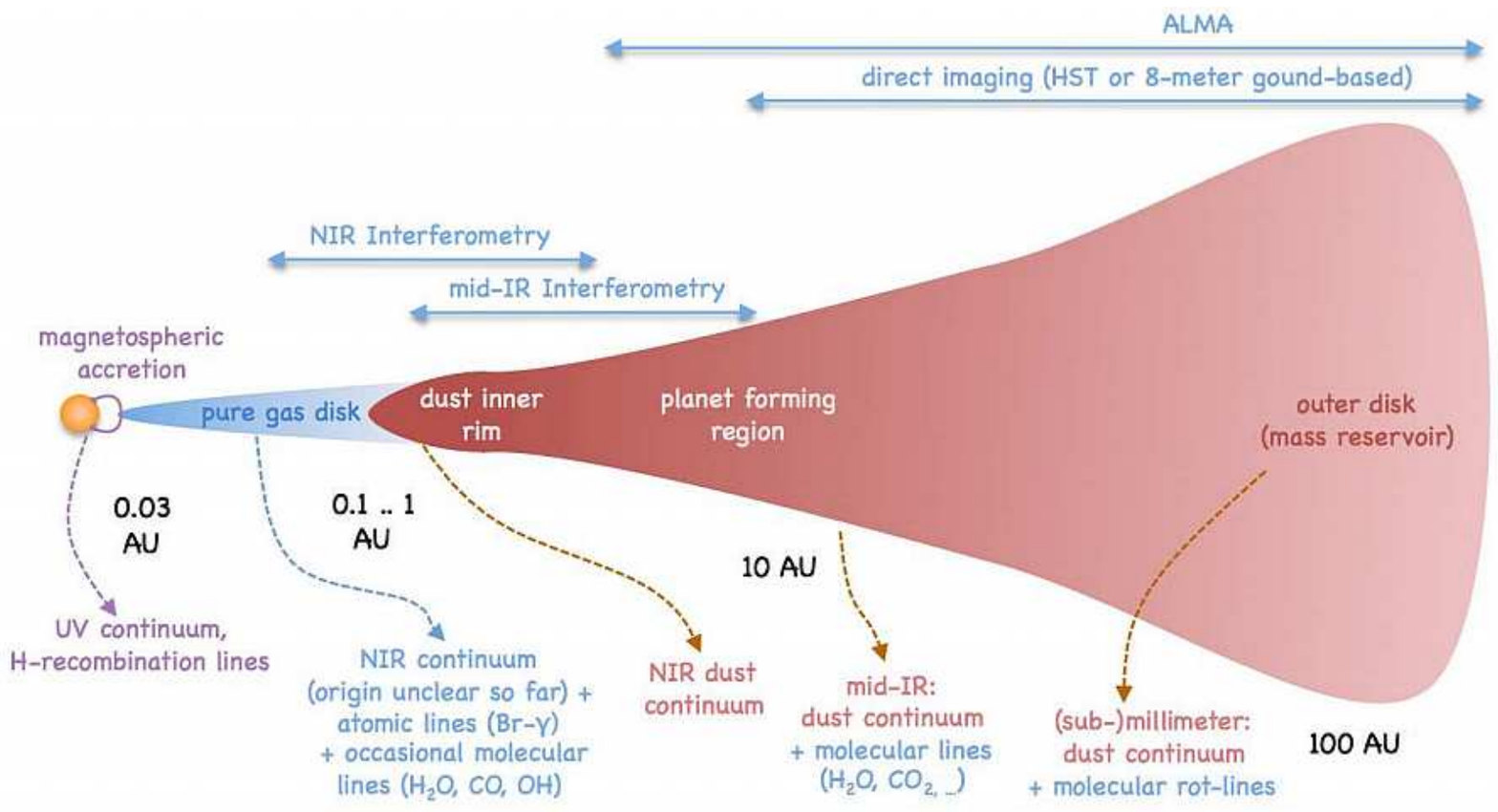}
  \caption{Schematic view of a protoplanetary disk. From Dullemond \& Monnier (2010).}
  \label{fig:disc}
\end{figure}

While optical direct imaging and (sub-)millimetric observations have revealed complex structures around young stellar
objects such as spirals, gaps, and holes at distances of few tens of astronomical units (AU), probing the innermost regions
where key processes for the star-disk-protoplanet(s) interactions are set requires a
very high angular resolution only provided by optical long-baseline interferometry (Figure~\ref{fig:disc}). Recently near-infrared
(spectro-)interferometry has directly probed the emission within this first AU and shown that these regions appear to be much more
complex than expected. Moreover all the complicated inner disk structures are strongly time variable on a timescale of weeks
to years.

Most of the continuum emission is due to the dust rim located at the dust evaporation front (at 1500~K) located at ~0.5-1~AU for HAeBe stars.
A gaseous disk extends further in, and material is accreted onto the star. Besides mass ejection through wind and/or jets are believed to originate
in this inner region. To probe the key mechanisms of our disk paradigm the distribution of the circumstellar gas and dust have to be studied. Multi-technique and multi-wavelength approaches are thus required to address the numerous open questions related to the puzzling environment of these objects.

The observations and modelling of gas lines to probe the geometry and physical conditions in these accretion and ejection mechanisms have been published since more than two decades (e.g. Hartmann 1982; Hartmann et al. 1990; Hartmann et al. 1994). Only recently it has been used through the technique of spectro-interferometry to spatially resolve the line. It has only been achieved on a limited number of stars, and mostly in the infrared range (e.g. across the H~I [Br$_\gamma$] line) (Benisty et al. 2010; Kraus et al. 2008; Eisner et al. 2014) with typical angular resolution of 0.5~AU. This corresponds also to the spatial extent of the inner rim of the dusty disk. Many more objects could be observed in the continuum (e.g. about 50 sources in the PIONIER Large Program (Berger et al. in prep.), and at very long baselines (Tannirkulam et al. 2008) and detailed modelling of these measurements have been published (e.g. Weigelt et al. 2011; Kraus et al. 2012; Ellerbroek et al. 2015). Improving the capability of visible interferometer and their ability to spatially resolve emission and absorption lines would be of strong interest as one would gain in angular resolution and probe complementary lines to the H~I [Br$_\gamma$] line.

\subsection{Accretion-ejection tracers}

\begin{figure}[t]
  \centering
  \includegraphics[angle=0,scale=0.37]{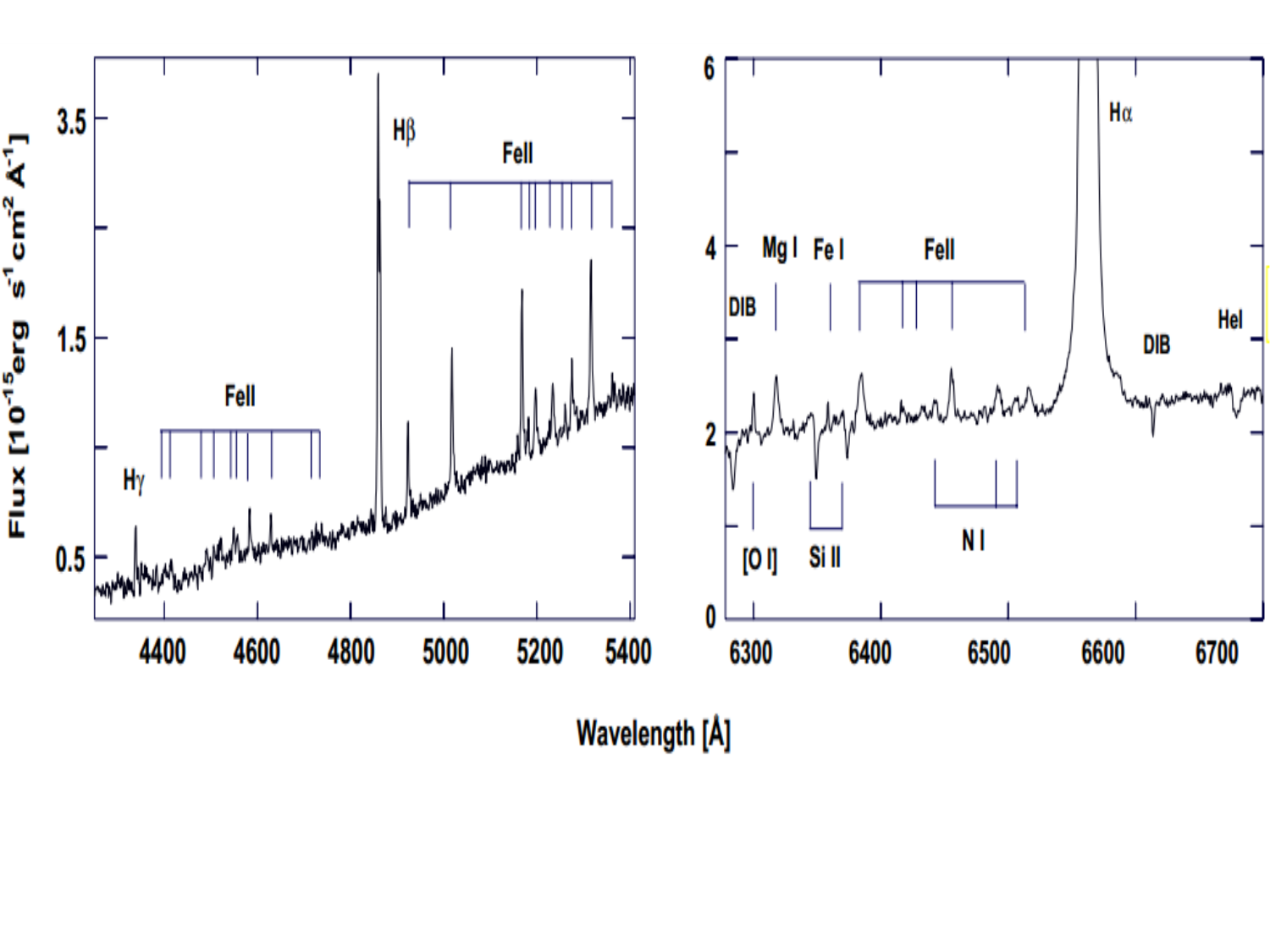}
  \caption{Spectrum of the Young Stellar Object ISOSS~J~20298+3559-IRS~1. From Krause et al. 2003.}
  \label{fig:spectrum}
\end{figure}

 \subsubsection{He\,I 10830\,nm}

 This line provides a unique diagnostic of kinematic motion in the regions close to the star. Its high opacity makes it a very sensitive probe to the geometry of the mechanisms at play. Observational studies have shown that the He I line is composite, with narrow and broad components that support an origin in magnetically controlled accretion from the disk onto the stellar surface but also in a hot wind (Beristain et al. 2001).  About 70\% of the surveyed Classical T~Tauri Stars (CTTS) show a sub-continuum blue-shifted absorption in the  He I 10830 line, in contrast with the hydrogen lines, where only 10\% of the CTTS have blue absorption (Edwards et al. 2003, 2006). The blue-shifted absorption below the continuum can be related to mass-loss through either a stellar or disk wind. The comparison with a survey of CTTS shows that the blue-shifted absorption are likely to be induced by a disk wind in 30\% of the stars, and by stellar winds in 60\%. On the other hand, the red absorption can be used as a probe of accreting material. Fischer et al. (2008) modelled this feature via scattering of stellar and veiling continua, assuming that the magnetospheric accretion is due to an azimuthally symmetric dipole. They derived accretion shock filling factors and sizes of the funnel flows. Kurosawa et al. (2011) simultaneously model the He~I~10830  with the Pa$\beta$, Br$_{\gamma}$ lines using non-LTE radiative transfer. Their model is consistent with the narrow blue shifted absorption component of the He I 10830 being caused by a disk wind, while the wider blueshfited absorption is caused by a bipolar stellar wind. However, these winds are still very compact (few stellar radii) and being able to resolve them spatially would require extremely high angular resolution.

\subsubsection{Recombination lines}

Besides the He\,I 10830\,nm line, there are several other promising lines (Figure~\ref{fig:spectrum}), such as the hydrogen recombination lines (H$\alpha$, H$\beta$, ...). These lines are considered accretion tracer (Hartmann et al.\ 1994), likely tracing structures on the stellar surface and in the magnetospheric accretion columns on scales of 3-5 stellar radii. Some commonly observed characteristics in the line profiles include: (1) slightly blueshifted emission peaks, (2) blueward asymmetries, and (3) redshifted absorption components and inverse P~Cygni profiles.  These features, as well as the Ultraviolet Balmer discontinuity, can be reproduced by combinations of accretion shock models and outflow (disk wind or stellar wind) models (e.g. Kurosawa et al.\ 2011). Assuming a typical T\,Tauri star (K7 type, R$_{\star}=1.85\,R_{\odot}$) in the Taurus star forming region (R$_{\star}=0.06$\,mas at $d=140$\,pc) and that the accretion columns extend to 3\,R$_{\star}$, we require an angular resolution better than 0.2\,mas to resolve the accretion geometry.  Also the Ca\,II infrared triplet (850/854/866\,nm) lines are considered tracers of magnetospheric infall in T\,Tauri stars (Azevedo et al.\ 2006).

\subsubsection{Forbidden lines}

The spectra of many pre-main-sequence stars show forbidden lines, for instance from [OI], [NII], [FeII], and [SII] (Figure~\ref{fig:spectrum}).  These lines trace low-density gas that is located on relatively large spatial scales.  Detailed imaging and integral field spectroscopy observations on individual objects have found that the line emission is sometimes distributed over several arcseconds (e.g.\ DG\,Tau: Podio et al.\ 2011; Bacciotti et al.\ 2002; RY\,Tau: Agra-Amboage et al.\ 2009). Therefore, most of the line emission in these tracers might be overresolved with interferometry, although there might also be contributions from a compact component.

\subsection{Looking for companions}

Young short period binaries (with a period less than a few tens days, and a separation of a few tenths AU) cannot support large circumstellar disks.
In these close binary systems stars orbit in a gap opened by tidal interactions inside a circumbinary disk. Evidence of an enhanced emission line activity close to periastron passages have been observed (DQ~Tau: Basri et al. 1997; UZ~Tau~E: Jensen et al. 2007). Such phenomena can be modelled by non-axisymmetric accretion (de Val-Borro et al. 2011). Interferometry will provide a critical test for such simulations.\\

\begin{figure}[h]
  \centering
  \includegraphics[angle=0,scale=0.25]{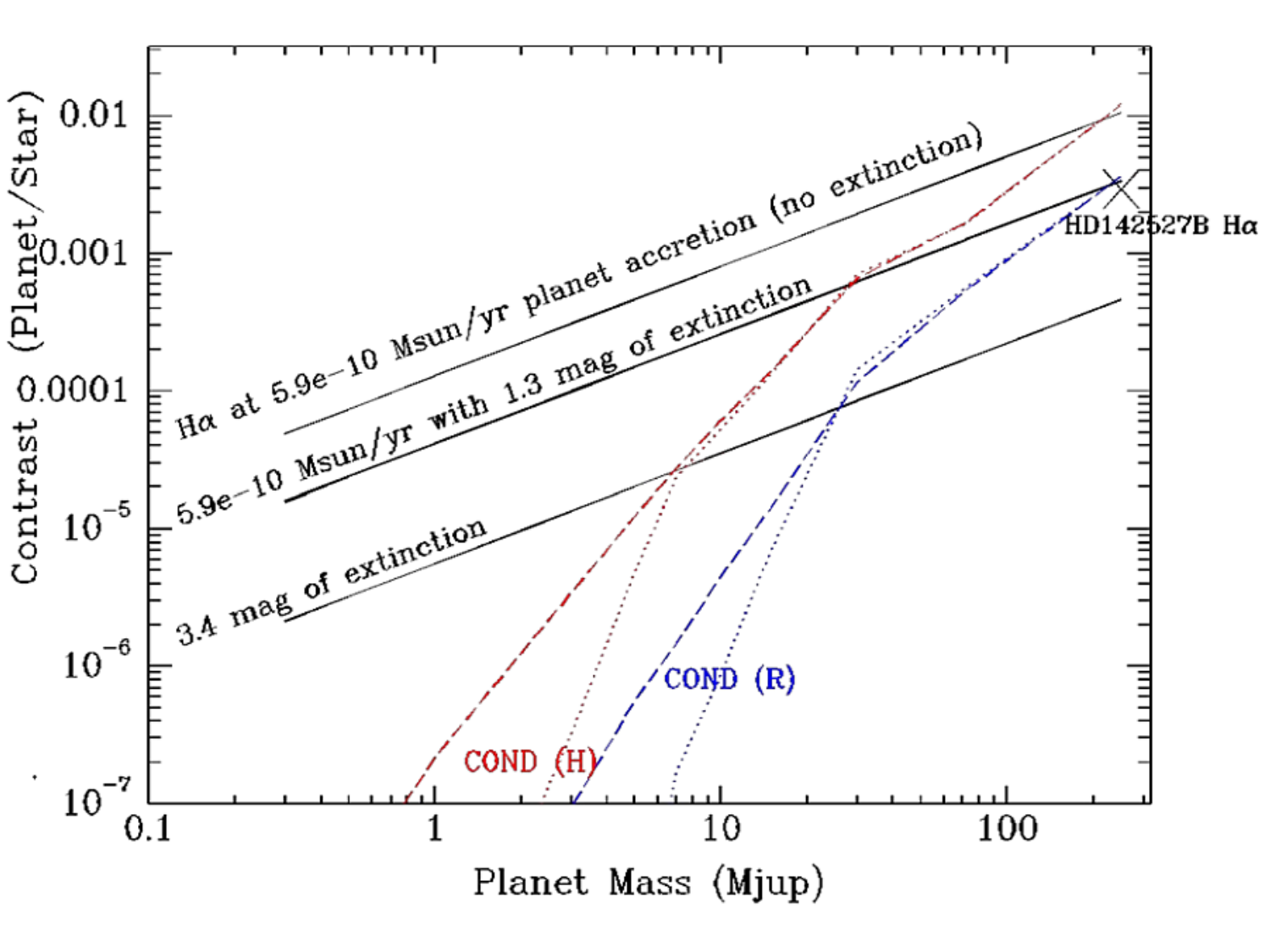}~\includegraphics[angle=0,scale=0.27]{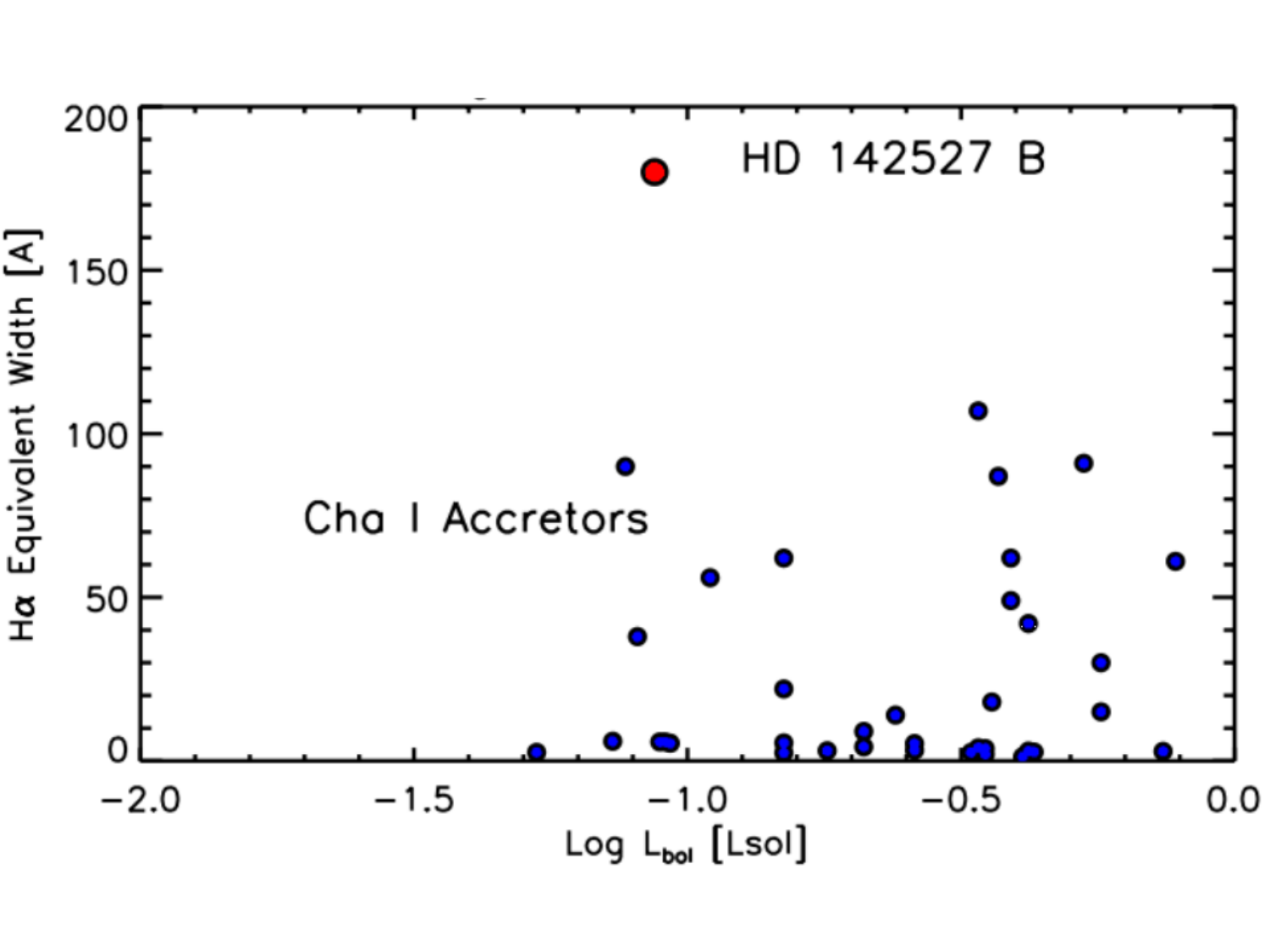}
  \caption{
    \textit{Left.} Predicted companion/star contrast ratio for the H$\alpha$-line (black solid lines,
    with different values of extinction) and the R-band and H-band continuum (blue and red lines),
    as function of companion mass. \textit{Right.} The companion (HD~142527B – red dot), inside the gap
    of a transitional disk, has significant accretion emission compared to younger low mass stars and brown dwarfs.
    (From Close et al.\ (2014)).
  }
  \label{fig:companion}
\end{figure}

Another interesting prospect for imaging in accretion-tracing lines could be to detect the signatures of low-mass companions that might be forming in the circumstellar disk. Close et al.\ (2014) argued that the companion/star contrast might be much more favourable in accretion-tracing lines than in visual or near-infrared continuum, in particular if one approaches the planetary-mass regime (Figure~\ref{fig:companion}). Assuming an accretion rate of $5.9 \times 10^{-10}$\,M$_{\odot}$/yr (as deduced for the 0.25\,M$_{\odot}$-mass companion around HD\,142527), then one could expect to detect a 260\,M$_{\rm Jupiter}$-mass brown dwarf at a contrast $10^{-3}$ and a 1\,M$_{\rm Jupiter}$-mass planet still at a contrast $10^{-4}$ (without extinction), although it should be noted that it is still highly uncertain whether the corresponding models can be extrapolated to the planetary-mass regime. Also, these young bodies should still be embedded in the circumstellar disk in order to allow significant accretion to occur, which should result in significant extinction.  Already a modest amount of extinction (e.g.\ 3.4\,mag) would lower the contrast to $5\times 10^{-5}$ (for the 260\,M$_{\rm Jupiter}$-mass brown dwarf) and $5\times 10^{-6}$ (for the Jupiter-mass object), respectively (see solid lines in Figure~\ref{fig:companion}).  This reduces the chances for detecting such objects. Anyway high-contrast imaging is required for this science case.\\

\subsection{Scattered light features}

Imaging in scattered light could reveal asymmetric structures that might be linked to planet formation, similar as seen in the outer disk of HD~141569 (Mouillet et al. 2001). Since the total integrated scattered light contributions are about 100 times fainter than the stellar flux, this will
require high-fidelity, high-contrast imaging. In that case, a polarimetric mode (like the ZIMPOL mode of SPHERE - Beuzit et al. 2014) or a nulling mode would be of strong interest.

\section{Summary of the High-Level Requirements}

\begin{figure}[h]
  \centering
  \includegraphics[angle=0,scale=0.25]{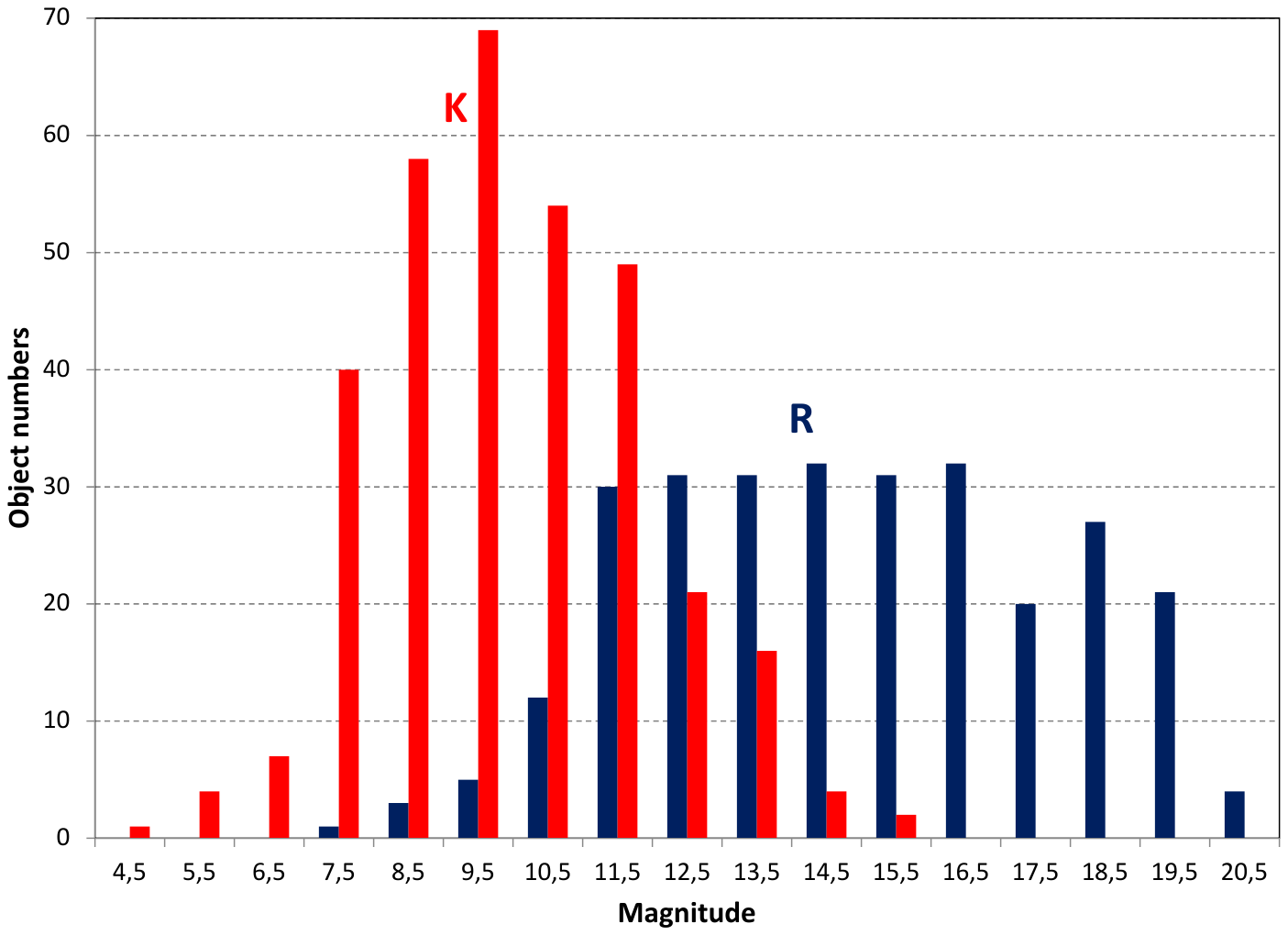}
  \caption{R$_{mag}$ histogram of Pre-Main Sequence stars in the Taurus-Auriga Star Forming Region (from Kenyon et al. (2008)). In Ophiuchus, K$_{mag}$s are similar but R$_{mag}$s are fainter because of greater extinction.}
  \label{fig:Vmag}
\end{figure}

The high-level requirements for each science case are summarized in Table~\ref{tab:specs}. The main challenge is related to the sensitivity in V due to the faintness of the YSO in this spectral range (Figure~\ref{fig:Vmag}). Given the equivalent width of the visible interesting lines (Table~\ref{tab:EW} a spectral resolution of a few thousands is required for studies across the H$\alpha$ line while the other lines require a spectral resolution of a few tens thousands.

\begin{table}[h]
\begin{center}
\label{tab:specs}
\begin{tabular}{cccccc}
  \hline
  {\tiny Science case} & {\tiny Baseline length} & {\tiny Spectral resolution} &{\tiny Accuracy} &{\tiny Imaging} & {\tiny Notes}\\
  \hline
 {\tiny Diameter of stars in SFR} &{\tiny unreachable} & & & & \\
 {\tiny Diameter of stars in NYMG} &{\tiny $>$~100~m} &{\tiny --} &{\tiny $<$~2\% on visibility} &{\tiny --} &\\
 {\tiny Dynamical masses} &{\tiny 300~m} &{\tiny --} &{\tiny $<$~2\% on visibility} & {\tiny x} & \\
 & & & {\tiny $<$~0.5$^\circ$ on the closure phase} & & {\tiny (1)}\\
 {\tiny Accretion-ejection} &{\tiny hectometric} & {\tiny a few 1000 (H$\alpha$)} & {\tiny $<$~2\% on visibility} & {\tiny x}\\
  & & {\tiny a few 10000 (other lines)} & & & \\
 & & & {\tiny $<$~1$^\circ$ on the closure phase} &&\\
 {\tiny Looking for companions} & {\tiny hectometric} & {\tiny a few 1000} & {\tiny $<$~0.001$^\circ$ on the closure phase} & {\tiny High-contrast} & {\tiny (2)}\\
 {\tiny Scattered light features} & {\tiny hectometric} & {\tiny --} & {\tiny $<$~0.1\% on visibility} & {\tiny High-contrast} \\
 &  &  & & {\tiny High-fidelity}& {\tiny (3)}\\
  \hline
\end{tabular}
\end{center}

{\footnotesize (1) A closure phase accuracy like $\pm$~0.3$^\circ$ leads to a contrast between stars of nearly 4-5 mag to be reached.}\\
{\footnotesize (2) The phase accuracy is the critical aspect and not the visibilities. To detect a companion with a contrast of 10$^{-5}$ a phase accuracy better than 0.001$^\circ$ is needed.}\\
{\footnotesize (3) Since the integrated scattered light contributions are likely 100-times fainter than the stellar flux an absolute visibility calibration better than 0.1\% is required to see some structures.}\\
\begin{center}
  \caption{High-level specifications for the different YSO science cases.}
\end{center}
\end{table}

\begin{table}[h]
\label{tab:EW}
\begin{center}
\begin{tabular}{ccc}
  \hline
  Line & Wavelengths (nm) & Equivalent width (${\AA}$) \\
  \hline
H$\alpha$ & 656.2 & 20 - 150 \\
Ca II triplet & 849.8 854.2 866.2 & 0.5 - 50 \\
He I & 667.8 706.5 & 0.5 - 2 \\
O I & 777.2 844.6 & 1.6 - 8 \\
$[$OI$]$ & 630 & $<$ 15\\
$[$SII$]$ & 673.1 & $<$ 5\\
$[$FeII$]$ & 715.5 & $<$ 2\\
  \hline
\end{tabular}
  \caption{Equivalent widths of visible spectral lines.}
\end{center}
\end{table}

\section{Conclusion}

The YSOs present a strong science case for interferometry in the visible. Interferometric imaging and spectroscopy will provide unique and complementary data for understanding star and planet formation. The techniques will probe the innermost regions of protoplanetary disks and will enable diameter measurements of stars still contracting to the main sequence. The science case is very challenging mainly because of the brightness of targets.

\chapter[Breaking the frontier to the cosmic distance
scale using visible interferometry of cepheids and eclipsing binaries]{Breaking the frontier to the cosmic distance
scale using visible interferometry of cepheids and eclipsing binaries\label{ch3}}

%%%%%%%%%%%%%%%%%%%%%%%%%%%%%%%%%%%%%%%%
%\usepackage{txfonts}
%%%%%%%%%%%%%%%%%%%%%%%%%%%%%%%%%%%%%%%%
%
\def\kms{km~s$^{-1}$}
\def\te{T_{\rm eff}}
\def\Mo{M_{\odot}}
\def\Lo{L_{\odot}}
\def\Ro{R_{\odot}}
\def\deg{\ensuremath{^\circ}}

\newcommand{\ep}[1]{\textcolor{red}{#1}}

\newcommand{\astroph}[1]{astro-ph/{#1}}
\newcommand{\AJ}[3]{{#1}, AJ, \vol{{#2}}, {#3}}
\newcommand{\ApJ}[3]{{#1}, ApJ, \vol{{#2}}, {#3}}
\newcommand{\ApJS}[3]{{#1}, ApJS, \vol{{#2}}, {#3}}
\newcommand{\ApJLett}[3]{{#1}, ApJ, \vol{{#2}}, {#3}}
\newcommand{\ApJL}[3]{{#1}, ApJ, \vol{{#2}}, {#3}}
\newcommand{\AnnRev}[3]{{#1}, Annual Review of Astron. and Astrophys.,
\vol{{#2}}, {#3}}
\newcommand{\AandA}[3]{{#1}, A\&A, \vol{{#2}}, {#3}}
\newcommand{\AandAsupp}[3]{{#1}, A\&A Supp.\rm, \vol{{#2}}, {#3}}
\newcommand{\AN}[3]{{#1}, Astron. Nachr.\rm, \vol{{#2}}, {#3}}
\newcommand{\MNRAS}[3]{{#1}, MNRAS\rm, \vol{{#2}}, {#3}}
\newcommand{\MmSAI}[3]{{#1}, Mem. Soc. Astron. Ital.\rm, \vol{{#2}},
{#3}}
\newcommand{\Mess}[3]{{#1}, The Messenger\rm, \vol{{#2}}, {#3}}
\newcommand{\Vistas}[3]{{#1}, Vistas Astr.\rm, \vol{{#2}}, {#3}}
\newcommand{\Observatory}[3]{{#1}, Observatory\rm, \vol{{#2}}, {#3}}
\newcommand{\PASP}[3]{{#1}, PASP\rm, \vol{{#2}}, {#3}}
\newcommand{\PASPC}[3]{{#1}, PASPC\rm, \vol{{#2}}, {#3}}
\newcommand{\PASJ}[3]{{#1}, PASJ\rm, \vol{{#2}}, {#3}}
\newcommand{\vol}[1]{{\mbox{#1}}}

\newcommand{\Teff}{\mbox{$T_{\mbox{\scriptsize eff}}\,$}}
\newcommand{\Tefffour}{\mbox{$T^{4}_{\mbox{\scriptsize eff}}\,$}}
\newcommand{\Teq}{\mbox{$T_{\mbox{\scriptsize eq}}\,$}}
\newcommand{\Sv}{\mbox{$S_{V}\,$}}
\newcommand{\BCv}{\mbox{$\mbox{B.C.}_{V}\,$}}
\newcommand{\geff}{\mbox{$g_{\mbox{\scriptsize eff}}\,$}}
\newcommand{\Vrot}{\mbox{$v_{\mbox{\scriptsize rot}}\,$}}
\newcommand{\mbol}{\mbox{$m_{\mbox{\scriptsize bol}}\,$}}
\newcommand{\mv}{\mbox{$m_{V}\,$}}
\newcommand{\mk}{\mbox{$m_{K}\,$}}
\newcommand{\Mbol}{\mbox{$M_{\mbox{\scriptsize bol}}\,$}}
\newcommand{\Mb}{\mbox{$M_{B}\,$}}
\newcommand{\Mv}{\mbox{$M_{V}\,$}}
\newcommand{\Mk}{\mbox{$M_{K}\,$}}
\newcommand{\Mh}{\mbox{$M_{H}\,$}}
\newcommand{\Mj}{\mbox{$M_{J}\,$}}
\newcommand{\Mi}{\mbox{$M_{I}\,$}}
\newcommand{\Mw}{\mbox{$M_{W}\,$}}
\newcommand{\Wvi}{\mbox{$W_{VI}\,$}}
\newcommand{\Wjk}{\mbox{$W_{JK}\,$}}
\newcommand{\Mrr}{\mbox{$M_{\mbox{\scriptsize RR}}\,$}}
\newcommand{\delMv}{\mbox{$\Delta
         M_{\mbox{\scriptsize V,TO}}^{\mbox{\scriptsize RR}}\,$}}
\newcommand{\meanMbol}{\mbox{$<\!\!M_{\mbox{\scriptsize bol}}\!\!>\,$}}
\newcommand{\meanMv}{\mbox{$<\!\!M_{V}\!\!>\,$}}
\newcommand{\meanMk}{\mbox{$<\!\!M_{K}\!\!>\,$}}
\newcommand{\Koff}{\mbox{$K_{\mbox{\scriptsize off}}\,$}}
\newcommand{\MvRR}{\mbox{$<M_{V}(RR)>\,$}}
\newcommand{\MkRR}{\mbox{$<M_{K}(RR)>\,$}}
\newcommand{\MbolRR}{\mbox{$<M_{\mbox{\scriptsize bol}}(RR)>\,$}}
\newcommand{\MvTO}{\mbox{$M_{V}(TO)\,$}}
\newcommand{\RRab}{\mbox{$RR{\mbox{\scriptsize ab}}\,$}}
\newcommand{\RRc}{\mbox{$RR{\mbox{\scriptsize c}}\,$}}
\newcommand{\RRd}{\mbox{$RR{\mbox{\scriptsize d}}\,$}}
\newcommand{\pmag}{\mbox{$\stackrel{\mbox{\scriptsize m}}{\textstyle
.}$}}
\newcommand{\mags}{\mbox{$^{\mbox{\scriptsize m}}$}}
\newcommand{\pday}{\mbox{$\stackrel{\mbox{d}}{\textstyle .}$}}
\newcommand{\pperiod}{\mbox{$\stackrel{p}{\textstyle .}$}}
\newcommand{\phour}{\mbox{$\stackrel{h}{\textstyle .}$}}
\newcommand{\hours}{\mbox{$^{h}$}}
\newcommand{\pmin}{\mbox{$\stackrel{m}{\textstyle .}$}}
\newcommand{\mins}{\mbox{$^{m}$}}
\newcommand{\psec}{\mbox{$\stackrel{s}{\textstyle .}$}}
\newcommand{\secs}{\mbox{$^{s}$}}
\newcommand{\parcsec}{\mbox{$\stackrel{\prime\prime}{\textstyle .}$}}
\newcommand{\arcsecs}{\mbox{$^{\prime\prime}$}}
\newcommand{\parcmin}{\mbox{$\stackrel{\prime}{\textstyle .}$}}
\newcommand{\arcmins}{\mbox{$^{\prime}$}}
\newcommand{\solar}{\mbox{$\odot$}}
\newcommand{\Rsolar}{\mbox{$R_{\odot}\,$}}
\newcommand{\Msolar}{\mbox{$\vec{M}_{\odot}\,$}}
\newcommand{\Lsolar}{\mbox{$L_{\odot}\,$}}
\newcommand{\Tsolar}{\mbox{$T_{\odot}\,$}}
\newcommand{\FeH}{\mbox{[Fe/H]}\,}
\newcommand{\OFe}{\mbox{[O/Fe]}\,}
\newcommand{\Mpc}{\mbox{Mpc}\,}
\newcommand{\prMpc}{\mbox{$\mbox{Mpc}^{-1}$}\,}
\newcommand{\loggzero}{\mbox{$\log (g_0)$}\ }
\newcommand{\logP}{\mbox{$\log P$}\ }
\newcommand{\mM}{\mbox{$(m-M)$}\ }
\newcommand{\mMzero}{\mbox{$(m-M)_0$}\ }
\newcommand{\thetaspec}{\mbox{$\theta_{\mbox{\scriptsize spec}}$}}
\newcommand{\thetaphot}{\mbox{$\theta_{\mbox{\scriptsize phot}}$}}
\newcommand{\Smin}{\mbox{$\Sigma_{\mbox{\scriptsize min}}$}\,}
\newcommand{\Ab}{\mbox{$\mbox{A}_{\mbox{\scriptsize B}}$}\,}
\newcommand{\vrad}{\mbox{$\mbox{v}_{\mbox{\scriptsize rad}}$}\,}
\newcommand{\hubble}{\mbox{$H_{\mbox{\scriptsize 0}}\:$}}
\newcommand{\phase}[2]{\mbox{[{#1};{#2}]}\,}
\newcommand{\magdex}{\mbox{mag~dex$^{-1}$}\,}
\newcommand{\thetaLD}{\mbox{ $\theta_{\mbox{\scriptsize LD}}$}\,}
\newcommand{\Fv}{\mbox{ $F_{\mbox{\scriptsize V}}$}\,}

\author{N.~Nardetto$^{1}$,
P.~Kervella$^{2}$,
 A.~Chiavassa$^{2}$,
A.~Gallenne$^{3}$,
A.~Merand$^{4}$}
 \author{D.~Mourard$^{1}$,
 J.~Breitfelder$^{2}$,
 }
 
\begin{center} 
$^{1}$~Laboratoire Lagrange, UMR 7293, CNRS, Observatoire de la C\^ote d'Azur, Universit\'e de Nice Sophia-Antipolis, France\\
$^{2}$~Observatoire de Paris-Meudon, LESIA, UMR 8109, 5 Place Jules Janssen, F-92195 Meudon Cedex, France\\
$^{3}$~Departamento de Astronom\'ia, Universidad de Concepci\'on, Casilla 160-C, Concepci\'on, Chile\\
$^{4}$~European Southern Observatory, Alonso de Cordova 3107, Casilla 19001, Santiago 19, Chile\\
\end{center}

\section{Introduction}\label{s_Introduction}
To understand the nature of dark energy and the global geometry of space, one of the best solution to date is proposed by the SHOES project (Supernovae, H0, for the Equation of State of dark energy), led by A. Riess, and the Carnegie Hubble Program, led by W. Freedman, who both derived the Hubble constant (Ho) with a $\simeq$3\% precision (including systematics) using basically two different anchors to the type 1a supernova relation (SN): the trigonometric parallaxes of few Galactic Cepheids and the distance to the Large Magellanic Cloud (LMC). Besides, the observation of extragalactic eclipsing binaries provides a new path to a very precise determination of the distance to galaxies in the local group. 
Using new generation of long-baseline visible interferometers (in north and south hemispheres), and also in the context of Gaia, we aim at improving the calibration of the first ladder of the distance scale (i.e. in the local group). This can be done by better understanding the physics of Cepheids and eclipsing binaries and also by improving a very fundamental tool in stellar physics, which is the surface-brigthness color relation (SBCR hereafter).   
These improved SBCR, once applied to Cepheids and eclipsing binaries will indeed break the frontier of the cosmic distance scale, which means (1) promote the Baade-Wesselink method as a even more robust tool to derive the distance of Milky Way and LMC Cepheids (which basically means to better understand also the physics of Cepheids thanks to Gaia), (2) derive the distance to LMC with a precision and accuracy of 1\% (1\% for eclipsing binary, probably 3\% for Cepheids), (3) derive the distance to M31 and M33 (and many other galaxies in the local Group) with a precision and accuracy of 5\%. 
We find that with a new generation of long-baseline visible interferometer (with limiting magnitude of  basically 10 with and an unprecedented precision on the visibility), we could reach indeed these objectives. The number of stars that could be observed in the north and south hemisphere and the expected precision is given for Cepheids and eclipsing binaries, respectively, and for each sub-projects. 
This cartography of the local universe relying on visible interferometry, once corrected from systematics (e.g. metallicity, ...) thanks to the cross-check with Gaia parallaxes, will be used by the SHOES and CHP projects to determine the Hubble constant with a precision and accuracy of 2\% or better shedding light on the nature of dark energy.

Two distance indicators are currently very powerful: namely Cepheids and eclipsing binaries.  We show in this white book how long-baseline and high precision visible interferometry can help to better understand the physics of Cepheids and also eclipsing binaries by using Gaia parallaxes, and how in turn, this would help to break the frontier to the cosmic distance scale in particular by providing a very precise distance to LMC and distant galaxies. This approach is particularly consistent and complementary with other on-going or future projects, such as Gaia, MATISSE. This work is part of the Araucaria Project \citep{gieren05_messenger, gp13}.

\section{The physics of Cepheids}

A powerful way of constraining the $PL$ relation is to use the Baade-Wesselink (BW) method of distance determination. The basic principle of the BW method is to compare the linear and angular size variation of a pulsating star in order to derive its distance through a simple division. The angular diameter is either derived by interferometry (hereafter IBW for Interferometric Baade-Wesselink method; see for instance \citep{kervella04a} or by using the InfraRed Surface Brightness (IRSB) relation  \citet{storm11a, storm11b}. However, when determining the linear radius variation of the Cepheid by spectroscopy, one has to use a conversion projection factor from radial to pulsation velocity. There is currently a consensus on the period-projection factor relation (hereafter $Pp$ relation) between hydrodynamical models \citep{nardetto04, nardetto07, nardetto09,  nardetto11b} and observational constraints based on infrared interferometry \citep{merand05} or Cepheids in Eclipsing Binaries \citep{pilecki13}. However, a 2$\sigma$ discrepancy is found when using the IRSB method, i.e. the surface-brightness relation \citet{storm11a, storm11b, nardetto11b}. \citet{riess11} adopted a conservative approach concerning the BW method: `{\it We have not made use of additional distance measures to Galactic Cepheids based on the BW method or stellar associations as they are much more uncertain than well-measured parallaxes, and the former appear to be under refinement due to uncertainties in their projection factors, as discussed by \citet{fouque07} and \citet{vl07}}" ({\it cf.} \citet{riess09a, riess09b}). 

Next generation of visible interferometers can greatly help in resolving such discrepancy in the projection factor.  On the one hand, it seems necessary to add additional direct constrains on the projection factor by increasing the number of Cepheids in binaries that could be resolved (Sect \ref{sscep_bin}) or by measuring for the first time the limb-darkening of Cepheids  (Sect \ref{sscep_LD}). On the other hand, it seems essential to test the surface-brightness relation of short-period Cepheids (where the p-factor discrepancy appears), by pushing the recent efforts made to characterize the CSEs of Cepheids (Sect \ref{sscep_CSE}). Another problem could come from the 'famous bump' of Cepheids which arises at the pulsation phase of minimum radius \citep{storm11a, fokin96}.  A monitoring of Cepheids with a homogeneous precision of few percents (in particular at minimum radius) is required. Moreover, such angular diameter curves of high precision will allow an application of the inverse BW method based on trigonometric parallaxes of Gaia  (Sect \ref{sscep_invGaia}). We will show also that high-spectral resolution visible interferometry can be a very robust tool to bring new constrain on the dynamical structure of Cepheids' atmosphere, but could also provide a novel method of distance determination (Sect \ref{sscep_HRA}).  The final objective is to have a better knowledge of the period-projection factor relation and promote the Baade-Wesselink as an even more robust tool for the distance scale.

%The synthesis concerning instrumental specification is given in Sect \ref{sscep_sum}

\subsection{The projection factor derived from Cepheids in binaries}\label{sscep_bin}

Several companion of Cepheids have been recently revealed by interferometry: V1334 Cyg \citep{gallenne13}, AX Cir and AW Per \citep{gallenne14}, see also \citep{gallenne14b}.The contrast between the companion and the Cepheid is rather high in H band (flux ratio between 30 and 100), where  current interferometric recombiners work. However, in the visible wavelength, the contrast is more favorable (flux ratio between 2 and 100).

With a visible interferometer of new generation it would be possible to detect and characterize more Cepheids in binary systems. Such detections are crucial for several reasons: (1) they bring new constrain on the mass of Cepheids and could help in resolving the Cepheid mass discrepancy \citep{neilson11,keller08, bono06}, (2) they allow a geometrical determination of the projection factor of Cepheids. 

\begin{table}[htbp] \caption{\label{Tab.ang} Cepheids in binaries}
\begin{center}
\begin{tabular}{lccc|ccccccccc}
\hline
\hline
N$_\star$	& 10-15 \\
spatial resolution 	&  0.3 mas \\
spectral resolution 	&  $R=3000$ \\			
temporal resolution	&  2-40 years \\
precision 	&	1\% even at low visibilities \\	
imaging	& 6T is best \\
\hline
\end{tabular}
\end{center}
%\begin{list}{}{}
%\item[$^{\mathrm{a}}$]  and/or minimum number of telescopes required
%\end{list}
\end{table}

\subsection{The limb-darkening of Cepheids as a constrain on the projection factor}\label{sscep_LD}

\begin{figure}[htbp]
\begin{center}
\resizebox{1.0\hsize}{!}{\includegraphics[clip=true]{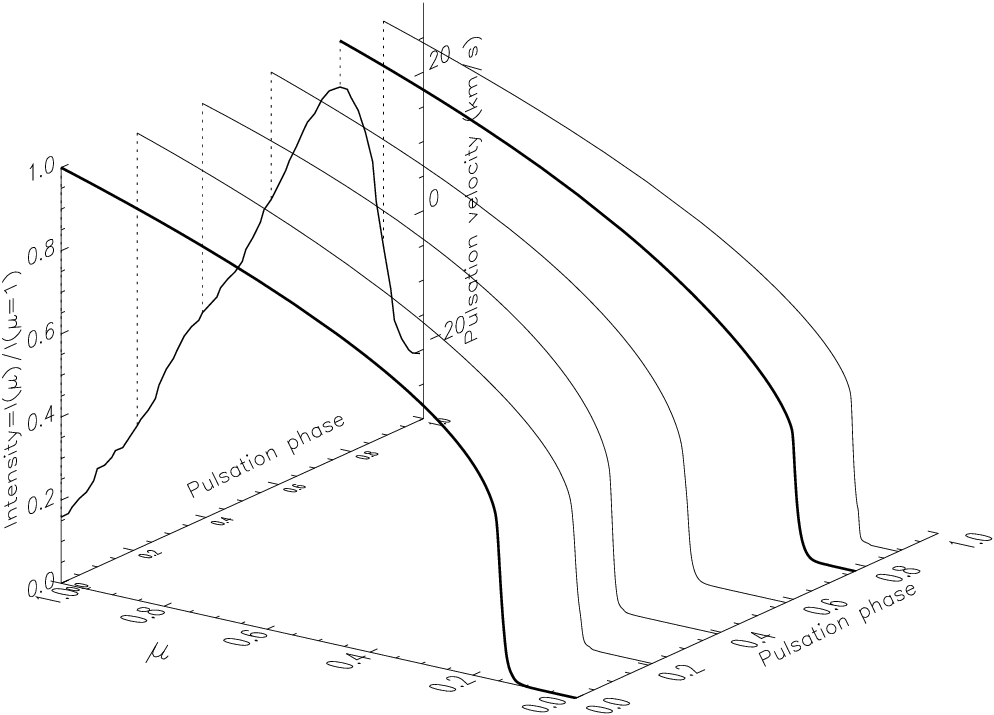}}
\end{center}
\caption{ A $3$D diagram that represents the normalized stellar disk
intensity distribution of $\delta$~Cep in the continuum as a function of $\mu$ (see
text) for different phases. The vertical plot (arbitrary unit)
represents the photospheric pulsation velocity as a function of the
phase (in the heliocentric frame). Two distributions (in
accentuated) are interesting because they correspond to extreme
cases in limb-darkening.} \label{Fig_Cint3D}
\end{figure}

\begin{figure}[htbp]
\begin{center}
\resizebox{1.0\hsize}{!}{\includegraphics[clip=true]{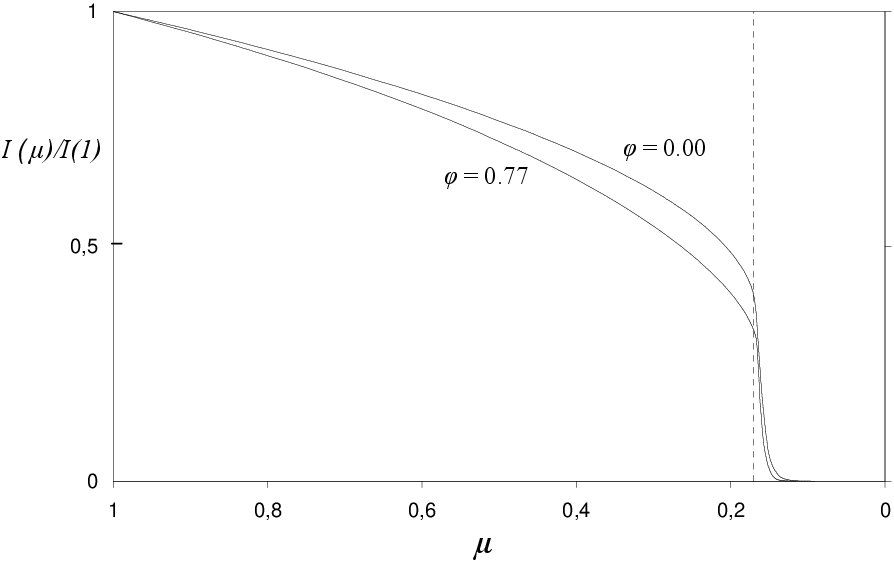}}
\end{center}
\caption{Details of the two intensity distributions for phases
$0.77$ and $0.00$ as a function of $\mu$ in the case of $\delta$~Cep. The profile corresponding
to the highest velocity at contraction ($\phi=0.77$) is the most
limb-darkened, while the profile corresponding to the highest
velocity at expansion ($\phi=0.0$) is the lowest limb-darkened. The
vertical dashed line represents the photosphere of the star defined
by $\tau_{\rm c}=2/3$. The photosphere corresponds to the same value
of $\mu$ for the two pulsation phases considered here.}
\label{Fig_Cint}
\end{figure}

Computation of limb-darkening profiles at different wavelengths and
pulsational phases is needed in the BW method not only to derive
the limb-darkened interferometric angular diameters, but also to determine the
geometric projection factor. However, theoretical results concerning
limb-darkening are currently inconsistent at an important level (see
for example Marengo et al. 2003 - Fig. 3, and \citet{nardetto06b}
- Fig 3). Figure~\ref{Fig_Cint3D} is derived from  \citet{nardetto06b} and shows the limb-darkening variation as a function of time based on the hydrodynamical model by \citet{fokin91}. Fig.~\ref{Fig_Cint} gives the detail of the continuum limb-darkening variation at maximum contraction and expansion velocities. From our theoretical simulations we find that a precision of 5\% on the visibility in the second lobe would give the average value of the  limb-darkening of the star at about 1\% of precision. But even better, 1\% on the visibility in the second lobe would provide the time-dependency of the limb-darkening of the star with a 5$\sigma$ detection. The precision of the projection factor and the physics of the photosphere of Cepheids would be greatly improved with such unique measurement.

\begin{table}[htbp] \caption{\label{Tab.ang} The limb-darkening of $\delta$~Cep and $\ell$~Car}
\begin{center}
\begin{tabular}{lccc|ccccccccc}
\hline
\hline
N$_\star$	& $\delta$ Cep and $\ell$~Car \\
spatial resolution 	&  0.3 mas \\
spectral resolution 	&  $R=3000$ \\			
temporal resolution	&  $P=5.36$d ($\delta$ Cep),  $P=35.5$d ($\ell$~Car) \\
precision 	&	1\% in the second lobe (V$ \simeq$ 10\%) \\	
imaging	& 3T is fine \\
\hline
\end{tabular}
\end{center}
\end{table}

\subsection{The circumstellar environment of Cepheids}\label{sscep_CSE}

Although Cepheids are fundamental distance indicators, the circumstellar envelopes (CSEs) we discovered around several Galactic Cepheids create a positive bias on their apparent luminosity (Figure 2; \citet{merand07}; \citet{gallenne12, gallenne13b}). Their potential impact on the calibration of the Period-Luminosity relation in different wavelength domains (visible, near-infrared, thermal infrared) is discussed in \citet{kervella13}. New observations of Cepheid CSEs in the thermal infrared domain using the VLTI/MIDI instrument (Fig. 3; \citet{gallenne13b}) indicate that these envelopes probably contain refractory dust grains (including alumina), but their nature is not yet fully understood. 

Recently, a bright CSE has been also discovered in the visible with VEGA around $\delta$~Cep (paper in preparation).If confirmed, a systematic study of theses CSEs in visible with next generation of long interferometers and also with MATISSE in infrared will be of high importance for the distance scale calibration. 

\begin{table}[htbp] \caption{\label{Tab.ang}The CSE of Cepheids}
\begin{center}
\begin{tabular}{lccc|ccccccccc}
\hline
\hline
N$_\star$	& about 20 \\
spatial resolution 	&  0.3 mas \\
spectral resolution 	&  $R=3000$ \\			
temporal resolution	&  several obs. per star is enough \\
precision 	&	1\% in the first lobe (V$ \simeq$ 50\%) \\	
imaging	& 3T is fine, but 6T better \\
\hline
\end{tabular}
\end{center}
\end{table}

\section{The robust BW method combined with Gaia: a constrain on the period-projection factor relation}\label{sscep_invGaia}

Gaia will provide the distances of one billion of stars in the Milky Way with an unprecedented precision. In the second release of Gaia in 2017, we should find the distance to 250 Cepheids (within 3 kpc) with precisions better than 2\%. Our aim is to apply the  {\it inverse} Baade-Wesselink to the whole sample of Galactic Cepheids observed by interferometry.

About 25 Cepheids have already been observed with VEGA/CHARA and FLUOR/CHARA (papers in preparation) but with inhomogeneous precision. A new generation of visible interferometer of high precision would allow to reach the 1 or 2\% precision for each individual determination of the angular diameter at a specific pulsation phase.  The most important is to concentrate on the minimum and maximum radii of the star. It is in particular well known that some unexplained discrepancies in the surface-brightness are found at minimum radius for some objects, probably not all Cepheids \citep{storm11a,storm11b}. Reaching a 1 or 2\% accuracy when the star has its smallest angular diameter is crucial to resolve this question. The CSEs and the limb-darkening should be also well defined in the analysis of the interferometric data. 

From the Galactic cepheid database \citep{fernie95} we found that 272 stars are observable from CHARA ($\delta > -25^{o}$) and 256 from VLTI ($\delta < -25^{o}$). If we select Cepheids brighter than $mV=10$, we have 105 Cepheids observable with CHARA and 147 with VLTI. If we consider a precision on the visibility of 5\% ($R=2500$), the amplification factor found in the PhD by Merand, and 20 measurements per star (at maximum and minimum radii), we find we could get a 2\% precision on the distance (or similarly 2\% on the projection factor) for 75 Cepheids with CHARA and 40 with VLTI. This would lead to a 2\% precision on the individual projection factors bringing exciting result on the period-projection factor relation. The current observational status of the \emph{Pp} can be found in \citet{storm11b}, their figure~6.

\begin{table}[htbp] \caption{\label{Tab.ang} The inverse BW method using Gaia}
\begin{center}
\begin{tabular}{lccc|ccccccccc}
\hline
\hline
N$_\star$	& 105 \\
spatial resolution 	&  0.3 mas \\
spectral resolution 	&  $R=3000$ \\			
temporal resolution	&  periods from 2 to 50 days \\
precision 	&	5\% on the visibility (V$ \simeq$ 50\%) \\	
imaging	& 3T is fine, but 6T better \\
\hline
\end{tabular}
\end{center}
\end{table}

\subsection{High spectral resolution interferometry of Cepheids as an indicator of distances?}\label{sscep_HRA}

\begin{figure}[htbp]
\begin{center}
\resizebox{1.0\hsize}{!}{\includegraphics[clip=true]{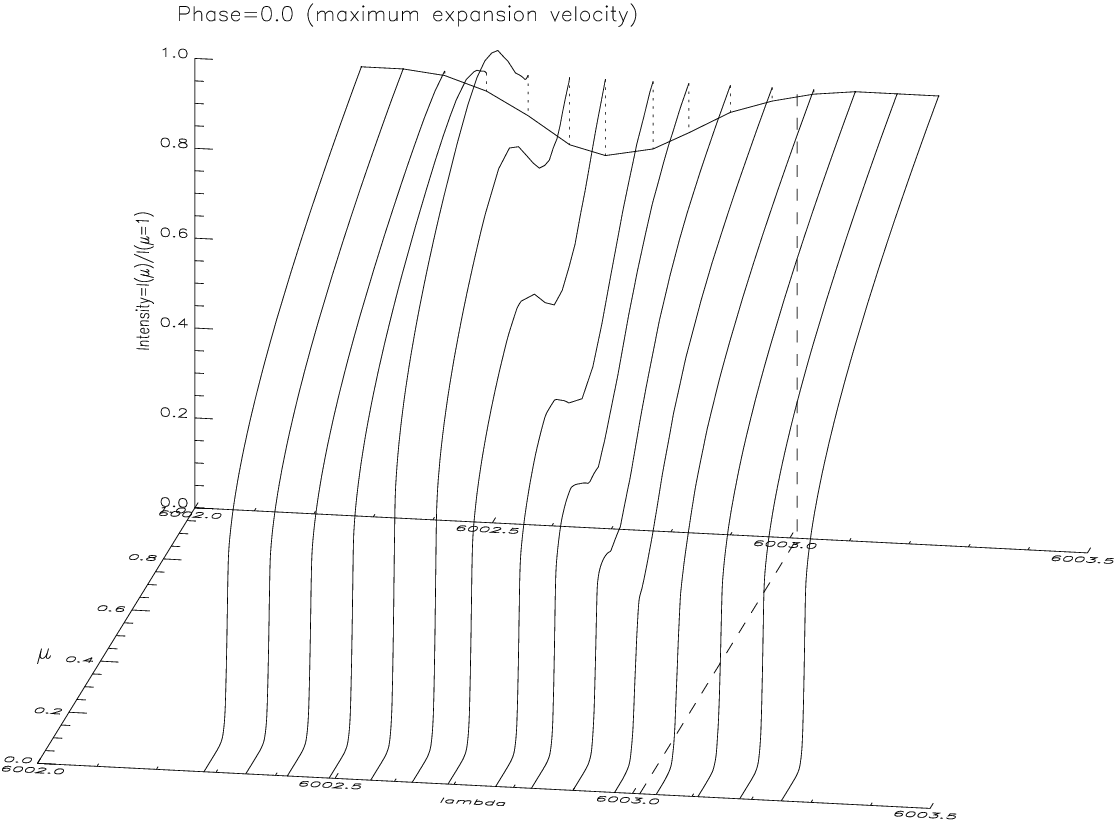}}
\end{center}
\caption{A 3D diagram that represents the spectral line profile
({Fe}{I} 6003.012 \AA) for the maximum expansion velocity ($
\phi=0.00$) with the corresponding intensity distributions. The
dashed line represents the reference wavelength in the stellar rest
frame. }
 \label{Fig_Rint3D_g}
\end{figure}

Spectro-interferometry within metallic or even H$\alpha$ lines of Cepheids is beyond the sensitivity of current interferometers. 
Nevertheless, with future visible interferometers, such observations will become possible. For instance, if we consider $\delta$~Cep, we find theoretically (using toy modeling) an S-signature photocenter displacement due mainly to pulsation, rotation (assumed to 8\kms), line depth (about 25\% of the continuum) and the angular diameter at a given pulsation phase. The largest S-signal is obtained at maximum radius (when the pulsation velocity is null) with a peak-to-peak value of 160$\mu$as when considering a spectral resolution of $R=60000$ (see Fig.~5.3 of the PhD of Nardetto). The second largest signal is obtained at minimum radius (when the pulsation velocity is null) with a peak-to-peak value of 145$\mu$as. The difference between these two situations is mainly due the angular diameter (all other parameters are the same) and of about 15$\mu$as (or about 10\%). For a 200m baseline, a wavelength of 1$\mu$m and a precision $\frac{\sigma(\phi)}{\phi}=10^{-3}$ on the phase measurements, one can reach a sensitivity of 1$\mu$as. If verified, spectro-interferometry would be a perfect tool to probe the dynamical structure of Cepheid' atmosphere (projection factor), but could also be used to apply a new method of distance determination (with a 15$\sigma$ detection or 5\% in precision on the individual angular diameter for only one measurement; actually it should be even more precise as all the S-signal signature has to be fitted). The only requirement is to extract without degeneracy the angular diameter from other parameters.  We emphasize that for such analysis, hydrodynamical models are ready \citep{nardetto06b} (Fig. \ref{Fig_Rint3D_g}).

We now anticipate the observation of LMC Cepheids (as presented in \citet{mourard06}).  We have used a sample of LMC Cepheids extracted from \citet{persson04}. This sample gives access to J, H and K magnitudes of about 90 cepheids with periods ranging from 3 to 48 days. In order to estimate angular diameter for each star of this sample, we have used empirical relations based on the (J,J-K) parameters. Half of the sample have a K magnitude between 10 and 12 and an angular diameter between 20 and 40$\mu$as. If we consider $\delta$~Cep as are reference, this values translate into an effect on the peak-to-peak S-signature in the differential phase of 2.5$\mu$as and 5$\mu$as, respectively. In order to get the distance (or angular diameter variation) we would need to detect an effect of 10\%, which means 0.25$\mu$as and 0.5$\mu$as, respectively. Thus, to derive the distance of LMC Cepheids with this technic a precision on the differential phase of at least 0.2$\mu$as would be necessary. 

\begin{table}[htbp] \caption{\label{Tab.ang} A new distance indicator using specto-interferometry of Cepheids}
\begin{center}
\begin{tabular}{lccc|ccccccccc}
\hline
\hline
N$_\star$	& 40 in MW ; 40 in LMC \\
spatial resolution 	&  0.3 mas \\
spectral resolution 	&  $R=60000$ \\			
temporal resolution	&  periods from 2 to 50 days \\
precision 	&	1$\mu$as for MW; 0.2$\mu$as for LMC \\	
imaging	& 3T is fine, but 6T better \\
\hline
\end{tabular}
\end{center}
\end{table}

%\subsection{Summary for Cepheids}\label{sscep_sum}
%A better understanding of the projection factor (Sect.~\ref{sscep_bin}, \ref{sscep_LD}, \ref{sscep_invGaia}) and the CSEs (Sect.~\ref{sscep_CSE}) will promote the BW method as an even more robust tool for the distance scale calibration. Besides, a high spectral resolution visible long baseline interferometer would be an unique tool to probe the dynamical structure of Cepheids' atmosphere and environment. And as explained in Sect. \ref{sscep_HRA} it could also allow to apply a novel method of distance determination. 

\section{Eclipsing binaries}

\subsection{The SBCR of early and late-type stars: distances in the local group}

The distance to the LMC, which provides the fiducial Cepheid PL relation constitutes the largest source of uncertainty in the whole process of constructing distance scale ladder \citet{freedman01}. 
%According to \citet{benedict02} and \citet{schaefer08}, current distance moduli for the LMC are in the
%range from 18.2 to 18.8 mag, which implies an almost incredible 30\% uncertainty on this fundamental number. Selecting only the modern and most reliable results, one still has uncertainty of about 10-15\%, which shows that most of the techniques used to measure the LMC distance and therefore the zero point of the whole distance scale calibrations are plagued by very substantial systematic errors. 
The Araucaria team led by \citet{gp13}  have detected a dozen very long-period (60-772d) eclipsing binary systems composed of intermediate-mass, late-type giants located in a quiet evolutionary phase on the helium-burning loop. By observing spectroscopically eight of these systems intensively over the past 8 yr, they could accurately measure the linear sizes of their components, while the angular sizes have been derived from the surface-brightness color-relation (SBCR). The LMC distance that we derive from these systems is accurate to 2.2\% and provides a firm base for a 3\% determination of the Hubble constant. However, the systematic uncertainty in the distance measurement comes mainly from the calibration of the surface-brightness color-relation. The root mean squared scatter in the current SBCR is 0.03 mag \citet{dibenedetto05}, which translates to an accuracy of 2\% in the respective angular diameters of the component stars. 

Some efforts are currently undertaken to achieve the 1\% precision on the relation. In this context, future interferometry combined with photometric campaigns could play a central role. The objective is not only to achieve 1\% but also to reach the intrinsic dispersion of the relation, which basically means to quantify the impact of stellar activity on the surface-brightness of the star. Late-type stars are not perfect blackbody, the magnetism, the convection, could play a role. For early-type, it is even more complex: rotation, wind, pulsation have probably an impact of up to 10\% on the SBCR relation \citep{challouf14}. Late-type stars are faint but extremely useful to derive the distance to LMC and SMC even if they are indeed difficult to detect. Conversely, early-type stars in eclipsing binaries could play a major role. These very bright systems are indeed very easily detected in the LMC and even in M31 and also in many other galaxies in the Local Group. Pushing the precision on the SBCR of early-type stars down to even 5\% would be of great interest.

\begin{table}[htbp] \caption{\label{Tab.ang} The SBCR of late-type stars: distance to LMC}
\begin{center}
\begin{tabular}{lccc|ccccccccc}
\hline
\hline
N$_\star$	&  $\simeq$ 200 \\
spatial resolution 	&  0.3 mas \\
spectral resolution 	&  $R=3000$ \\			
temporal resolution	&  no particular constrain \\
precision 	&	1\% or better on $\theta$ \\	
imaging	& 3T is fine, but 6T better \\
\hline
\end{tabular}
\end{center}
\end{table}

%\subsection{The SBCR of early-type stars: distance to M31, M33}

%Our aim is to open a new path to the Hubble constant by using new anchors in the calibration of the cosmic distance scale, which are the M31 and M33 galaxies. M31 and M33 are the nearest and most suitable Local Group galaxies for calibrating the extragalactic distance scale. However, they present a much greater observational challenge than LMC. Distances are now known to no better than 10\%-15\%, as there are discrepancies of 0.2-0.3 mag between various distance indicators (e.g., \citet{benedict02}; Fig. 8). For such purpose, we propose to develop a distance indicator based on early-type eclipsing binaries. 
%With a new generation instrument in visible, we expect to reach a 3\% precision and accuracy. 

%\begin{table}[htbp] \caption{\label{Tab.ang} The SBCR of early-type stars: distance to M31, M33}
%\begin{center}
%\begin{tabular}{lccc|ccccccccc}
%\hline
%\hline
%N$_\star$	& 40 in MW ; 40 in LMC \\
%spatial resolution 	&  0.3 mas \\
%spectral resolution 	&  $R=60000$ \\			
%temporal resolution	&  periods from 2 to 50 days \\
%precision 	&	1$\mu$as for MW; 0.2$\mu$as for LMC \\	
%imaging	& 3T is fine, but 6T better \\
%\hline
%\end{tabular}
%\end{center}
%\end{table}

\subsection{The distance of Galactic eclipsing binaries: a cross-check for the method}

With new generation visible long-baseline interferometry it could be possible to measure of orbital separation of a large sample of Galactic eclipsing binaries, about 50 systems (on-going project actually). Then, we could compare several types of distances:
\begin{itemize}
\item the distance derived from the interferometric measurement of the separation (combined with spectroscopy)
\item the distance derived from the classical approach: photometry, spectroscopy are used to derive the individual radii and are combined with the angular diameter obtained from SBCR relation
\item if the individual angular diameter of the system are resolved by interferometry, once combined with individual radii (classical approach), it provides another estimate of the distance
\item distance from trigonometric parallaxes (Gaia)
\end{itemize}  
These different approaches will greatly help in resolving some possible systematics in the method. 

\begin{table}[htbp] \caption{\label{Tab.ang} The distance of Galactic EBs: a cross-check for the method}
\begin{center}
\begin{tabular}{lccc|ccccccccc}
\hline
\hline
N$_\star$	& 50 eclipsing binary systems \\
spatial resolution 	&  0.3 mas \\
spectral resolution 	&  $R=3000$ \\			
temporal resolution	&  to be defined\\
precision 	&	1\% on the visibilities \\	
imaging	& 6T better \\
\hline
\end{tabular}
\end{center}
\end{table}

\section{Conclusion}

For more than one decade now, visible interferometry has played (and continue to play) a central for the distance scale calibration. 
We have shown that a next generation interferometer in the visible - with high precision measurements, 6 telescopes with baselines of 300m, but also high spectral resolution - would be extremely important to reach the 1\% precision and accuracy on the distances in our Galaxy and beyond, but also to understand the physics of Cepheids and eclipsing binaries. Such improvements are crucial to reduce the precision and systematics on the Hubble constant.

\chapter[Massive Stars]{Massive Stars\label{ch4}}

 \author{Ph. Stee$^{1}$,
A.~Meilland$^{1}$,
F.~Millour$^{1}$,
M.~Borges Fernandes$^{2}$,
A.~Carciofi$^{3}$,
W-J~de~Wit$^{4}$,
A.~Domiciano de Souza Jr$^{1}$,
D.~Faes$^{5,1}$,
N.~Kostogryz$^{6}$,
 N.~Nardetto$^{1}$,
 J.~Zorec$^{7}$
 }

\begin{center} 
$^{1}$~Laboratoire Lagrange, UMR 7293, CNRS, Observatoire de la C\^ote d'Azur, Universit\'e de Nice Sophia-Antipolis,
06304 Nice, France\\
$^{2}$~Observat\'orio Nacional, Rua General Jos\'e Cristino 77, 20921-400 S\~ao Cristov\~ao, Rio de Janeiro, Brazil\\
$^{3}$~Instituto de Astronomia, Geof\'isica e Ciencias Atmosf\'ericas, Universidade de S\~ao Paulo, Rua do Mat\~ao 1226, Cidade Universit\'aria, 05508-090, S\~ao Paulo, SP, Brazil\\
$^{4}$~ESO - European Organisation for Astronomical Research in the Southern Hemisphere, Chile\\
$^{5}$~Instituto de Astronomia, Geof\'isica e Ci\^encias Atmosf\'ericas, Universidade de S\~ao Paulo, Rua do Mat\~ao 1226, Cidade Universit\'aria, 05508-900, S\~ao Paulo, SP, Brazil\\
$^{6}$~Kiepenheuer-Institut f\"ur Sonnenphysik (KIS), Sch\"oneckstrasse 6, 79104, Freiburg, Germany\\
$^{7}$~Sorbonne Universit\'es, UPMC Universit\'e Paris 6 et CNRS, UMR7095 Institut d'Astrophysique de Paris, F-75014 Paris, France.\\
\end{center}

\section{Massive multiple stars}
The mass of the most massive stars has long been investigated (see e.g. Banerjee \& Kroupa 2012 and references therein). There have been even debates on the so-called ै supermassive starsै (e.g., Schmidt- Kaler and Feitzinger 1981; Cassinelli et al. 1981), and R136a in the 30 Doradus nebula was suggested to be a single supermassive star with a mass of 1000 to 3000 M$_{\odot}$. Speckle interferometric observations resolved R136a for the first time and demonstrated that R136a consists of eight dominant stars (a1 to a8) within 1 arcsec (Weigelt \& Baier 1985). SINFONI spectroscopy of R136a (Schnurr et al. 2009) showed that the three brightest stars a1 to a3 are WN5h stars.\\

Recently, Massey et al. (2002), Crowther et al. (2010), Schnurr et al. (2009) and others performed detailed studies of the brightest members of several young clusters, which may be among the most massive stars known (suggesting very high initial masses of up to 300 M$_{\odot}$  for the brightest, young stars in R136, Crowther et al. 2010) under the assumption that these stars are single. However, one cannot rule out yet that these stars might be binaries, because speckle or adaptive optics-assisted imaging lack the highest angular resolution necessary to resolve them into binaries. For such systems, radial velocity measurements are also extremely difficult given the line broadening due to the emerging winds. Because of the implications for massive star formation and evolution, it is then of foremost interest to establish conclusively whether they are binary or single stars (see Sana et al. 2012 and Sana et al. 2013). Indeed, binary evolution may have drastic consequences on the evolution of high-mass stars, via mass transfer or possibly mergers, hence on the progenitors of core-collapsed supernovae. Establishing the distribution of binary separations and of mass ratios is therefore essential to constrain the influence of mass loss and mass transfer on the evolution of high-mass binaries and to ascertain the origin and properties of their remnants.\\

Observing stars in distant clusters was however too challenging for the VLTI up to now, even for the most luminous ones. Indeed, the currently offered limiting magnitude of AMBER is K=7.5 mag on the UTs (under regular conditions, seeing 0.8ै€).\\

AMBER has already demonstrated spectro-astrometric capabilities. Spectro-astrometry is a technique aiming at going beyond the classical diffraction limit of a telescope, when emission lines are available in the selected targets. Spectro-astrometry has long been used for studies of binaries (see e.g. Baines et al. 2006, Wheelwright et al. 2010). By computing the photocenter shift of the point-spread function as a function of wavelength, it is possible to detect binaries whose components have different spectral types, or trace Doppler shift-induced photocenter in rotating disks of Be stars. With interferometry in the visible, this means going beyond the classically accepted resolution limit of 1 mas (in the V band), by making use of the wavelength-differential phase, which can be compared to the photocenter in the case of unresolved objects. Spectro-astrometry on single-dish telescopes has demonstrated in the past the detection of photocenter shifts up to 1000 times smaller than the seeing limit (see e.g. Oudmaijer et al. 2008).
With spectro-interferometry using AMBER, the proven photocenter-shift detection limit using differential phases is as small as 10 $\mu$arc-seconds (see e.g. Le Bouquin et al. 2009, Domiciano de Souza et al. 2012). The use of differential phases over spectral lines allows us to probe binaries well-below the technical resolution limit. For example, given the current errors on differential phases at low flux, of the order of 2 $\sigma$, this translates into a potential separation limit of 0.1 mas for a WR+O system (see an example in Fig. 4). This technique works well for 1:1 flux ratio and different spectral types of the components. Hence, it works best with WR+O stars, or emission-line stars such as LBVs or sgB[e] stars.
In the LMC, the detectable binary separation threshold would shrink from 100 AU to 5 AU. These improved limits, made available by an intermediate spectral resolution, will provide a significant overlap with radial velocity (RV) binary detections. Hence, a comparison of our data with the available RV measurements will enable a direct assessment of the biases from RV programs on the massive binary populations.

\begin{figure}
\center{\includegraphics[width=0.4\textwidth, angle =270]{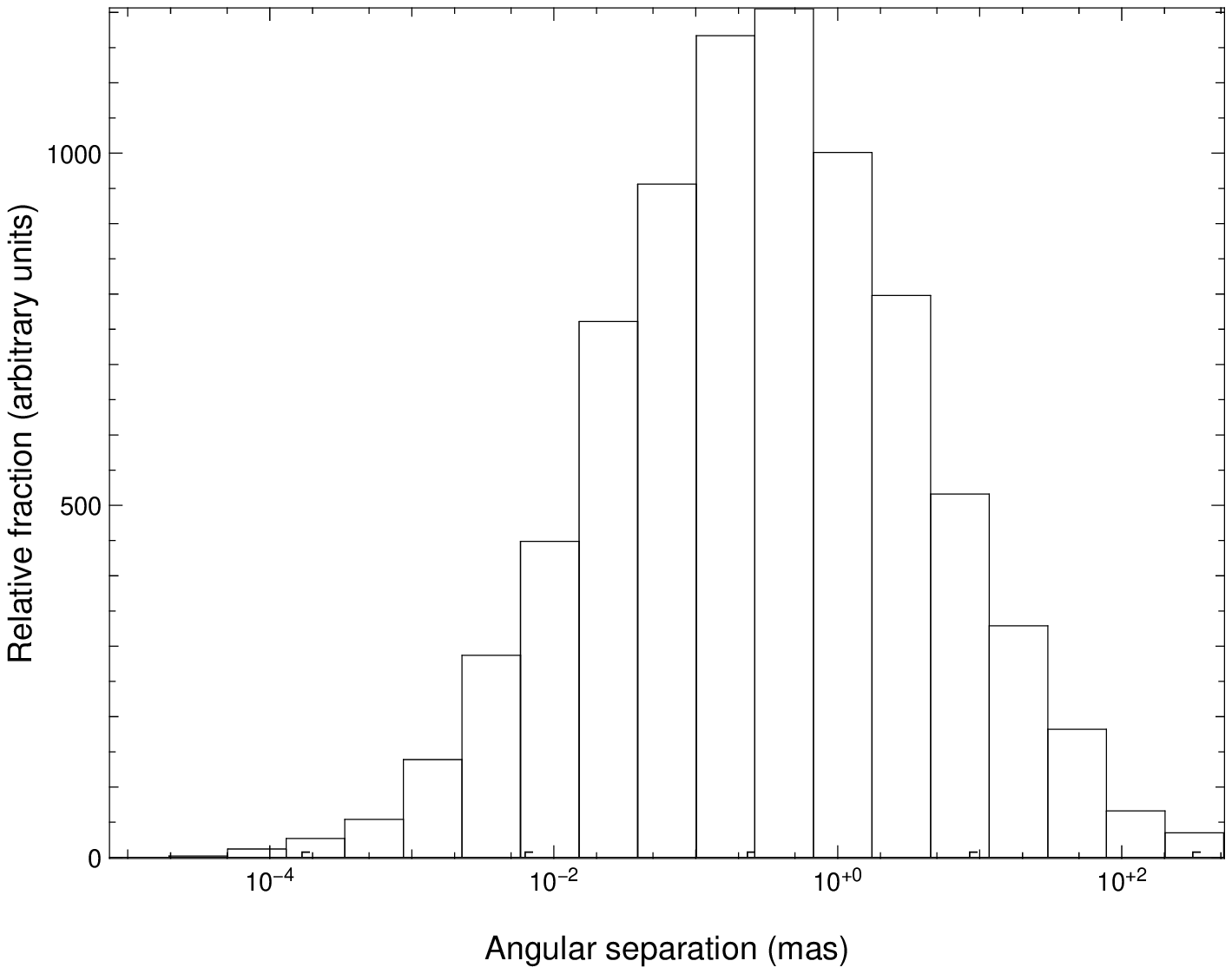}}
\caption{\label{bina_mass}Simulation of a purely theoretic population of 100 massive binary stars, with the semimajor axis having the same distribution as in the Bates et al. 2009 paper (see their figure 18). We used random orbital parameters and masses ranging from 60 to 90 solar masses. The  figure shows the resulting one-epoch binary separation for a 49\,kpc-placed cluster. The maximum of the distribution is perfectly suited for interferometric studies in the visible.}
\end{figure}

\section{Physics of active massive stars}

The nature of the circumstellar environment (CE) of active hot stars is central to understand many issues related to these objects. As a natural tracer of mass-loss, the study of this material is crucial to improve massive star evolution models. Moreover, the effect of rotation on the non-spherical distribution of mass-loss for these luminous objects is an active subject of debate. This is especially for classical Be stars which are known to be the fastest-rotating non-degenerated stars. To make progress in the understanding of these objects and the role of rotation in mass-loss processes, one needs to accurately determine the structure of their CE. By providing simultaneously high spatial and spectral resolution, interferometry in the visible is particularly well suited for the study of these objects.\\

\subsection{Stellar rotation among Be stars}
Stellar rotation is a key in our understanding of both mass-loss and evolution of massive stars (Maeder \& Meynet 2003). The fast rotation, often accompanied by anisotropic winds seems to be a major physical effect to lose angular momentum. It may also leads to a situation where the star reaches its ै€œcriticalै€ velocity and the matter is then no more bound to the star and can easily be ejected in the CE. Moreover, it is now well established that at low metallicity, stellar winds are less efficient and thus cannot expel enough angular momentum for the fastest rotators which implies that the first generation of massive stars may be all critical rotators. Classical Be stars are known to be the fastest class of non-degenerated stars, even if their rotational rate is still highly debated: Fr\'emat et al. (2005) found  $\Omega/\Omega_c$ $\sim$0.88, Zorec et al. (2013) 0.92$<$ $\Omega/\Omega_c$  $<$0.95, and Cranmer (2005) found clues of a dependence of the Be stars rotational rate with their rotational rate. Consequently, it is not clear if Be stars can be considered as a homogeneous group of stars in term of mass-ejection processes (Stee \& Meilland 2009).\\

Spectro-interferometry is the most suitable technique to test the various hypotheses on the mass-ejection processes by constraining both the CE geometry and kinematics. For example, the first VLTI/AMBER observations of Be stars published in Meilland et al. (2007) or Carciofi et al. (2009) have provided evidence that the disks surrounding these objects are fully dominated by rotation, finally ruling out the hypothesis of slowly expanding wind due to the bi-stability mechanism introduced by Lamers \& Pauldrach (1991).\\

After these early successes we have also initiated a first spectro-interferometric survey on the brightest Be stars with the VLTI/AMBER completed by VLTI/MIDI (mid-IR) and CHARA/VEGA (visible) observations. The observations published in (Meilland et al. 2008, 2009, 2011, 2012, 2013 ; Delaa et al. 2011 ; Stee et al. 2012) allow us to start a statistical study of the CE geometry (in R, V, and N bands) and the kinematics (using H$\alpha$, Br$\gamma$, and HeI 2.05 $\mu$m lines). We have found that all the CE were well modeled by geometrically thin disk in Kelperian rotation. We also found evidence of the N-band emission compactness (see Fig.1) that may be linked either to disk truncation (Chesneau et al. 2005) or isothermal disk (Jones et al. 2004 ; 2009). Moreover, we strongly constrain the disk inclination angle and thus the stellar rotation itself. We found a rotation rate of $\Omega/\Omega_c$= 0.95 $\pm$ 0.02 which is slightly higher than the previous estimate by Fr\'emat et al.(2005) but compatible with Zorec et al. (2012) estimation. Consequently, Be stars seem to rotate very close to their critical velocity. However despite these successes in modeling the Be disc kinematics, its physical structure i.e. radial and vertical density distribution, ionization, and temperature, and their dependency on the stellar parameters remains poorly known. Thus, in order to put strong constants on these physical parameters we need at least a spectral resolution in the visible of 20000
which corresponds to a spectral resolution of about 0.3 $\AA$, i.e. 15 km s$^{-1}$ to have enough spectral channels to accurately study the envelope kinematics. Regarding the spatial resolution, since most of the stellar photospheres of Be stars are smaller than 0.5 mas, we need at least baselines of $\sim$ 300m, as already provided by the CHARA array,  but if we want to "see" some photospheric structures baselines of $\sim$ 1000m are needed. 

\subsection{Measuring differential rotation on the surface of Bn stars}
As already outlined in the previous section and even though $\frac{\Omega}{\Omega_{c}}$  can be very high it still underestimates the actual stellar rotation, which is more clearly evidenced by the $\eta$ parameter. $\eta$ is defined as the ratio of the centrifugal and the gravitational force at the stellar equator. Thus, the measurement of $\eta$ is crucial if we would like to know how close are Be stars to the critical rotation. Unfortunately, Be stars can have photospheric spectral lines marred by emission/absorptions due to their circumstellar disc. Bn stars, which are nearly as fast rotators as Be stars, don't have spectra perturbed by circumstellar matter, so that the study of their apparent geometry can be carried out more properly and reliably.\\

\begin{figure}
\center{\includegraphics[width=0.5\textwidth]{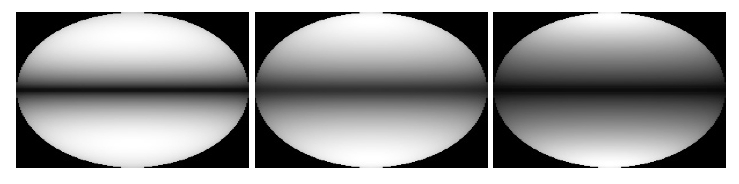}}
\caption{\label{delaa}Brightness distribution on the stellar disk in the continuum at $\lambda$ = 4437 $\AA$, near the He I 4471 $\AA$ line, calculated for a star rotating at $\eta$ = 0.9 and for different values of $\alpha$, where $\alpha$ is the constant that parameterizes the degree of differential rotation (from left to right): -1.0, 0.0 and 1.5, seen at an inclination angle of 90$^o$. From \citet{delaa2013}.}
\end{figure}

Moreover, due to the rapid rotation, the surface geometry of Bn and Be stars is highly deformed. The surface angular velocity can be dependent on the stellar latitude. Then,  not  only  the  centrifugal  force  acting  on  the  stellar  surface,  but also the effective temperature distribution (gravity darkening effect) should depend on the particular surface rotation law. Up to now, these effects have been measured only for few stars (\citet{domiciano2003}, \citet{monnier2007}, \citet{zhao2009}, \citet{che2011}) and different values of the $\beta$ exponent of the gravity darkening law were deduced using image reconstruction technique (\citet{monnier2007}, \citet{zhao2009}, \citet{che2011}).  \citet{delaa2013} has demonstrated that a differential rotation at the surface of fast rotating star may affect the brightness distribution of the stellar disc and so, also modify the value of the $\beta$ exponent. 

In all these measurements it was assumed that a genuine constant gravity-darkening $\beta$ exponent exists. Actually, $\beta$ is a function of the stellar co-latitude
and on the particular differential rotation law. Using nothing but the simplest Maunder's approximation to sketch the stellar surface differential rotation, $\Omega(\theta)=$ $\Omega_e(1+\alpha\cos^2\theta)$, it can be shown that $\beta=\beta(\alpha,\theta)$, which consequently forces that every empirical determined $\beta$ exponent is dependent also on the inclination-angle i of the stellar rotation axis.
This implies, that a value of $\beta$ different from 0.25 (for a fully a radiative an barotropic envelope; \citet{vonzeipel1924}) can also be due to a non-conservative rotational law in the outermost stellar layers (Zorec et al., 2011) mixed with aspect angle effects. Such $\beta$ values must be constrained by spectrally resolved interferometric measurements, because nonetheless they carry precious information on the surface rotation law that may be unscrambled using differential interferometry through the  spectral line differential phases and spectroscopic models of stellar atmospheres with differential rotation. This approach is useful, since the latitude dependent angular velocity introduces  peculiar  characteristics  to  the  spectral  lines and spectral line differential  phases, which are sensitive to the stellar inclination angle and to the surface angular velocity law \citep{delaa2013}. 

\begin{figure}[]
\centerline{\includegraphics{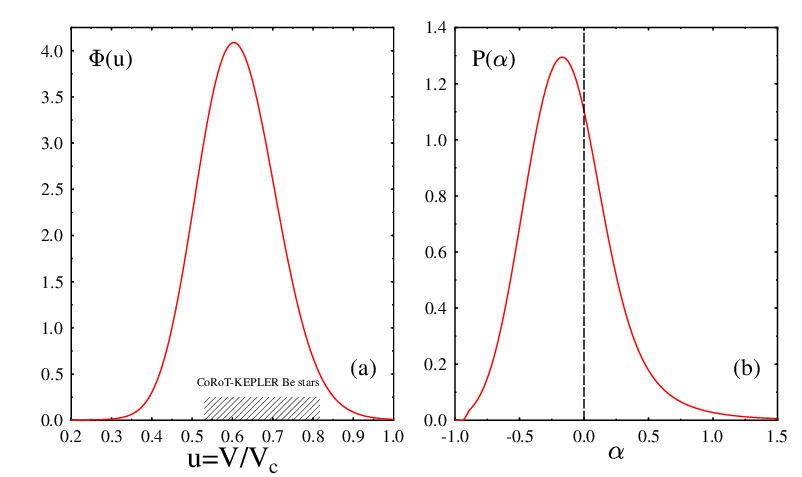}} 
\caption{\label{alph} (a) Distribution $\Phi(u)$ of the equatorial velocity ratios $u=V_{\rm e}/V_{\rm c}$ of classical Be stars corrected for micro-turbulence, presence of binary components, and having Maunder's
like surface differential rotation. The shaded box indicate the $V/V_{\rm c}-$ interval of up to now studied CoRoT and KEPLER Be stars; (b) Occurrence probability function $P(\alpha)$ of the differential rotation parameter $\alpha$.}
\end{figure}

  Statistical studies based on the distribution of true velocity ratios $V/V_{\rm c}$ of classical Be stars, corrected for induced effects on the measured values of $V\!\sin i$ due to the gravity darkening effect, macro-turbulence and presence of close binary components, produce the probability distribution of the differential rotation parameter $\alpha$ in Maunder's expression for $\Omega(\theta)$ shown in Fig.~\ref{alph}, which suggests that $-0.5 < \alpha < +0.3$ is the probable interval of values of $\alpha$ characterizing the surface rotation of Be stars and possibly in other fast rotators, like Bn stars.

 In  order  to  constrain  this  differential  rotation  empirically we  definitely need an interferometric instrument in the visible since it the only way to combine high spatial resolution (for our purpose we need to reach $\sim$ 0.2 mas) and spectral resolution as high as 100000 in order to reach 0.1 \AA\ in spectral lines sensitive to this effect as HeI lines.  With a spectral resolution of 20000 it is still already possible to access to the differential rotation if we are able to obtain measurements with a radial velocity precision of $15 < \Delta V < 30$  km~s$^{-1}$ and a phase shit precision of $2^o < \Delta\phi < 5^o$ as described in \citep{delaa2013}.

\subsection{Physics of B[e] stars}
Stars with the B[e] phenomenon are B-type stars that present in the optical spectrum, strong Balmer lines, and also permitted and forbidden low-excitation lines of singly ionized or neutral metals, e.g. Fe\,{\sc ii} and O\,{\sc i}, all of them in emission. In addition, there is also an excess in the near- and mid-IR, due to circumstellar dust (\citet{Conti}). 

This group of stars is actually very heterogenous, with objects in different evolutionary stages, such as low and high mass evolved stars (supergiants and compact planetary nebula with the B[e] phenomenon), intermediate-mass pre-main sequence objects (Herbig Ae/B[e] stars) and symbiotic stars (\citet{Lamers}). However, most of the stars with the B[e] phenomenon in the Milky Way is still unclassified with respect to their nature, mainly due to the absence of realiable stellar parameters and especially distances, being called as unclassified B[e] stars. These objects are usually embedded in dense circumstellar material. Consequently, their optical spectra are strongly polluted with circumstellar emission, while photospheric lines are rare or absent, making difficult a proper stellar classification.

One of the main questions related to these objects is how such different stars may have similar spectral characteristics. The most probable answer seems to be linked to the presence of circumstellar disks, as proposed originally by \cite{Zickgraf} for B[e] supergiants. Due to the structure and kinematics of these disks, they would be the ideal environment not only for the origin of forbidden lines and dust emission, and consequently the B[e] phenomenon, but also for molecular emission. Recent papers that studied the CO emission in these objects revealed that dense material, in a disk or ring, is in Keplerian rotation, rather than in expansion (\citet{Oksala, Kraus2013,  Muratore}). Of course, the mechanisms for the formation of these disks may be different for each class of stars with the B[e] phenomenon, and there are still many open questions related to their exact properties (shape, density, composition, temperature, etc.).

\citet{Miroshnichenko} suggested that most of the unclassified B[e] stars would actually be non-supergiants objects in binary systems, and a new class of objects was proposed, the FS CMa stars (name of its template object). However, the presence of a companion could not be confirmed for several of these objects yet, and this suggestion deserves future investigation. In addition, based on recent papers (\citet{Millour2009, Millour2011, Kraus2012a, Kraus2012b}), we cannot discard a possible link between binarity and the B[e] phenomenon for B[e] supergiants.

Based on this, long baseline interferometry, associated to the state-of-art 3D radiative transfer codes, like HDUST\footnote{HDUST is a three-dimensional Monte Carlo radiative transfer code that combines the full non-local thermodynamic equilibrium (NLTE) treatment of the radiative transfer in gaseous media, with a very general treatment for circumstellar dust grains (\citet{Carciofi}).}, is definitely a powerful tool to search for both companions and disks around stars with the B[e] phenomenon and describe their physical characteristics.\\

An interesting interferometric study of a B[e] star has been done on HD 62623 (3 Puppis, HR 2996) which is the brightest star showing the B[e] phenomenon. Its distance was estimated to be 700 $\pm$ 150 pc and its radius R$_{\star}$ = 55 $\pm$10 R$_{\odot}$ (Bittar et al. 2001). It exhibits radial velocity variations with a period of 137.7 days which where interpreted as the presence of an unseen companion (Lambert 1988).
MIDI observations (Meilland et et al. 2010) showed that the dust around HD 62623 is concentrated in a disc-like structure which extends to a hundred mas in the N band. Using radiative transfer modeling combining dust (computed using MC3D code by Wolf 2003) and gas (SIMECA, Stee et al. 1994) the authors constrained the disc structure, i.e. mass, dust inner radius, stratification, and gas contribution in the N-band (See Fig 2). They also constrained the object inclination angle and rotational velocity and showed that the star was rotating too fast according to its evolutionary stage but too slow to explain the break of symmetry of its environment. HD 62623 was then observed during four full nights using AMBER in HR mode centered on the Br$\gamma$ emission line (Millour et al. 2011). These observations were used to reconstructed narrow-spectral-channel images through Br$\gamma$. The reconstructed images not only confirmed the model deduced from the MIDI observations, but also allowed to significantly constrain the gas geometry and kinematics, showing that it was concentrated in a Keplerian rotating disc. Using additional spectroscopic observations from Plets et al. (1995), the authors concluded that the gas and dust were part of a same at least partly stratified Keplerian disc, similar in many aspects to the one found around young stars. Nevertheless, as for most B[e] stars, the origin and formation mechanism of HD62623 disc remains unsettled, even if the most credible hypotheses imply effects from the putative companion. Consequently, to progress in the understanding of binary and rotation effects on evolution of massive stars we propose to image B[e] stars in the visible which will require at least three quadruplets, i.e. 18 visibilities measurements with phase closure to provide a decent (u,v) plan coverage. Since in the visible we are probing the inner part of the gaseous disk, the spatial as well as spectral requirements are globally similar to the ones for Be stars, i.e. respectively 0.1 mas and 20000 for the spatial and spectral resolution. 

However, a strong limitation is, in general, the low sensitivity of the interferometers, making possible the observation of only a few bright objects. For example, considering the list of stars with the B[e] phenomenon (\citet{Lamers}) and instruments, like AMBER and MIDI, we have the following:

\begin{itemize}
 
\item VLTI/AMBER: only 8 objects can be observed by ATs and 24 by UTs

\item VLTI/MIDI: only 12 objects can be observed by ATs (those ones with an emission at 12 $\mu m$ higher than 20 Jy) 

\end{itemize}

Despite this limitation, different papers, based mainly on interferometric data, have confirmed the presence of gaseous and dusty circumstellar disks for eight objects: HD\,50138, see Fig.\,1 (\citet{Borges, Ellerbroek}), CPD-57$^o$2874 (\citet{Domiciano de Souza}), MWC\,300 (\citet{Wang}), CPD-52$^o$9243 (\citet{Cidale}), HD\,62623 (\citet{Meilland}), HD\,85567 (\citet{Wheelwright2013, Vural}), HD\,45677 (Kanaan et al., in preparation), and GG Car (Seriacopi et al., in preparation). In addition, for three B[e] supergiants candidates the presence of a companion was discovered or confirmed by interferometry: HD\,327083 (\citet{Wheelwright2012a, Wheelwright2012b}), CD-4211721 (\citet{Kraus2012a, Kraus2012b}), and HD\,62623 (\citet{Millour2011}).   

\begin{figure}[!tbh]
\centering   
\includegraphics[width=0.5\textwidth]{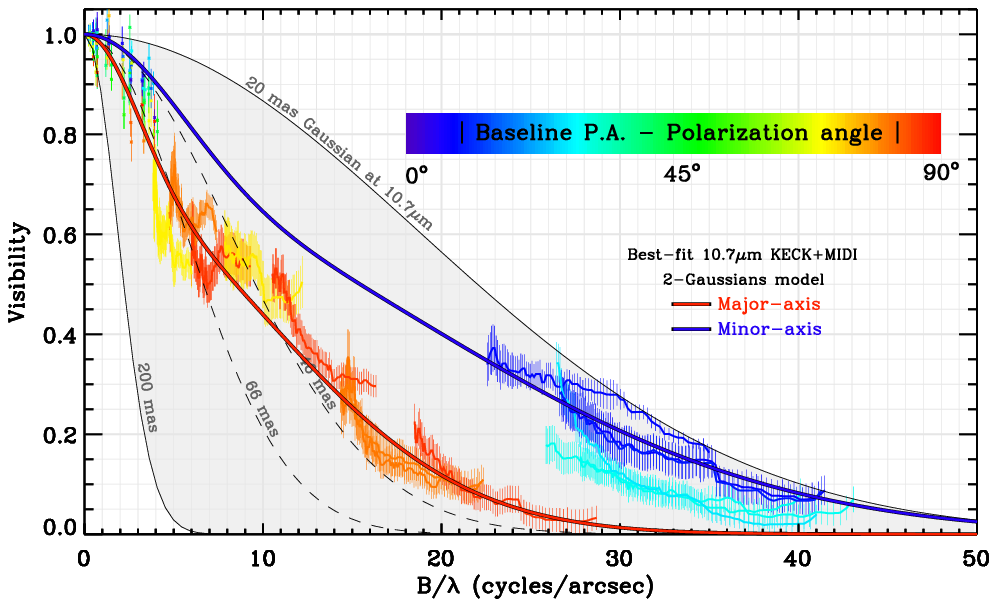}
\caption{VLTI/MIDI N-band spectrally resolved visibilities plotted as a function of the spatial frequency (B/$\lambda$). Colors indicate the baselines position angle with respect to the polarization measurement. The color scale (i.e., blue, cyan, green, yellow, orange, and red) goes from blue for baselines parallel to the polarization angle to red for the perpendicular ones. The dots with error bars indicate the 10.7$\mu$m Keck measurements. The thin solid and dashed lines represent the visibilities from Gaussian disks with different FWHM. The thick red solid line represents the model fitting aligned to the PA of the major axes of the 2-Gaussian distribution and the blue one to the PA of the minor axes, confirming the presence of a dusty disk around HD\,50138 (extracted from \cite{Borges}). }
\label{spfvis}
\end{figure}

All the previous cited works are in the IR wavelengths, but the importance of interferometry in the optical range is also remarkable. We would like to highlight the work made by VEGA, which is attached to CHARA interferometer. It is an ideal instrument to measure wind sizes and compare them with the optical continuum forming regions.

However, for the study of stars with the B[e] phenomenon, only two objects brighter than $M_V$ = 8 can be observed by VEGA. For one of them, the young massive spectroscopic binary MWC\,361, \citet{Benisty} could spatially resolve the H$\alpha$ line emission region that can be in a disk-wind, which is enhanced by the periastron passage of the companion, due to gravitational perturbations.

It is clear that new similar instruments, especially at the southern hemisphere, are definitely  necessary for the study of a larger sample of B[e] stars. This will provide us:

\begin{itemize}
\item good quality imaging of the circumstellar environment and wind structures that can be modelled by modern 3D codes;

\item spatial and spectral resolution of the close circumstellar gas environment, where the B[e] phenomenon is formed and the identification of companions. In addition, interferometric measurements of different spectral lines in the visible (H, He, etc.) will allow us to better obtain velocity fields and temperature distribution in the disks;

\item temporal resolution to confirm the effects of short and long-time variations in the close atmosphere and wind structure caused by binary interactions or even by pulsations, as seen for HD\,50138 (\citet{Borges}).

\end{itemize}

Long baseline interferometry, associated to the state-of-art 3D radiative transfer codes, is the ideal tool to provide information concerning the geometry and inclination of the circumstellar matter of stars with the B[e] phenomenon. Up to now several papers are based on IR data. However, interferometry in the visible can also be a very important source of information. It is clear that only a multi-wavelength study may, in a near future, allow us:

\begin{itemize}
\item (i) to comprehend the nature of some unclassified B[e] stars;

\item (ii) to include the B[e] phenomenon in the evolutionary tracks;

\item (iii) to understand phases of the stellar evolution that are not well understood yet.

\end{itemize}

\begin{figure}
\center{\includegraphics[width=0.4\textwidth]{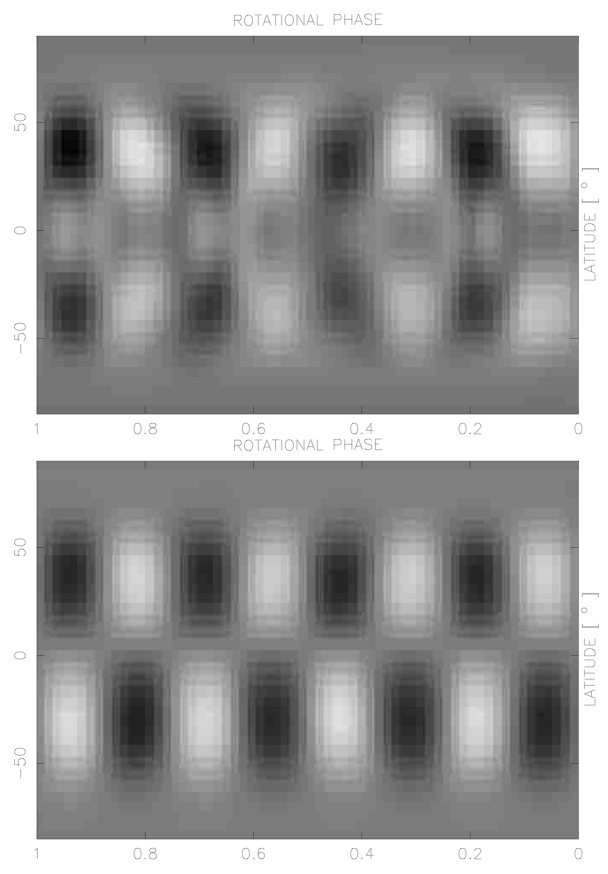}}
\caption{\label{jankov}Top: Maximum Entropy reconstruction of a Mercator projection of the visible surface from normalized flux spectra from stellar surface brightness perturbation due to the non-radial pulsation m=4, l=5 mode on a star tilted at i=85$^o$. One notices a characteristic mirroring due to the north-south ambiguity. Bottom: Maximum Entropy reconstruction from photocenter shifts parallel to rotation. The north-south ambiguity is removed. From \citet{jankov2001}.}
\end{figure}

\subsection{Direct detection of Non Radial Pulsations of massive stars.}
The study of the surface of massive rapidly rotating stars with strong non-radial oscillations using Differential Speckle Interferometry (DSI) technique was already proposed by \citet{vakili1991}. The application of this technique to the study of surface structure is particularly well justified for the stars at high inclinations  \citep{petrov1988}. \citet{jankov2001} treated explicitly the case of non-radial stellar pulsations, for which the cancellation of opposite sign temperature or velocity fields introduces difficulties, and showed that interferometric constraint introduces the crucial improvement. In fact, the photocenter shift provides the first order moment of the spatial brightness distribution and (comparing to the zero order moment spectroscopic information) the corresponding stellar regions are reinforced by weighting with the coordinate parallel or orthogonal to rotation. Consequently, the modes that are cancelled in flux spectrum should appear in the spectrally resolved photocenter shift data. Of course, the correct detection and identification of modes present in the star is crucial for a credible asteroseismological analysis.

Following this technique, \citet{jankov2001}  propose to detect non-radial pulsations by differential interferometry using the dynamic spectra of photocenter shift variability characterized by bumps traveling from blue to red within the spectral lines. The theoretical estimation of expected signal-to-noise ratios in differential speckle interferometry \citep{chelli1989} demonstrated the practical applicability of the technique to a wide number of sources. For instance, the high precision CHARA/VEGA or VLTI/AMBER measurements of differential fringe phase corresponding to $\sim$ 10 $\mu$as photocenter shift (standard limit), makes measurements feasible for the low order pulsation of 20 kms$^{-1}$ and considering a star at an intermediate inclination (i = 45$^o$), which should yield a signal of about $\sim$ 25 $\mu$as for the typical angular diameter of $\eta$ Cen ($\sim$ 0.5 mas). Moreover, for $\eta$ Cen $i  \sim$ 70$^o$ and, since the star is tilted at high inclinations, the expected signals, calculated by \citet{jankov2001}, should be even stronger, particularly for low order pulsations, making the feasibility better. For that purpose,visible interferometric observations are mandatory since this technics requires baselines late enough to (partly) resolve the stellar surface (on the order of $\sim$ 500m) and spectral resolution of at least 20000 in order to have sufficient spectral channels across photospheric lines. A good (u,v) plane coverage is mendatory.\\

Up to now, interferometric studies of active hot and massive stars have produced the largest contributions of referred papers using the VLTI/AMBER and CHARA/VEGA instruments. This is certainly due to the fact that they are bright (many sources have m$_{v}$ $<$ 7) with diameters well suited for current interferometric baselines, i.e. central stars with diameters $\sim$ 0.5 mas and circumstellar disk of a few mas. These properties will certainly also benefit from a future visible interferometric instrument.

\chapter[Evolved stars, Planetary Nebulae, $\delta$~Scuti, and RR~Lyrae  as seen by a visible interferometer]{Evolved stars, Planetary Nebulae, $\delta$~Scuti, and RR~Lyrae  as seen by a visible interferometer\label{ch5}}
  
   \author{C. Paladini$^{1}$,
          A. Chiavassa$^{2}$,
         K. Ohnaka$^{3}$,
               N. Fabas$^{4, 5}$}
        \author{ M. Hillen$^{6}$,
          N. Nardetto$^{2}$,
      	  van Winckel H.$^{6}$,  
        Wittkowski M.$^{7}$,    
        M. Ireland$^{8}$,
        J.D. Monnier$^{9}$
        }

        \begin{center} 
$^{1}$~IInstitut of Astronomy and Astrophysics, Universit\'e Libre de Bruxelles, CP226, Boulevard du Triomphe, B-1050 Brussels, Belgium\\
$^{2}$~Laboratoire Lagrange, UMR7293, Universit\'e de Nice Sophia-Antipolis, CNRS, Observatoire de la C\^ote d'Azur, Nice, France\\
$^{3}$~Max-Planck-Institut f\"ur Radioastronomie, Auf dem HŸgel 69, 53121 Bonn, Germany\\
$^{4}$~Instituto de Astrof\'isica de Canarias, E-38205 La Laguna, S/C de Tenerife, Spain\\
$^{5}$~Departamento de Astrof\'isica, Facultad de F\'isica, Universidad de la Laguna, Tenerife, Spain\\
$^{6}$~ Instituut voor Sterrenkunde, University of Leuven, Celestijnenlaan 200D, B-3001 Leuven, Belgium\\
$^{7}$~European Southern Observatory, Karl-Schwarzschild Str. 2, 85748 Garching bei Muenchen, Germany\\
$^{8}$~Research School of Astronomy and Astrophysics, Australian National University, Canberra, Australian Capital Territory 2611, Australia\\
$^{9}$~Department of Astronomy, University of Michigan, 941 Dennison Building, Ann Arbor, Michigan 48109, USA.\\
\end{center}

\section{Asymptotic Giant Branch Stars}
The AGB is the late evolutionary stage of low to intermediate mass stars.
During this phase the atmosphere of the star is dramatically affected by several dynamical processes such as pulsation, dust formation,
and stellar wind. The latter is a crucial process for galactic and extra-galactic astrophysics: through the mass-loss process all the material that was synthesized during
the life-time of the star via nuclear processes is returned to the interstellar medium. 

There have been few attempt to study AGBs in the visible with interferometry.
The observations were mainly limited to measure stellar parameters, determine the diameter variation over a pulsation period, 
and/or across a few spectral wavelengths ranges \citep{koechlin1985, quirrenbach1992, quirrenbach1993, quirrenbach1994}.

%\cite{quirrenbach1994} measured the diameters of three carbon stars using the MarkIII interferometer
%\footnote{the MarkIII Optical Interferometer was located on Mt. Wilson near Los Angeles, CA.
%The instrument was operated by the Remote Sensing Division of the Naval Research Laboratiry, and it was decommissioned in 1992.}. The measurements were carried at 0.712, 0.754, and 0%.8~$\mu$m. The increase of diameter from one wavelength to the other was smaller than the one measured for O-rich AGB stars.
%This is due to the different molecular opacities that characterize the oxygen and carbon-rich chemistry (TiO vs. C$_2$ in the specific). 
%Temporal variability was also observed and stellar parameters derived.

The visible spectrum of AGBs is populated by several molecular and spectral bands.
According to the chemistry, an oxygen-rich star will exhibit TiO (already mentioned in Sect.~\ref{tio}),
H$_2$O, VO; an AGB star with $0.5\geq$~C/O~$\leq 1$ (also called S-type star) will show ZrO and LaO bands beside the TiO.
Carbon rich stars will have CN, C$_2$, and other carbon-rich molecules (easily observable at $\lambda\geq6600$, $\lambda\sim7000 - 7200$ and beyond $\lambda 7900~\AA$).
Despite the strong molecular blending, many atomic lines can be measured. Beside the classic CaII and H$\alpha$, other lines
of interest are lithium (four absorption lines 4603, 6104, 6708, and $8126~\AA$), technetium (quite weak and detectable towards the blue 
4238.19, 4262.27 and 4297.06$~\AA$), rubidium (7800.3~$\AA$), and more in general all the s-process lines.

This plethora of lines is used for abundances measurements, but how do the pulsation, non-LTE effects, inhomogeneities, 
and shock-wave propagation affect these measurements? A visible interfererometer will help to address part of these questions
in the following ways: (i) monitoring programs (three images within one year 
during selected pulsation phases on a few selected targets will constrain the time-scale of the life of the inhomogeneities; (ii) monitoring imaging through a pulsation cycle
will enable to assess the effect of pulsation on the stellar atmosphere. A few observations (~6 to 10 visibilities) 
covering different spatial frequencies for a sample of targets with various chemistry and variability type
will help constraining the molecular stratification of the available model atmospheres. 

Beside these general questions there are also some very specific science cases that would benefit from the
existence of a visible interferometer that provides: medium-to-high spectral resolution, 
imaging capabilities, and if possible a polarimetry mode.
The science cases are discussed in the following Sect.s~\ref{dust}, \ref{magneticfield}, \ref{chromosphere}, \ref{tio}, and \ref{shockfront} while a summary of the requirements
is presented in Sect.~\ref{AGBrequirements}.

\subsection{The dust distribution around AGB stars}
\label{dust}
The atmosphere of an AGB star is cool enough to allow formation of various dust species.
Dust plays a crucial role in the stellar wind, although there are some cases where the dynamics is not completely understood.
The dust-driven stellar wind in small particle approximation can successfully explain the observations of 
carbon-rich stars \citep{gautschy-loidl2004, nowotny2005, sacuto2011}, while it fails for the O-rich case.
\cite{hoefner2008} suggested that in the case of oxygen-rich atmospheres, the wind is driven by  scattering on micron-size  Fe-free silicate dust grains. 
These models were tested vs. interferometric observations of the O-rich variable RT~Vir by \cite{sacuto2013}, while 
the proof of the existence of such large grains arrived with the SAM-NACO observations presented by \cite{norris2012}.
The authors used the interferometric (aperture masking) technique to
reach the angular resolution required to observe close to the stellar surface ($\sim 10$~mas). 
This technique was combined with polarimetry that allows to distinguish between the contribution coming from the stellar surface and the 
one coming from a dust shell. Through modelling one can establish the dust grain size, composition, distribution and
importance of clumpiness.

The gain of going to the visible to study the dust distribution around AGBs is that the scattering from dust contribution becomes more important.
The challenge is the fact that dusty stars become fainter. There are around 200 objects that could be observed from Paranal (selection criteria DEC$<+10$) with
an average V-band magnitude $\sim7.45$ mag.
A few spectral channels spread over the visible should be enough to perform the observations.
One could choose the same ones available on other instruments mounted on single-dish telescopes to be able
to cover different spatial scales. VLT/EFOSC2 for example covers the H$\alpha$ filter (6577 $\AA$), a Str\"omgren
y filter (5482 $\AA$), and narrow filters (50~$\AA$) centred on the resonance lines of Na (5894$\AA$).
The size of the photosphere in the visible will be in average $<10$~mas,
but the dust component is expected to be more extended, especially for the nearby objects.
To be able to reconstruct the distribution of grain size, one will need both long and very short configurations.
Because of the complexity of the AGB environment, six telescope is the minimum number of apertures required. 
Single epoch observations need to be performed within maximum 1 week to avoid 
to be affected by the temporal variability of the star.

Adding a polarimetry mode in the visible might be a big challenge
but the science case presented for AGBs can be extended also
to other dust-enshrouded objects like K-Giant stars, post-AGBs (Sect.~\ref{postagb.sect}) 
and protoplanetary discs.

\subsection{Mapping magnetic fields and velocity fields over the surface of stars}
\label{magneticfield}
%(K. Ohnaka \& N. Fabas)
The generation mechanism of magnetic fields in cool evolved stars and their 
possible connections to the mass loss are not well understood.  
We expect the study of starspots to be the key of understanding stellar 
dynamos \citep{Berdyugina2005}.  
Interferometry combined with high spectral resolution provides the spatially 
resolved spectrum at each position over the surface of stars, as demonstrated 
by the recent velocity-resolved VLTI/AMBER imaging of the RSGs Betelgeuse and 
Antares \citep{ohnaka2011,ohnaka2013}\footnote{see also movies at\\ 
http://www3.mpifr-bonn.mpg.de/staff/kohnaka/alfori2.html
http://www3.mpifr-bonn.mpg.de/staff/kohnaka/alfsco2.html}. 
This allows us to map the velocity field, opening a new window to observe 
stars like the Sun.  
The spatially resolved spectra of regions with strong local 
magnetic fields (e.g., starspots) would show the Zeeman splitting, which 
enables us to map the magnetic fields over the surface of stars.  
The combined mapping of the velocity fields and magnetic fields would 
bring stellar astrophysics to an entirely new dimension.

\subsection{Spatially resolving the chromosphere and its dynamics}
\label{chromosphere}
%(K. Ohnaka)
The heating mechanism of the chromosphere in cool evolved stars and its 
role in the mass loss is little understood.  Recent observations suggest the 
coexistence of the hot chromospheric plasma and cool molecular gas within 
several stellar radii, forming inhomogeneous structures \citep[e.g.,][]{harper2001,harper2006}.  
\cite{berio2011} probed the chromosphere 
of K giants in the H${\alpha}$and Ca~II lines with CHARA/VEGA.  
We propose to image the chromosphere and derive its physical properties by 
interferometry of the H alpha and Ca II lines.   High resolution imaging would 
help to understand the heating mechanism of the chromosphere, its role in 
the mass loss, and why it can coexist with the molecular component.

\subsection{Probing the MOLsphere with TiO lines}
\label{tio}
%(K. Ohnaka)
The MOLsphere is quasi-static, dense, warm molecular layers extending to 
several stellar radii and exists not only in Mira stars but also in normal red 
giants and red supergiants.  Recent interferometric observations have 
succeeded in spatially resolving or imaging the MOLsphere and demonstrated 
the presence of outer atmospheres much more extended than hydrostatic 
photospheric models \citep[e.g.][and references therein]{ohnaka2013}. 
Nevertheless there are observational data challenging the presence of 
the MOLsphere \citep{ryde2014}.  
For example, the current MOLsphere models predict the 
TiO bands in the visible to be too strong compared to the observed spectra, 
suggesting that we do not yet understand the structure of the outer atmosphere 
very well.  Spatially resolving the stars in the TiO lines helps to better 
understand the structure of the outer atmosphere.  Given that TiO may serve as 
seed nuclei for dust formation, it is important to know the physical 
properties of TiO layers for understanding the dust formation.

\subsection{Geometrical characterization of shock waves in Mira stars }
\label{shockfront}
Radially pulsating AGB stars (a.k.a. Mira stars) are
characterized by strong emission in the Balmer lines linked to
the propagation of a strong hypersonic radiative shock waves.
Strong signatures in linear polarization in these lines have been
detected \citep{fabas2011} and this result points toward global
non-sphericity of the shock since the polarization of the disk-integrated flux does not cancel out.
Spectro-interferometry in the visible would help establish the shape and time evolution of
these shock waves, all of this in comparison to
the shape and evolution of the photosphere. Indeed, we think that, while it
propagates outwards, the shock front breaks down and is at the
origin of the formation of clumps in the circumstellar envelope.
The requirements for this kind of observations should allow to
resolve the lower atmosphere of a given Mira star, so 30 mas
should be a good upper value for spatial resolution and, of course, a high spectral resolution is needed ($>30,000$) in order
to study individual lines. Since the pulsation period can be as
long as a year, a time sampling of one week around and after
the maximum light would be good. One triplet of telescopes is
sufficient in order to get visibilities and closure phases, which
means imaging is not necessary. The Mira stars to be observed
typically have a magnitude $V<7$ at maximum light.

\subsection{Requirements for the science cases of AGB stars}
\label{AGBrequirements}
%Claudia
AGBs are very luminous and have extended atmosphere, therefore they become naturally the first objects to be observed with infrared interferometry.
When moving to the visible though one has to be careful. In fact many evolved AGBs will be
faint because of dust obscuration. Typical magnitudes for the most evolved Mira variables will vary between 8 and 13 mag in the V-band.
One has also to take into account the fact that Mira variable will change their brightness of up to 9 magnitude within one year.
Imaging programs will have to take this into account, and a proper number of telescopes (6+)
or a fast system to change configuration (in the current VLTI scheme: 3 configurations within 2 weeks maximum) are needed.

Well-studied cool evolved stars like Betelgeuse, omi Cet, etc, have angular 
diameters larger than 30 mas.  At 0.6~$\mu$m, they are completely resolved 
out already at 8 m.  For imaging, it is necessary to sample the first lobe, therefore baselines as short as 4 m
are needed.  
A diffraction-limited imager on a single-dish telescope may help, but 
only if the single-dish imager's spectral resolution matches with that of the 
interferometer. 
For imaging of stars with angular diameters of $10$--$20$~mas, a spatial 
resolution of $\sim1$--$2$~mas is required.  There are $\sim30$ stars with $10$--$20$~mas 
with $V < 4$--$5$.   Time interval of 1 month would be useful for probing time 
variations.   Precision: $5$--$10\%$ in visibility, $2$--$3$ deg in CP. 

A spectral resolution of a few thousands ($>10\,000$) is enough to resolve the molecular bands,
while we need a spectral resolution of $> 40\,000$ to resolve the atomic line profiles. 
To detect the Zeeman splitting (Sect.\ref{magneticfield}) the spectral resolution of at least $50\,000$ (better $10^5$) is required.

\section{Red Supergiant Stars}
Massive evolved stars with masses between roughly 10 and 25~M$_{\odot}$ spend some time as
red supergiant (RSG) stars being the largest stars in the universe. They have effective temperatures, $T_{\rm eff}$, ranging from
     3\,450 to 4\,100\,K, luminosities of 20\,000 to 300\,000\,L$_\odot$
     and radii up to 1\,500\,R$_\odot$ \citep{2005ApJ...628..973L,2010NewAR..54....1L}. Their luminosities place them among the brightest
stars, visible to very large distances. These stars exhibit variations in integrated brightness,
surface features, and the depths, shapes, and Doppler shifts of
spectral lines; as a consequence, stellar parameters and abundances are difficult to determine. 

 RSGs eject massive amounts of mass back to the interstellar medium 
\citep[ 3$\times10^{-4}$ -- 2$\times10^{-7}$ - M$_\odot$/yr][]{2010A&A...523A..18D} with an unidentified process that may be related to acoustic waves and radiation pressure on molecules \citep{2007A&A...469..671J}, or to the dissipation of Alfv\'en waves from magnetic field, recently discovered on RSGs \citep{2010A&A...516L...2A,2010MNRAS.408.2290G}, as early suggested by \cite{1984ApJ...284..238H, 1989A&A...209..198P, 1997A&A...325..709C}. The understanding of the physics of the convective envelope is crucial for these stars that contribute extensively to the dust and chemical enrichment of the galaxies. Moreover, the mass-loss significantly affects the evolution of massive stars, and it is a key to understanding the progenitors of core-collapse supernovae. The effects of convection and non-radial waves can be represented by numerical multi-dimensional time-dependent radiation hydrodynamics (RHD) simulations with realistic input physics carried out with CO$^5$BOLD code. These simulations have been used extensively employed to predict and interpret interferometric observations for evolved stars such as, e.g., for AGBs \citep{2010A&A...511A..51C} and for RSGs \citep{2010A&A...515A..12C}. 

There is a number of multiwavelength imaging examples of RSGs because of their high luminosity and large angular diameter. Several research teams \citep{1990MNRAS.245P...7B, 1992MNRAS.257..369W, 1997MNRAS.285..529T, 1997MNRAS.291..819W, 2000MNRAS.315..635Y} detected time-variable inhomogeneities on the surface of $\alpha$~Ori with WHT and COAST; \cite{2009A&A...508..923H} reported a reconstructed image in the H-band with two large spots; \cite{2009A&A...503..183O,2011A&A...529A.163O,2014A&A...568A..17O}
detected possible convective motions in CO line formation layers for $\alpha$~Ori and Antares. \cite{1997MNRAS.285..529T} reported bright spots on the surface of the supergiants $\alpha$~Her and $\alpha$~SCO using WHT, and \cite{2010A&A...511A..51C} on VX~Sgr using VLTI/AMBER.

\subsection{Requirements}\label{rsgsect}

The main objectives that we need to recover from visible interferometric images/observables are: 
\begin{itemize}
\item the size of the granulation cells from the large scale granule to the small-medium scale convection-related structures; 
\item the surface intensity contrast;
\item the velocity field and the timescale of the convective motions.
\end{itemize}

These objectives are crucial:
\begin{itemize}
\item to understand the photosphere and dynamics of RSGs where about half of the atomic elements are produced;
\item to explain the (unknown) mass-loss process;
\item to constrain model atmospheres that will be used to determine accurate abundances in distant galaxies from complex spectra.
\end{itemize}
 
In Section~\ref{tio}, we indicated several molecular and atomic lines predominant in evolved stars. Several spectral resolutions should be used to attack the different related problems either for RSG and AGB stars:

\begin{itemize}
\item Low resolution, $R=\lambda/\Delta\lambda=30$. This resolution will help to characterize the size granulation pattern in terms of intensity surface contrast and spatial coverage of the surface.
\item Medium resolution, $R=\lambda/\Delta\lambda=3000$. This resolution will help to study the temperature stratification in the atmosphere. Observations at wavelengths in a molecular band and in the close pseudo-continuum probe different atmospheric depths, and thus layers at different temperatures that highly constraint the limb-darkening effect to be compared to 3D RHD simulations.
\item High resolution, $R=\lambda/\Delta\lambda>30000$. This resolution will help to study the cinematic in the spectral lines. In fact, these stars have velocities up to 20 ? 30 km/s in the optical thin region that strongly affect the line broadening.
\end{itemize} 
 
A last important point concerns the temporal and the wavelength variability. Monochromatic intensity maps from 3D RHD simulations are characterized by strong wavelength dependence that changes the appearance of the stellar surface. In average, the brightest areas exhibit an intensity $\sim$50 times or larger than the dark ones in the optical and up to $\sim$10 times in the infrared. Probing very narrow wavelength filters close to particular optical spectral line centers may increase these values. The importance of the spectral resolution is shown in Fig.~14 and 15 and related Section~5.4 of \cite{2009A&A...506.1351C}. Moreover, the temporal time scales of the granulation pattern differ also with respect to the spectral region probed: with, e.g., the optical region characterized by short-lived (a few weeks to a few months) small-scale structures \citep{2011A&A...528A.120C}; and the near infrared characterized by large cells evolving over years \citep{2009A&A...506.1351C}.
The temporal monitoring is then a key point, images must be carried out every month (or less) in the optical region for all the spectral resolution options.
 
 \begin{figure*}
   \centering
   \includegraphics[width=1.\textwidth]{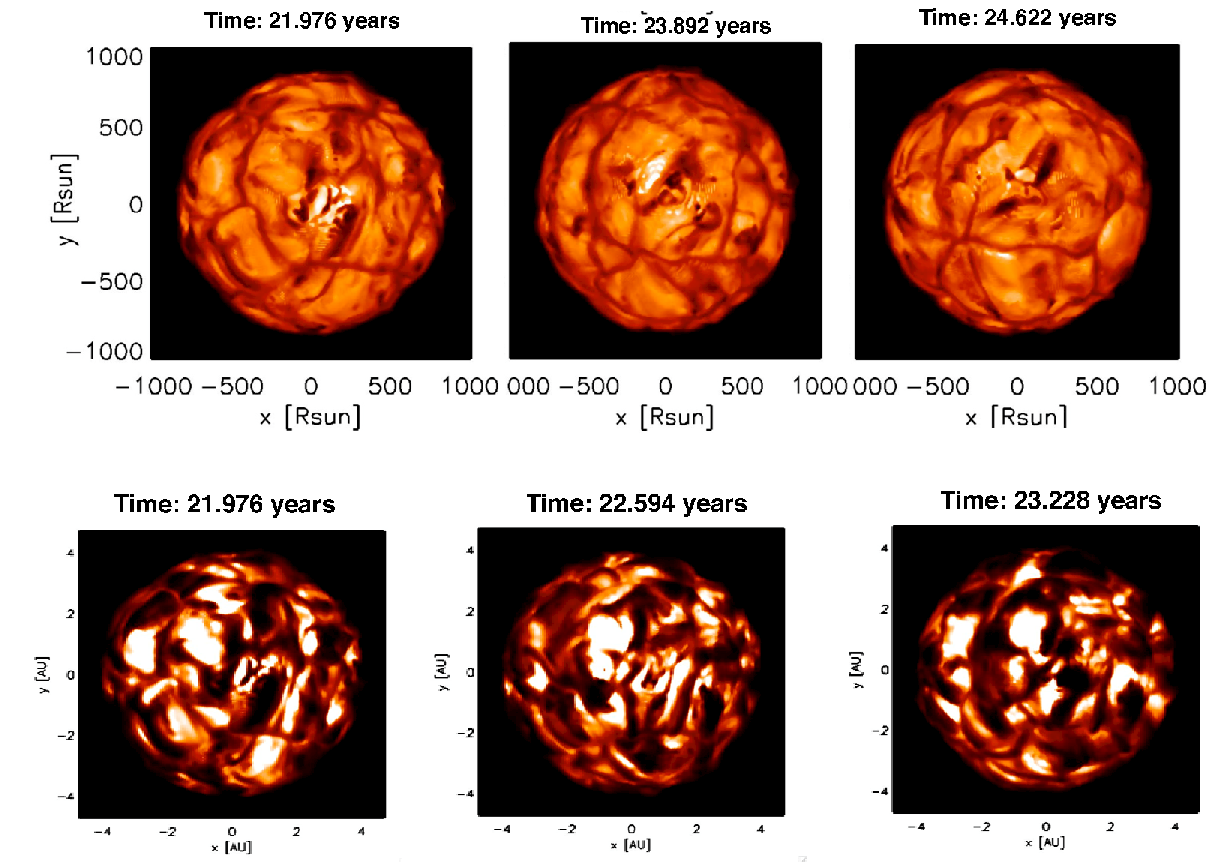}
   \caption{Synthetic maps (linear intensity) of a RSG simulations at different stellar time in the H band \citep[top row,][]{2009A&A...506.1351C} and in the optical region \citep[bottom row,][]{2011A&A...528A.120C}.}
              \label{chiavassafig}%
    \end{figure*}
 
 \section{Post-AGB}
\label{postagb.sect}
Optically-bright post-AGB stars are commonly found in binary systems with separations of typically 1~AU and that are
surrounded by rather compact, stable circumbinary disks \citep{2014MNRASKamath,2006AAdeRuyter}. 
These disks are very similar to the protoplanetary disks (PPDs) around young stars. However, the formation 
history of post-AGB disks is clearly different from PPDs, as is the further evolution of the illuminating star.
Optically thick and dusty disks are very complex, which, in the post-AGB case, is added by the complication 
that the central object is an (interacting) binary. To answer the relevant questions concerning the origin, evolution
and fate of these systems, therefore, requires a multiwavelength approach, combined with a high angular resolving power.  
A visible interferometric instrument is highly needed to extend the constraints on the parameter space of the models 
beyond what the ALMA and the second-generation instruments on the VLTI are able to provide. Visible interferometry 
can provide unique additional insights and constraints.
There are, in particular, two fundamental questions that can be addressed by such an instrument.

\begin{figure*}
   \centering
   \includegraphics[ width = \hsize,bb= 6 6 957 247]{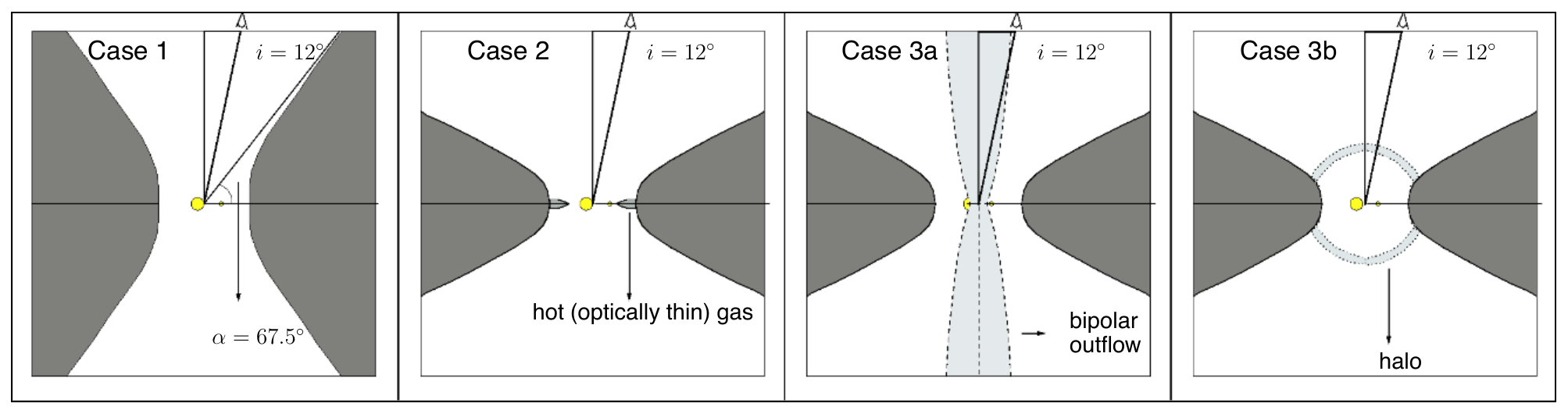}
   \caption{The various proposed configurations to explain the existing visible interferometric data for the 89 Her post-AGB binary. This system 
   is seen pole-on so that the various possibilities cannot be distinguished based on the existing data, although the modeling of \citet{2014AAHillen}
   suggests that the first option of a highly scattering inner disk rim does not work. More systems need to be observed to put this 
   intriguing result in a bigger perspective. For more background info concerning 89 Herculis, see \citet{2013AAHillen} and \citet{2014AAHillen}.}
   \label{figure:postAGB}
\end{figure*}

\subsection{Locating the continuum scattered light around post-AGB binaries}
Recently, using the visible instruments on the NPOI 
and the CHARA Array, \citet{2013AAHillen} spatially resolved a surprisingly large fraction of circumstellar optical light  
around the bright prototypical post-AGB binary 89 Her. Between 35 and 40\% of all optical flux received from this pole-on system 
is light scattered by the inner region, either from the inner rim of the disk or from a region extended in the vertical direction along the line of 
sight (see Fig.~\ref{figure:postAGB}). Based on radiative transfer disk models, \citet{2014AAHillen} claim that the former option can be excluded: 
such a large amount of optical light cannot exclusively be scattered by the disk and this particular system 
must therefore have an optically thin dust component in an outflow as well. 
Do all the post-AGB binary+disk systems have such a scattered light contribution to the stellar part of the spectral energy distribution? What 
fraction of the scattered light is coming from the disk and what is coming from the environment, including possible outflows? 
Are the input data to our models or some of the basic modeling assumptions/practices incorrect? 
More precise observations (5-10\% in visibility, 1-3$^\circ$ in CP) with a better uv-coverage 
(at least 4T combination), such as can be provided easily with 
the VLTI baselines, are needed to fully characterize the scattering source in this system. 
A polarimetric mode would be highly advantageous for this kind of study as well, since different 3D-distributions of the material
can yield a variety of polarimetric signatures whereas the resulting intensity projections on the sky may be similar.
Although scattering can be studied in the near-IR, additional observations in the visible wavelength domain give much stronger 
constraints on the material properties and the spatial distribution of the scattering material.
With a visible instrument on the VLTI the light scattered in post-AGB binary systems could be exploited in 
full detail, which would significantly reduce the extensive parameter space involved in modeling these objects. 
About 60 Galactic sources with $V<12$~mag are accessible to the VLTI. Extrapolating from 
the 89 Her results, a spatial resolution of 1~mas and a low spectral resolution (preferentially over a wide spectral range) is sufficient 
to characterize a scattered-light component in the majority of these systems.

\subsection{Tracing the inner gas streams in post-AGB binaries: accretion disk and/or jet?} \label{postAGB2}
Although outflows detected in optical continuum light are currently rare among post-AGB binaries, 89 Her and 
the Red Rectangle being the only known examples, outflows have been detected in these systems with a variety of other techniques. 
For many objects, the CO sub-mm rotational lines have an outflowing component in addition to the rotating disk 
\citep{2013AABujarrabalB}, which in some cases has been resolved with the PdBI \citep{2007AABujarrabal}, and recently ALMA
\citep{2013AABujarrabalC}. Jet-like outflows emanating from the neighborhood of the companion star have also been inferred 
on the basis of temporal spectroscopic monitoring campaigns \citep[e.g.,][]{2012AAGorlova,2013MNRASThomas}. Although 
feasible to detect jets in this way, to fully charactize these structures requires a more direct imaging method. 
This is of great importance because the link between the ``inner outflows'' detected in the optical and the large-scale 
outflows seen in CO is not at all understood, nor is the driving mechanism for either outflow or the mass reservoir from which they 
emanate. Is there an accretion disk around the secondary? Is this disk being fed by accretion from the circumbinary disk or from the 
post-AGB star? Are there more objects with a source morphology as complex as the Red Rectangle? Will
these post-AGB binaries ever become planetary nebulae, which are by default asymmetric \citep{2002ARAABalick}? 
Many questions remain about these objects and can only be resolved by constraining their outflow properties.
The number of sources and required spatial resolution is the same as in the previous science case. In this case, however, a 
spectral resolution of at least 10\,000 is needed, to spectrally resolve the Halpha line.

\subsection{Requirements}
About $60$ sources with V $< 12--13$ can be observed from Paranal.
The spectral resolution needed for the first science case is 3-30, but more important is the spectral range (up to B?).
The required spatial resolution is about 1 AU at a distance of 1 kpc, or 1 mas, to fully resolve the nearest sources.
Visibility precision: $5\%$, closure phase precision: 1-3$^\circ$.
Imaging is required, because we want to trace the scattered light and the source morphology can be very complex: how much 
of it is coming from the circumbinary disk, from a jet-like outflow, a halo or yet another morphology?
Imaging with four telescopes is possible, but six telescopes would be better, given that RV Tauri stars pulsate on
timescales of $\sim$15 days.

A spectral resolution of 10 000 is required for the second science case (Sect.~\ref{postAGB2}). 
The focus is on Halpha (and perhaps the Ca triplet).
The required spatial resolution is the same as for the first science case.
Visibility precision: differential measurements at the $<5\%$ level are needed.
Imaging is required for this science case, given the complex morphologes that are to be expected. 
Six telescopes would be preferred, given the pulsations and source complexity, although a 2-telescope temporal monitoring
would already give interesting constraints.

\section{The projection factor of $\delta$ Scuti and RR Lyrae stars}

The Baade-Wesselink (BW) method has been widely used to calibrate the period-luminosity relation of Cepheids \citep{fouque07, storm11a, storm11b} with important implications for the extragalactic distance scale and
cosmology \citep{riess11}. It has been applied for Cepheids using interferometry \citep{kervella04a, davis09} and photometry \citep{gieren05}, but also for RR Lyrae stars \citep{fernley94} and $\delta$~Scuti stars \citep{wilson93, milone94,  wilson98}. Concerning $\delta$ Scuti stars a radial pulsation is required to apply the {\it classical} BW method. Many High-Amplitude $\delta$-Scuti (hereafter HADS) stars show radial fundamental mode of pulsation (combined sometimes with other harmonics). These are the best candidates to apply the BW analysis. However, few studies have been done in this field and the parallaxes of Hipparcos currently remains the only geometric method that has been used to calibrate their \emph{PL} relations \citep{petersen98}.

The basic principle of the BW method is to compare the linear and
angular size {\it variation} of a pulsating star in order to derive
its distance through a simple division. The key point is that
photometric measurements lead to angular diameters corresponding to the {\it
photospheric} layer, while the linear stellar radius variation is
deduced by spectroscopy, i.e., based on line-forming regions which
form higher in the atmosphere. Thus, radial velocities
$V_{\mathrm{rad}}$, which are derived from line profiles, include
the integration in two directions: over the stellar surface through
limb-darkening and over the atmospheric layers through velocity
gradients. All these phenomena are currently merged in one specific
quantity, generally considered constant with time: the projection
factor $p$, defined as $V_{\mathrm{puls}}=pV_{\mathrm{rad}}$, where
$V_{\mathrm{puls}}$ is defined as the {\it photospheric} pulsation
velocity \citep{nardetto04}. $V_{\mathrm{puls}}$ is then
integrated with time to derive the photospheric radius
variation.

\begin{figure}[htbp]
\begin{center}
\resizebox{1.0\hsize}{!}{\includegraphics[clip=true]{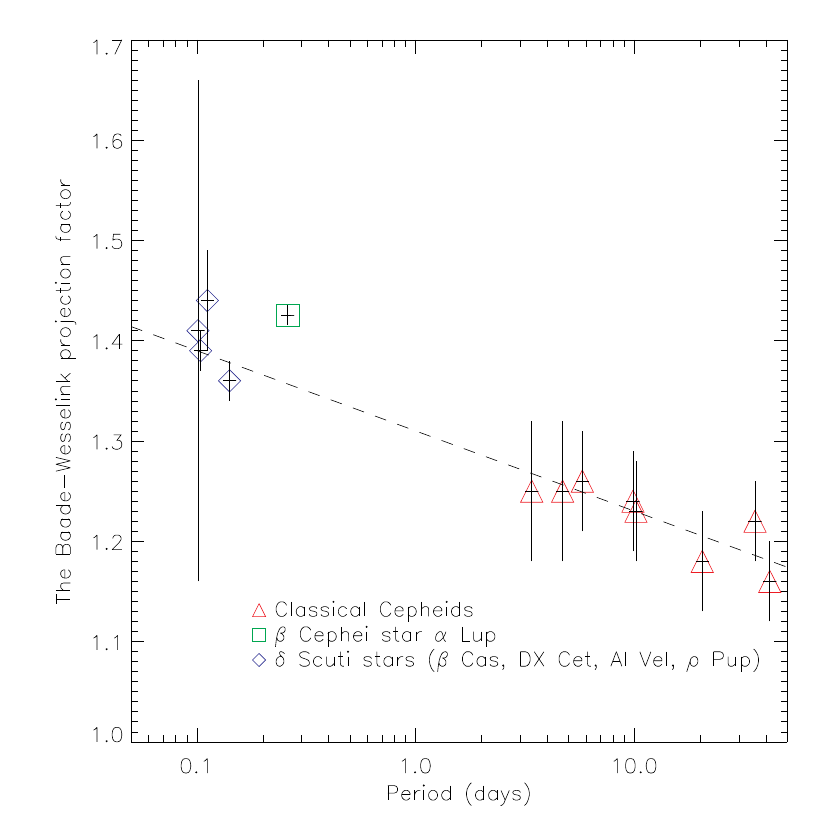}}
\end{center}
\caption{The Baade-Wesselink projection factor as a function of the period for different kinds of pulsating stars. The $\delta$~Scuti stars indicated as blue diamonds are, by increasing period: $\beta$~Cas, DX~Cet, AI~Vel and $\rho$~Pup.} \label{Fig2}
\end{figure}

The projection factor appears widely in the literature for Cepheids. Up to recently, a
constant geometric projection factor was commonly used for all kinds
of pulsating stars (generally $p=1.36$, \citet{burki82}). Using high
precision spectroscopic measurements of 10 southern hemisphere Galactic
Cepheids with HARPS, we measure directly the velocity gradient
within the atmosphere of Cepheids and derive semi-theoretical
$p$-factors. We discovered a relation between the period of Cepheids
and the projection factor. We also showed that if one uses a
constant $p$-factor ($p=1.36$) to derive the \emph{PL} relation,
errors of $0.10$ and $0.03$ on the slope and zero-point of the
\emph{PL} relation can be introduced. This means that distances can
be overestimated by 10\% for long-period Cepheids \citep{nardetto07, nardetto09}.

The aim of this study is to derive the expected period-projection factor relation of RR Lyrae and High-Amplitude $\delta$ Scuti stars (HADS) by applying the inverse BW method. HADS are generally supposed to pulsate radially. However, weak non-radial pulsation could be detected using long baseline visible interferometry. 
In that case, the Baade-Wesselink method, if revised, can constitute an interesting tool to distinguish radial and non-radial modes of pulsation. \citet{balona79a, balona79b, balona79c} and \citet{stamford81} give a $p$-factor estimates connected to the non-radial mode of the star, while the classical approach done for Cepheids have been recently applied to several $\delta$-Scuti with dominant radial mode \citet{guiglion13, nardetto14}. The low and medium spectral resolutions (reported in Section~\ref{rsgsect}) are the best modes of observation in order to derive the angular diameter variation of pulsating stars with high precision.

\begin{figure}[htbp]
\begin{center}
\resizebox{1.0\hsize}{!}{\includegraphics[clip=true]{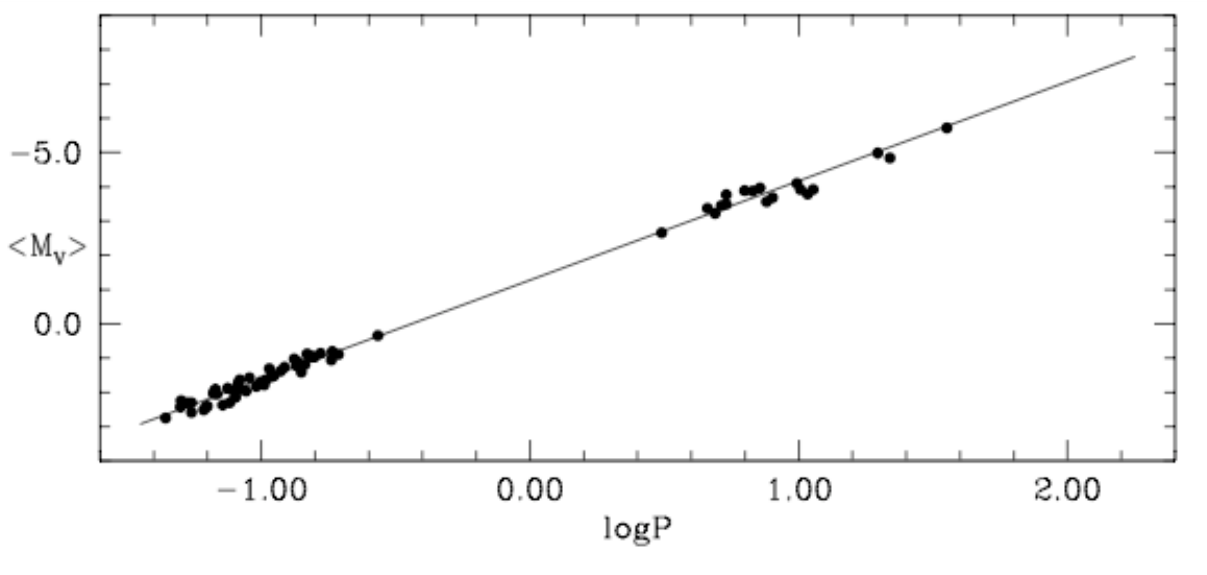}}
\end{center}
\caption{ Period-Luminosity relation of Galactic $\delta$~Scuti (lower magnitude) and Cepheid (larger magnitude) variable stars (figure from Mc Namara et al. 2007).  The PL relation of Cepheids combined with the PL relation of $\delta$~Scuti stars would provide an interesting tool for the extragalactic distance scale.} \label{Fig2}
\end{figure}

Once the projection factor derived, the photometric version of the BW method can be applied to a wider sample of fainter $\delta$~Scuti and RR Lyrae stars, and, with the combination of Gaia parallaxes, this can help to provide a robust period-luminosity relation. Deriving such relation is extremely interesting for different reasons. First, \citep{fernie92} and then \citep{mcnamara07} showed that the PL relation of $\delta$~Scuti stars and Cepheids seems to be consistent in slope and zero-point (Fig.~{\ref{Fig2}). Second, Cepheids are
supergiant stars and are relatively rare in space, while $\delta$-Scuti stars are dwarf or near-dwarf stars and are numerous in the solar neighborhood. Many are so nearby that they have negligible or very small reddening. Third, SX Ph stars can be found in Globular Clusters, that helps also to constrain the PL relation of RR Lyrae  \citep{mcnamara07}. The PL relation of Cepheids combined with the PL relation of $\delta$~Scuti stars would provide an interesting tool to the extragalactic distance scale. And if we show that the BW method can be used to calibrate the PL relation of $\delta$~Scuti using a specific p-factor (and in a consistent way compared to trigonometric measurements) it will open a new and promising field of investigation. 

\begin{table}[htbp] \caption{\label{Tab.ang} HADS and RR Lyrae}
\begin{center}
\begin{tabular}{lccc|ccccccccc}
\hline
\hline
N$_\star$	& $\simeq$ 10 \\
spatial resolution 	& $\theta$ from 0.1 to 0.25 mas \\
spectral resolution 	&  $R=3000$ \\			
temporal resolution	&  10 minutes (high sensitivity required) \\
precision 	&	1\% at low visibilities \\	
imaging	& 3T is fine, 6T is best \\
\hline
\end{tabular}
\end{center}
%\begin{list}{}{}
%\item[$^{\mathrm{a}}$]  and/or minimum number of telescopes required
%\end{list}
\end{table}

\section{Pulsating stars in binaries}

Another very interesting aspect is to study pulsating stars in binaries in order to (1) measure their orbital solution with interferometry and (2) derive their mass, which put important constrains on the evolutionary models. it concerns not only  $\delta$ Scuti stars, but also RR Lyrae, $\beta$ Cepheids, $\gamma$ Dor and RoAp stars.  From Garcia et al. (1995), Stankov et al. (2005) and Matthews et al. (1999), together with the Galactic Variable Catalogue, we found about 140 objects to be observed with $mV <7$ and 200 objects with $mV <9$. The exact orbital separation still have to be calculated to better characterize how many objects could be resolved by a 300 meters baseline interferometer.

\chapter[Interacting binaries]{Interacting binaries\label{ch6}}

 \author{N.~Blind$^{1}$,
H. Boffin$^{2}$,
M. Borges Fernandes$^{3}$,
A. Gallenne$^{4}$,
C. Guerrero Pena$^{5}$,
M. Hillen$^{6}$,
A. Labeyrie$^{7}$,
T. Marsh$^{8}$,
M. Simon$^{9}$,
F. Milllour$^{1}$,
N. Nardetto$^{1}$,
D. Steeghs$^{8}$,
T. ten Brummelaar$^{10}$,
 }

\begin{center} 
$^{1}$~Max Planck Institute for Extraterrestrial Physics, Giessenbachstrasse, 85741, Garching, Germany\\
$^{2}$~ESO, Alonso de C\'ordova 3107, Casilla 19001, Santiago 19, Chile\\
$^{3}$~Observat\'orio Nacional, Rua General Jos\'e Cristino 77, 20921-400 S\~ao Cristov\~ao, Rio de Janeiro, Brazil\\
$^{4}$~Universidad de Concepci\'on, Departamento de Astronom\'ia, Casilla 160-C, Concepci\'on, Chile\\
$^{5}$~Instituto de Astronom\'ia UNAM, Apdo. Postal 70-264, Cd. Universitaria, 04510 M\'exico D.F., M\'exico\\
$^{6}$~Instituut voor Sterrenkunde, KU Leuven, 3001, Leuven, Belgium\\
$^{7}$~Collge de France, 11, place Marcelin Berthelot, F-75231 Paris Cedex 05, France\\
$^{8}$~Department of Physics, University of Warwick, Coventry CV4 7AL, UK\\
$^{9}$~Department of Physics and Astronomy, Stony Brook University, Stony Brook, NY 11794, USA\\
$^{10}$~CHARA Array, Mount Wilson Observatory, 91023, Mount Wilson, CA, USA\\
\end{center} 

\section{Binaries}
It can be safely said that binary stars are critical for astrophysics, and especially interacting binaries. This is by no means a gratuitous statement.
First of all, stars tend to form and live in binaries, with the fraction of binary stars being higher for more massive stars: the binary fraction increases from
25--30 \% for M dwarfs and 54$\pm$2\% for solar-like stars, to more than 70\% for B and O stars \citep{Raghavan2010, Chini2012}.
Moreover, a significant fraction of these binary stars will interact in one way or another during their evolution. For example, \citet{Sana2012} found that 
more than 70\% of all massive stars will exchange mass with a companion, leading to a binary merger in one-third of the cases. 

\noindent Interactions in a binary system can be of different kind:
\begin{itemize}
\item Tidal interaction: stars will tidally interact, leading to synchronisation of the stellar rotation with the orbital motion, and then later, to the circularisation of the orbit.
The typical timescales for tidal synchronisation in convective stars is $10^4 P^4$ years (where $P$ is in days), that is, about $10^8$ years for a system with an orbital period of 10 days, while circularisation occurs on timescales of 
$10^6 P^{16/3}$ years, meaning a timescale of $10^{11}$ years for a 10-d period system. The circularisation period, i.e. the orbital period for which the orbit is circularised, will thus depend on the age of a star and 
is therefore about a few days for solar-like stars in our neighborhood or in open clusters.  
As a star evolves towards the red giant and asymptotic giant branches, its radius increases by factor of ten to hundreds, leading to much stronger tidal effects. The circularisation period
for red giants in open clusters is, for example, of several hundred days \citet{Mermilliod2007}.
\item Mass transfer by wind: all stars loose mass by stellar wind, with mass-loss varying between 10$^{-14}$ M$_\odot$yr$^{-1}$ for solar-like stars to 10$^{-6}$--10$^{-4}$  M$_\odot$yr$^{-1}$ for the most evolved stars. When a star is in a binary system, its wind will be partially intercepted by the companion, leading to mass exchange. The simplest formalism of such accretion is by Hoyle \& Lyttleton (1944) and Bondi \& Hoyle (1950), but more sophisticated simulations in binary systems have shown that things are more complicated, with the formation of spirals \citep{Boffin1994, Theuns1996, Nagae2004}. Such spirals have now been detected, e.g. by ALMA around the star R Scl \citet{Maercker2012}. When the two stars have strong winds, these will collide, creating shocks, which will emit X-rays. This is particularly the case in massive binaries.
\item Roche-lobe overflow: When a star fills its Roche lobe -- a critical region defining its zone of gravitational influence (see Fig.~\ref{fig:binary}) -- it is no more possible for it to keep its material, and matter flows towards its companion. This is the typical mass loss happening in Algols. 
\item Angular momentum loss: mass loss and mass transfer lead to angular momentum loss and transfer, but one can also have angular momentum loss without any mass loss. This is the case when one of the stars has a strong enough magnetic field, such as in cataclysmic variables, or if the stars are very close and degenerate, in which case, there will be emission of gravitational radiation. The angular momentum variations lead to changes in the orbital elements and in some cases merging of the two stars.
\end{itemize}

\begin{figure}[htbp]
   \centering
   \includegraphics[width=1.8in]{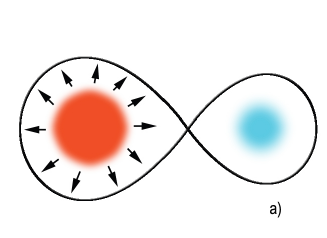}\\
    \includegraphics[width=1.8in]{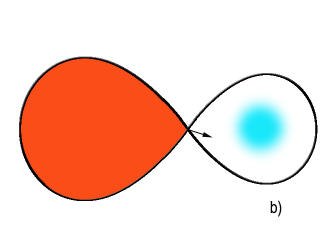}  % requires the graphicx package
   \caption{The two main cases of mass transfer in a binary system: by wind accretion (a) or by Roche lobe overflow (b).}
   \label{fig:binary}
\end{figure}

The mass transfer in a binary system will affect the chemical composition of the companion, which can for example become polluted in processed elements like in Algols or in Barium stars. It also leads to a mass increase of the companion, with sometimes strong consequences, such as blue straggler stars or Type Ia supernovae. Finally, mass transfer has also an impact of the evolution of the orbital separation. 

Because of the various possible interactions, and the various sizes and mass-loss of stars in the different evolutionary stages, interacting binaries are happening on many different scales. 
On one extreme, the system SDSSJ010657.39-100003.3 \cite(Killic2011} is a detached binary comprising two white dwarfs in orbit around each other with an orbital period of 39 minutes: one of the white dwarf is tidally deformed, and because the system is losing angular momentum through gravitational  radiation, the two stars will merge in 37 Myr. This is by no means the binary systems with the shortest orbital period: to our knowledge, this 
title is bestowed on the AM Canum Venaticorum star HM Cancri, with an orbital period of 321 seconds \citet{Roelofs2010}. This system also contains two white dwarfs, with one of the two transferring mass to the other by Roche lobe overflow. 
At the other extreme, one of the most remarkable star in our Galaxy, $\eta$~Car, is a binary system with an orbital period of 5.5 years and a rather eccentric orbit. Every passage  of the companion star at the perihelion is leading to drastic effects \citet{Mehner2015}. In the 1840's, $\eta$~Car underwent a gigantic outburst, making it the second brightest star in the night sky, after Sirius. During this outburst, the star lost about 10 M$_\odot$, creating the magnificent homunculus nebula. The likely explanation of this outburst lies in the binary nature of the system.

Binary stars are thus critical as they are the clues to understand a large variety of astrophysical phenomena, some playing an important role in chemical enrichment of galaxies, such as Type Ia supernovae, short gamma-ray bursts, novae, barium stars and other peculiar red giants, symbiotic stars, blue straggler stars, etc. Moreover, as we have seen, some stars in binary systems will merge, so that even if a star appear single now, it may not have been always the case, and the previous binarity may be the cause for some odd physical properties of the stars. This could explain some blue stragglers or some B subdwarfs, among others. Binary stars are also crucial as they allow us to determine uniquely, and in a model-independent way, the masses and radii of stars. It is thus critical to be able to study them in details.

For solar-like stars, Duquennoy \& Mayor (1991) showed that the period distribution is log normal between 1 day and 10 million years, with a mean at $\log P=4.8$ days, and 
a standard deviation in $\log $ of 2.3. Thus, various techniques will be needed to probe these different scales. The longest periods are often found as astrometric or visual binaries, while for the shortest periods, one need to rely on spectroscopy or photometry (spectroscopic binaries, eclipsing binaries, W Uma stars, etc.). The advance in technology has allowed to change also the application of a given technique to various systems.

Currently the number of systems resolved as visual binaries is biased towards long period systems: this is because traditionally, it was only possible to discern visual binaries with a separation of 0.3'' or more. The use of adaptive optics on the largest telescopes as well as the Hubble Spatial Telescope has allowed us to go to smaller separation, that is, the diffraction limit of the telescopes, typically 0.05'' or 50 mas. With interferometry, one can go down to much smaller separations even, down to about 1 mas. This thus allows one to problem systems with short orbital period and significant radial-velocity variations, that is, we can combine spectroscopic and visual orbits and get most information on the system, in particular the masses. As a typical distance of 100 pc, 0.3'' corresponds to a separation of 30 AU and orbital periods of about 100 years; 0.05'' to about 5 AU and 10 years, while 1 mas, corresponds to 0.1 AU and a period of a few days. For cases more typical of massive stars, at a distance of 1.5 kpc, these numbers become, respectively, 450 AU	and 2000 yrs, 75 AU and 100 yrs, and 1.5 yrs and 0.3 yrs. Thus interferometry really allows us to reach a new class of systems, those most likely to interact.
We note that with techniques such as astro-tomography, it is even possible to go down to the micro-arcsecond scale, but this is limited to a certain kind of binary systems \citet{Boffin2001}.

We identify three major interests of using interferometry for the different types of binaries:
\begin{itemize}
\item Finding a companion to a star;
\item Determine the fundamental parameters of stars, such as mass and radius;
\item Study the kinematics of gas, accretion discs, etc., in interacting systems.
\end{itemize}

Interferometry has already been used for a variety of binary stars, such as double-lined stars, symbiotic stars, post-AGB stars, binary cepheids, massive stars and their various flavours, such as B[e] and luminous blue variable stars, as well as in young binaries. 
As a testimony of the ubiquity of binary stars in all fields of stellar astrophysics, some of these topics are covered in previous chapters in this book: 
young binaries is covered in  Chapter 3 by Perraut et al., while eclipsing binaries and binary cepheids are covered in Chapter 4 by Nardetto et al.,  binarity of massive stars and the use of spectro-astrometry in Chapter 5 by Stee et al., while discs in binary post-AGB stars as well as pulsations in binaries are covered in Chapter 6 by Paladini et al. Here, we will present some results obtained with interferometry, not presented elsewhere in this book.

We also refer to the review of \citet{Stee2012} that present  results between 2007 and 2011 based on interferometric observations of binaries and multiple systems, as well as to the work by \citet{Millour2014}, which used the VLT Interferometer to study the close environment of two post-merger stellar systems.

\section{Finding a companion}

The presence of a companion will be betrayed by its signature in the interferometric signal. If the difference in magnitude between the two objects is less than 6, i.e. the contrast in flux is about 200 or less, then current interferometers such as PIONIER on ESO's Very Large Telescope can detect the companion, using visibilities and closure phases provided there is an error of 0.5 degrees or less on the latter \citet{LeBouquin2011}. Once the companion has been detected, it is then possible to follow the relative orbital motion of the two components, providing a visual orbit, which coupled with radial velocities measurements lead to the measurement of the masses of the two stars (see next section). 

Working at high spectral resolution  allows one to probe the smallest interacting systems by spectro-astrometry. While the giant is in generally dominating the continuum emission in the visible domain, the faint companion is situated in the energetic accretion zone presenting strong emission line like H$\alpha$. The photocentre of the binary, centred on the giant companion in the continuum, shifts towards the accretor in the line, which could be measured by help of differential phases. A simple calculation and the experience gained on VEGA show that reaching precision of order 10 $\mu$as or better is well within reach, and is therefore able to probe the most compact semi-detached systems.

An example of such detection is given by the discovery of a companion to the luminous blue variable (LBV), HR Car. 
As mentioned earlier, the best-known LBV, $\eta$~Car, is thought to have lost about 10~M$_\odot$ during its great outburst in the 1840's. There is no firmly established mechanism to explain such a mass loss, but the discovery that $\eta$~Car is a five-year highly eccentric binary puts focus on possible binarity-induced mechanism for the giant outbursts, and prompted binarity searches among LBVs. So far, however, while several wide LBV binaries were identified, no system similar to $\eta$~Car (relatively close \& eccentric) has been found. The LBV HR Car was observed as part of the OHANA sample (spectrally resolved 3-beam interferometry of Br $\gamma$; see Rivinius et al., 2014). The OHANA data obtained for HR Car showed a clear and temporally stable (over the several months of observations) phase signature across the blue part of the emission line, but little to no visibility signature. This marks a photocentre displacement of the emission line with respect to the continuum, which is hard to explain with the more-or-less symmetric, but variable wind of a single supergiant star. These data hinted at the possibility for HR Car to be a binary, but more observations were needed to confirm this. HR Car was observed with the four 1.8-metre Auxiliary Telescopes of ESO's Very Large Telescope Interferometer, using the PIONIER instrument in the $H$-band. The visibilities and closure phases obtained at the two epochs (Rivinius et al. 2014) clearly reveal asymmetries that cannot be due to a single, spherical object. In fact, two sources are clearly detected with a flux ratio of 6, and about 2 mas apart (Boffin et al. 2015, in prep.). This would correspond to an orbital period of the order of 10 years and would make HR Car a binary similar to $\eta$~Car.

\section{Determining accurate masses}
Mass is the most crucial input in stellar internal structure modelling. It predominantly influences the luminosity of a star and, therefore, its lifetime. Unfortunately, the mass of a star can generally only be determined when the star belongs to a binary system. Therefore, modelling stars with extremely accurate masses (better than 1 \%), in different ranges of masses, would allow to firmly anchor the models of the more loosely constrained single stars.
At present, accurate masses are still rare, however: in a recent paper, \citet{Torres2010} have listed the non-interacting systems with masses more accurate than 3\%. They found 95 eclipsing binaries and 23 astrometric binaries. Among these systems, the simultaneously astrometric and spectroscopic binaries provide the less model-dependent masses, since the derivation of their masses only relies on the interpretation of their positions and radial velocities by the Kepler's laws. In addition, these systems are less likely to have been affected by mass transfer, since their periods are much longer than those of the eclipsing binaries. However, among the 23 astrometric systems, only 14 components have masses more accurate than 1\%. The majority of them are on the main sequence, two are subgiant stars, one is a F9-type giant and only the components of Capella are giant stars as late as G-type.
Jean-Louis Halbwachs and colleagues have started \citep{Halbwachs2009, Halbwachs2014} a survey of of about 70 double-lined spectroscopic binaries (SB2), using the high-resolution spectrograph Sophie (Haute-Provence Observatory) and Hermes (Roque de los Muchachos Observatory), and with the aim to determine the minimum masses of the components, $M_1 \sin^3 i$ and $M_2 \sin^3 i$, with an accuracy better than 1\%. Accurate masses will be obtained when the orbital inclinations, $ i$, will be derived from astrometric measurements. The binaries of the sample are expected to have large photocentric orbits that will be observed with the Gaia satellite, and merging the radial velocities (RVs) and the astrometric measurements from Gaia will provide accurate masses and also magnitudes in the Gaia G photometric band. Combined with the Gaia exquisite parallaxes, this will provide both masses and luminosities for each component. Despite the high quality that is expected for the Gaia measurements, it is known from the reduction of the Hipparcos satellite (ESA, 1997 and \citep{vanLeeuwen2007}) that the large space astrometric surveys may be prone to systematic errors. Moreover, concerning double stars, a good orbit determination with Gaia depends on the measurements at various epochs, which rely ultimately on the application of the PSF calibration, considering the object as a single star. Binaries may then be sensitive to small differences between the position given by the PSF and the position of the actual photocentre. Therefore, an independent derivation of the masses of some stars of the sample will be necessary in order to validate any future results. Interferometry, and PIONIER in particular, is quite suitable for that: it  observes the relative orbits instead of the photocentric ones, and this compensates the fact that the accuracy of the instrument is a bit worse than that of Gaia.

\begin{figure}[htbp]
   \centering
   \includegraphics[width=10cm]{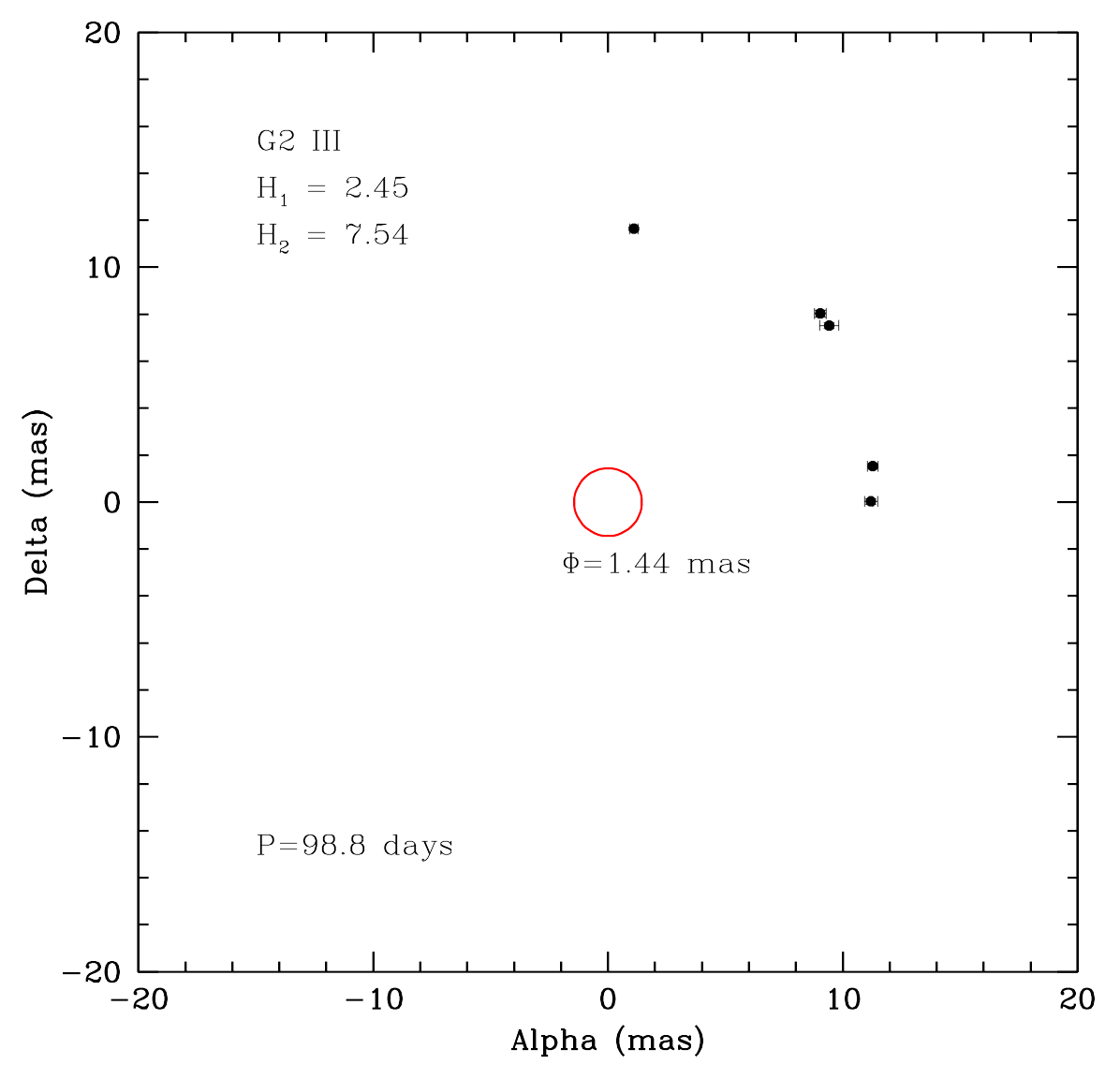} % requires the graphicx package
   \caption{Relative positions of the two stars in one system studied with PIONIER. The primary is a G2 III giant star whose diameter could be derived, and the position of the companion is shown for various epochs, covering about a quarter of the full 98-d orbit. With more data points, it should be possible to derive precisely the orbit of the two stars, leading to individual masses with an error at the 1\% level. From Halbwachs et al., in prep.}
   \label{fig:hdorbit}
\end{figure}

For systems which are double-lined spectroscopic binaries, that is, for which we already have the mass ratio, as well as the orbital period and eccentricity, using interferometry to make it a visual binary, allows us to determine the remaining parameters, i.e. the inclination, position angle of the node line, and, thereby, the individual masses. This thus allows obtaining precise stellar masses. As an example of what can be achieved, we show in Fig.~\ref{fig:hdorbit} an example obtained with PIONIER of a binary with a period of 98.8 days and containing a red giant (G2 III) and a fainter companion -- the luminosity ratio in the $H$-band is more than 5 magnitudes, that is a factor 100 in flux. In this case, the giant is rather bright ($H$=2.4, $V$=3.9). Because the companion is most likely a F star, in the visible, the luminosity contrast would be smaller, hence, the companion would be easier to observe. Here, the giant is also resolved and one can also estimate its diameter.

\section{Probing the mass transfer mechanism}
Symbiotic stars show a composite spectrum, composed of the absorption features of a cool giant, in addition to strong hydrogen and helium emission lines, linked to the presence of a hot star and a nebula. It is now well established that such a ``symbiosis'' is linked to the fact that these stars are active binary systems, with orbital periods between a hundred days and several years.

In such systems, 
the red giant is losing mass that is partly transferred to the accreting companion -- either a main sequence (MS) star or a white dwarf (WD).
One of the main questions related to symbiotic stars is how the mass transfer takes place: by stellar wind? Through Roche-lobe overflow (RLOF)? Or through some intermediate process? 
Answering this question requires  being able to compare the radii of the stars to their Roche lobe radius (which depends on the separation and the mass ratio). However, determining the radius of the red giant in symbiotic systems is not straightforward, and there has always been some controversy surrounding this. 
Optical interferometry is currently the only available technique that can achieve this. It allows determining the size and the distortion of the giant star, and in some cases, the orbital parameters of the system, without any a priori on their characteristics.

\begin{figure}[h]
   \centering
   \includegraphics[width=10cm]{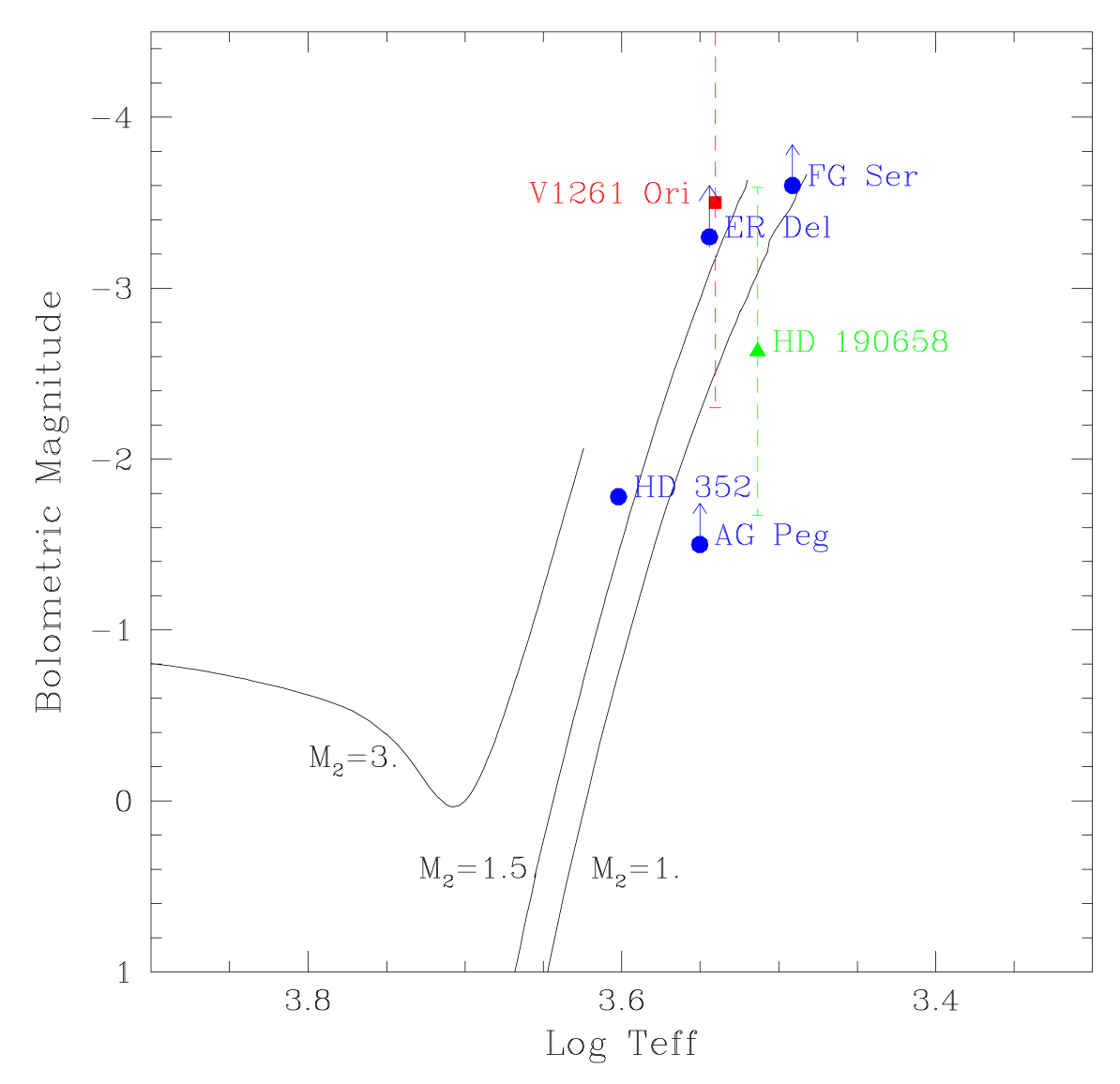} % requires the graphicx package
   \caption{Hertzsprung-Russell diagram of the symbiotic and related stars studied by \citet{Boffin2014a}, based on their estimate of the radius made with interferometry. This shows that such stars are quite normal giants, and if, as thought, they have larger mass-loss rates than normal giants, the origin of this must be in their binary nature.}
   \label{fig:hrsymbio}
\end{figure}

Recently, Boffin et al. (2014a) have observed with PIONIER several symbiotic and related stars, measuring their diameters -- in the range of 0.6 to 2.3 milli-arcseconds (mas) -- and thereby assessing for the first time in an independent way the filling factor of the Roche lobe of the mass-losing giants. For the three stars with the shortest orbital periods (i.e. HD 352, FG Ser and HD 190658), they found that the giants are filling (or are close to) their Roche lobe, consistent with the fact that these objects present ellipsoidal variations in their light curve. In the case of the symbiotic star FG Ser, there are some indications that the diameter is changing by 13\% over the course of 41 days, while observations of HD352 are, seemingly, indicative of an elongation at the level of 10\%. Such deformations need, however, still to be confirmed with more precise interferometric observations and compared with the outcome of light-curve modelling. The other three studied stars (V1261 Ori, ER Del, and AG Peg) have filling factors in the range 0.2 to 0.55, i.e., the star is well within its Roche lobe, and for these the mass transfer must take place by wind.

The detailed analysis of these stars also allows Boffin et al. to place the systems in an HR-diagram (Fig.~\ref{fig:hrsymbio}), showing that red giants in symbiotic systems are rather normal and obey similar relations between colour, spectral type, temperature, luminosity, and radius. We refer the reader to \citet{Boffin2014b} for further discussion.

\section{A case in point: SS Lep}

SS Leporis is composed of an evolved M giant and an A star in a 260-day orbit, and presents the most striking effect of mass transfer, the ``Algol'' paradox. That is, the M giant is less massive than the A star. The A star is unusually luminous and surrounded by an expanding shell, certainly as the result of accretion. The observation of regular outbursts and of ultraviolet activity from the A star shell are further hints that mass transfer is currently ongoing. Additionally, a large circumbinary dusty disc surrounds the binary system, implying that the mass transfer process is non-conservative.  
SS Lep was observed during the commissioning of the PIONIER visitor instrument at the VLTI \citet{LeBouquin 2011}. Because it combines four telescopes (in our cases, the four 1.8-m Auxiliary Telescopes), PIONIER provides six visibilities and four closure phases simultaneously, which allows reliable image reconstructions. It was possible to detect the two components of SS Lep as they moved across their orbit and measure the diameter of the red giant in the system ($\sim$2.2 milli-arcseconds). By reconstructing the visual orbit and combining it with the previous spectroscopic one, it was possible to well constrain the parameters of the two stars. The M giant is found to have a mass of 1.3 M$_\odot$, while the less evolved A star has a mass twice as large: a clear mass reversal must have taken place, with more than 0.7 M$_\odot$ having been transferred from one star to the other. Such results also indicate that the M giant only fills around 85$\pm$3\% of its Roche lobe, which means, that the mass transfer in this system is most likely by a wind and not by Roche lobe overflow. However, as the M giant is currently thought to be on the early-AGB phase, where mass loss is still very small, it is difficult to understand how it could have lost so much mass in a few million years (the time spent on the AGB), unless one invokes the Companion-Reinforced Attrition Process (CRAP) of \citet{Tout1988}, i.e. assume that the tidal force of the companion dramatically increases the wind mass loss. Although this allows in principle to explain the current state of SS Lep without resorting to a Roche-lobe overflow, this still needs to be proven by a more detailed study. This would require more sensitivity and higher spatial resolution, which can be achieved by going in the visible.

\section{Going to the Visible}
We are now briefly considering what can be done if we have an interferometer working in the visible.
In Tab.~\ref{tab:distance}, we show the 
 absolute magnitude in the $V$ and $H$ bands for stars on the main-sequence and on the giant branch, as well as distances we can reach assuming that with PIONIER we can currently study objects with about $H$=8, while it is hoped that one can reach $V$=10 in the visible. 
 
% Requires the booktabs if the memoir class is not being used
\begin{table}[htbp]
   \centering
   \tbl{Distances reached in $V$ and $H$}
   %\topcaption{Table captions are better up top} % requires the topcapt package
  { \begin{tabular}{@{} lrrrr @{}} % Column formatting, @{} suppresses leading/trailing space
      \toprule
       Sp. Type & M$_V$ & M$_H$ & D$_V$ (pc) & D$_H$ (pc)\\
    \toprule
O V&--5&-4&10,000 &2,500\\
B V&--3&--2.9&4,000&1,500\\
A V&1.5&1.3&500&215\\
G V&5&3.5&100&80\\
M V&9&4.8&16&44\\
WD&12&11.8&4&1.7\\
K III&0.5&--2.1&800&1000\\
M IIII&--0.6&--4.8&1300&3700\\
      \toprule
   \end{tabular}}
   \label{tab:distance}
\end{table}

It is clear from this table that going to the visible, with such limiting magnitudes, would allow us to reach more distant objects of early spectral types than currently possible, a similar number of objects for early giants (K-type), while slightly less of the reddest objects, that is M stars. However, one need to convolve this with the fact, that in the visible, the magnitude contrast between a M star and its companion would be higher, and it would thus be easier to detect the companion.
It is also noteworthy that these numbers indicate that for O stars,  one can thus reach the Bulge and thus have access to many objects. 

If we could improve on current interferometers, and achieve a resolution of 0.1 mas, then a system at 10 kpc with such a separation, would imply a physical separation of 1 AU. This is an interesting separation range to probe, as it is exactly those of interest for, e.g., LBVs. 

More generally, one can estimate for the different spectral classes, the range of orbital periods and mass ratios we could detect, if we assume we can observe objects up to $V$=10 with a resolution of 0.1 mas: \\
\pagebreak

\begin{table}[h]
\begin{tabular}{lll}
\multicolumn{3}{l}{\bf O V stars}\\
D=150 pc		& P $\sim$ 6 -- 1000 days & q $>$ 0.04 \\
D=1 kpc 		& P $>$ 100 d & q $>$ 0.1 \\
D=10 kpc		& P $>$ 9 yrs & q $\sim$ 1\\
\multicolumn{3}{l}{\bf A V stars}\\
D=10 pc		& P $<$ 100 d & q $>$ 0.1\\
D=100 pc		& P  $\sim$ 1 d to 10 yrs & q $>$ 0.25\\
D=500 pc		& P  $\sim$ 8 d to 100 yrs & q $\sim$ 1\\
\multicolumn{3}{l}{\bf G V stars}\\
D=10 pc		& P $<$ 70d & q $> $0.2\\
D=50 pc		& P $< $2 yrs&  q $>$ 0.5\\
D=100 pc		& P $<$ 6 yrs & q $\sim$ 1\\
\end{tabular}
\end{table}

The above numbers indicate that one could probe, for example, at least 250 Algol systems,
leading to a dramatic improvement in the understanding of such systems. This would be similar for all kinds of stars.
It is thus clear that the range of parameters that can be probed would be amazingly wide, and would provide much constraints for stellar evolution and binary processes!

\chapter[Imaging, Technics and the FRIEND prototype]{Imaging, Technics and the FRIEND prototype\label{ch7}}

 \author{F.~Millour$^{1}$,
 Ph. Berio$^{1}$,
 Labeyrie Antoine$^{2}$,
Mourard Denis$^{1}$,
Schneider Jean$^{3}$,
Soulez Ferreol$^{4}$,
Tallon-Bosc Isabelle$^{4}$,
Tallon Michel$^{4}$,
Ten Brummelaar Theo$^{5}$,
Thiebaut Eric$^{4}$
 }

\begin{center} 
$^{1}$~Laboratoire Lagrange, UMR 7293, CNRS, Observatoire de la C\^ote d'Azur, Universit\'e de Nice Sophia-Antipolis,
Nice, France\\
$^{2}$~Collge de France, 11, place Marcelin Berthelot, F-75231 Paris Cedex 05, France\\
$^{3}$~Paris Observatory Ð LUTh, 5 place Jules Janssen, 92190 Meudon, France\\
$^{4}$~Centre de Recherche Astronomique de Lyon, CNRS, Observatoire de Lyon, France\\
$^{5}$~CHARA Array, Mount Wilson Observatory, 91023, Mount Wilson, CA, USA\\
\end{center}

 \section{The CHARA Array}
 The VEGA and PAVO beam combiners already give access to visible wavelengths and CHARA will provide adaptive optics with 25\% Strehl at 700 nm (80 \% in H band) with 1m telescopes (Berio et al. 2014). An important limitation for imaging purposes is the limitation of short baselines on the CHARA array. Adding one telescope would be an interesting idea to boost its imaging capabilities. The forthcoming adaptative optics on the 1m telescopes will certainly help to increase the limiting magnitude.
 
 \begin{figure}
\center{\includegraphics[width=0.7\textwidth]{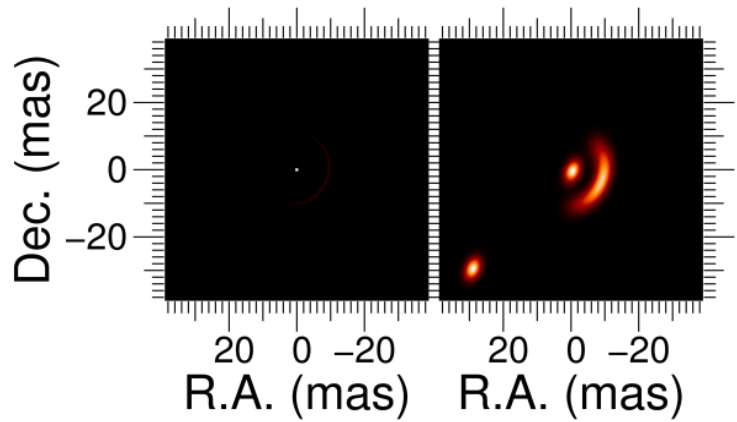}}
\caption{\label{Image_Be}Images of a B[e] star model made of a point source (the star) and a skewed ring (the dusty disk). Left: the image and Right: the image convolved with a resolution matching the one of the interferometer.}
\end{figure}

 \begin{figure}
\includegraphics[width=0.7\textwidth]{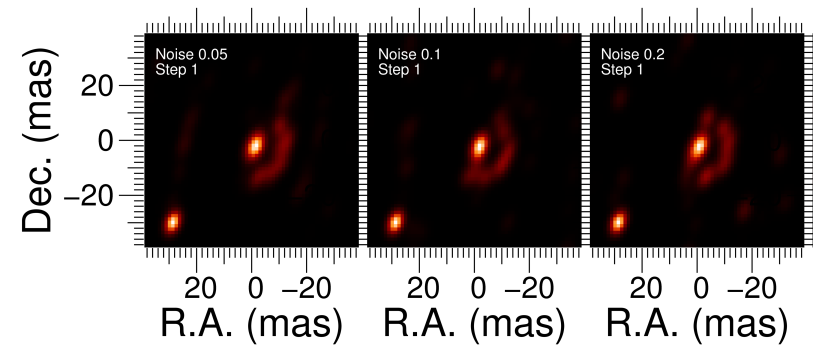}
\includegraphics[width=0.7\textwidth]{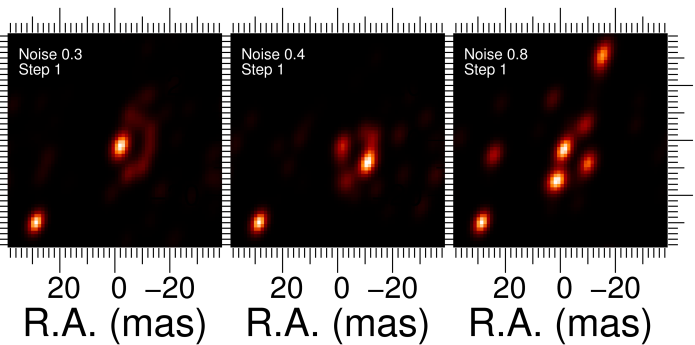}
\caption{\label{Image_Be_noise}Image reconstruction attempts by simulating a 3-nights observation of the above model with one measurement every hour. The noise level of the observables is raised progressively from top-left to bottom-right. The interferometric beam is displayed on the bottom-left corner of each image.}
\end{figure}

 \begin{figure}
\includegraphics[width=0.4\textwidth]{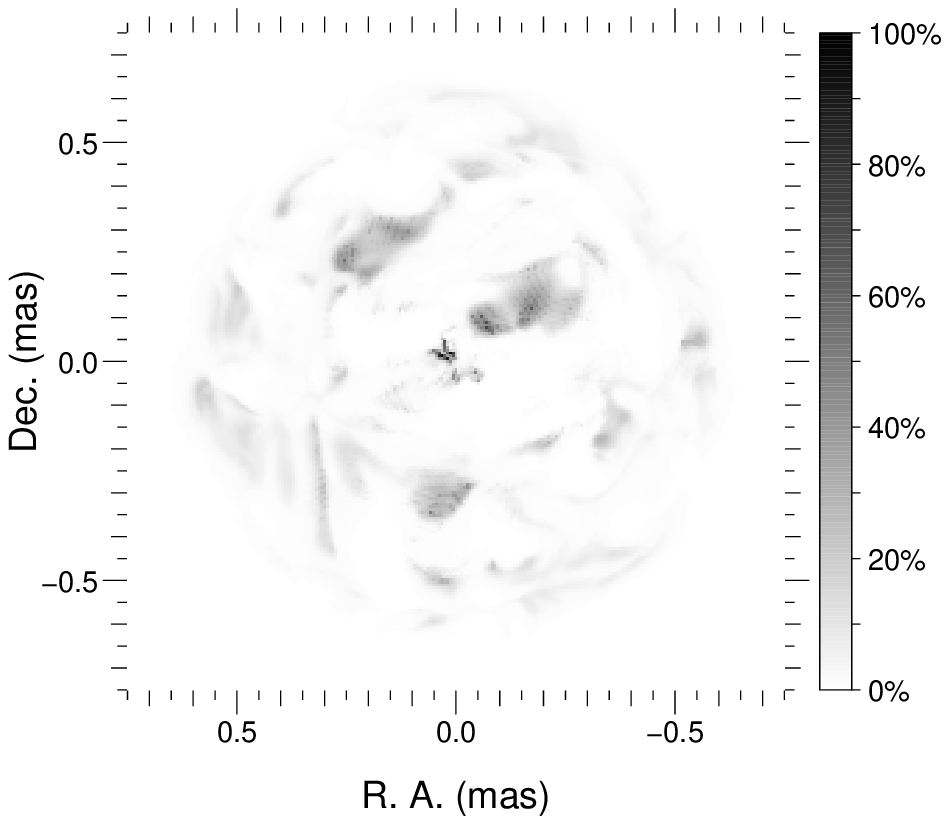}
\includegraphics[width=0.4\textwidth]{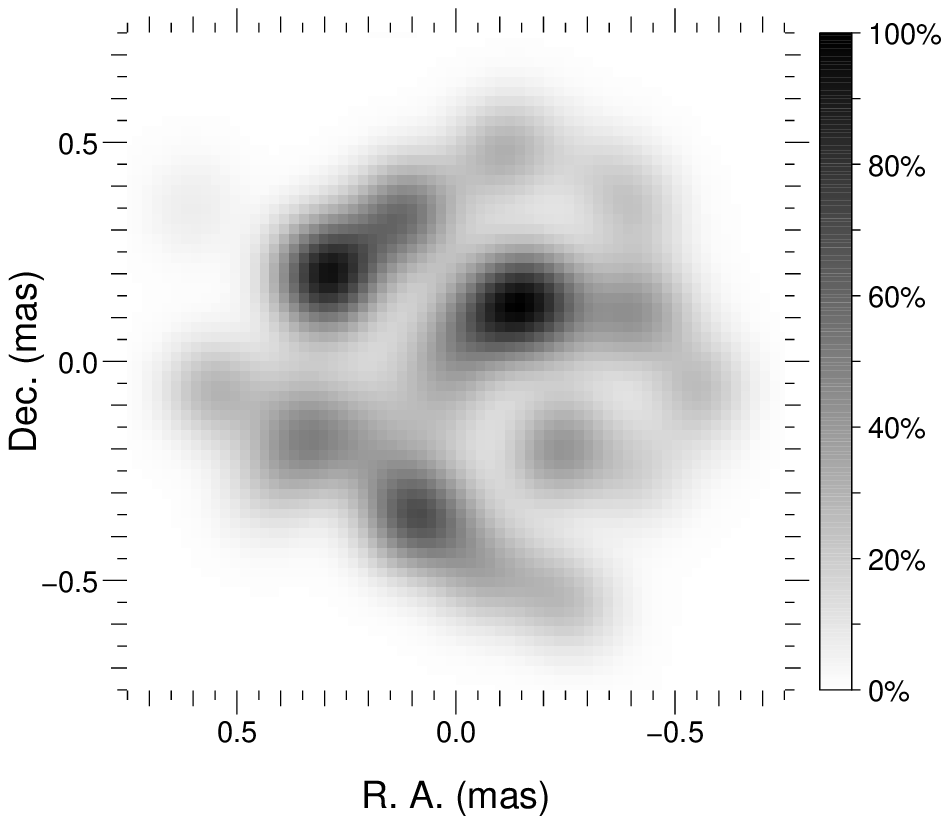}
\includegraphics[width=0.4\textwidth]{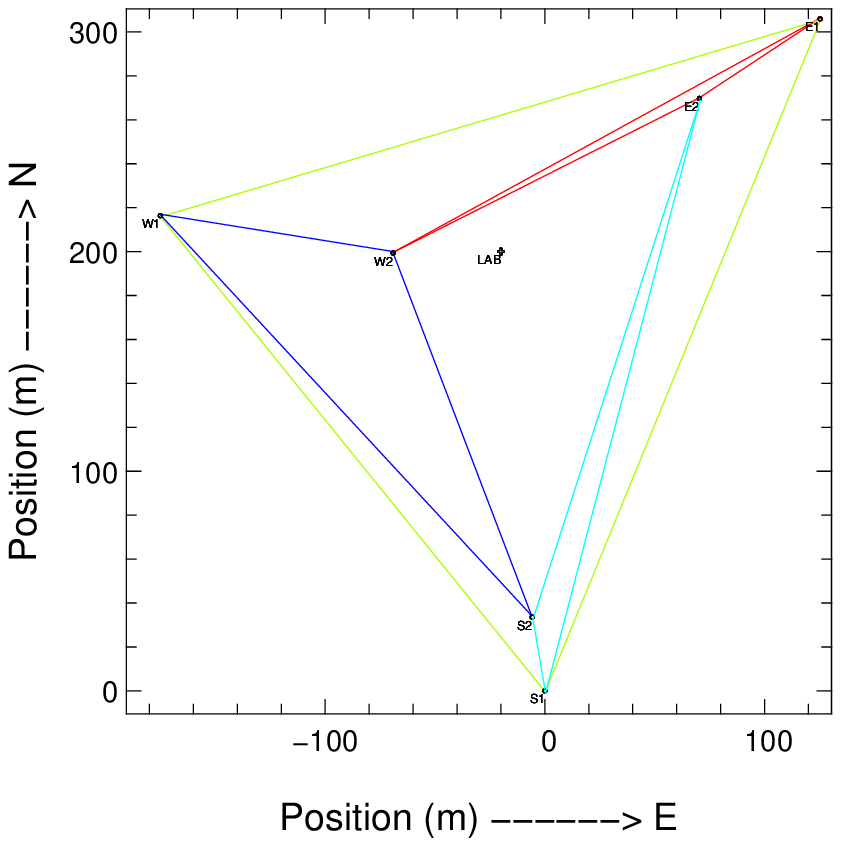}
\includegraphics[width=0.4\textwidth]{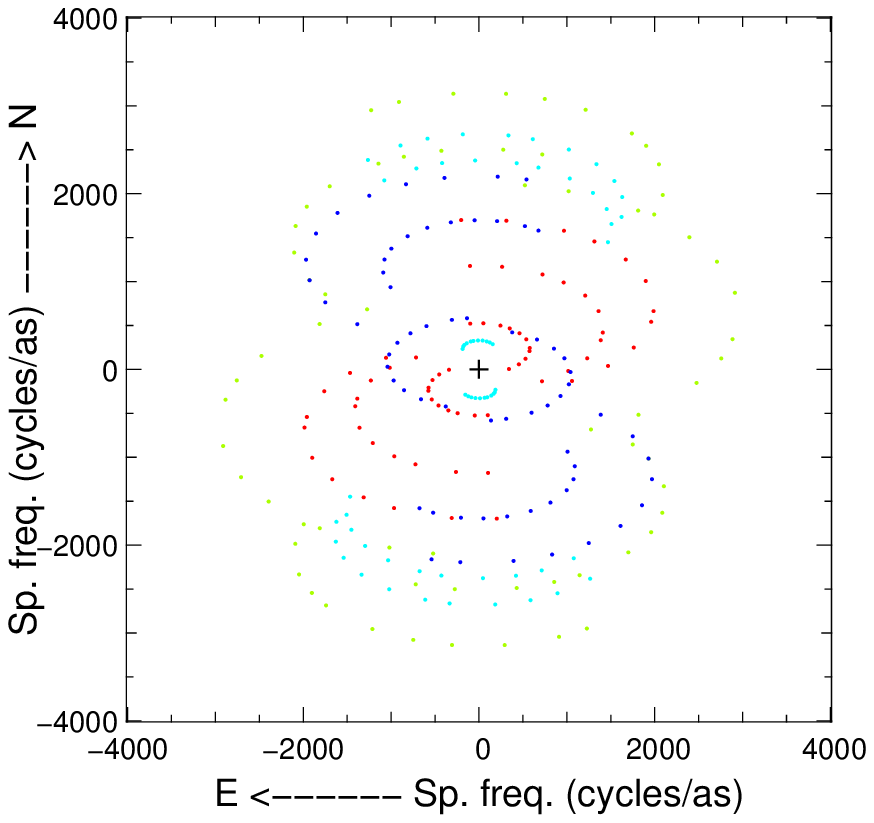}
\caption{\label{betelgeuse}(Top Left) Model of Betelgeuse in the visible (from A. Chiavassa) scaled to a diameter of 3 mas. (Top right) Image reconstruction of synthetic data made by simulating a visible instrument using the previous model with the CHARA array and four sets of 3 telescopes. (Bottom Left) The stations used for the simulation of a 3 telescope visible instrument at the CHARA array. (Bottom Right) (u,v) plane of the simulated observations.}
\end{figure}

  \section{The VLTI Array}
 The VLTI has larger telescopes but no visible beam combiner up to now. Adaptive optics will be available on the Auxiliary Telescopes (AT) thanks to the NAOMI project (Aller-Charpentier et al. 2014). The required Strehl ratio is 30 \% in the H band. Using Marechal relationship, stating that the Strehl ratio is approximately $S=exp({S_{f}}^2$) where S$_{f}$ is the phase residual after adaptive optics correction and holds for S$_{f}$ larger than $\sim$ 20\%, this Strehl ratio translates to 0.1\% at 700nm. This means that the height of the core of the image is approximatively the same as the halo. At this level, the use of Marechal approximation is questionable and this regime deserves numerical simulations to get reliable values. Adaptive optics is already available on the Unit Telescopes (UT). It is advertised for a 50\% Strehl ration in the K band, i.e. again 0.1\% at 700nm.\\
 
 The VLTI is not yet ready for visible light combination, and a few technical tweaks need to be implemented to enable UBV bands (RIJ do go through already):
 \begin{itemize}
 \item The MACAO dichroic would need to be changed and we are wondering if one would be able to select between different dichroic/semi-reflecting plates?
 \item The guiding strategy should also be modified.
  \end{itemize}

  \begin{figure}
\includegraphics[width=0.7\textwidth]{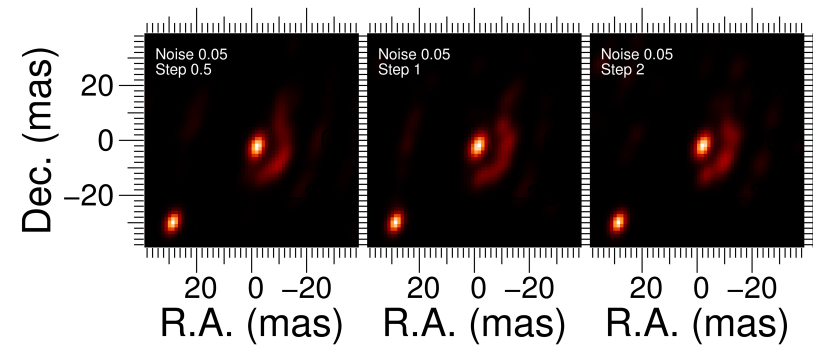}
\includegraphics[width=0.7\textwidth]{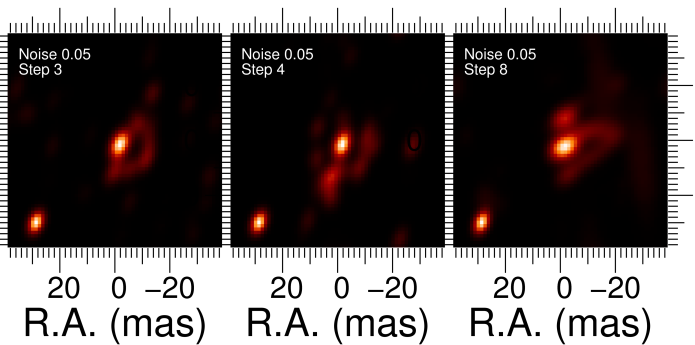}
\caption{Image reconstruction attempts by simulating a 3-nights observation of the above model with a given data noise level. The (u,v) plane filling is lowered progressively from top-left to bottom-right. The interferometric beam is displayed on the bottom-left corner of each image.}
\label{Be_uvplane}
\end{figure}
 
For the MACAO system, one possibility could be to use the infrared wavefront sensors of GRAVITY. We need to assess how would that improve visible Strehl ratio compared to a couple visible wave front sensor with dichroic or semi?reflecting plates.

\section{Polychromatic data}
There is already some existing tools thanks to the JMMC to prepare interferometric observations and make some model fitting such as ASPRO, LITPro and searchCal. On the other side, a new visible instrument will provide differential visibilities and differential phases, as do VEGA, AMBER and MIDI today but current tools (LITpro, MIRA, BSMEM) used squared visibilities and closure phases only. Some developments have occurred to take chromatic information into account, e.g. SPIDAST (Cruzalebes et al. 2013) or fitomatic (Millour et al. 2009, 2011a, 2011b). The POLCA project has successfully developed tools for handling polychromatic data. In image reconstruction, new algorithms are now able to reconstruct full 3D polychromatic maps from all available interferometric data (Soulez \& Thi\'ebaut 2013; Schutz, A. et al. 2014). Project POLCA has also allowed improvements in LITpro model fitting software, by including for instance VEGA differential visibilities taking nutrimental chromatic artifacts into account (Tallon et al. 2014).\\

Spectral resolution is also important for helping regularizations in image reconstruction, for instance when there is a large continuum with various spectral lines. PIONIER instrument was missing the spectral information which would have been better for data interpretation.  VEGA has spectroscopic capabilities but is limited due to saturation effects on the detector. Varying spectral resolution as a function of baseline length is a powerfull method in some hyperspectral remote sensing studies since high frequency information position is not going to change much depending on high frequency baseline. The spectrum of the object is also quite important for image reconstruction.

\section{Optimal number of telescopes to use and baselines length}
A possibility would be many telescopes with few delay lines that can be switched quickly wheres another option would be a continuously following of one baseline along time with a high frequency of recording. A minimum is certainly 4 telescopes and different configurations to make imaging. Experience at CHARA with MIRC shows that 6 telescopes is a strict minimum for snapshot imaging. In the same direction we
recall that the Plateau de Bure interferometer of IRAM was only able to make direct images when the telescope network was extended to 6 antennas  within 10 years and that actually the NOEMA project is extended the array to 12 antennas. Regarding the baselines length, short baseline are really mandatory for imaging purposes in order to fix features in the field of view. Longest baselines are not always usable due to SNR issues, and short baselines are needed to get the low frequencies. This pushes the need of additional telescopes. Often, the problem is that there is a gap between speckle data and interferometric data, and this is sometimes where most the information resides.

\section{Recombination scheme}
 PAVO uses holes for spatial filtering whereas VEGA uses speckle data. None of them is able to get closure phases and for VEGA the issue is coming from the detector. Nevertheless, we have learned for instance from MIRC that closure phases are very important for image reconstruction (ref ?). 

  \section{The FRIEND prototype}
 In the next 2 or 3 years, the two major interferometric arrays, VLTI and CHARA, will equip their telescopes of 1.8m and 1m respectively with Adaptive Optics (AO hereafter) systems. This improvement will permit to apply with a reasonable efficiency in the visible domain, the principle of spatial filtering with single mode fibers demonstrated in the near-infrared. It will clearly open new astrophysical fields by taking benefit of an improved sensitivity and state-of-the-art precision and accuracy on interferometric observables. To prepare this future possibility, we started the development of a demonstrator called FRIEND (Fibered and spectrally Resolved Interferometric Experiment - New Design). FRIEND \citep{berio14} combines the beams coming from 3 telescopes after injection in single mode optical fibers and provides some spectral capabilities for characterization purposes as well as photometric channels. It operates in the R spectral band (from $600nm$ to $750nm$) and uses the world's fastest and more sensitive analogic detector OCAM2.  \\
 
 From a technical point-of-view, the main goal of FRIEND is to validate spectrally-resolved interferometric observations in the visible in the case of partial correction by AO. In terms of astrophysics, a first and non-exhaustive analysis has confirmed the need to reach a 10th limiting magnitude at least in the case of low spectral resolution observations and to keep high spectral resolution capabilities. So, we decided to use the well-known multi-axial 'all-in-one' beam recombination scheme. The FRIEND performance strongly depends on the quality of the AO correction. We used  a simulator developped for the AO of CHARA to estimate the coupling into a single mode fiber at $\lambda=700nm$. We considered one seeing condition (effective $r_0=12cm$ and $t_0=10ms$ at $\lambda=0.5\mu m$ ) corresponding to the 80th percentile of the summer seeing at Mount Wilson. We considered also two conditions of AO corrections: {\it Tip-Tilt only} and {\it Full AO}. We found a mean coupling of $3\%$ and $25\%$ for the {\it Tip-Tilt only} and {\it Full AO} conditions, respectively.
The FRIEND performance has been derived at $R=2600$. 

\begin{figure}[h]
\begin{center}
\begin{tabular}{c}
\includegraphics[scale=0.5]{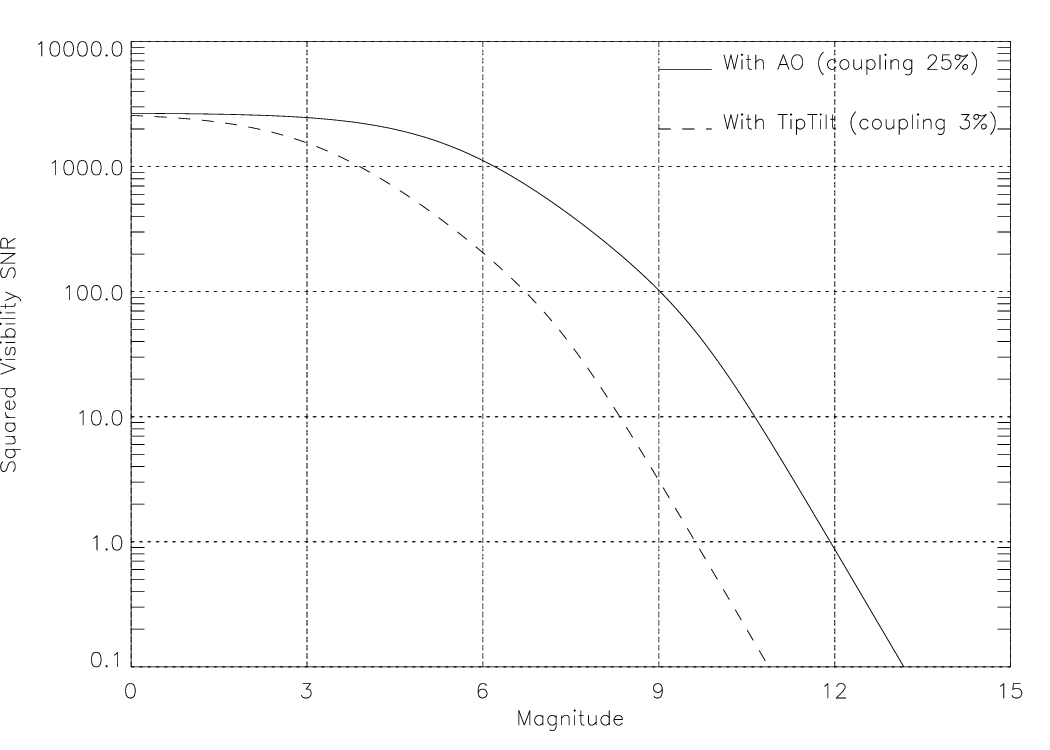}
\end{tabular}
\end{center}
\caption[]{
\label{fig9} Expected SNR of the squared visibility with FRIEND in the medium spectral resolution mode for an unresolved star. Used parameters: $\lambda=700nm$, $\Delta\lambda=30nm$, $R=2600$, $DIT=10ms$, $QE=90\%$, $M=30000$ (i.e. 5 minutes of observation), $V_{inst}=0.9$ and $V_*=1$.}
\end{figure} 

Figure \ref{fig9} shows the expected SNR of squared visibility with respect to the star magnitude for two conditions of AO correction in the case of an unresolved star.
From a theoretical point-of-view (i.e. considering photon and detector noise only), it clearly appears that a 10th limiting magnitude\footnote{we define the limiting magnitude as the magnitude corresponding to $SNR_{V^2}=10$} could be reached by averaging the visibility estimates over a spectral band of $30nm$ in the case of {\it Full AO} correction. This value is well within our specifications. In the case of {\it Tip-Tilt only} correction, the limiting magnitude could not exceed the 8th magnitude.

Tests on sky at the focus of the CHARA interferometer have been successfully done in December 2014. The real tests of FRIEND in AO correction conditions could not be done before the Phase II AO system where a fast deformable mirror is implemented to close the loop for {\it Full AO} correction. Meanwhile, FRIEND will remain installed in the focal laboratory of CHARA. Depending on the first tests and more especially on the data quality in the case of {\it Tip-Tilt only} correction, we plan to carry out dedicated astrophysical programs with FRIEND in 2015.

\chapter[AGNs]{AGNs\label{ch8}}

 \author{R.Petrov$^{1}$,
 S.~Rakshit$^{1}$
 E.~Martin$^{2}$,
M.~Kishimoto$^{3}$,
A.~Marconi$^{4}$,
A.~Meilland$^{1}$,
F.~Millour$^{1}$,
G.~Weigelt$^{5}$
 }

\begin{center} 
$^{1}$~Laboratoire Lagrange, UMR 7293, CNRS, Observatoire de la C\^ote d'Azur, Universit\'e de Nice Sophia-Antipolis,
Nice, France\\
$^{2}$~INTA-CSIC Centro de Astrobiolog\'ia, 28850 Torrej\'on de Ardoz, Madrid, Spain\\
$^{3}$~Department of Physics, Faculty of Science, Kyoto Sangyo University, Kamigamo-motoyama, Kita-ku, Kyoto 603-8555, Japan\\
$^{4}$~Dipartimento di Fisica e Astronomia, Universit\`a di Firenze, via Sansone 1, 50019 Sesto Fiorentino, Firenze, Italy\\
$^{5}$~Max-Planck-Institut f\"ur Radioastronomie Auf dem H\"ugel 69, 53121 Bonn, Germany\\
\end{center}

\underline{Foreword}: This text reflects discussions in late 2014 and early 2015 about the interest of a new interferometric instrument or a new interferometer operating in the visible for the study of Active Galactic Nuclei. However it has not been iterated among all participants and is substantially biased toward the personal views of the redactors 

\section{Active Galactic Nuclei}
Active Galactic Nuclei (AGN) are extremely bright sources powered by the accretion of material on a central super massive black hole (SMBH). They emit more than 1/5 of the electromagnetic power in the universe and a majority of galaxies might host a central BH triggering some level of nuclei activity. AGNs can be considered as important contributors and markers for the global history of mass accretion and galaxy evolution in the Universe. If well understood, they could be used as standard candles for the evaluation of cosmological distances at redshifts $z>3$. Quasars make the current reference grid for the calibration of GAIA astrometry but, at the tens of $\mu$as accuracy of GAIA, the structure of these sources could have a significant impact on the definition of their GAIA photocenter.
We have an unified model of AGNs (\citet{Antonucci1993}, \citet{Urry1995}) that is increasingly considered as over simplistic but still offers a simple and useful structure to discuss typical sizes and magnitudes and therefore to contribute to a first evaluation of the potential of visible long baseline interferometry. This model features a very compact accretion disk (AD) around the central SMBH, a Broad Line Region (BLR) composed of high velocity gas clouds producing broad emission lines when this central region is not obscured by the clumpy dust torus located after the dust sublimation radius. This dust torus (DT) collimates the light from the central source that can ionize some narrow line region (NLR) lower velocity gas clouds placed in the non-obscured double cone and producing narrow emission lines. When an AGN is close to equator-on, the dust torus shields the BLR, that can be detected only in polarized light reprocessed by the more far away NLR clouds. Some AGNs emit high velocity jets near their rotation axis. A strong radio emission can be present in a cone around the rotation axis and produce radio-loud AGNs if the line of sight is close enough to their axis.
In the following we discuss the possibility and the interest of observing some of these features with a long baseline interferometer operated in the visible.

\section{Accretion Disks}
Imaging accretion disks would be decisive to understand and model the accretion and jet launching mechanisms. However, the largest accretion disks are smaller than 1 microarcsecond and even simple evaluations of equivalent size, inclination and thickness would require baselines larger than 50 km for a visible interferometer. Actual imaging would need typically 250 km baselines for a resolution of 0.2 microarcseconds.  This seems quite unrealistic for ground-based interferometers.
However, the observation of the BLR and of the innermost parts of the dust torus can strongly constrain the spatial distribution of light emitted by the central accretion disk. The BLR, and maybe the inner NLR, can give accurate estimates of the mass of the BH. This should allow accurate modeling of the accretion disk.

\section{Broad Line Region}

The BLR is currently the best tracer of the characteristics of the central source and of the transport of material toward and away from the accretion disk. The spatial and the brightness distribution of BLR clouds are supposed to be dominated by the luminosity distribution of the central source and by its interaction with the inner parts of the dust torus. The kinematic of the BLR is supposed to be dominated by the mass of the central SMBH. Time resolved spectro-photometric techniques, called reverberation mapping (RM), provide estimates of the BLR and inner dust rim sizes, and constraints on the BLR morphology. RM is based on the study of the time needed for a feature in a broad emission line, or in the dust torus, to echo a variation of light near the central source. The time-velocity echo diagrams are claimed (Peterson, 1993) to be unique signatures of the BLR morphology, but they are often difficult to obtain with a sufficient quality for an image reconstruction or a full model fitting, mainly because a perfect RM observation should be continuous, well sampled and "unlimited" in time, which is seldom possible from ground based observations. Quite often, RM is limited to a typical lag $\tau_{\mathrm{RM}}$ between the continuum light curve and the global time variation of an emission line. The combination of this RM linear size $\tau_{\mathrm{RM}}$ with the line broadening $\Delta V$ attributed to the global velocity field yields a virial mass estimate:

\begin{equation}
M_{\mathrm{BH}}=f_M  (c \tau_{\mathrm{RM}} \Delta V^2)/G
\end{equation}

This is used to produce mass-luminosity curves for quasars with dispersion of the order of 0.44 dex, i.e. 275\% \citet{Kaspi2007}.

The typical BLR size $c\tau_{\mathrm{RM}}$ can is related to the luminosity $L$ of the AGN by:
\begin{equation}
R_{\mathrm{BLR}}=c\tau_{\mathrm{RM}}= f_L L^{1/2} 
\end{equation}

This yields size-luminosity laws, with a dispersion that can be reduced down to 0.19 dex, or even 0.13 dex, for low luminosity quasars (Bentz, 2013). Reducing the dispersion to 0.05 dex (12\%) for a larger range of luminosity would transform QSOs in decisive cosmological standard candles (Bentz, 2013), much more numerous than type I supernova and observable at least up to $z=3$.

The combination of the linear equivalent size\footnote{The concept of equivalent size depends on the geometry of the sources and some distributions have no characteristic size. Here the "equivalent sizes" refer to measures: the centroid of the cross-correlation between continuum and line light curves for $\tau_{\mathrm{RM}}$ and the measured visibility for $\Lambda$.} $c\tau_{\mathrm{RM}}$ with the equivalent angular size $\Lambda$  (defined as $\Lambda=\frac{\lambda}{B} \sqrt{1-V}$ from the visibility $V$ measured in the line for a partially unresolved source), yield a distance estimate:
\begin{equation}
D_{\mathrm{BLR}}=f_D \alpha \frac{(c\tau_{\mathrm{RM}})}{\Lambda} 
\end{equation}

where $\alpha$ is a unit conversion constant. This is the base of the quasar parallax direct distance measurement method proposed by \citep{Elvis2002}. This could be used to check and calibrate the RM Quasars distance measurements. Interferometric observations of $\Lambda$ are limited to a handful of targets with the 2nd generation of VLTI instruments GRAVITY and MATISSE. However, spectro-astrometry with an ELT would allow measuring equivalent angular size for QSOs in the $z=2$ to $z=3$ range, at least for the most likely geometry dominated by Keplerian rotation in a moderately open disk (Petrov 2013, 2015).

\begin{figure}
        \centering
       \includegraphics[width=10cm, height=7cm]{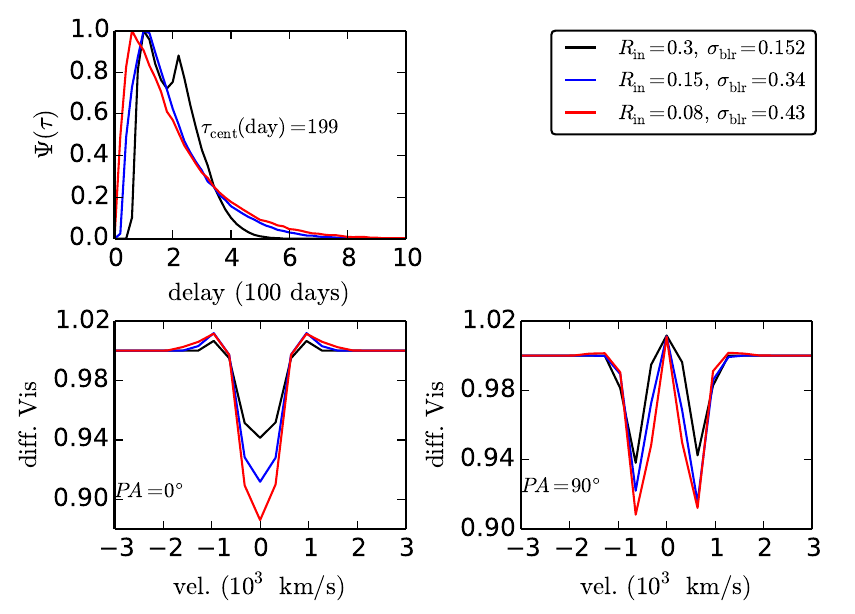}
       \caption{\small{RM 1D response function (upper panel) and visibility in the $\parallel$ (lower-left panel) and the $\perp$ (lower-right panel) baselines for different BLR geometries that produce same $\tau_{\mathrm{cent}}$ but different visibilities.}}\label{Fig:same_tcent}
  \end{figure}

The fudge factors $f_M$, $f_D$ and $f_L$ strongly depend from the geometry of the source. \citep{Rakshit2015} has shown that $f_M$ and $f_D$ can change by more than a factor 4 just with the inclination, opening angle and local-to-global velocity ratios without changing the RM measures $\tau_{\mathrm{RM}}$ and $\Delta V$. Obtaining spectrally resolved images of BLRs through emission lines would allow computing very accurately these fudge factors. Unfortunately, a real image of the largest BLRs, such as the 3C273 one, requires a resolution of at least 0.1 mas, i.e. a baseline of 5 km in the visible. Imaging an handful of very high luminosity BLRs would be decisive but with a limited impact on the general understanding of Quasar mechanism and on the calibration of the size-luminosity and mass luminosity laws that would make Quasars global cosmological tools. However, in \citep{Rakshit2015} we showed that measure of the visibility, differential visibility and differential phases, that are accessible on many quite unresolved sources constrain very strongly these fudge factors, typically from few hundreds \% with RM alone to better than 20\% from RM and optical interferometry together, if the optical interferometry data has a decent SNR, of at least 5. Figure \ref{Fig:same_tcent} shows different models produce same $\tau_{\mathrm{RM}}$ or same RM BLR size but interferometric differential visibility and phase signals show different signatures allowing to constrain BLR model having a few points in the emission line. Note that it is important to obtain all measures to fully constrain the model and strongly reduce the fudge factor uncertainty:

\begin{itemize}

\item The absolute visibility in the continuum yields the size of the inner rim of the dust torus and is used as a reference for differential observations in the line. The accuracy on absolute visibility is mostly limited by the accuracy of instrumental visibility calibration, which requires an active stabilization of the interferometric fringes. In that respect the availability of a fringe tracker operating up to magnitude 14 is mandatory to achieve the full VLTI potential with GRAVITY, MATISSE as well as with specialized AGN instruments such as OASIS\footnote{OASIS: Optimized AGN Spectro-Interferometric Sensor is a simple visitor instrument (or visitor module installed on AMBER) that optimizes the resolution, sampling and data processing for BLR observations. It would allow a gain in accuracy of about a factor 10 over GRAVITY, which is optimized for different science goals.}.

\item The differential visibility in the line yields the BLR size and intensity distribution, if the absolute visibility is well known, from absolute visibility or RM measurements. In the K band, the accuracy on the differential visibility is mostly limited by detector noise, and needs instruments designed to minimize the number of pixel read. The differential visibility is accessible on quite unresolved sources, but in K band, the typical limit to be able to measure a differential visibility is $K \simeq 11.5$ and only 10 to 20 targets are accessible with the VLTI even with the best instrument+FT combination.

\item The differential phase constrains the BLR size, geometry and kinematics, but is quite model dependent if measured alone, or combined only with RM data. It gives access to the highest super resolution factor and hence to the larger number of targets, up to 60 targets with the VLTI at its best.

\item A natural extension of the interferometric differential phase is the spectro-astrometry photocenter measurements, which measures photocenter variations with wavelength just as the differential phase. With an ELT, we should be able to measure BLR photocenter displacements up to z=3 for high luminosity QSOs similar to 3C273 or z=2 for lower luminosity targets such as 3C120 (Petrov, 2013, 2015). This needs to control the instrument systematics down to 1/1000 pixels, which has been shown possible by the laboratory experiment DIAMS (Abe, 2012).

\end{itemize}

The current strategy to observe AGNs in the near and mid infrared with the VLTI, which is currently the single optical interferometer allowing a significant program in that domain can be very roughly sketched as follows:

\begin{itemize}

\item MATISSE/VLTI will provide images for 2 to 4 of AGNs, providing a better understanding of the dust torus structure. In addition, it will provide equivalent sizes and global structural parameters for maybe two dozens of targets. This assumes an external FT operating at least at K=13. MATISSE will also provide constraints on the torus geometry and size for all targets for which we try to obtain BLR information.

\item MATISSE and GRAVITY will provide torus equivalent sizes and global structural parameters for maybe two dozens targets. The angular torus size can be combined with IR reverberation mapping to obtain "dust parallax" distance measurements \citet{Hoenig2014}. This is easy in the K band, because the K band structure seems simpler, but for a number or targets strongly limited by the VLTI resolution in the K band. The L, M and N bands would give access to more targets but would require more modeling. These distance measurements will complement, check and calibrate the “BLR parallax” measurements discussed next.

\item GRAVITY, and to a lesser extent MATISSE will make differential observations of BLRs up to K$\sim$10.5 (GRAVITY FT), which represents about 10 targets. With an external FT, this number goes up to 20 targets with GRAVITY and 60 targets with an optimized module for BLR observations in the K band such as OASIS. 

\item The AMBER+ observations of 3C273 (\citep{Petrov2012}, \citep{Rakshit2015}) showed that the BLR extends beyond the inner rim of the dust torus. Thus GRAVITY should be able to image the gas in or above the dust torus for a few objects.

\item All VLTI observations of AGNs should be complemented by RM measures.

\item This might be enough to try finding the relationship between the RM fudge factors and spectro-photometry measures such as the luminosity and the line profile. However, with GRAVITY alone, we cover a small range of luminosities. The situation improves with an external FT and improves even further with the specialized module OASIS. The accuracy of distance and mass measurements will be good only for a dozen targets, allowing poor calibration of the size-luminosity and mass-luminosity laws.

\end{itemize}

Observations in the visible would bring a decisive improvement, if the interferometer can be made sensitive enough in that spectral band. First we will gain in resolution by a factor 4. Second we will be allowed to use Balmer lines (for low z sources) that are 3 to 20 times stronger than the Paschen and Bracket lines available in the near IR. The combination of these two effects would yield a significant gain in the accuracy of quasar parallax distance measurements, as well as in the other parameters measures such as the masses. It will also strongly enhance the possibility to image gas in direct relation with the dust torus.

\begin{figure}
        \centering
       \includegraphics[width=6cm, height=5cm]{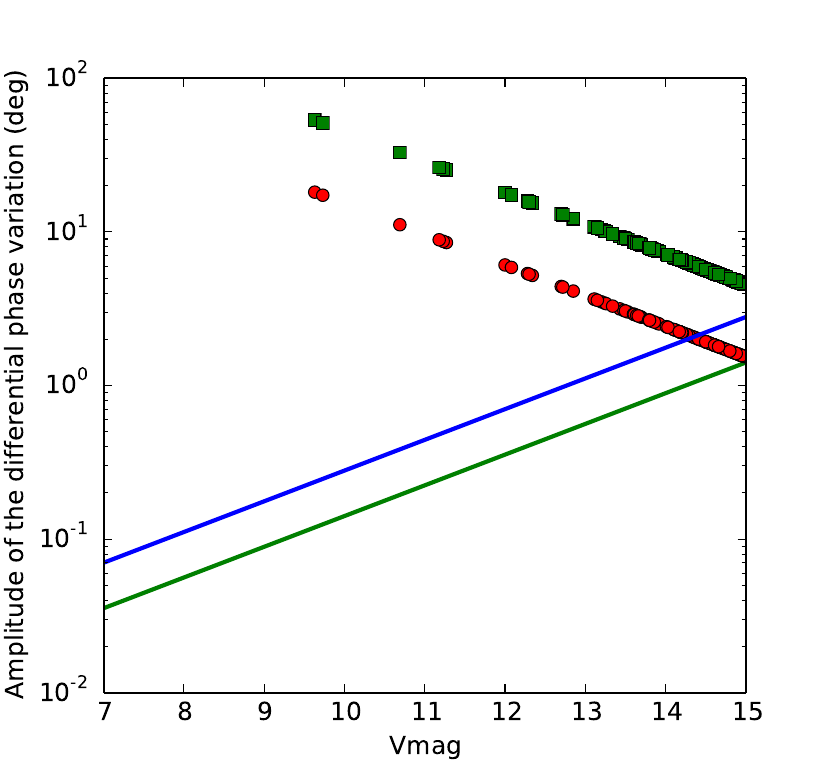}
       \includegraphics[width=6cm, height=5cm]{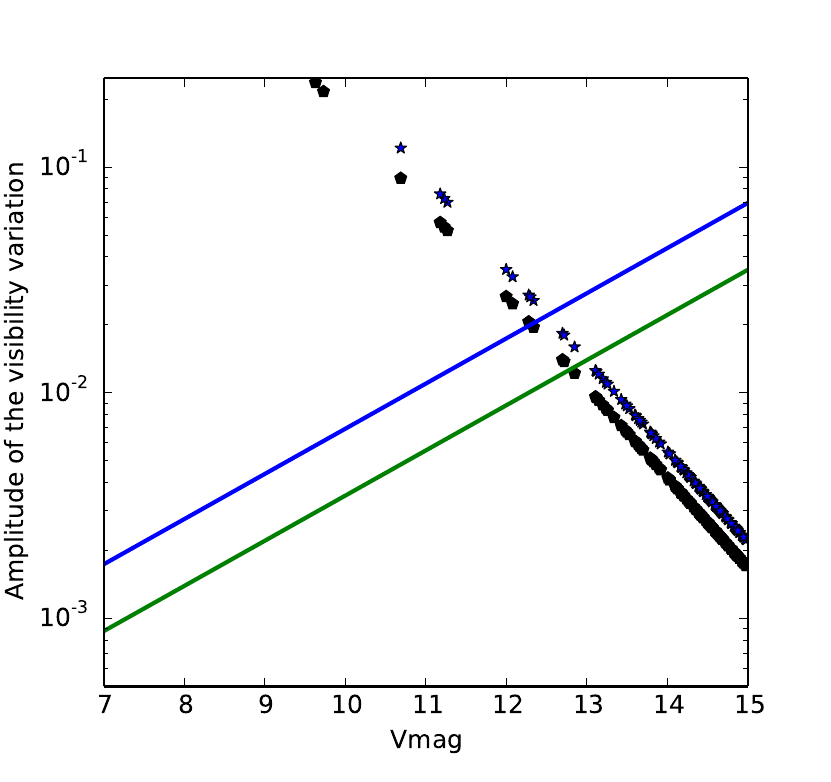}
       \caption[Feasibility in visible]{\small{Feasibility of BLR observation with a visible mode of VLTI (considering H$\alpha$ line). The solid curves represent differential phase (left) and differential visibility (right) for a visible mode with UTs (green) and ATs (blue). Each symbol represents one Sy1 AGN observable at Paranal. UTs at VLTI with Strehl ratio 0.1 could reach up to $V=15$ allowing to observe more than 130 targets, while ATs could reach up to $V=14$ if we have Strehl ratio 0.5 in the visible. Credit: \citet{Rakshit2015}.}}\label{Fig:VISIBLE}
  \end{figure}

Another major advantage of visible observations is to allow a direct combination with RM observations that have so far been made almost exclusively in the visible, without the need to use models to correct the relative scales in the visible and in the IR, nor having to launch very long term IR RM campaigns.
The condition for visible observation of BLRs with the VLTI to give a significant contribution is to be able to observe with a sufficient limiting magnitude in V. Figure \ref{Fig:VISIBLE} shows a very preliminary results of a feasibility study of BLR observation in visible using VLTI from \citet{Rakshit2015}, extrapolated from his work for the K band. It shows that:

\begin{itemize}

\item At V=14, we would obtain visibility, differential visibility and differential phase for all the GRAVITY targets (without FT), but with an improved distance accuracy by typically a factor 10. Such a magnitude could be achieved with the ATs if they have adaptive optics system providing a Strehl ratio of 0.5 in the (red) visible. In a few cases we will obtain images of the gas in or above the dust torus.

\item At V=15, we will be able to observe more than 120 quasars. For all K band VLTI targets we should have substantial distance and mass accuracy improvements. The exact gains need to be assessed but the very first estimate is that we would have more than 50 targets with Quasars parallax more accurate than 10\%. Such a magnitude can be achieved on the UTs if they are equipped with an AO system providing a Strehl ratio of 10\% in the red visible.

\end{itemize}

\section{Narrow Line Region}

In radio-quiet AGN, the NLR is the most extended structure associated with the black hole's mass accretion and ejection mechanisms. It ranges from kiloparsec scales well below down the resolution limit of even the largest optical telescopes. 

The emission in the NLR is dominated by high-ionization optical lines. The most prominent line with the largest equivalent width is the [OIII] line at 5007 $\AA$. It is widely used as a tracer for the radiative power output of the AGN. Interestingly, the NLR gas as traced by the emission lines is predominantly outflowing with velocities of several 100 km/s, thus providing direct energy feedback from the AGN to the galaxy. The emission lines are rather well understood in terms of photo-ionization of gas clouds by the central AGN. Indeed, it has been suggested that the sizes of the NLR structure in different galaxies scales with the square root of the AGN luminosity in a similar way as the BLR and torus sizes. This indicates rather simple, radiation-driven physics that lead to the observed emission.

While the NLR is clearly extended, the dominant part of the flux is originating from an unresolved region even in nearby objects such as NGC1068 or Circinus. Therefore, we currently do not know the physical condition that lead to the launch of the outflow associated with the NLR. Possibilities range from association to hydrodynamic winds in the accretion disk to evaporating clouds in the BLR or torus region. Pinning down the origin provides access to constrain density, ionization parameter, and temperature in the launching region. It will also allow us to determine the origin of the outflowing gas and its connection to the other structures in the AGN unification scheme.

The resolution requirements on an optical interferometer to study the NLR are modest. In terms of spatial resolution it has similar requirements as for studying the BLR. The degree of resolved emission, and its corresponding density and mass, will make it possible to locate the launching region. The width of typical narrow emission lines from the nuclear source dictate the required spectral resolution. Taking the [OIII] line as reference and assuming typical line widths of 300-500 km/s, a spatial resolution in the V-band range of $3000-5000$ will offer enough resolution to map the kinematic structure on sub-parsec scales. So, as for the BLR, the key condition is to achieve a sensitivity in the V=13 to V=15 range. For V between 12 and 15, the main target being the $13-14$ range.

\section{Dust Torus}

The innermost parts of the dust torus are targets for MATISSE in the L, M and N bands and to a lesser extent for GRAVITY in the K band continuum. The contribution of the torus to the visible continuum is expected to be faint, but has not been actually evaluated. In particular the BLR and torus gas could have a contribution to the visible continuum. This point still has to be investigated. The potential benefit is the gain in resolution as well as constrains on the physical parameters of the gas.
On the other hand we have already underlined the complementarity between gas observations, that are particularly interesting in the visible and torus observations, as these features interact and are both shaped by the central source intensity distribution. Thus it is highly desirable that all targets observed by MATISSE and GRAVITY be also observed in the visible.
We have already discussed the "dust parallax" distance measurements that would be a decisive corner stone to Quasar distance measurements \citet{Hoenig2014}. A significant torus contribution in the visible continuum could extend the domain of that technique.

\section{Specifications of Visible Interferometry for AGN Science}

\begin{tabular}{|p{2cm}|p{8cm}|p{4cm}|}\hline
Facility &	Application	 & Condition \\
\hline
VLTI (B=135-200 m) & &	\\	\hline
$V=13$ &	A few NLR observations. Imaging the gas near the torus for a few sources.	& Strehl$>$0.2 in V on the ATs \\ \hline
$V=14$ &	BLR and NLR: Observing the GRAVITY-MATISSE (without external FT) targets. Fudge factors constraints and improved accuracy on mass and distance measurements on 10-20 targets	& Strehl$>$0.5 in V on the ATs\\ \hline
$V=15$ &	Fudge factors, mass and distance measurements on $>$120 targets in a large luminosity range. Grand unification based on fudge factor constrained by spectro-photometric measures. Critical to calibrate RM and RM+spectro-astrometry mass and distance measurements. & 	Strehl$>$0.1 in V on the UTs \\ \hline
New generation interferometer, PFI
(B$>$10km) & 		& \\ \hline
$V=13$ &	Imaging BLRs &	Strehl$>$0.2 2m class telescope or Strehl$>$0.8 1m class telescope\\ \hline

$V=14$ &	Imaging BLRs &	Strehl$>$0.5 2m class telescope\\ \hline

$V=15$ &	Very accurate and distance measurements up to z~2 with much less model dependence than from RM alone or RM+spectro-astrometry	& Strehl$>$0.1 8m class telescope
Strehl$>$0.4 4m class telescope\\ \hline

E-ELT spectro-astrometry  &	RM+spectro astrometry yield direct distance measurements up to z=3 (3C273 L) and z=2 (3C120 L), IF spectro-interferometric measures allowed constraining the fudge factors as a function of luminosity and line width. &	Spectro-astrometry systematics controlled down to 1/1000 pixel \\ \hline

\end{tabular}

\bibliographystyle{ws-book-har}    % Bibliography: Author-Date system
\bibliography{visi_interf}      % pls. call your database here

\begin{thebibliography}{22}
\expandafter\ifx\csname natexlab\endcsname\relax\def\natexlab#1{#1}\fi

\bibitem[{{Balona} \& {Stobie}(1979{\natexlab{a}})}]{balona79c}
{Balona}, L.~A. \& {Stobie}, R.~S. 1979{\natexlab{a}}, \mnras, 187, 217

\bibitem[{{Balona} \& {Stobie}(1979{\natexlab{b}})}]{balona79b}
{Balona}, L.~A. \& {Stobie}, R.~S. 1979{\natexlab{b}}, \mnras, 189, 649

\bibitem[{{Balona} \& {Stobie}(1979{\natexlab{c}})}]{balona79a}
{Balona}, L.~A. \& {Stobie}, R.~S. 1979{\natexlab{c}}, \mnras, 189, 659

\bibitem[{{Burki} {et~al.}(1982){Burki}, {Mayor}, \& {Benz}}]{burki82}
{Burki}, G., {Mayor}, M., \& {Benz}, W. 1982, \aap, 109, 258

\bibitem[{{Davis} {et~al.}(2009){Davis}, {Jacob}, {Robertson}, {Ireland},
  {North}, {Tango}, \& {Tuthill}}]{davis09}
{Davis}, J., {Jacob}, A.~P., {Robertson}, J.~G., {et~al.} 2009, \mnras, 394,
  1620

\bibitem[{{Fernley}(1994)}]{fernley94}
{Fernley}, J. 1994, \aap, 284, L16

\bibitem[{{Fouqu{\'e}} {et~al.}(2007){Fouqu{\'e}}, {Arriagada}, {Storm},
  {Barnes}, {Nardetto}, {M{\'e}rand}, {Kervella}, {Gieren}, {Bersier},
  {Benedict}, \& {McArthur}}]{fouque07}
{Fouqu{\'e}}, P., {Arriagada}, P., {Storm}, J., {et~al.} 2007, \aap, 476, 73

\bibitem[{{Gieren} {et~al.}(2005){Gieren}, {Storm}, {Barnes}, {Fouqu{\'e}},
  {Pietrzy{\'n}ski}, \& {Kienzle}}]{gieren05}
{Gieren}, W., {Storm}, J., {Barnes}, III, T.~G., {et~al.} 2005, \apj, 627, 224

\bibitem[{{Guiglion} {et~al.}(2013){Guiglion}, {Nardetto}, {Mathias},
  {Domiciano de Souza}, {Poretti}, {Rainer}, {Fokin}, {Mourard}, \&
  {Gieren}}]{guiglion13}
{Guiglion}, G., {Nardetto}, N., {Mathias}, P., {et~al.} 2013, \aap, 550, L10

\bibitem[{{Kervella} {et~al.}(2004){Kervella}, {Nardetto}, {Bersier},
  {Mourard}, \& {Coud{\'e} du Foresto}}]{kervella04a}
{Kervella}, P., {Nardetto}, N., {Bersier}, D., {Mourard}, D., \& {Coud{\'e} du
  Foresto}, V. 2004, \aap, 416, 941

\bibitem[{{Milone} {et~al.}(1994){Milone}, {Wilson}, {Fry}, \&
  {Schiller}}]{milone94}
{Milone}, E.~F., {Wilson}, W.~J.~F., {Fry}, D.~J.~I., \& {Schiller}, S.~J.
  1994, \pasp, 106, 1120

\bibitem[{{Nardetto} {et~al.}(2004){Nardetto}, {Fokin}, {Mourard}, {Mathias},
  {Kervella}, \& {Bersier}}]{nardetto04}
{Nardetto}, N., {Fokin}, A., {Mourard}, D., {et~al.} 2004, \aap, 428, 131

\bibitem[{{Nardetto} {et~al.}(2009){Nardetto}, {Gieren}, {Kervella},
  {Fouqu{\'e}}, {Storm}, {Pietrzynski}, {Mourard}, \& {Queloz}}]{nardetto09}
{Nardetto}, N., {Gieren}, W., {Kervella}, P., {et~al.} 2009, \aap, 502, 951

\bibitem[{{Nardetto} {et~al.}(2007){Nardetto}, {Mourard}, {Mathias}, {Fokin},
  \& {Gillet}}]{nardetto07}
{Nardetto}, N., {Mourard}, D., {Mathias}, P., {Fokin}, A., \& {Gillet}, D.
  2007, \aap, 471, 661

\bibitem[{{Nardetto} {et~al.}(2014){Nardetto}, {Poretti}, {Rainer}, {Guiglion},
  {Scardia}, {Schmid}, \& {Mathias}}]{nardetto14}
{Nardetto}, N., {Poretti}, E., {Rainer}, M., {et~al.} 2014, \aap, 561, A151

\bibitem[{{Petersen} \& {Hog}(1998)}]{petersen98}
{Petersen}, J.~O. \& {Hog}, E. 1998, \aap, 331, 989

\bibitem[{{Riess} {et~al.}(2011){Riess}, {Macri}, {Casertano}, {Lampeitl},
  {Ferguson}, {Filippenko}, {Jha}, {Li}, \& {Chornock}}]{riess11}
{Riess}, A.~G., {Macri}, L., {Casertano}, S., {et~al.} 2011, \apj, 730, 119

\bibitem[{{Stamford} \& {Watson}(1981)}]{stamford81}
{Stamford}, P.~A. \& {Watson}, R.~D. 1981, \apss, 77, 131

\bibitem[{{Storm} {et~al.}(2011{\natexlab{a}}){Storm}, {Gieren}, {Fouqu{\'e}},
  {Barnes}, {Pietrzy{\'n}ski}, {Nardetto}, {Weber}, {Granzer}, \&
  {Strassmeier}}]{storm11a}
{Storm}, J., {Gieren}, W., {Fouqu{\'e}}, P., {et~al.} 2011{\natexlab{a}}, \aap,
  534, A94

\bibitem[{{Storm} {et~al.}(2011{\natexlab{b}}){Storm}, {Gieren}, {Fouqu{\'e}},
  {Barnes}, {Soszy{\'n}ski}, {Pietrzy{\'n}ski}, {Nardetto}, \&
  {Queloz}}]{storm11b}
{Storm}, J., {Gieren}, W., {Fouqu{\'e}}, P., {et~al.} 2011{\natexlab{b}}, \aap,
  534, A95

\bibitem[{{Wilson} {et~al.}(1993){Wilson}, {Milone}, \& {Fry}}]{wilson93}
{Wilson}, W.~J.~F., {Milone}, E.~F., \& {Fry}, D.~J.~I. 1993, \pasp, 105, 809

\bibitem[{{Wilson} {et~al.}(1998){Wilson}, {Milone}, {Fry}, \& {van
  Leeuwen}}]{wilson98}
{Wilson}, W.~J.~F., {Milone}, E.~F., {Fry}, D.~J.~I., \& {van Leeuwen}, J.
  1998, \pasp, 110, 433

\end{thebibliography}


\begin{thebibliography}{216}
\newcommand{\enquote}[1]{#1}
\providecommand{\natexlab}[1]{#1}
\providecommand{\url}[1]{\texttt{#1}}
\providecommand{\urlprefix}{URL }
\providecommand{\eprint}{eprint }
\expandafter\ifx\csname urlstyle\endcsname\relax
  \providecommand{\doi}[1]{doi:\discretionary{}{}{}#1}\else
  \providecommand{\doi}{doi:\discretionary{}{}{}\begingroup
  \urlstyle{rm}\Url}\fi

\bibitem[{{Antonucci}(1993)}]{Antonucci1993}
{Antonucci}, R. (1993). \enquote{{Unified models for active galactic nuclei and
  quasars},} \emph{\araa} \textbf{31}, pp. 473--521,
  \doi{10.1146/annurev.aa.31.090193.002353}.

\bibitem[{{Auri{\`e}re} \emph{et~al.}(2010){Auri{\`e}re}, {Donati},
  {Konstantinova-Antova}, {Perrin}, {Petit} and
  {Roudier}}]{2010A&A...516L...2A}
{Auri{\`e}re}, M., {Donati}, J., {Konstantinova-Antova}, R., {Perrin}, G.,
  {Petit}, P.,  and {Roudier}, T. (2010). \enquote{{The magnetic field of
  Betelgeuse: a local dynamo from giant convection cells?}} \emph{\aap}
  \textbf{516}, p.~L2, \doi{10.1051/0004-6361/201014925}, \eprint{1005.4845}.

\bibitem[{{Babcock}(1960)}]{1960ApJ...132..521B}
{Babcock}, H.~W. (1960). \enquote{{The 34-KILOGAUSS Magnetic Field of HD
  215441.}} \emph{\apj} \textbf{132}, p. 521, \doi{10.1086/146960}.

\bibitem[{{Baines} \emph{et~al.}(2014){Baines}, {Armstrong}, {Schmitt},
  {Benson}, {Zavala} and {van Belle}}]{Baines2014}
{Baines}, E.~K., {Armstrong}, J.~T., {Schmitt}, H.~R., {Benson}, J.~A.,
  {Zavala}, R.~T.,  and {van Belle}, G.~T. (2014). \enquote{{Navy Precision
  Optical Interferometer Measurements of 10 Stellar Oscillators},} \emph{\apj}
  \textbf{781}, 90, \doi{10.1088/0004-637X/781/2/90}.

\bibitem[{{Balick} and {Frank}(2002)}]{2002ARAABalick}
{Balick}, B. and {Frank}, A. (2002). \enquote{{Shapes and Shaping of Planetary
  Nebulae},} \emph{\araa} \textbf{40}, pp. 439--486,
  \doi{10.1146/annurev.astro.40.060401.093849}.

\bibitem[{{Balona} and {Stobie}(1979{\natexlab{a}})}]{balona79c}
{Balona}, L.~A. and {Stobie}, R.~S. (1979{\natexlab{a}}). \enquote{{Application
  of the Wesselink method to a non-radially oscillating star},} \emph{\mnras}
  \textbf{187}, pp. 217--222.

\bibitem[{{Balona} and {Stobie}(1979{\natexlab{b}})}]{balona79b}
{Balona}, L.~A. and {Stobie}, R.~S. (1979{\natexlab{b}}). \enquote{{The effect
  of radial and non-radial stellar oscillations on the light, colour and
  velocity variations},} \emph{\mnras} \textbf{189}, pp. 649--658.

\bibitem[{{Balona} and {Stobie}(1979{\natexlab{c}})}]{balona79a}
{Balona}, L.~A. and {Stobie}, R.~S. (1979{\natexlab{c}}). \enquote{{Wesselink
  radii of double-mode cepheids},} \emph{\mnras} \textbf{189}, pp. 659--666.

\bibitem[{{Benisty} \emph{et~al.}(2013){Benisty}, {Perraut}, {Mourard}, {Stee},
  {Lima}, {Le Bouquin}, {Borges Fernandes}, {Chesneau}, {Nardetto},
  {Tallon-Bosc}, {McAlister}, {Ten Brummelaar}, {Ridgway}, {Sturmann},
  {Sturmann}, {Turner}, {Farrington} and {Goldfinger}}]{Benisty}
{Benisty}, M., {Perraut}, K., {Mourard}, D., {Stee}, P., {Lima}, G.~H.~R.~A.,
  {Le Bouquin}, J.~B., {Borges Fernandes}, M., {Chesneau}, O., {Nardetto}, N.,
  {Tallon-Bosc}, I., {McAlister}, H., {Ten Brummelaar}, T., {Ridgway}, S.,
  {Sturmann}, J., {Sturmann}, L., {Turner}, N., {Farrington}, C.,  and
  {Goldfinger}, P.~J. (2013). \enquote{{Enhanced H$_{α}$ activity at
  periastron in the young and massive spectroscopic binary HD 200775},}
  \emph{\aap} \textbf{555}, A113, \doi{10.1051/0004-6361/201219893},
  \eprint{1306.0390}.

\bibitem[{{Berdyugina}(2005)}]{Berdyugina2005}
{Berdyugina}, S.~V. (2005). \enquote{{Starspots: A Key to the Stellar Dynamo},}
  \emph{Living Reviews in Solar Physics} \textbf{2}, p.~8,
  \doi{10.12942/lrsp-2005-8}.

\bibitem[{{Berio} \emph{et~al.}(2014{\natexlab{a}}){Berio}, {Bresson},
  {Clausse}, {Mourard}, {Dejonghe}, {Duthu}, {Lagarde}, {Meilland}, {Perraut},
  {Tallon-Bosc}, {Nardetto}, {Spang}, {Bailet}, {Marcotto}, {Chesneau}, {Stee},
  {Feautrier}, {Balard} and {Gach}}]{berio2014}
{Berio}, P., {Bresson}, Y., {Clausse}, J.~M., {Mourard}, D., {Dejonghe}, J.,
  {Duthu}, A., {Lagarde}, S., {Meilland}, A., {Perraut}, K., {Tallon-Bosc}, I.,
  {Nardetto}, N., {Spang}, A., {Bailet}, C., {Marcotto}, A., {Chesneau}, O.,
  {Stee}, P., {Feautrier}, P., {Balard}, P.,  and {Gach}, J.~L.
  (2014{\natexlab{a}}). \enquote{{Long baseline interferometry in the visible:
  the FRIEND project},} in \emph{Society of Photo-Optical Instrumentation
  Engineers (SPIE) Conference Series}, \emph{Society of Photo-Optical
  Instrumentation Engineers (SPIE) Conference Series}, Vol. 9146, p.~16,
  \doi{10.1117/12.2054890}.

\bibitem[{{Berio} \emph{et~al.}(2014{\natexlab{b}}){Berio}, {Bresson},
  {Clausse}, {Mourard}, {Dejonghe}, {Duthu}, {Lagarde}, {Meilland}, {Perraut},
  {Tallon-Bosc}, {Nardetto}, {Spang}, {Bailet}, {Marcotto}, {Chesneau}, {Stee},
  {Feautrier}, {Balard} and {Gach}}]{berio14}
{Berio}, P., {Bresson}, Y., {Clausse}, J.~M., {Mourard}, D., {Dejonghe}, J.,
  {Duthu}, A., {Lagarde}, S., {Meilland}, A., {Perraut}, K., {Tallon-Bosc}, I.,
  {Nardetto}, N., {Spang}, A., {Bailet}, C., {Marcotto}, A., {Chesneau}, O.,
  {Stee}, P., {Feautrier}, P., {Balard}, P.,  and {Gach}, J.~L.
  (2014{\natexlab{b}}). \enquote{{Long baseline interferometry in the visible:
  the FRIEND project},} in \emph{Society of Photo-Optical Instrumentation
  Engineers (SPIE) Conference Series}, \emph{Society of Photo-Optical
  Instrumentation Engineers (SPIE) Conference Series}, Vol. 9146, p.~16,
  \doi{10.1117/12.2054890}.

\bibitem[{{Berio} \emph{et~al.}(2011){Berio}, {Merle}, {Th{\'e}venin},
  {Bonneau}, {Mourard}, {Chesneau}, {Delaa}, {Ligi}, {Nardetto}, {Perraut},
  {Pichon}, {Stee}, {Tallon-Bosc}, {Clausse}, {Spang}, {McAlister}, {ten
  Brummelaar}, {Sturmann}, {Sturmann}, {Turner}, {Farrington} and
  {Goldfinger}}]{berio2011}
{Berio}, P., {Merle}, T., {Th{\'e}venin}, F., {Bonneau}, D., {Mourard}, D.,
  {Chesneau}, O., {Delaa}, O., {Ligi}, R., {Nardetto}, N., {Perraut}, K.,
  {Pichon}, B., {Stee}, P., {Tallon-Bosc}, I., {Clausse}, J.~M., {Spang}, A.,
  {McAlister}, H., {ten Brummelaar}, T., {Sturmann}, J., {Sturmann}, L.,
  {Turner}, N., {Farrington}, C.,  and {Goldfinger}, P.~J. (2011).
  \enquote{{Chromosphere of K giant stars. Geometrical extent and spatial
  structure detection},} \emph{\aap} \textbf{535}, A59,
  \doi{10.1051/0004-6361/201117479}, \eprint{1109.5476}.

\bibitem[{{Boffin} \emph{et~al.}(2014{\natexlab{a}}){Boffin}, {Hillen},
  {Berger}, {Jorissen}, {Blind}, {Le Bouquin}, {Miko{\l}ajewska} and
  {Lazareff}}]{Boffin2014a}
{Boffin}, H.~M.~J., {Hillen}, M., {Berger}, J.~P., {Jorissen}, A., {Blind}, N.,
  {Le Bouquin}, J.~B., {Miko{\l}ajewska}, J.,  and {Lazareff}, B.
  (2014{\natexlab{a}}). \enquote{{Roche-lobe filling factor of
  mass-transferring red giants: the PIONIER view},} \emph{\aap} \textbf{564},
  A1, \doi{10.1051/0004-6361/201323194}, \eprint{1402.1798}.

\bibitem[{{Boffin} \emph{et~al.}(2014{\natexlab{b}}){Boffin}, {Pourbaix},
  {Mu{\v z}i{\'c}}, {Ivanov}, {Kurtev}, {Beletsky}, {Mehner}, {Berger},
  {Girard} and {Mawet}}]{Boffin2014b}
{Boffin}, H.~M.~J., {Pourbaix}, D., {Mu{\v z}i{\'c}}, K., {Ivanov}, V.~D.,
  {Kurtev}, R., {Beletsky}, Y., {Mehner}, A., {Berger}, J.~P., {Girard}, J.~H.,
   and {Mawet}, D. (2014{\natexlab{b}}). \enquote{{Possible astrometric
  discovery of a substellar companion to the closest binary brown dwarf system
  WISE J104915.57-531906.1},} \emph{\aap} \textbf{561}, L4,
  \doi{10.1051/0004-6361/201322975}, \eprint{1312.1303}.

\bibitem[{{Boffin} \emph{et~al.}(2001){Boffin}, {Steeghs} and
  {Cuypers}}]{Boffin2001}
{Boffin}, H.~M.~J., {Steeghs}, D.,  and {Cuypers}, J. (eds.) (2001).
  \emph{{Astrotomography}}, \emph{Lecture Notes in Physics, Berlin Springer
  Verlag}, Vol. 573.

\bibitem[{{Boffin} \emph{et~al.}(1994){Boffin}, {Theuns} and
  {Jorissen}}]{Boffin1994}
{Boffin}, H.~M.~J., {Theuns}, T.,  and {Jorissen}, A. (1994). \enquote{{Wind
  Accretion in Binary Systems: 3D SPH Simulations},} \emph{\memsai}
  \textbf{65}, p. 1199.

\bibitem[{{Bono} \emph{et~al.}(2006){Bono}, {Caputo} and {Castellani}}]{bono06}
{Bono}, G., {Caputo}, F.,  and {Castellani}, V. (2006). \enquote{{Stellar
  pulsation and evolution: a stepping-stone to match reality.}} \emph{\memsai}
  \textbf{77}, p. 207.

\bibitem[{{Borges Fernandes} \emph{et~al.}(2011){Borges Fernandes}, {Meilland},
  {Bendjoya}, {Domiciano de Souza}, {Niccolini}, {Chesneau}, {Millour},
  {Spang}, {Stee} and {Kraus}}]{Borges}
{Borges Fernandes}, M., {Meilland}, A., {Bendjoya}, P., {Domiciano de Souza},
  A., {Niccolini}, G., {Chesneau}, O., {Millour}, F., {Spang}, A., {Stee}, P.,
  and {Kraus}, M. (2011). \enquote{{The Galactic unclassified B[e] star HD
  50138. II. Interferometric constraints on the close circumstellar
  environment},} \emph{\aap} \textbf{528}, A20,
  \doi{10.1051/0004-6361/201015602}, \eprint{1101.4964}.

\bibitem[{{Boyajian} \emph{et~al.}(2013){Boyajian}, {von Braun}, {van Belle},
  {Farrington}, {Schaefer}, {Jones}, {White}, {McAlister}, {ten Brummelaar},
  {Ridgway}, {Gies}, {Sturmann}, {Sturmann}, {Turner}, {Goldfinger} and
  {Vargas}}]{2013ApJ...771...40B}
{Boyajian}, T.~S., {von Braun}, K., {van Belle}, G., {Farrington}, C.,
  {Schaefer}, G., {Jones}, J., {White}, R., {McAlister}, H.~A., {ten
  Brummelaar}, T.~A., {Ridgway}, S., {Gies}, D., {Sturmann}, L., {Sturmann},
  J., {Turner}, N.~H., {Goldfinger}, P.~J.,  and {Vargas}, N. (2013).
  \enquote{{Stellar Diameters and Temperatures. III. Main-sequence A, F, G, and
  K Stars: Additional High-precision Measurements and Empirical Relations},}
  \emph{\apj} \textbf{771}, 40, \doi{10.1088/0004-637X/771/1/40},
  \eprint{1306.2974}.

\bibitem[{{Bruntt} \emph{et~al.}(2010){Bruntt}, {Kervella}, {M{\'e}rand},
  {Brand{\~a}o}, {Bedding}, {ten Brummelaar}, {Coud{\'e} du Foresto}, {Cunha},
  {Farrington}, {Goldfinger}, {Kiss}, {McAlister}, {Ridgway}, {Sturmann},
  {Sturmann}, {Turner} and {Tuthill}}]{2010A&A...512A..55B}
{Bruntt}, H., {Kervella}, P., {M{\'e}rand}, A., {Brand{\~a}o}, I.~M.,
  {Bedding}, T.~R., {ten Brummelaar}, T.~A., {Coud{\'e} du Foresto}, V.,
  {Cunha}, M.~S., {Farrington}, C., {Goldfinger}, P.~J., {Kiss}, L.~L.,
  {McAlister}, H.~A., {Ridgway}, S.~T., {Sturmann}, J., {Sturmann}, L.,
  {Turner}, N.,  and {Tuthill}, P.~G. (2010). \enquote{{The radius and
  effective temperature of the binary Ap star {$\beta$} CrB from CHARA/FLUOR
  and VLT/NACO observations},} \emph{\aap} \textbf{512}, A55,
  \doi{10.1051/0004-6361/200913405}, \eprint{0912.3215}.

\bibitem[{{Bruntt} \emph{et~al.}(2008){Bruntt}, {North}, {Cunha},
  {Brand{\~a}o}, {Elkin}, {Kurtz}, {Davis}, {Bedding}, {Jacob}, {Owens},
  {Robertson}, {Tango}, {Gameiro}, {Ireland} and
  {Tuthill}}]{2008MNRAS.386.2039B}
{Bruntt}, H., {North}, J.~R., {Cunha}, M., {Brand{\~a}o}, I.~M., {Elkin},
  V.~G., {Kurtz}, D.~W., {Davis}, J., {Bedding}, T.~R., {Jacob}, A.~P.,
  {Owens}, S.~M., {Robertson}, J.~G., {Tango}, W.~J., {Gameiro}, J.~F.,
  {Ireland}, M.~J.,  and {Tuthill}, P.~G. (2008). \enquote{{The fundamental
  parameters of the roAp star {$\alpha$} Circini},} \emph{\mnras} \textbf{386},
  pp. 2039--2046, \doi{10.1111/j.1365-2966.2008.13167.x}, \eprint{0803.1518}.

\bibitem[{{Bujarrabal} \emph{et~al.}(2013{\natexlab{a}}){Bujarrabal},
  {Alcolea}, {Van Winckel}, {Santander-Garc{\'{\i}}a} and
  {Castro-Carrizo}}]{2013AABujarrabalB}
{Bujarrabal}, V., {Alcolea}, J., {Van Winckel}, H., {Santander-Garc{\'{\i}}a},
  M.,  and {Castro-Carrizo}, A. (2013{\natexlab{a}}). \enquote{{Extended
  rotating disks around post-AGB stars},} \emph{\aap} \textbf{557}, A104,
  \doi{10.1051/0004-6361/201322015}, \eprint{1307.1975}.

\bibitem[{{Bujarrabal} \emph{et~al.}(2013{\natexlab{b}}){Bujarrabal},
  {Castro-Carrizo}, {Alcolea}, {Van Winckel}, {S{\'a}nchez Contreras},
  {Santander-Garc{\'{\i}}a}, {Neri} and {Lucas}}]{2013AABujarrabalC}
{Bujarrabal}, V., {Castro-Carrizo}, A., {Alcolea}, J., {Van Winckel}, H.,
  {S{\'a}nchez Contreras}, C., {Santander-Garc{\'{\i}}a}, M., {Neri}, R.,  and
  {Lucas}, R. (2013{\natexlab{b}}). \enquote{{ALMA observations of the Red
  Rectangle, a preliminary analysis},} \emph{\aap} \textbf{557}, L11,
  \doi{10.1051/0004-6361/201322232}, \eprint{1307.5959}.

\bibitem[{{Bujarrabal} \emph{et~al.}(2007){Bujarrabal}, {van Winckel}, {Neri},
  {Alcolea}, {Castro-Carrizo} and {Deroo}}]{2007AABujarrabal}
{Bujarrabal}, V., {van Winckel}, H., {Neri}, R., {Alcolea}, J.,
  {Castro-Carrizo}, A.,  and {Deroo}, P. (2007). \enquote{{The nebula around
  the post-AGB star 89 Herculis},} \emph{\aap} \textbf{468}, pp. L45--L48,
  \doi{10.1051/0004-6361:20066969}, \eprint{arXiv:astro-ph/0703718}.

\bibitem[{{Burki} \emph{et~al.}(1982){Burki}, {Mayor} and {Benz}}]{burki82}
{Burki}, G., {Mayor}, M.,  and {Benz}, W. (1982). \enquote{{The peculiar
  classical Cepheid HR 7308},} \emph{\aap} \textbf{109}, pp. 258--270.

\bibitem[{{Buscher} \emph{et~al.}(1990){Buscher}, {Baldwin}, {Warner} and
  {Haniff}}]{1990MNRAS.245P...7B}
{Buscher}, D.~F., {Baldwin}, J.~E., {Warner}, P.~J.,  and {Haniff}, C.~A.
  (1990). \enquote{{Detection of a bright feature on the surface of
  Betelgeuse},} \emph{\mnras} \textbf{245}, pp. 7P--11P.

\bibitem[{{Carciofi} and {Bjorkman}(2006)}]{Carciofi}
{Carciofi}, A.~C. and {Bjorkman}, J.~E. (2006). \enquote{{Non-LTE Monte Carlo
  Radiative Transfer. I. The Thermal Properties of Keplerian Disks around
  Classical Be Stars},} \emph{\apj} \textbf{639}, pp. 1081--1094,
  \doi{10.1086/499483}, \eprint{astro-ph/0511228}.

\bibitem[{{Casagrande} \emph{et~al.}(2014){Casagrande}, {Portinari}, {Glass},
  {Laney}, {Silva Aguirre}, {Datson}, {Andersen}, {Nordstr{\"o}m}, {Holmberg},
  {Flynn} and {Asplund}}]{casagrande14}
{Casagrande}, L., {Portinari}, L., {Glass}, I.~S., {Laney}, D., {Silva
  Aguirre}, V., {Datson}, J., {Andersen}, J., {Nordstr{\"o}m}, B., {Holmberg},
  J., {Flynn}, C.,  and {Asplund}, M. (2014). \enquote{{Towards stellar
  effective temperatures and diameters at 1 per cent accuracy for future
  surveys},} \emph{\mnras} \textbf{439}, pp. 2060--2073,
  \doi{10.1093/mnras/stu089}, \eprint{1401.3754}.

\bibitem[{{Casagrande} \emph{et~al.}(2010){Casagrande}, {Ram{\'{\i}}rez},
  {Mel{\'e}ndez}, {Bessell} and {Asplund}}]{casagrande10}
{Casagrande}, L., {Ram{\'{\i}}rez}, I., {Mel{\'e}ndez}, J., {Bessell}, M.,  and
  {Asplund}, M. (2010). \enquote{{An absolutely calibrated T$_{eff}$ scale from
  the infrared flux method. Dwarfs and subgiants},} \emph{\aap} \textbf{512},
  A54, \doi{10.1051/0004-6361/200913204}, \eprint{1001.3142}.

\bibitem[{{Casagrande} \emph{et~al.}(2011){Casagrande}, {Sch{\"o}nrich},
  {Asplund}, {Cassisi}, {Ram{\'{\i}}rez}, {Mel{\'e}ndez}, {Bensby} and
  {Feltzing}}]{casagrande11}
{Casagrande}, L., {Sch{\"o}nrich}, R., {Asplund}, M., {Cassisi}, S.,
  {Ram{\'{\i}}rez}, I., {Mel{\'e}ndez}, J., {Bensby}, T.,  and {Feltzing}, S.
  (2011). \enquote{{New constraints on the chemical evolution of the solar
  neighbourhood and Galactic disc(s). Improved astrophysical parameters for the
  Geneva-Copenhagen Survey},} \emph{\aap} \textbf{530}, A138,
  \doi{10.1051/0004-6361/201016276}, \eprint{1103.4651}.

\bibitem[{{Challouf} \emph{et~al.}(2014){Challouf}, {Nardetto}, {Mourard},
  {Graczyk}, {Aroui}, {Chesneau}, {Delaa}, {Pietrzy{\'n}ski}, {Gieren}, {Ligi},
  {Meilland}, {Perraut}, {Tallon-Bosc}, {McAlister}, {ten Brummelaar},
  {Sturmann}, {Sturmann}, {Turner}, {Farrington}, {Vargas} and
  {Scott}}]{challouf14}
{Challouf}, M., {Nardetto}, N., {Mourard}, D., {Graczyk}, D., {Aroui}, H.,
  {Chesneau}, O., {Delaa}, O., {Pietrzy{\'n}ski}, G., {Gieren}, W., {Ligi}, R.,
  {Meilland}, A., {Perraut}, K., {Tallon-Bosc}, I., {McAlister}, H., {ten
  Brummelaar}, T., {Sturmann}, J., {Sturmann}, L., {Turner}, N., {Farrington},
  C., {Vargas}, N.,  and {Scott}, N. (2014). \enquote{{Improving the surface
  brightness-color relation for early-type stars using optical
  interferometry},} \emph{ArXiv e-prints} \eprint{1409.1351}.

\bibitem[{{Che} \emph{et~al.}(2011){Che}, {Monnier}, {Zhao}, {Pedretti},
  {Thureau}, {M{\'e}rand}, {ten Brummelaar}, {McAlister}, {Ridgway}, {Turner},
  {Sturmann} and {Sturmann}}]{che2011}
{Che}, X., {Monnier}, J.~D., {Zhao}, M., {Pedretti}, E., {Thureau}, N.,
  {M{\'e}rand}, A., {ten Brummelaar}, T., {McAlister}, H., {Ridgway}, S.~T.,
  {Turner}, N., {Sturmann}, J.,  and {Sturmann}, L. (2011). \enquote{{Colder
  and Hotter: Interferometric Imaging of {$\beta$} Cassiopeiae and {$\alpha$}
  Leonis},} \emph{\apj} \textbf{732}, 68, \doi{10.1088/0004-637X/732/2/68},
  \eprint{1105.0740}.

\bibitem[{{Chelli}(1989)}]{chelli1989}
{Chelli}, A. (1989). \enquote{{The phase problem in optical interferometry -
  Error analysis in the presence of photon noise},} \emph{\aap} \textbf{225},
  pp. 277--290.

\bibitem[{{Chiavassa} \emph{et~al.}(2012){Chiavassa}, {Bigot}, {Kervella},
  {Matter}, {Lopez}, {Collet}, {Magic} and {Asplund}}]{2012A&A...540A...5C}
{Chiavassa}, A., {Bigot}, L., {Kervella}, P., {Matter}, A., {Lopez}, B.,
  {Collet}, R., {Magic}, Z.,  and {Asplund}, M. (2012).
  \enquote{{Three-dimensional interferometric, spectrometric, and planetary
  views of Procyon},} \emph{\aap} \textbf{540}, A5,
  \doi{10.1051/0004-6361/201118652}, \eprint{1201.3264}.

\bibitem[{{Chiavassa} \emph{et~al.}(2010{\natexlab{a}}){Chiavassa}, {Collet},
  {Casagrande} and {Asplund}}]{2010A&A...524A..93C}
{Chiavassa}, A., {Collet}, R., {Casagrande}, L.,  and {Asplund}, M.
  (2010{\natexlab{a}}). \enquote{{Three-dimensional hydrodynamical simulations
  of red giant stars: semi-global models for interpreting interferometric
  observations},} \emph{\aap} \textbf{524}, A93,
  \doi{10.1051/0004-6361/201015507}, \eprint{1009.1745}.

\bibitem[{{Chiavassa} \emph{et~al.}(2010{\natexlab{b}}){Chiavassa}, {Haubois},
  {Young}, {Plez}, {Josselin}, {Perrin} and {Freytag}}]{2010A&A...515A..12C}
{Chiavassa}, A., {Haubois}, X., {Young}, J.~S., {Plez}, B., {Josselin}, E.,
  {Perrin}, G.,  and {Freytag}, B. (2010{\natexlab{b}}). \enquote{{Radiative
  hydrodynamics simulations of red supergiant stars. II. Simulations of
  convection on Betelgeuse match interferometric observations},} \emph{\aap}
  \textbf{515}, p. A12, \doi{10.1051/0004-6361/200913907}, \eprint{1003.1407}.

\bibitem[{{Chiavassa} \emph{et~al.}(2010{\natexlab{c}}){Chiavassa}, {Lacour},
  {Millour}, {Driebe}, {Wittkowski}, {Plez}, {Thi{\'e}baut}, {Josselin},
  {Freytag}, {Scholz} and {Haubois}}]{2010A&A...511A..51C}
{Chiavassa}, A., {Lacour}, S., {Millour}, F., {Driebe}, T., {Wittkowski}, M.,
  {Plez}, B., {Thi{\'e}baut}, E., {Josselin}, E., {Freytag}, B., {Scholz}, M.,
  and {Haubois}, X. (2010{\natexlab{c}}). \enquote{{VLTI/AMBER
  spectro-interferometric imaging of VX Sagittarii's inhomogenous outer
  atmosphere},} \emph{\aap} \textbf{511}, p. A51,
  \doi{10.1051/0004-6361/200913288}, \eprint{0911.4422}.

\bibitem[{{Chiavassa} \emph{et~al.}(2014){Chiavassa}, {Ligi}, {Magic},
  {Collet}, {Asplund} and {Mourard}}]{2014A&A...567A.115C}
{Chiavassa}, A., {Ligi}, R., {Magic}, Z., {Collet}, R., {Asplund}, M.,  and
  {Mourard}, D. (2014). \enquote{{Planet transit and stellar granulation
  detection with interferometry. Using the three-dimensional stellar atmosphere
  Stagger-grid simulations},} \emph{\aap} \textbf{567}, A115,
  \doi{10.1051/0004-6361/201323207}, \eprint{1404.7049}.

\bibitem[{{Chiavassa} \emph{et~al.}(2011){Chiavassa}, {Pasquato}, {Jorissen},
  {Sacuto}, {Babusiaux}, {Freytag}, {Ludwig}, {Cruzal{\`e}bes}, {Rabbia},
  {Spang} and {Chesneau}}]{2011A&A...528A.120C}
{Chiavassa}, A., {Pasquato}, E., {Jorissen}, A., {Sacuto}, S., {Babusiaux}, C.,
  {Freytag}, B., {Ludwig}, H.-G., {Cruzal{\`e}bes}, P., {Rabbia}, Y., {Spang},
  A.,  and {Chesneau}, O. (2011). \enquote{{Radiative hydrodynamic simulations
  of red supergiant stars. III. Spectro-photocentric variability, photometric
  variability, and consequences on Gaia measurements},} \emph{\aap}
  \textbf{528}, p. A120, \doi{10.1051/0004-6361/201015768}, \eprint{1012.5234}.

\bibitem[{{Chiavassa} \emph{et~al.}(2009){Chiavassa}, {Plez}, {Josselin} and
  {Freytag}}]{2009A&A...506.1351C}
{Chiavassa}, A., {Plez}, B., {Josselin}, E.,  and {Freytag}, B. (2009).
  \enquote{{Radiative hydrodynamics simulations of red supergiant stars. I.
  interpretation of interferometric observations},} \emph{\aap} \textbf{506},
  pp. 1351--1365, \doi{10.1051/0004-6361/200911780}, \eprint{0907.1860}.

\bibitem[{{Chini} \emph{et~al.}(2012){Chini}, {Hoffmeister}, {Nasseri}, {Stahl}
  and {Zinnecker}}]{Chini2012}
{Chini}, R., {Hoffmeister}, V.~H., {Nasseri}, A., {Stahl}, O.,  and
  {Zinnecker}, H. (2012). \enquote{{A spectroscopic survey on the multiplicity
  of high-mass stars},} \emph{\mnras} \textbf{424}, pp. 1925--1929,
  \doi{10.1111/j.1365-2966.2012.21317.x}, \eprint{1205.5238}.

\bibitem[{{Cidale} \emph{et~al.}(2012){Cidale}, {Borges Fernandes},
  {Andruchow}, {Arias}, {Kraus}, {Chesneau}, {Kanaan}, {Cur{\'e}}, {de Wit} and
  {Muratore}}]{Cidale}
{Cidale}, L.~S., {Borges Fernandes}, M., {Andruchow}, I., {Arias}, M.~L.,
  {Kraus}, M., {Chesneau}, O., {Kanaan}, S., {Cur{\'e}}, M., {de Wit}, W.~J.,
  and {Muratore}, M.~F. (2012). \enquote{{Observational constraints for the
  circumstellar disk of the B[e] star CPD-52 9243},} \emph{\aap} \textbf{548},
  A72, \doi{10.1051/0004-6361/201220120}.

\bibitem[{{Conti}(1997)}]{Conti}
{Conti}, P.~S. (1997). \enquote{{B[e] Stars: What Are They and Why Are They At
  This Workshop?}} in A.~{Nota} and H.~{Lamers} (eds.), \emph{Luminous Blue
  Variables: Massive Stars in Transition}, \emph{Astronomical Society of the
  Pacific Conference Series}, Vol. 120, p. 161.

\bibitem[{{Creevey} \emph{et~al.}(2014){Creevey}, {Th{\'e}venin}, {Berio},
  {Heiter}, {von Braun}, {Mourard}, {Bigot}, {Boyajian}, {Kervella}, {Morel},
  {Pichon}, {Chiavassa}, {Nardetto}, {Perraut}, {Meilland}, {Mc Alister}, {ten
  Brummelaar}, {Sturmann}, {Sturmann} and {Turner}}]{creevey14}
{Creevey}, O., {Th{\'e}venin}, F., {Berio}, P., {Heiter}, U., {von Braun}, K.,
  {Mourard}, D., {Bigot}, L., {Boyajian}, T.~S., {Kervella}, P., {Morel}, P.,
  {Pichon}, B., {Chiavassa}, A., {Nardetto}, N., {Perraut}, K., {Meilland}, A.,
  {Mc Alister}, H.~A., {ten Brummelaar}, T.~A., {Sturmann}, C.~F.~J.,
  {Sturmann}, L.,  and {Turner}, N. (2014). \enquote{{Benchmark stars for Gaia:
  fundamental properties of the Population II star HD140283 from
  interferometric, spectroscopic and photometric data},} \emph{ArXiv e-prints}
  \eprint{1410.4780}.

\bibitem[{{Creevey} \emph{et~al.}(2007){Creevey}, {Monteiro}, {Metcalfe},
  {Brown}, {Jim{\'e}nez-Reyes} and {Belmonte}}]{cre07}
{Creevey}, O.~L., {Monteiro}, M.~J.~P.~F.~G., {Metcalfe}, T.~S., {Brown},
  T.~M., {Jim{\'e}nez-Reyes}, S.~J.,  and {Belmonte}, J.~A. (2007).
  \enquote{{The Complementary Roles of Interferometry and Asteroseismology in
  Determining the Mass of Solar-Type Stars},} \emph{\apj} \textbf{659}, pp.
  616--625, \doi{10.1086/512097}, \eprint{astro-ph/0702270}.

\bibitem[{{Creevey} \emph{et~al.}(2013){Creevey}, {Th{\'e}venin}, {Basu},
  {Chaplin}, {Bigot}, {Elsworth}, {Huber}, {Monteiro} and
  {Serenelli}}]{creevey13}
{Creevey}, O.~L., {Th{\'e}venin}, F., {Basu}, S., {Chaplin}, W.~J., {Bigot},
  L., {Elsworth}, Y., {Huber}, D., {Monteiro}, M.~J.~P.~F.~G.,  and
  {Serenelli}, A. (2013). \enquote{{A large sample of calibration stars for
  Gaia: log g from Kepler and CoRoT fields},} \emph{\mnras} \textbf{431}, pp.
  2419--2432, \doi{10.1093/mnras/stt336}, \eprint{1302.7158}.

\bibitem[{{Creevey} \emph{et~al.}(2012){Creevey}, {Th{\'e}venin}, {Boyajian},
  {Kervella}, {Chiavassa}, {Bigot}, {M{\'e}rand}, {Heiter}, {Morel}, {Pichon},
  {Mc Alister}, {ten Brummelaar}, {Collet}, {van Belle}, {Coud{\'e} du
  Foresto}, {Farrington}, {Goldfinger}, {Sturmann}, {Sturmann} and
  {Turner}}]{creevey12}
{Creevey}, O.~L., {Th{\'e}venin}, F., {Boyajian}, T.~S., {Kervella}, P.,
  {Chiavassa}, A., {Bigot}, L., {M{\'e}rand}, A., {Heiter}, U., {Morel}, P.,
  {Pichon}, B., {Mc Alister}, H.~A., {ten Brummelaar}, T.~A., {Collet}, R.,
  {van Belle}, G.~T., {Coud{\'e} du Foresto}, V., {Farrington}, C.,
  {Goldfinger}, P.~J., {Sturmann}, J., {Sturmann}, L.,  and {Turner}, N.
  (2012). \enquote{{Fundamental properties of the Population II fiducial stars
  <ASTROBJ>HD 122563</ASTROBJ> and <ASTROBJ>Gmb 1830</ASTROBJ> from CHARA
  interferometric observations},} \emph{\aap} \textbf{545}, A17,
  \doi{10.1051/0004-6361/201219651}, \eprint{1207.5954}.

\bibitem[{{Cuntz}(1997)}]{1997A&A...325..709C}
{Cuntz}, M. (1997). \enquote{{Chromospheric velocity fields in {$\alpha$}
  Orionis (M2 Iab) generated by stochastic shocks.}} \emph{\aap} \textbf{325},
  pp. 709--713.

\bibitem[{{Davis} \emph{et~al.}(2009){Davis}, {Jacob}, {Robertson}, {Ireland},
  {North}, {Tango} and {Tuthill}}]{davis09}
{Davis}, J., {Jacob}, A.~P., {Robertson}, J.~G., {Ireland}, M.~J., {North},
  J.~R., {Tango}, W.~J.,  and {Tuthill}, P.~G. (2009). \enquote{{Observations
  of the pulsation of the Cepheid l Car with the Sydney University Stellar
  Interferometer},} \emph{\mnras} \textbf{394}, pp. 1620--1630,
  \doi{10.1111/j.1365-2966.2009.14433.x}, \eprint{0812.4791}.

\bibitem[{{De Beck} \emph{et~al.}(2010){De Beck}, {Decin}, {de Koter},
  {Justtanont}, {Verhoelst}, {Kemper} and {Menten}}]{2010A&A...523A..18D}
{De Beck}, E., {Decin}, L., {de Koter}, A., {Justtanont}, K., {Verhoelst}, T.,
  {Kemper}, F.,  and {Menten}, K.~M. (2010). \enquote{{Probing the mass-loss
  history of AGB and red supergiant stars from CO rotational line profiles. II.
  CO line survey of evolved stars: derivation of mass-loss rate formulae},}
  \emph{\aap} \textbf{523}, A18, \doi{10.1051/0004-6361/200913771},
  \eprint{1008.1083}.

\bibitem[{{de Ruyter} \emph{et~al.}(2006){de Ruyter}, {van Winckel}, {Maas},
  {Lloyd Evans}, {Waters} and {Dejonghe}}]{2006AAdeRuyter}
{de Ruyter}, S., {van Winckel}, H., {Maas}, T., {Lloyd Evans}, T., {Waters},
  L.~B.~F.~M.,  and {Dejonghe}, H. (2006). \enquote{{Keplerian discs around
  post-AGB stars: a common phenomenon?}} \emph{\aap} \textbf{448}, pp.
  641--653, \doi{10.1051/0004-6361:20054062}, \eprint{arXiv:astro-ph/0601578}.

\bibitem[{{Delaa} \emph{et~al.}(2013){Delaa}, {Zorec}, {Domiciano de Souza},
  {Mourard}, {Perraut}, {Stee}, {Fr{\'e}mat}, {Monnier}, {Kraus}, {Che},
  {B{\'e}rio}, {Bonneau}, {Clausse}, {Challouf}, {Ligi}, {Meilland},
  {Nardetto}, {Spang}, {McAlister}, {ten Brummelaar}, {Sturmann}, {Sturmann},
  {Turner}, {Farrington} and {Goldfinger}}]{delaa2013}
{Delaa}, O., {Zorec}, J., {Domiciano de Souza}, A., {Mourard}, D., {Perraut},
  K., {Stee}, P., {Fr{\'e}mat}, Y., {Monnier}, J., {Kraus}, S., {Che}, X.,
  {B{\'e}rio}, P., {Bonneau}, D., {Clausse}, J.~M., {Challouf}, M., {Ligi}, R.,
  {Meilland}, A., {Nardetto}, N., {Spang}, A., {McAlister}, H., {ten
  Brummelaar}, T., {Sturmann}, J., {Sturmann}, L., {Turner}, N., {Farrington},
  C.,  and {Goldfinger}, P.~J. (2013). \enquote{{Spectrally resolved
  interferometric observations of {$\alpha$} Cephei and physical modeling of
  fast rotating stars},} \emph{\aap} \textbf{555}, A100,
  \doi{10.1051/0004-6361/201220689}.

\bibitem[{{Di Benedetto}(2005)}]{dibenedetto05}
{Di Benedetto}, G.~P. (2005). \enquote{{Predicting accurate stellar angular
  diameters by the near-infrared surface brightness technique},} \emph{\mnras}
  \textbf{357}, pp. 174--190, \doi{10.1111/j.1365-2966.2005.08632.x}.

\bibitem[{{Domiciano de Souza} \emph{et~al.}(2003){Domiciano de Souza},
  {Kervella}, {Jankov}, {Abe}, {Vakili}, {di Folco} and
  {Paresce}}]{domiciano2003}
{Domiciano de Souza}, A., {Kervella}, P., {Jankov}, S., {Abe}, L., {Vakili},
  F., {di Folco}, E.,  and {Paresce}, F. (2003). \enquote{{The spinning-top Be
  star Achernar from VLTI-VINCI},} \emph{\aap} \textbf{407}, pp. L47--L50,
  \doi{10.1051/0004-6361:20030786}, \eprint{astro-ph/0306277}.

\bibitem[{{Eisenhauer} \emph{et~al.}(2011){Eisenhauer}, {Perrin}, {Brandner},
  {Straubmeier}, {Perraut}, {Amorim}, {Sch{\"o}ller}, {Gillessen}, {Kervella},
  {Benisty}, {Araujo-Hauck}, {Jocou}, {Lima}, {Jakob}, {Haug}, {Cl{\'e}net},
  {Henning}, {Eckart}, {Berger}, {Garcia}, {Abuter}, {Kellner}, {Paumard},
  {Hippler}, {Fischer}, {Moulin}, {Villate}, {Avila}, {Gr{\"a}ter}, {Lacour},
  {Huber}, {Wiest}, {Nolot}, {Carvas}, {Dorn}, {Pfuhl}, {Gendron}, {Kendrew},
  {Yazici}, {Anton}, {Jung}, {Thiel}, {Choquet}, {Klein}, {Teixeira}, {Gitton},
  {Moch}, {Vincent}, {Kudryavtseva}, {Str{\"o}bele}, {Sturm}, {F{\'e}dou},
  {Lenzen}, {Jolley}, {Kister}, {Lapeyr{\`e}re}, {Naranjo}, {Lucuix},
  {Hofmann}, {Chapron}, {Neumann}, {Mehrgan}, {Hans}, {Rousset}, {Ramos},
  {Suarez}, {Lederer}, {Reess}, {Rohloff}, {Haguenauer}, {Bartko}, {Sevin},
  {Wagner}, {Lizon}, {Rabien}, {Collin}, {Finger}, {Davies}, {Rouan},
  {Wittkowski}, {Dodds-Eden}, {Ziegler}, {Cassaing}, {Bonnet}, {Casali},
  {Genzel} and {Lena}}]{eisenhauer2011}
{Eisenhauer}, F., {Perrin}, G., {Brandner}, W., {Straubmeier}, C., {Perraut},
  K., {Amorim}, A., {Sch{\"o}ller}, M., {Gillessen}, S., {Kervella}, P.,
  {Benisty}, M., {Araujo-Hauck}, C., {Jocou}, L., {Lima}, J., {Jakob}, G.,
  {Haug}, M., {Cl{\'e}net}, Y., {Henning}, T., {Eckart}, A., {Berger}, J.-P.,
  {Garcia}, P., {Abuter}, R., {Kellner}, S., {Paumard}, T., {Hippler}, S.,
  {Fischer}, S., {Moulin}, T., {Villate}, J., {Avila}, G., {Gr{\"a}ter}, A.,
  {Lacour}, S., {Huber}, A., {Wiest}, M., {Nolot}, A., {Carvas}, P., {Dorn},
  R., {Pfuhl}, O., {Gendron}, E., {Kendrew}, S., {Yazici}, S., {Anton}, S.,
  {Jung}, Y., {Thiel}, M., {Choquet}, {\'E}., {Klein}, R., {Teixeira}, P.,
  {Gitton}, P., {Moch}, D., {Vincent}, F., {Kudryavtseva}, N., {Str{\"o}bele},
  S., {Sturm}, S., {F{\'e}dou}, P., {Lenzen}, R., {Jolley}, P., {Kister}, C.,
  {Lapeyr{\`e}re}, V., {Naranjo}, V., {Lucuix}, C., {Hofmann}, R., {Chapron},
  F., {Neumann}, U., {Mehrgan}, L., {Hans}, O., {Rousset}, G., {Ramos}, J.,
  {Suarez}, M., {Lederer}, R., {Reess}, J.-M., {Rohloff}, R.-R., {Haguenauer},
  P., {Bartko}, H., {Sevin}, A., {Wagner}, K., {Lizon}, J.-L., {Rabien}, S.,
  {Collin}, C., {Finger}, G., {Davies}, R., {Rouan}, D., {Wittkowski}, M.,
  {Dodds-Eden}, K., {Ziegler}, D., {Cassaing}, F., {Bonnet}, H., {Casali}, M.,
  {Genzel}, R.,  and {Lena}, P. (2011). \enquote{{GRAVITY: Observing the
  Universe in Motion},} \emph{The Messenger} \textbf{143}, pp. 16--24.

\bibitem[{{Elkin} \emph{et~al.}(2010){Elkin}, {Mathys}, {Kurtz}, {Hubrig} and
  {Freyhammer}}]{2010MNRAS.402.1883E}
{Elkin}, V.~G., {Mathys}, G., {Kurtz}, D.~W., {Hubrig}, S.,  and {Freyhammer},
  L.~M. (2010). \enquote{{A rival for Babcock's star: the extreme 30-kG
  variable magnetic field in the Ap star HD75049},} \emph{\mnras} \textbf{402},
  pp. 1883--1891, \doi{10.1111/j.1365-2966.2009.16015.x}, \eprint{0908.0849}.

\bibitem[{{Ellerbroek} \emph{et~al.}(2014){Ellerbroek}, {Podio}, {Dougados},
  {Cabrit}, {Sitko}, {Sana}, {Kaper}, {de Koter}, {Klaassen}, {Mulders},
  {Mendigut{\'{\i}}a}, {Grady}, {Grankin}, {van Winckel}, {Bacciotti},
  {Russell}, {Lynch}, {Hammel}, {Beerman}, {Day}, {Huelsman}, {Werren},
  {Henden} and {Grindlay}}]{Ellerbroek}
{Ellerbroek}, L.~E., {Podio}, L., {Dougados}, C., {Cabrit}, S., {Sitko}, M.~L.,
  {Sana}, H., {Kaper}, L., {de Koter}, A., {Klaassen}, P.~D., {Mulders}, G.~D.,
  {Mendigut{\'{\i}}a}, I., {Grady}, C.~A., {Grankin}, K., {van Winckel}, H.,
  {Bacciotti}, F., {Russell}, R.~W., {Lynch}, D.~K., {Hammel}, H.~B.,
  {Beerman}, L.~C., {Day}, A.~N., {Huelsman}, D.~M., {Werren}, C., {Henden},
  A.,  and {Grindlay}, J. (2014). \enquote{{Relating jet structure to
  photometric variability: the Herbig Ae star HD 163296},} \emph{\aap}
  \textbf{563}, A87, \doi{10.1051/0004-6361/201323092}, \eprint{1401.3744}.

\bibitem[{{Elvis} and {Karovska}(2002)}]{Elvis2002}
{Elvis}, M. and {Karovska}, M. (2002). \enquote{{Quasar Parallax: A Method for
  Determining Direct Geometrical Distances to Quasars},} \emph{\apjl}
  \textbf{581}, pp. L67--L70, \doi{10.1086/346015}, \eprint{astro-ph/0211385}.

\bibitem[{{Fabas} \emph{et~al.}(2011){Fabas}, {L{\`e}bre} and
  {Gillet}}]{fabas2011}
{Fabas}, N., {L{\`e}bre}, A.,  and {Gillet}, D. (2011). \enquote{{Shock-induced
  polarized hydrogen emission lines in the Mira star o Ceti},} \emph{\aap}
  \textbf{535}, A12, \doi{10.1051/0004-6361/201117748}, \eprint{1109.6500}.

\bibitem[{{Fernie}(1992)}]{fernie92}
{Fernie}, J.~D. (1992). \enquote{{A new approach to the Cepheid
  period-luminosity law - Delta Scuti stars as small Cepheids},} \emph{\aj}
  \textbf{103}, pp. 1647--1651, \doi{10.1086/116179}.

\bibitem[{{Fernie} \emph{et~al.}(1995){Fernie}, {Evans}, {Beattie} and
  {Seager}}]{fernie95}
{Fernie}, J.~D., {Evans}, N.~R., {Beattie}, B.,  and {Seager}, S. (1995).
  \enquote{{A Database of Galactic Classical Cepheids},} \emph{Information
  Bulletin on Variable Stars} \textbf{4148}, p.~1.

\bibitem[{{Fernley}(1994)}]{fernley94}
{Fernley}, J. (1994). \enquote{{A revision to the absolute magnitudes of RR
  Lyrae stars from the Baade-Wesselink method},} \emph{\aap} \textbf{284}, pp.
  L16--L18.

\bibitem[{{Fokin}(1991)}]{fokin91}
{Fokin}, A.~B. (1991). \enquote{{On hydrogen line formation in the atmospheres
  of pulsating giants},} \emph{\mnras} \textbf{250}, pp. 258--269.

\bibitem[{{Fokin} \emph{et~al.}(1996){Fokin}, {Gillet} and
  {Breitfellner}}]{fokin96}
{Fokin}, A.~B., {Gillet}, D.,  and {Breitfellner}, M.~G. (1996).
  \enquote{{Pulsating motions and turbulence in {$\delta$} Cephei.}}
  \emph{\aap} \textbf{307}, pp. 503--515.

\bibitem[{{Fouqu{\'e}} \emph{et~al.}(2007){Fouqu{\'e}}, {Arriagada}, {Storm},
  {Barnes}, {Nardetto}, {M{\'e}rand}, {Kervella}, {Gieren}, {Bersier},
  {Benedict} and {McArthur}}]{fouque07}
{Fouqu{\'e}}, P., {Arriagada}, P., {Storm}, J., {Barnes}, T.~G., {Nardetto},
  N., {M{\'e}rand}, A., {Kervella}, P., {Gieren}, W., {Bersier}, D.,
  {Benedict}, G.~F.,  and {McArthur}, B.~E. (2007). \enquote{{A new calibration
  of Galactic Cepheid period-luminosity relations from B to K bands, and a
  comparison to LMC relations},} \emph{\aap} \textbf{476}, pp. 73--81,
  \doi{10.1051/0004-6361:20078187}, \eprint{0709.3255}.

\bibitem[{{Freedman} \emph{et~al.}(2001){Freedman}, {Madore}, {Gibson},
  {Ferrarese}, {Kelson}, {Sakai}, {Mould}, {Kennicutt}, {Ford}, {Graham},
  {Huchra}, {Hughes}, {Illingworth}, {Macri} and {Stetson}}]{freedman01}
{Freedman}, W.~L., {Madore}, B.~F., {Gibson}, B.~K., {Ferrarese}, L., {Kelson},
  D.~D., {Sakai}, S., {Mould}, J.~R., {Kennicutt}, R.~C., Jr., {Ford}, H.~C.,
  {Graham}, J.~A., {Huchra}, J.~P., {Hughes}, S.~M.~G., {Illingworth}, G.~D.,
  {Macri}, L.~M.,  and {Stetson}, P.~B. (2001). \enquote{{Final Results from
  the Hubble Space Telescope Key Project to Measure the Hubble Constant},}
  \emph{\apj} \textbf{553}, pp. 47--72, \doi{10.1086/320638},
  \eprint{astro-ph/0012376}.

\bibitem[{{Gallenne} \emph{et~al.}(2012){Gallenne}, {Kervella} and
  {M{\'e}rand}}]{gallenne12}
{Gallenne}, A., {Kervella}, P.,  and {M{\'e}rand}, A. (2012). \enquote{{Thermal
  infrared properties of classical and type II Cepheids. Diffraction limited 10
  {$\mu$}m imaging with VLT/VISIR},} \emph{\aap} \textbf{538}, A24,
  \doi{10.1051/0004-6361/201117307}, \eprint{1111.7215}.

\bibitem[{{Gallenne} \emph{et~al.}(2014{\natexlab{a}}){Gallenne}, {Kervella},
  {M{\'e}rand}, {Evans}, {Girard}, {Gieren} and
  {Pietrzy{\'n}ski}}]{gallenne14b}
{Gallenne}, A., {Kervella}, P., {M{\'e}rand}, A., {Evans}, N.~R., {Girard},
  J.~H.~V., {Gieren}, W.,  and {Pietrzy{\'n}ski}, G. (2014{\natexlab{a}}).
  \enquote{{Searching for visual companions of close Cepheids. VLT/NACO lucky
  imaging of Y Oph, FF Aql, X Sgr, W Sgr, and {$\eta$} Aql},} \emph{\aap}
  \textbf{567}, A60, \doi{10.1051/0004-6361/201423872}, \eprint{1406.0493}.

\bibitem[{{Gallenne} \emph{et~al.}(2014{\natexlab{b}}){Gallenne}, {M{\'e}rand},
  {Kervella}, {Breitfelder}, {Le Bouquin}, {Monnier}, {Gieren}, {Pilecki} and
  {Pietrzy{\'n}ski}}]{gallenne14}
{Gallenne}, A., {M{\'e}rand}, A., {Kervella}, P., {Breitfelder}, J., {Le
  Bouquin}, J.-B., {Monnier}, J.~D., {Gieren}, W., {Pilecki}, B.,  and
  {Pietrzy{\'n}ski}, G. (2014{\natexlab{b}}). \enquote{{Multiplicity of
  Galactic Cepheids from long-baseline interferometry. II. The Companion of AX
  Circini revealed with VLTI/PIONIER},} \emph{\aap} \textbf{561}, L3,
  \doi{10.1051/0004-6361/201322883}, \eprint{1312.1950}.

\bibitem[{{Gallenne} \emph{et~al.}(2013{\natexlab{a}}){Gallenne}, {M{\'e}rand},
  {Kervella}, {Chesneau}, {Breitfelder} and {Gieren}}]{gallenne13b}
{Gallenne}, A., {M{\'e}rand}, A., {Kervella}, P., {Chesneau}, O.,
  {Breitfelder}, J.,  and {Gieren}, W. (2013{\natexlab{a}}). \enquote{{Extended
  envelopes around Galactic Cepheids. IV. T Monocerotis and X Sagittarii from
  mid-infrared interferometry with VLTI/MIDI},} \emph{\aap} \textbf{558}, A140,
  \doi{10.1051/0004-6361/201322257}, \eprint{1309.0854}.

\bibitem[{{Gallenne} \emph{et~al.}(2013{\natexlab{b}}){Gallenne}, {Monnier},
  {M{\'e}rand}, {Kervella}, {Kraus}, {Schaefer}, {Gieren}, {Pietrzy{\'n}ski},
  {Szabados}, {Che}, {Baron}, {Pedretti}, {McAlister}, {ten Brummelaar},
  {Sturmann}, {Sturmann}, {Turner}, {Farrington} and {Vargas}}]{gallenne13}
{Gallenne}, A., {Monnier}, J.~D., {M{\'e}rand}, A., {Kervella}, P., {Kraus},
  S., {Schaefer}, G.~H., {Gieren}, W., {Pietrzy{\'n}ski}, G., {Szabados}, L.,
  {Che}, X., {Baron}, F., {Pedretti}, E., {McAlister}, H., {ten Brummelaar},
  T., {Sturmann}, J., {Sturmann}, L., {Turner}, N., {Farrington}, C.,  and
  {Vargas}, N. (2013{\natexlab{b}}). \enquote{{Multiplicity of Galactic
  Cepheids from long-baseline interferometry. I. CHARA/MIRC detection of the
  companion of V1334 Cygni},} \emph{\aap} \textbf{552}, A21,
  \doi{10.1051/0004-6361/201321091}, \eprint{1302.1817}.

\bibitem[{{Gastine} \emph{et~al.}(2013){Gastine}, {Morin}, {Duarte}, {Reiners},
  {Christensen} and {Wicht}}]{2013A&A...549L...5G}
{Gastine}, T., {Morin}, J., {Duarte}, L., {Reiners}, A., {Christensen}, U.~R.,
  and {Wicht}, J. (2013). \enquote{{What controls the magnetic geometry of M
  dwarfs?}} \emph{\aap} \textbf{549}, L5, \doi{10.1051/0004-6361/201220317},
  \eprint{1212.0136}.

\bibitem[{{Gautschy-Loidl} \emph{et~al.}(2004){Gautschy-Loidl}, {H{\"o}fner},
  {J{\o}rgensen} and {Hron}}]{gautschy-loidl2004}
{Gautschy-Loidl}, R., {H{\"o}fner}, S., {J{\o}rgensen}, U.~G.,  and {Hron}, J.
  (2004). \enquote{{Dynamic model atmospheres of AGB stars. IV. A comparison of
  synthetic carbon star spectra with observations},} \emph{\aap} \textbf{422},
  pp. 289--306, \doi{10.1051/0004-6361:20035860}.

\bibitem[{{Gieren} \emph{et~al.}(2005{\natexlab{a}}){Gieren}, {Pietrzynski},
  {Bresolin}, {Kudritzki}, {Minniti}, {Urbaneja}, {Soszynski}, {Storm},
  {Fouque}, {Bono}, {Walker} and {Garcia}}]{gieren05_messenger}
{Gieren}, W., {Pietrzynski}, G., {Bresolin}, F., {Kudritzki}, R.-P., {Minniti},
  D., {Urbaneja}, M., {Soszynski}, I., {Storm}, J., {Fouque}, P., {Bono}, G.,
  {Walker}, A.,  and {Garcia}, J. (2005{\natexlab{a}}). \enquote{{Measuring
  Improved Distances to Nearby Galaxies: The Araucaria Project},} \emph{The
  Messenger} \textbf{121}, pp. 23--28.

\bibitem[{{Gieren} \emph{et~al.}(2005{\natexlab{b}}){Gieren}, {Storm},
  {Barnes}, {Fouqu{\'e}}, {Pietrzy{\'n}ski} and {Kienzle}}]{gieren05}
{Gieren}, W., {Storm}, J., {Barnes}, T.~G., III, {Fouqu{\'e}}, P.,
  {Pietrzy{\'n}ski}, G.,  and {Kienzle}, F. (2005{\natexlab{b}}).
  \enquote{{Direct Distances to Cepheids in the Large Magellanic Cloud:
  Evidence for a Universal Slope of the Period-Luminosity Relation up to Solar
  Abundance},} \emph{\apj} \textbf{627}, pp. 224--237, \doi{10.1086/430496},
  \eprint{astro-ph/0503637}.

\bibitem[{{Gorlova} \emph{et~al.}(2012){Gorlova}, {Van Winckel}, {Gielen},
  {Raskin}, {Prins}, {Pessemier}, {Waelkens}, {Fr{\'e}mat}, {Hensberge},
  {Dumortier}, {Jorissen} and {Van Eck}}]{2012AAGorlova}
{Gorlova}, N., {Van Winckel}, H., {Gielen}, C., {Raskin}, G., {Prins}, S.,
  {Pessemier}, W., {Waelkens}, C., {Fr{\'e}mat}, Y., {Hensberge}, H.,
  {Dumortier}, L., {Jorissen}, A.,  and {Van Eck}, S. (2012).
  \enquote{{Time-resolved spectroscopy of BD+46{\deg}442: Gas streams and jet
  creation in a newly discovered evolved binary with a disk},} \emph{\aap}
  \textbf{542}, A27, \doi{10.1051/0004-6361/201118727}, \eprint{1204.3004}.

\bibitem[{{Grunhut} \emph{et~al.}(2010){Grunhut}, {Wade}, {Hanes} and
  {Alecian}}]{2010MNRAS.408.2290G}
{Grunhut}, J.~H., {Wade}, G.~A., {Hanes}, D.~A.,  and {Alecian}, E. (2010).
  \enquote{{Systematic detection of magnetic fields in massive, late-type
  supergiants},} \emph{\mnras} \textbf{408}, pp. 2290--2297,
  \doi{10.1111/j.1365-2966.2010.17275.x}, \eprint{1006.5891}.

\bibitem[{{Guiglion} \emph{et~al.}(2013){Guiglion}, {Nardetto}, {Mathias},
  {Domiciano de Souza}, {Poretti}, {Rainer}, {Fokin}, {Mourard} and
  {Gieren}}]{guiglion13}
{Guiglion}, G., {Nardetto}, N., {Mathias}, P., {Domiciano de Souza}, A.,
  {Poretti}, E., {Rainer}, M., {Fokin}, A., {Mourard}, D.,  and {Gieren}, W.
  (2013). \enquote{{Understanding the dynamical structure of pulsating stars:
  The Baade-Wesselink projection factor of the {$\delta$} Scuti stars AI
  Velorum and {$\beta$} Cassiopeiae},} \emph{\aap} \textbf{550}, L10,
  \doi{10.1051/0004-6361/201220780}, \eprint{1301.2475}.

\bibitem[{{Halbwachs}(2009)}]{Halbwachs2009}
{Halbwachs}, J.-L. (2009). \enquote{{Local effects in astrometric binary
  orbits: perspective transformation and light-travel time},} \emph{\mnras}
  \textbf{394}, pp. 1075--1084, \doi{10.1111/j.1365-2966.2009.14406.x},
  \eprint{0812.3224}.

\bibitem[{{Halbwachs} \emph{et~al.}(2014){Halbwachs}, {Arenou}, {Pourbaix},
  {Famaey}, {Guillout}, {Lebreton}, {Salomon}, {Tal-Or}, {Ibata} and
  {Mazeh}}]{Halbwachs2014}
{Halbwachs}, J.-L., {Arenou}, F., {Pourbaix}, D., {Famaey}, B., {Guillout}, P.,
  {Lebreton}, Y., {Salomon}, J.-B., {Tal-Or}, L., {Ibata}, R.,  and {Mazeh}, T.
  (2014). \enquote{{Masses of the components of SB2 binaries observed with Gaia
  - I. Selection of the sample and mass ratios of 20 new SB2s discovered with
  Sophie},} \emph{\mnras} \textbf{445}, pp. 2371--2377,
  \doi{10.1093/mnras/stu1838}, \eprint{1409.1384}.

\bibitem[{{Harper} and {Brown}(2006)}]{harper2006}
{Harper}, G.~M. and {Brown}, A. (2006). \enquote{{Electron Density and
  Turbulence Gradients within the Extended Atmosphere of the M Supergiant
  Betelgeuse ({$\alpha$} Orionis)},} \emph{\apj} \textbf{646}, pp. 1179--1202,
  \doi{10.1086/505073}.

\bibitem[{{Harper} \emph{et~al.}(2001){Harper}, {Brown} and {Lim}}]{harper2001}
{Harper}, G.~M., {Brown}, A.,  and {Lim}, J. (2001). \enquote{{A Spatially
  Resolved, Semiempirical Model for the Extended Atmosphere of {$\alpha$}
  Orionis (M2 Iab)},} \emph{\apj} \textbf{551}, pp. 1073--1098,
  \doi{10.1086/320215}.

\bibitem[{{Hartmann} and {Avrett}(1984)}]{1984ApJ...284..238H}
{Hartmann}, L. and {Avrett}, E.~H. (1984). \enquote{{On the extended
  chromosphere of Alpha Orionis},} \emph{\apj} \textbf{284}, pp. 238--249,
  \doi{10.1086/162402}.

\bibitem[{{Haubois} \emph{et~al.}(2009){Haubois}, {Perrin}, {Lacour},
  {Verhoelst}, {Meimon}, {Mugnier}, {Thi{\'e}baut}, {Berger}, {Ridgway},
  {Monnier}, {Millan-Gabet} and {Traub}}]{2009A&A...508..923H}
{Haubois}, X., {Perrin}, G., {Lacour}, S., {Verhoelst}, T., {Meimon}, S.,
  {Mugnier}, L., {Thi{\'e}baut}, E., {Berger}, J.~P., {Ridgway}, S.~T.,
  {Monnier}, J.~D., {Millan-Gabet}, R.,  and {Traub}, W. (2009).
  \enquote{{Imaging the spotty surface of <ASTROBJ>Betelgeuse</ASTROBJ> in the
  H band},} \emph{\aap} \textbf{508}, pp. 923--932,
  \doi{10.1051/0004-6361/200912927}, \eprint{0910.4167}.

\bibitem[{{Hillen} \emph{et~al.}(2014){Hillen}, {Menu}, {Van Winckel}, {Min},
  {Gielen}, {Wevers}, {Mulders}, {Regibo} and {Verhoelst}}]{2014AAHillen}
{Hillen}, M., {Menu}, J., {Van Winckel}, H., {Min}, M., {Gielen}, C., {Wevers},
  T., {Mulders}, G.~D., {Regibo}, S.,  and {Verhoelst}, T. (2014). \enquote{{An
  interferometric study of the post-AGB binary 89 Herculis. II. Radiative
  transfer models of the circumbinary disk},} \emph{\aap} \textbf{568}, A12,
  \doi{10.1051/0004-6361/201423749}, \eprint{1405.1960}.

\bibitem[{{Hillen} \emph{et~al.}(2013){Hillen}, {Verhoelst}, {Van Winckel},
  {Chesneau}, {Hummel}, {Monnier}, {Farrington}, {Tycner}, {Mourard}, {ten
  Brummelaar}, {Banerjee} and {Zavala}}]{2013AAHillen}
{Hillen}, M., {Verhoelst}, T., {Van Winckel}, H., {Chesneau}, O., {Hummel},
  C.~A., {Monnier}, J.~D., {Farrington}, C., {Tycner}, C., {Mourard}, D., {ten
  Brummelaar}, T., {Banerjee}, D.~P.~K.,  and {Zavala}, R.~T. (2013).
  \enquote{{An interferometric study of the post-AGB binary 89 Herculis. I.
  Spatially resolving the continuum circumstellar environment at optical and
  near-IR wavelengths with the VLTI, NPOI, IOTA, PTI, and the CHARA Array},}
  \emph{\aap} \textbf{559}, A111, \doi{10.1051/0004-6361/201321616},
  \eprint{1308.6715}.

\bibitem[{{H{\"o}fner}(2008)}]{hoefner2008}
{H{\"o}fner}, S. (2008). \enquote{{Winds of M-type AGB stars driven by
  micron-sized grains},} \emph{\aap} \textbf{491}, pp. L1--L4,
  \doi{10.1051/0004-6361:200810641}.

\bibitem[{{H{\"o}nig} \emph{et~al.}(2014){H{\"o}nig}, {Watson}, {Kishimoto} and
  {Hjorth}}]{Hoenig2014}
{H{\"o}nig}, S.~F., {Watson}, D., {Kishimoto}, M.,  and {Hjorth}, J. (2014).
  \enquote{{A dust-parallax distance of 19 megaparsecs to the supermassive
  black hole in NGC 4151},} \emph{\nat} \textbf{515}, pp. 528--530,
  \doi{10.1038/nature13914}, \eprint{1411.7032}.

\bibitem[{{Huber} \emph{et~al.}(2012){Huber}, {Ireland}, {Bedding},
  {Brand{\~a}o}, {Piau}, {Maestro}, {White}, {Bruntt}, {Casagrande},
  {Molenda-{\.Z}akowicz}, {Silva Aguirre}, {Sousa}, {Barclay}, {Burke},
  {Chaplin}, {Christensen-Dalsgaard}, {Cunha}, {De Ridder}, {Farrington},
  {Frasca}, {Garc{\'{\i}}a}, {Gilliland}, {Goldfinger}, {Hekker}, {Kawaler},
  {Kjeldsen}, {McAlister}, {Metcalfe}, {Miglio}, {Monteiro}, {Pinsonneault},
  {Schaefer}, {Stello}, {Stumpe}, {Sturmann}, {Sturmann}, {ten Brummelaar},
  {Thompson}, {Turner} and {Uytterhoeven}}]{huber12}
{Huber}, D., {Ireland}, M.~J., {Bedding}, T.~R., {Brand{\~a}o}, I.~M., {Piau},
  L., {Maestro}, V., {White}, T.~R., {Bruntt}, H., {Casagrande}, L.,
  {Molenda-{\.Z}akowicz}, J., {Silva Aguirre}, V., {Sousa}, S.~G., {Barclay},
  T., {Burke}, C.~J., {Chaplin}, W.~J., {Christensen-Dalsgaard}, J., {Cunha},
  M.~S., {De Ridder}, J., {Farrington}, C.~D., {Frasca}, A., {Garc{\'{\i}}a},
  R.~A., {Gilliland}, R.~L., {Goldfinger}, P.~J., {Hekker}, S., {Kawaler},
  S.~D., {Kjeldsen}, H., {McAlister}, H.~A., {Metcalfe}, T.~S., {Miglio}, A.,
  {Monteiro}, M.~J.~P.~F.~G., {Pinsonneault}, M.~H., {Schaefer}, G.~H.,
  {Stello}, D., {Stumpe}, M.~C., {Sturmann}, J., {Sturmann}, L., {ten
  Brummelaar}, T.~A., {Thompson}, M.~J., {Turner}, N.,  and {Uytterhoeven}, K.
  (2012). \enquote{{Fundamental Properties of Stars Using Asteroseismology from
  Kepler and CoRoT and Interferometry from the CHARA Array},} \emph{\apj}
  \textbf{760}, 32, \doi{10.1088/0004-637X/760/1/32}, \eprint{1210.0012}.

\bibitem[{{Ireland} \emph{et~al.}(2008){Ireland}, {M{\'e}rand}, {ten
  Brummelaar}, {Tuthill}, {Schaefer}, {Turner}, {Sturmann}, {Sturmann} and
  {McAlister}}]{Ireland2008}
{Ireland}, M.~J., {M{\'e}rand}, A., {ten Brummelaar}, T.~A., {Tuthill}, P.~G.,
  {Schaefer}, G.~H., {Turner}, N.~H., {Sturmann}, J., {Sturmann}, L.,  and
  {McAlister}, H.~A. (2008). \enquote{{Sensitive visible interferometry with
  PAVO},} in \emph{Society of Photo-Optical Instrumentation Engineers (SPIE)
  Conference Series}, Vol. 7013, \doi{10.1117/12.788386}.

\bibitem[{{Jankov} \emph{et~al.}(2003){Jankov}, {Domiciano de Souza}, {Stehle},
  {Vakili}, {Perraut-Rousselet} and {Chesneau}}]{2003SPIE.4838..587J}
{Jankov}, S., {Domiciano de Souza}, A., Jr., {Stehle}, C., {Vakili}, F.,
  {Perraut-Rousselet}, K.,  and {Chesneau}, O. (2003).
  \enquote{{Interferometric-Doppler imaging of stellar surface abundances},} in
  W.~A. {Traub} (ed.), \emph{Interferometry for Optical Astronomy II},
  \emph{Society of Photo-Optical Instrumentation Engineers (SPIE) Conference
  Series}, Vol. 4838, pp. 587--593.

\bibitem[{{Jankov} \emph{et~al.}(2001){Jankov}, {Vakili}, {Domiciano de Souza}
  and {Janot-Pacheco}}]{jankov2001}
{Jankov}, S., {Vakili}, F., {Domiciano de Souza}, A., Jr.,  and
  {Janot-Pacheco}, E. (2001). \enquote{{Interferometric-Doppler imaging of
  stellar surface structure},} \emph{\aap} \textbf{377}, pp. 721--734,
  \doi{10.1051/0004-6361:20011047}.

\bibitem[{{Jofr{\'e}} \emph{et~al.}(2014){Jofr{\'e}}, {Heiter}, {Soubiran},
  {Blanco-Cuaresma}, {Worley}, {Pancino}, {Cantat-Gaudin}, {Magrini},
  {Bergemann}, {Gonz{\'a}lez Hern{\'a}ndez}, {Hill}, {Lardo}, {de Laverny},
  {Lind}, {Masseron}, {Montes}, {Mucciarelli}, {Nordlander}, {Recio Blanco},
  {Sobeck}, {Sordo}, {Sousa}, {Tabernero}, {Vallenari} and {Van Eck}}]{jofre14}
{Jofr{\'e}}, P., {Heiter}, U., {Soubiran}, C., {Blanco-Cuaresma}, S., {Worley},
  C.~C., {Pancino}, E., {Cantat-Gaudin}, T., {Magrini}, L., {Bergemann}, M.,
  {Gonz{\'a}lez Hern{\'a}ndez}, J.~I., {Hill}, V., {Lardo}, C., {de Laverny},
  P., {Lind}, K., {Masseron}, T., {Montes}, D., {Mucciarelli}, A.,
  {Nordlander}, T., {Recio Blanco}, A., {Sobeck}, J., {Sordo}, R., {Sousa},
  S.~G., {Tabernero}, H., {Vallenari}, A.,  and {Van Eck}, S. (2014).
  \enquote{{Gaia FGK benchmark stars: Metallicity},} \emph{\aap} \textbf{564},
  A133, \doi{10.1051/0004-6361/201322440}, \eprint{1309.1099}.

\bibitem[{{Josselin} and {Plez}(2007)}]{2007A&A...469..671J}
{Josselin}, E. and {Plez}, B. (2007). \enquote{{Atmospheric dynamics and the
  mass loss process in red supergiant stars},} \emph{\aap} \textbf{469}, pp.
  671--680, \doi{10.1051/0004-6361:20066353}, \eprint{0705.0266}.

\bibitem[{{Kamath} \emph{et~al.}(2014){Kamath}, {Wood} and {Van
  Winckel}}]{2014MNRASKamath}
{Kamath}, D., {Wood}, P.~R.,  and {Van Winckel}, H. (2014). \enquote{{Optically
  visible post-AGB/RGB stars and young stellar objects in the Small Magellanic
  Cloud: candidate selection, spectral energy distributions and spectroscopic
  examination},} \emph{\mnras} \textbf{439}, pp. 2211--2270,
  \doi{10.1093/mnras/stt2033}, \eprint{1402.5954}.

\bibitem[{{Kaspi} \emph{et~al.}(2007){Kaspi}, {Brandt}, {Maoz}, {Netzer},
  {Schneider} and {Shemmer}}]{Kaspi2007}
{Kaspi}, S., {Brandt}, W.~N., {Maoz}, D., {Netzer}, H., {Schneider}, D.~P.,
  and {Shemmer}, O. (2007). \enquote{{Reverberation Mapping of High-Luminosity
  Quasars: First Results},} \emph{\apj} \textbf{659}, pp. 997--1007,
  \doi{10.1086/512094}, \eprint{astro-ph/0612722}.

\bibitem[{{Keller}(2008)}]{keller08}
{Keller}, S.~C. (2008). \enquote{{Cepheid Mass loss and the
  Pulsation-Evolutionary Mass Discrepancy},} \emph{\apj} \textbf{677}, pp.
  483--487, \doi{10.1086/529366}, \eprint{0801.1342}.

\bibitem[{{Kervella} \emph{et~al.}(2013){Kervella}, {Gallenne} and
  {M{\'e}rand}}]{kervella13}
{Kervella}, P., {Gallenne}, A.,  and {M{\'e}rand}, A. (2013).
  \enquote{{Circumstellar envelopes of Cepheids: a possible bias affecting the
  distance scale?}} in R.~{de Grijs} (ed.), \emph{IAU Symposium}, \emph{IAU
  Symposium}, Vol. 289, pp. 157--160, \doi{10.1017/S1743921312021291}.

\bibitem[{{Kervella} \emph{et~al.}(2004{\natexlab{a}}){Kervella}, {Nardetto},
  {Bersier}, {Mourard} and {Coud{\'e} du Foresto}}]{kervella04a}
{Kervella}, P., {Nardetto}, N., {Bersier}, D., {Mourard}, D.,  and {Coud{\'e}
  du Foresto}, V. (2004{\natexlab{a}}). \enquote{{Cepheid distances from
  infrared long-baseline interferometry. I. VINCI/VLTI observations of seven
  Galactic Cepheids},} \emph{\aap} \textbf{416}, pp. 941--953,
  \doi{10.1051/0004-6361:20031743}, \eprint{astro-ph/0311525}.

\bibitem[{{Kervella} \emph{et~al.}(2004{\natexlab{b}}){Kervella},
  {Th{\'e}venin}, {Di Folco} and {S{\'e}gransan}}]{ker04}
{Kervella}, P., {Th{\'e}venin}, F., {Di Folco}, E.,  and {S{\'e}gransan}, D.
  (2004{\natexlab{b}}). \enquote{{The angular sizes of dwarf stars and
  subgiants. Surface brightness relations calibrated by interferometry},}
  \emph{\aap} \textbf{426}, pp. 297--307, \doi{10.1051/0004-6361:20035930},
  \eprint{astro-ph/0404180}.

\bibitem[{{Kochukhov} \emph{et~al.}(2007){Kochukhov}, {Adelman}, {Gulliver} and
  {Piskunov}}]{2007NatPh...3..526K}
{Kochukhov}, O., {Adelman}, S.~J., {Gulliver}, A.~F.,  and {Piskunov}, N.
  (2007). \enquote{{Weather in stellar atmosphere revealed by the dynamics of
  mercury clouds in {$\alpha$} Andromedae},} \emph{Nature Physics} \textbf{3},
  pp. 526--529, \doi{10.1038/nphys648}, \eprint{0705.4469}.

\bibitem[{{Kochukhov} and {Wade}(2010)}]{2010A&A...513A..13K}
{Kochukhov}, O. and {Wade}, G.~A. (2010). \enquote{{Magnetic Doppler imaging of
  {$\alpha$}$^{2}$ Canum Venaticorum in all four Stokes parameters. Unveiling
  the hidden complexity of stellar magnetic fields},} \emph{\aap} \textbf{513},
  A13, \doi{10.1051/0004-6361/200913860}, \eprint{1002.0025}.

\bibitem[{{Koechlin} and {Rabbia}(1985)}]{koechlin1985}
{Koechlin}, L. and {Rabbia}, Y. (1985). \enquote{{Stellar diameter measurements
  with the CERGA optical interferometer - Recent developments and results},}
  \emph{\aap} \textbf{153}, pp. 91--98.

\bibitem[{{Kraus} \emph{et~al.}(2013){Kraus}, {Oksala}, {Nickeler}, {Muratore},
  {Borges Fernandes}, {Aret}, {Cidale} and {de Wit}}]{Kraus2013}
{Kraus}, M., {Oksala}, M.~E., {Nickeler}, D.~H., {Muratore}, M.~F., {Borges
  Fernandes}, M., {Aret}, A., {Cidale}, L.~S.,  and {de Wit}, W.~J. (2013).
  \enquote{{Molecular emission from GG Carinae's circumbinary disk},}
  \emph{\aap} \textbf{549}, A28, \doi{10.1051/0004-6361/201220442},
  \eprint{1211.5149}.

\bibitem[{{Kraus} \emph{et~al.}(2012{\natexlab{a}}){Kraus}, {Calvet},
  {Hartmann}, {Hofmann}, {Kreplin}, {Monnier} and {Weigelt}}]{Kraus2012a}
{Kraus}, S., {Calvet}, N., {Hartmann}, L., {Hofmann}, K.-H., {Kreplin}, A.,
  {Monnier}, J.~D.,  and {Weigelt}, G. (2012{\natexlab{a}}). \enquote{{On the
  Nature of the Herbig B[e] Star Binary System V921 Scorpii: Discovery of a
  Close Companion and Relation to the Large-scale Bipolar Nebula},}
  \emph{\apjl} \textbf{746}, L2, \doi{10.1088/2041-8205/746/1/L2},
  \eprint{1201.2420}.

\bibitem[{{Kraus} \emph{et~al.}(2012{\natexlab{b}}){Kraus}, {Calvet},
  {Hartmann}, {Hofmann}, {Kreplin}, {Monnier} and {Weigelt}}]{Kraus2012b}
{Kraus}, S., {Calvet}, N., {Hartmann}, L., {Hofmann}, K.-H., {Kreplin}, A.,
  {Monnier}, J.~D.,  and {Weigelt}, G. (2012{\natexlab{b}}). \enquote{{On the
  Nature of the Herbig B[e] Star Binary System V921 Scorpii: Geometry and
  Kinematics of the Circumprimary Disk on Sub-AU Scales},} \emph{\apj}
  \textbf{752}, 11, \doi{10.1088/0004-637X/752/1/11}, \eprint{1204.1969}.

\bibitem[{{Kuschnig} \emph{et~al.}(1999){Kuschnig}, {Ryabchikova}, {Piskunov},
  {Weiss} and {Gelbmann}}]{1999A&A...348..924K}
{Kuschnig}, R., {Ryabchikova}, T.~A., {Piskunov}, N.~E., {Weiss}, W.~W.,  and
  {Gelbmann}, M.~J. (1999). \enquote{{Multi element Doppler imaging of AP
  stars. I. He, Mg, Si, CR and Fe surface distribution for CU Virginis},}
  \emph{\aap} \textbf{348}, pp. 924--932.

\bibitem[{{Lamers} \emph{et~al.}(1998){Lamers}, {Zickgraf}, {de Winter},
  {Houziaux} and {Zorec}}]{Lamers}
{Lamers}, H.~J.~G.~L.~M., {Zickgraf}, F.-J., {de Winter}, D., {Houziaux}, L.,
  and {Zorec}, J. (1998). \enquote{{An improved classification of B[e]-type
  stars},} \emph{\aap} \textbf{340}, pp. 117--128.

\bibitem[{{Le Bouquin} \emph{et~al.}(2011){Le Bouquin}, {Berger}, {Lazareff},
  {Zins}, {Haguenauer}, {Jocou}, {Kern}, {Millan-Gabet}, {Traub}, {Absil},
  {Augereau}, {Benisty}, {Blind}, {Bonfils}, {Bourget}, {Delboulbe},
  {Feautrier}, {Germain}, {Gitton}, {Gillier}, {Kiekebusch}, {Kluska},
  {Knudstrup}, {Labeye}, {Lizon}, {Monin}, {Magnard}, {Malbet}, {Maurel},
  {M{\'e}nard}, {Micallef}, {Michaud}, {Montagnier}, {Morel}, {Moulin},
  {Perraut}, {Popovic}, {Rabou}, {Rochat}, {Rojas}, {Roussel}, {Roux},
  {Stadler}, {Stefl}, {Tatulli} and {Ventura}}]{lebouquin2011}
{Le Bouquin}, J.-B., {Berger}, J.-P., {Lazareff}, B., {Zins}, G., {Haguenauer},
  P., {Jocou}, L., {Kern}, P., {Millan-Gabet}, R., {Traub}, W., {Absil}, O.,
  {Augereau}, J.-C., {Benisty}, M., {Blind}, N., {Bonfils}, X., {Bourget}, P.,
  {Delboulbe}, A., {Feautrier}, P., {Germain}, M., {Gitton}, P., {Gillier}, D.,
  {Kiekebusch}, M., {Kluska}, J., {Knudstrup}, J., {Labeye}, P., {Lizon},
  J.-L., {Monin}, J.-L., {Magnard}, Y., {Malbet}, F., {Maurel}, D.,
  {M{\'e}nard}, F., {Micallef}, M., {Michaud}, L., {Montagnier}, G., {Morel},
  S., {Moulin}, T., {Perraut}, K., {Popovic}, D., {Rabou}, P., {Rochat}, S.,
  {Rojas}, C., {Roussel}, F., {Roux}, A., {Stadler}, E., {Stefl}, S.,
  {Tatulli}, E.,  and {Ventura}, N. (2011). \enquote{{PIONIER: a 4-telescope
  visitor instrument at VLTI},} \emph{\aap} \textbf{535}, A67,
  \doi{10.1051/0004-6361/201117586}, \eprint{1109.1918}.

\bibitem[{{Leinert} \emph{et~al.}(2003){Leinert}, {Graser}, {Przygodda},
  {Waters}, {Perrin}, {Jaffe}, {Lopez}, {Bakker}, {B{\"o}hm}, {Chesneau},
  {Cotton}, {Damstra}, {de Jong}, {Glazenborg-Kluttig}, {Grimm}, {Hanenburg},
  {Laun}, {Lenzen}, {Ligori}, {Mathar}, {Meisner}, {Morel}, {Morr}, {Neumann},
  {Pel}, {Schuller}, {Rohloff}, {Stecklum}, {Storz}, {von der L{\"u}he} and
  {Wagner}}]{leinert2003}
{Leinert}, C., {Graser}, U., {Przygodda}, F., {Waters}, L.~B.~F.~M., {Perrin},
  G., {Jaffe}, W., {Lopez}, B., {Bakker}, E.~J., {B{\"o}hm}, A., {Chesneau},
  O., {Cotton}, W.~D., {Damstra}, S., {de Jong}, J., {Glazenborg-Kluttig},
  A.~W., {Grimm}, B., {Hanenburg}, H., {Laun}, W., {Lenzen}, R., {Ligori}, S.,
  {Mathar}, R.~J., {Meisner}, J., {Morel}, S., {Morr}, W., {Neumann}, U.,
  {Pel}, J.-W., {Schuller}, P., {Rohloff}, R.-R., {Stecklum}, B., {Storz}, C.,
  {von der L{\"u}he}, O.,  and {Wagner}, K. (2003). \enquote{{MIDI - the 10
  $\backslash$mum instrument on the VLTI},} \emph{\apss} \textbf{286}, pp.
  73--83, \doi{10.1023/A:1026158127732}.

\bibitem[{{Levesque}(2010)}]{2010NewAR..54....1L}
{Levesque}, E.~M. (2010). \enquote{{The physical properties of red
  supergiants},} \emph{\nar} \textbf{54}, pp. 1--12,
  \doi{10.1016/j.newar.2009.10.002}, \eprint{0911.4720}.

\bibitem[{{Levesque} \emph{et~al.}(2005){Levesque}, {Massey}, {Olsen}, {Plez},
  {Josselin}, {Maeder} and {Meynet}}]{2005ApJ...628..973L}
{Levesque}, E.~M., {Massey}, P., {Olsen}, K.~A.~G., {Plez}, B., {Josselin}, E.,
  {Maeder}, A.,  and {Meynet}, G. (2005). \enquote{{The Effective Temperature
  Scale of Galactic Red Supergiants: Cool, but Not As Cool As We Thought},}
  \emph{\apj} \textbf{628}, pp. 973--985, \doi{10.1086/430901},
  \eprint{arXiv:astro-ph/0504337}.

\bibitem[{{Ligi} \emph{et~al.}(2014){Ligi}, {Mourard}, {Lagrange}, {Perraut}
  and {Chiavassa}}]{Ligi2014}
{Ligi}, R., {Mourard}, D., {Lagrange}, A.-M., {Perraut}, K.,  and {Chiavassa},
  A. (2014). \enquote{{On the characterization of transiting exoplanets and
  magnetic spots with optical interferometry},} \emph{ArXiv e-prints}
  \eprint{1410.5333}.

\bibitem[{{Ligi} \emph{et~al.}(2015){Ligi}, {Mourard}, {Lagrange}, {Perraut}
  and {Chiavassa}}]{Ligi2015}
{Ligi}, R., {Mourard}, D., {Lagrange}, A.-M., {Perraut}, K.,  and {Chiavassa},
  A. (2015). \enquote{{Transiting exoplanets and magnetic spots characterized
  with optical interferometry},} \emph{\aap} \textbf{574}, A69,
  \doi{10.1051/0004-6361/201424013}, \eprint{1410.5333}.

\bibitem[{Ligi \emph{et~al.}(2013)Ligi, Mourard, Nardetto and
  Clausse}]{Ligi2013}
Ligi, R., Mourard, D., Nardetto, N.,  and Clausse, J.-M. (2013). \enquote{The
  operation of vega/chara: From the scientific idea to the final products,}
  \emph{Journal of Astronomical Instrumentation} \textbf{02}, 02, p. 1340003,
  \doi{10.1142/S2251171713400035},
  \eprint{http://www.worldscientific.com/doi/pdf/10.1142/S2251171713400035}.

\bibitem[{{Lopez} \emph{et~al.}(2014){Lopez}, {Lagarde}, {Jaffe}, {Petrov},
  {Sch{\"o}ller}, {Antonelli}, {Beckman}, {B{\'e}rio}, {Bettonvil}, {Graser},
  {Millour}, {Robbe-Dubois}, {Venema}, {Wolf}, {Bristow}, {Glindemann},
  {Gonzalez}, {Lanz}, {Henning}, {Weigelt}, {Ag{\'o}cs}, {Augereau},
  {{\'A}vila}, {Bailet}, {Behrend}, {Berger}, {von Boekel}, {Bonhomme},
  {Bourget}, {Brast}, {Bresson}, {Clausse}, {Chesneau}, {Cs{\'e}p{\'a}ny},
  {Connot}, {Crida}, {Danchi}, {Delbo}, {Delplancke}, {Dominik}, {Dugu{\'e}},
  {Elswijk}, {Fante{\"i}}, {Finger}, {Gabasch}, {Girard}, {Girault}, {Gitton},
  {Glazenborg}, {Gont{\'e}}, {Guitton}, {Guniat}, {De Haan}, {Haguenauer},
  {Hanenburg}, {Heininger}, {Hofmann}, {Hogerheijde}, {ter Horst}, {Hron},
  {Hughes}, {Ives}, {Jakob}, {Jask{\'o}}, {Jolley}, {Kragt}, {K{\"o}hler},
  {Kroener}, {Kroes}, {Labadie}, {Laun}, {Lehmitz}, {Leinert}, {Lizon},
  {Lucuix}, {Marcotto}, {Martinache}, {Matter}, {Martinot-Lagarde}, {Mauclert},
  {Mehrgan}, {Meilland}, {Mellein}, {M{\'e}nardi}, {Menut}, {Meisenheimer},
  {Morel}, {Mosoni}, {Navarro}, {Neumann}, {Nussbaum}, {Ottogalli}, {Palsa},
  {Panduro}, {Pantin}, {Percheron}, {Duc}, {Pott}, {Pozna}, {Przygodda},
  {Richichi}, {Rigal}, {Rupprecht}, {Schertl}, {Stegmeier}, {Thiam}, {Tromp},
  {Vannier}, {Vakili}, {van Belle}, {Wagner} and {Woillez}}]{lopez2014}
{Lopez}, B., {Lagarde}, S., {Jaffe}, W., {Petrov}, R., {Sch{\"o}ller}, M.,
  {Antonelli}, P., {Beckman}, U., {B{\'e}rio}, P., {Bettonvil}, F., {Graser},
  U., {Millour}, F., {Robbe-Dubois}, S., {Venema}, L., {Wolf}, S., {Bristow},
  P., {Glindemann}, A., {Gonzalez}, J.-C., {Lanz}, T., {Henning}, T.,
  {Weigelt}, G., {Ag{\'o}cs}, T., {Augereau}, J.-C., {{\'A}vila}, G., {Bailet},
  C., {Behrend}, J., {Berger}, J.-P., {von Boekel}, R., {Bonhomme}, S.,
  {Bourget}, P., {Brast}, R., {Bresson}, Y., {Clausse}, J.~M., {Chesneau}, O.,
  {Cs{\'e}p{\'a}ny}, G., {Connot}, C., {Crida}, A., {Danchi}, W.~C., {Delbo},
  M., {Delplancke}, F., {Dominik}, C., {Dugu{\'e}}, M., {Elswijk}, E.,
  {Fante{\"i}}, Y., {Finger}, G., {Gabasch}, A., {Girard}, P., {Girault}, V.,
  {Gitton}, P., {Glazenborg}, A., {Gont{\'e}}, F., {Guitton}, F., {Guniat}, S.,
  {De Haan}, M., {Haguenauer}, P., {Hanenburg}, H., {Heininger}, M., {Hofmann},
  K.-H., {Hogerheijde}, M., {ter Horst}, R., {Hron}, J., {Hughes}, Y., {Ives},
  D., {Jakob}, G., {Jask{\'o}}, A., {Jolley}, P., {Kragt}, J., {K{\"o}hler},
  R., {Kroener}, T., {Kroes}, G., {Labadie}, L., {Laun}, W., {Lehmitz}, M.,
  {Leinert}, C., {Lizon}, J.~L., {Lucuix}, C., {Marcotto}, A., {Martinache},
  F., {Matter}, A., {Martinot-Lagarde}, G., {Mauclert}, N., {Mehrgan}, L.,
  {Meilland}, A., {Mellein}, M., {M{\'e}nardi}, S., {Menut}, J.~L.,
  {Meisenheimer}, K., {Morel}, S., {Mosoni}, L., {Navarro}, R., {Neumann}, U.,
  {Nussbaum}, E., {Ottogalli}, S., {Palsa}, R., {Panduro}, J., {Pantin}, E.,
  {Percheron}, I., {Duc}, T.~P., {Pott}, J.-U., {Pozna}, E., {Przygodda}, F.,
  {Richichi}, A., {Rigal}, F., {Rupprecht}, G., {Schertl}, D., {Stegmeier}, J.,
  {Thiam}, L., {Tromp}, N., {Vannier}, M., {Vakili}, F., {van Belle}, G.,
  {Wagner}, K.,  and {Woillez}, J. (2014). \enquote{{MATISSE status report and
  science forecast},} in \emph{Society of Photo-Optical Instrumentation
  Engineers (SPIE) Conference Series}, \emph{Society of Photo-Optical
  Instrumentation Engineers (SPIE) Conference Series}, Vol. 9146, p.~0,
  \doi{10.1117/12.2056419}.

\bibitem[{{L{\"u}ftinger} \emph{et~al.}(2010){L{\"u}ftinger}, {Kochukhov},
  {Ryabchikova}, {Piskunov}, {Weiss} and {Ilyin}}]{2010A&A...509A..71L}
{L{\"u}ftinger}, T., {Kochukhov}, O., {Ryabchikova}, T., {Piskunov}, N.,
  {Weiss}, W.~W.,  and {Ilyin}, I. (2010). \enquote{{Magnetic Doppler imaging
  of the roAp star HD 24712},} \emph{\aap} \textbf{509}, A71,
  \doi{10.1051/0004-6361/200811545}, \eprint{0910.5556}.

\bibitem[{{Maercker} \emph{et~al.}(2012){Maercker}, {Mohamed}, {Vlemmings},
  {Ramstedt}, {Groenewegen}, {Humphreys}, {Kerschbaum}, {Lindqvist},
  {Olofsson}, {Paladini}, {Wittkowski}, {de Gregorio-Monsalvo} and
  {Nyman}}]{Maercker2012}
{Maercker}, M., {Mohamed}, S., {Vlemmings}, W.~H.~T., {Ramstedt}, S.,
  {Groenewegen}, M.~A.~T., {Humphreys}, E., {Kerschbaum}, F., {Lindqvist}, M.,
  {Olofsson}, H., {Paladini}, C., {Wittkowski}, M., {de Gregorio-Monsalvo}, I.,
   and {Nyman}, L.-A. (2012). \enquote{{Unexpectedly large mass loss during the
  thermal pulse cycle of the red giant star R Sculptoris},} \emph{\nat}
  \textbf{490}, pp. 232--234, \doi{10.1038/nature11511}, \eprint{1210.3030}.

\bibitem[{{Maestro} \emph{et~al.}(2013){Maestro}, {Che}, {Huber}, {Ireland},
  {Monnier}, {White}, {Kok}, {Robertson}, {Schaefer}, {ten Brummelaar} and
  {Tuthill}}]{2013MNRAS.434.1321M}
{Maestro}, V., {Che}, X., {Huber}, D., {Ireland}, M.~J., {Monnier}, J.~D.,
  {White}, T.~R., {Kok}, Y., {Robertson}, J.~G., {Schaefer}, G.~H., {ten
  Brummelaar}, T.~A.,  and {Tuthill}, P.~G. (2013). \enquote{{Optical
  interferometry of early-type stars with PAVO@CHARA - I. Fundamental stellar
  properties},} \emph{\mnras} \textbf{434}, pp. 1321--1331,
  \doi{10.1093/mnras/stt1092}, \eprint{1306.5937}.

\bibitem[{{McAlister} \emph{et~al.}(2012){McAlister}, {ten Brummelaar},
  {Ridgway}, {Gies}, {Sturmann}, {Sturmann}, {Turner}, {Schaefer}, {Boyajian},
  {Farrington}, {Goldfinger} and {Webster}}]{McAlister2012}
{McAlister}, H.~A., {ten Brummelaar}, T.~A., {Ridgway}, S.~T., {Gies}, D.~R.,
  {Sturmann}, J., {Sturmann}, L., {Turner}, N.~H., {Schaefer}, G.~H.,
  {Boyajian}, T.~S., {Farrington}, C.~D., {Goldfinger}, P.~J.,  and {Webster},
  L. (2012). \enquote{{Recent technical and scientific highlights from the
  CHARA Array},} in \emph{Society of Photo-Optical Instrumentation Engineers
  (SPIE) Conference Series}, Vol. 8445, \doi{10.1117/12.926452}.

\bibitem[{{McNamara} \emph{et~al.}(2007){McNamara}, {Clementini} and
  {Marconi}}]{mcnamara07}
{McNamara}, D.~H., {Clementini}, G.,  and {Marconi}, M. (2007). \enquote{{A
  {$\delta$} Scuti Distance to the Large Magellanic Cloud},} \emph{\aj}
  \textbf{133}, pp. 2752--2763, \doi{10.1086/513717},
  \eprint{astro-ph/0702107}.

\bibitem[{{Mehner} \emph{et~al.}(2015){Mehner}, {Davidson}, {Humphreys},
  {Walter}, {Baade}, {de Wit}, {Martin}, {Ishibashi}, {Rivinius}, {Martayan},
  {Ruiz} and {Weis}}]{Mehner2015}
{Mehner}, A., {Davidson}, K., {Humphreys}, R.~M., {Walter}, F.~M., {Baade}, D.,
  {de Wit}, W.~J., {Martin}, J., {Ishibashi}, K., {Rivinius}, T., {Martayan},
  C., {Ruiz}, M.~T.,  and {Weis}, K. (2015). \enquote{{Eta Carinae's 2014.6
  spectroscopic event: Clues to the long-term recovery from its Great
  Eruption},} \emph{\aap} \textbf{578}, A122,
  \doi{10.1051/0004-6361/201425522}, \eprint{1504.04940}.

\bibitem[{{Meilland} \emph{et~al.}(2010){Meilland}, {Kanaan}, {Borges
  Fernandes}, {Chesneau}, {Millour}, {Stee} and {Lopez}}]{Meilland}
{Meilland}, A., {Kanaan}, S., {Borges Fernandes}, M., {Chesneau}, O.,
  {Millour}, F., {Stee}, P.,  and {Lopez}, B. (2010). \enquote{{Resolving the
  dusty circumstellar environment of the A[e] supergiant HD 62623 with the
  VLTI/MIDI},} \emph{\aap} \textbf{512}, A73,
  \doi{10.1051/0004-6361/200913640}, \eprint{0912.1954}.

\bibitem[{{M{\'e}rand} \emph{et~al.}(2007){M{\'e}rand}, {Aufdenberg},
  {Kervella}, {Foresto}, {ten Brummelaar}, {McAlister}, {Sturmann}, {Sturmann}
  and {Turner}}]{merand07}
{M{\'e}rand}, A., {Aufdenberg}, J.~P., {Kervella}, P., {Foresto}, V.~C.~d.,
  {ten Brummelaar}, T.~A., {McAlister}, H.~A., {Sturmann}, L., {Sturmann}, J.,
  and {Turner}, N.~H. (2007). \enquote{{Extended Envelopes around Galactic
  Cepheids. III. Y Ophiuchi and {$\alpha$} Persei from Near-Infrared
  Interferometry with CHARA/FLUOR},} \emph{\apj} \textbf{664}, pp. 1093--1101,
  \doi{10.1086/518597}, \eprint{0704.1825}.

\bibitem[{{M{\'e}rand} \emph{et~al.}(2005){M{\'e}rand}, {Kervella}, {Coud{\'e}
  du Foresto}, {Ridgway}, {Aufdenberg}, {ten Brummelaar}, {Berger}, {Sturmann},
  {Sturmann}, {Turner} and {McAlister}}]{merand05}
{M{\'e}rand}, A., {Kervella}, P., {Coud{\'e} du Foresto}, V., {Ridgway}, S.~T.,
  {Aufdenberg}, J.~P., {ten Brummelaar}, T.~A., {Berger}, D.~H., {Sturmann},
  J., {Sturmann}, L., {Turner}, N.~H.,  and {McAlister}, H.~A. (2005).
  \enquote{{The projection factor of {$\delta$} Cephei. A calibration of the
  Baade-Wesselink method using the CHARA Array},} \emph{\aap} \textbf{438}, pp.
  L9--L12, \doi{10.1051/0004-6361:200500139}.

\bibitem[{{Mermilliod} \emph{et~al.}(2007){Mermilliod}, {Andersen}, {Latham}
  and {Mayor}}]{Mermilliod2007}
{Mermilliod}, J.-C., {Andersen}, J., {Latham}, D.~W.,  and {Mayor}, M. (2007).
  \enquote{{Red giants in open clusters. XIII. Orbital elements of 156
  spectroscopic binaries},} \emph{\aap} \textbf{473}, pp. 829--845,
  \doi{10.1051/0004-6361:20078007}.

\bibitem[{{Metcalfe} \emph{et~al.}(2014){Metcalfe}, {Creevey}, {Do{\u g}an},
  {Mathur}, {Xu}, {Bedding}, {Chaplin}, {Christensen-Dalsgaard}, {Karoff},
  {Trampedach}, {Benomar}, {Brown}, {Buzasi}, {Campante}, {{\c C}elik},
  {Cunha}, {Davies}, {Deheuvels}, {Derekas}, {Di Mauro}, {Garc{\'{\i}}a},
  {Guzik}, {Howe}, {MacGregor}, {Mazumdar}, {Montalb{\'a}n}, {Monteiro},
  {Salabert}, {Serenelli}, {Stello}, {Stesacute}, {licki}, {Suran},
  {Y{\i}ld{\i}z}, {Aksoy}, {Elsworth}, {Gruberbauer}, {Guenther}, {Lebreton},
  {Molaverdikhani}, {Pricopi}, {Simoniello} and {White}}]{metcalfe14}
{Metcalfe}, T.~S., {Creevey}, O.~L., {Do{\u g}an}, G., {Mathur}, S., {Xu}, H.,
  {Bedding}, T.~R., {Chaplin}, W.~J., {Christensen-Dalsgaard}, J., {Karoff},
  C., {Trampedach}, R., {Benomar}, O., {Brown}, B.~P., {Buzasi}, D.~L.,
  {Campante}, T.~L., {{\c C}elik}, Z., {Cunha}, M.~S., {Davies}, G.~R.,
  {Deheuvels}, S., {Derekas}, A., {Di Mauro}, M.~P., {Garc{\'{\i}}a}, R.~A.,
  {Guzik}, J.~A., {Howe}, R., {MacGregor}, K.~B., {Mazumdar}, A.,
  {Montalb{\'a}n}, J., {Monteiro}, M.~J.~P.~F.~G., {Salabert}, D., {Serenelli},
  A., {Stello}, D., {Stesacute}, {licki}, M., {Suran}, M.~D., {Y{\i}ld{\i}z},
  M., {Aksoy}, C., {Elsworth}, Y., {Gruberbauer}, M., {Guenther}, D.~B.,
  {Lebreton}, Y., {Molaverdikhani}, K., {Pricopi}, D., {Simoniello}, R.,  and
  {White}, T.~R. (2014). \enquote{{Properties of 42 Solar-type Kepler Targets
  from the Asteroseismic Modeling Portal},} \emph{\apjs} \textbf{214}, 27,
  \doi{10.1088/0067-0049/214/2/27}, \eprint{1402.3614}.

\bibitem[{{Michaud}(1970)}]{1970ApJ...160..641M}
{Michaud}, G. (1970). \enquote{{Diffusion Processes in Peculiar a Stars},}
  \emph{\apj} \textbf{160}, p. 641, \doi{10.1086/150459}.

\bibitem[{{Miglio} \emph{et~al.}(2013){Miglio}, {Chiappini}, {Morel},
  {Barbieri}, {Chaplin}, {Girardi}, {Montalb{\'a}n}, {Valentini}, {Mosser},
  {Baudin}, {Casagrande}, {Fossati}, {Aguirre} and {Baglin}}]{miglio13}
{Miglio}, A., {Chiappini}, C., {Morel}, T., {Barbieri}, M., {Chaplin}, W.~J.,
  {Girardi}, L., {Montalb{\'a}n}, J., {Valentini}, M., {Mosser}, B., {Baudin},
  F., {Casagrande}, L., {Fossati}, L., {Aguirre}, V.~S.,  and {Baglin}, A.
  (2013). \enquote{{Galactic archaeology: mapping and dating stellar
  populations with asteroseismology of red-giant stars},} \emph{\mnras}
  \textbf{429}, pp. 423--428, \doi{10.1093/mnras/sts345}, \eprint{1211.0146}.

\bibitem[{{Millour} \emph{et~al.}(2009){Millour}, {Chesneau}, {Borges
  Fernandes}, {Meilland}, {Mars}, {Benoist}, {Thi{\'e}baut}, {Stee}, {Hofmann},
  {Baron}, {Young}, {Bendjoya}, {Carciofi}, {Domiciano de Souza}, {Driebe},
  {Jankov}, {Kervella}, {Petrov}, {Robbe-Dubois}, {Vakili}, {Waters} and
  {Weigelt}}]{Millour2009}
{Millour}, F., {Chesneau}, O., {Borges Fernandes}, M., {Meilland}, A., {Mars},
  G., {Benoist}, C., {Thi{\'e}baut}, E., {Stee}, P., {Hofmann}, K.-H., {Baron},
  F., {Young}, J., {Bendjoya}, P., {Carciofi}, A., {Domiciano de Souza}, A.,
  {Driebe}, T., {Jankov}, S., {Kervella}, P., {Petrov}, R.~G., {Robbe-Dubois},
  S., {Vakili}, F., {Waters}, L.~B.~F.~M.,  and {Weigelt}, G. (2009).
  \enquote{{A binary engine fuelling HD 87643's complex circumstellar
  environment. Determined using AMBER/VLTI imaging},} \emph{\aap} \textbf{507},
  pp. 317--326, \doi{10.1051/0004-6361/200811592}, \eprint{0908.0227}.

\bibitem[{{Millour} \emph{et~al.}(2014){Millour}, {Chesneau}, {Meilland} and
  {Nardetto}}]{Millour2014}
{Millour}, F., {Chesneau}, O., {Meilland}, A.,  and {Nardetto}, N. (2014).
  \enquote{{Optical Interferometry and Adaptive Optics of Bright Transients},}
  in P.~R. {Wozniak}, M.~J. {Graham}, A.~A. {Mahabal},  and R.~{Seaman} (eds.),
  \emph{The Third Hot-wiring the Transient Universe Workshop}, pp. 177--183,
  \eprint{1407.5153}.

\bibitem[{{Millour} \emph{et~al.}(2011){Millour}, {Meilland}, {Chesneau},
  {Stee}, {Kanaan}, {Petrov}, {Mourard} and {Kraus}}]{Millour2011}
{Millour}, F., {Meilland}, A., {Chesneau}, O., {Stee}, P., {Kanaan}, S.,
  {Petrov}, R., {Mourard}, D.,  and {Kraus}, S. (2011). \enquote{{Imaging the
  spinning gas and dust in the disc around the supergiant A[e] star HD 62623},}
  \emph{\aap} \textbf{526}, A107, \doi{10.1051/0004-6361/201016193},
  \eprint{1012.2957}.

\bibitem[{{Milone} \emph{et~al.}(1994){Milone}, {Wilson}, {Fry} and
  {Schiller}}]{milone94}
{Milone}, E.~F., {Wilson}, W.~J.~F., {Fry}, D.~J.~I.,  and {Schiller}, S.~J.
  (1994). \enquote{{Studies of large-amplitude Delta Scuti variables. 2: DY
  Herculis},} \emph{\pasp} \textbf{106}, pp. 1120--1133, \doi{10.1086/133488}.

\bibitem[{{Miroshnichenko}(2007)}]{Miroshnichenko}
{Miroshnichenko}, A.~S. (2007). \enquote{{Toward Understanding the B[e]
  Phenomenon. I. Definition of the Galactic FS CMa Stars},} \emph{\apj}
  \textbf{667}, pp. 497--504, \doi{10.1086/520798}.

\bibitem[{{Monnier} \emph{et~al.}(2007){Monnier}, {Zhao}, {Pedretti},
  {Thureau}, {Ireland}, {Muirhead}, {Berger}, {Millan-Gabet}, {Van Belle}, {ten
  Brummelaar}, {McAlister}, {Ridgway}, {Turner}, {Sturmann}, {Sturmann} and
  {Berger}}]{monnier2007}
{Monnier}, J.~D., {Zhao}, M., {Pedretti}, E., {Thureau}, N., {Ireland}, M.,
  {Muirhead}, P., {Berger}, J.-P., {Millan-Gabet}, R., {Van Belle}, G., {ten
  Brummelaar}, T., {McAlister}, H., {Ridgway}, S., {Turner}, N., {Sturmann},
  L., {Sturmann}, J.,  and {Berger}, D. (2007). \enquote{{Imaging the Surface
  of Altair},} \emph{Science} \textbf{317}, pp. 342--,
  \doi{10.1126/science.1143205}, \eprint{0706.0867}.

\bibitem[{{Monnier} \emph{et~al.}(2008){Monnier}, {Zhao}, {Pedretti},
  {Thureau}, {Ireland}, {Muirhead}, {Berger}, {Millan-Gabet}, {Van Belle}, {ten
  Brummelaar}, {McAlister}, {Ridgway}, {Turner}, {Sturmann}, {Sturmann},
  {Berger}, {Tannirkulam} and {Blum}}]{monnier08}
{Monnier}, J.~D., {Zhao}, M., {Pedretti}, E., {Thureau}, N., {Ireland}, M.,
  {Muirhead}, P., {Berger}, J.-P., {Millan-Gabet}, R., {Van Belle}, G., {ten
  Brummelaar}, T., {McAlister}, H., {Ridgway}, S., {Turner}, N., {Sturmann},
  L., {Sturmann}, J., {Berger}, D., {Tannirkulam}, A.,  and {Blum}, J. (2008).
  \enquote{{Imaging the surface of Altair and a MIRC update},} in \emph{Society
  of Photo-Optical Instrumentation Engineers (SPIE) Conference Series},
  \emph{Society of Photo-Optical Instrumentation Engineers (SPIE) Conference
  Series}, Vol. 7013, p.~2, \doi{10.1117/12.789754}.

\bibitem[{{Mourard} \emph{et~al.}(2011){Mourard}, {B{\'e}rio}, {Perraut},
  {Ligi}, {Blazit}, {Clausse}, {Nardetto}, {Spang}, {Tallon-Bosc}, {Bonneau},
  {Chesneau}, {Delaa}, {Millour}, {Stee}, {Le Bouquin}, {ten Brummelaar},
  {Farrington}, {Goldfinger} and {Monnier}}]{mourard11}
{Mourard}, D., {B{\'e}rio}, P., {Perraut}, K., {Ligi}, R., {Blazit}, A.,
  {Clausse}, J.~M., {Nardetto}, N., {Spang}, A., {Tallon-Bosc}, I., {Bonneau},
  D., {Chesneau}, O., {Delaa}, O., {Millour}, F., {Stee}, P., {Le Bouquin},
  J.~B., {ten Brummelaar}, T., {Farrington}, C., {Goldfinger}, P.~J.,  and
  {Monnier}, J.~D. (2011). \enquote{{Spatio-spectral encoding of fringes in
  optical long-baseline interferometry. Example of the 3T and 4T recombining
  mode of VEGA/CHARA},} \emph{\aap} \textbf{531}, A110,
  \doi{10.1051/0004-6361/201116976}.

\bibitem[{{Mourard} \emph{et~al.}(2009{\natexlab{a}}){Mourard}, {Clausse},
  {Marcotto}, {Perraut}, {Tallon-Bosc}, {B{\'e}rio}, {Blazit}, {Bonneau},
  {Bosio}, {Bresson}, {Chesneau}, {Delaa}, {H{\'e}nault}, {Hughes}, {Lagarde},
  {Merlin}, {Roussel}, {Spang}, {Stee}, {Tallon}, {Antonelli}, {Foy},
  {Kervella}, {Petrov}, {Thiebaut}, {Vakili}, {McAlister}, {ten Brummelaar},
  {Sturmann}, {Sturmann}, {Turner}, {Farrington} and {Goldfinger}}]{mourard09}
{Mourard}, D., {Clausse}, J.~M., {Marcotto}, A., {Perraut}, K., {Tallon-Bosc},
  I., {B{\'e}rio}, P., {Blazit}, A., {Bonneau}, D., {Bosio}, S., {Bresson}, Y.,
  {Chesneau}, O., {Delaa}, O., {H{\'e}nault}, F., {Hughes}, Y., {Lagarde}, S.,
  {Merlin}, G., {Roussel}, A., {Spang}, A., {Stee}, P., {Tallon}, M.,
  {Antonelli}, P., {Foy}, R., {Kervella}, P., {Petrov}, R., {Thiebaut}, E.,
  {Vakili}, F., {McAlister}, H., {ten Brummelaar}, T., {Sturmann}, J.,
  {Sturmann}, L., {Turner}, N., {Farrington}, C.,  and {Goldfinger}, P.~J.
  (2009{\natexlab{a}}). \enquote{{VEGA: Visible spEctroGraph and polArimeter
  for the CHARA array: principle and performance},} \emph{\aap} \textbf{508},
  pp. 1073--1083, \doi{10.1051/0004-6361/200913016}.

\bibitem[{{Mourard} \emph{et~al.}(2009{\natexlab{b}}){Mourard}, {Clausse},
  {Marcotto}, {Perraut}, {Tallon-Bosc}, {B{\'e}rio}, {Blazit}, {Bonneau},
  {Bosio}, {Bresson}, {Chesneau}, {Delaa}, {H{\'e}nault}, {Hughes}, {Lagarde},
  {Merlin}, {Roussel}, {Spang}, {Stee}, {Tallon}, {Antonelli}, {Foy},
  {Kervella}, {Petrov}, {Thiebaut}, {Vakili}, {McAlister}, {ten Brummelaar},
  {Sturmann}, {Sturmann}, {Turner}, {Farrington} and
  {Goldfinger}}]{Mourard2009}
{Mourard}, D., {Clausse}, J.~M., {Marcotto}, A., {Perraut}, K., {Tallon-Bosc},
  I., {B{\'e}rio}, P., {Blazit}, A., {Bonneau}, D., {Bosio}, S., {Bresson}, Y.,
  {Chesneau}, O., {Delaa}, O., {H{\'e}nault}, F., {Hughes}, Y., {Lagarde}, S.,
  {Merlin}, G., {Roussel}, A., {Spang}, A., {Stee}, P., {Tallon}, M.,
  {Antonelli}, P., {Foy}, R., {Kervella}, P., {Petrov}, R., {Thiebaut}, E.,
  {Vakili}, F., {McAlister}, H., {ten Brummelaar}, T., {Sturmann}, J.,
  {Sturmann}, L., {Turner}, N., {Farrington}, C.,  and {Goldfinger}, P.~J.
  (2009{\natexlab{b}}). \enquote{{VEGA: Visible spEctroGraph and polArimeter
  for the CHARA array: principle and performance},} \emph{\aap} \textbf{508},
  pp. 1073--1083, \doi{10.1051/0004-6361/200913016}.

\bibitem[{{Mourard} and {Nardetto}(2006)}]{mourard06}
{Mourard}, D. and {Nardetto}, N. (2006). \enquote{{Prospects for direct
  distance determination of LMC Cepheids by differential interferometry},}
  \emph{\memsai} \textbf{77}, p. 555.

\bibitem[{{Muratore} \emph{et~al.}(2015){Muratore}, {Kraus}, {Oksala}, {Arias},
  {Cidale}, {Borges Fernandes} and {Liermann}}]{Muratore}
{Muratore}, M.~F., {Kraus}, M., {Oksala}, M.~E., {Arias}, M.~L., {Cidale}, L.,
  {Borges Fernandes}, M.,  and {Liermann}, A. (2015). \enquote{{Evidence of the
  Evolved Nature of the B[e] Star MWC 137},} \emph{\aj} \textbf{149}, 13,
  \doi{10.1088/0004-6256/149/1/13}, \eprint{1409.7550}.

\bibitem[{{Nagae} \emph{et~al.}(2004){Nagae}, {Oka}, {Matsuda}, {Fujiwara},
  {Hachisu} and {Boffin}}]{Nagae2004}
{Nagae}, T., {Oka}, K., {Matsuda}, T., {Fujiwara}, H., {Hachisu}, I.,  and
  {Boffin}, H.~M.~J. (2004). \enquote{{Wind accretion in binary stars. I. Mass
  accretion ratio},} \emph{\aap} \textbf{419}, pp. 335--343,
  \doi{10.1051/0004-6361:20040070}, \eprint{astro-ph/0403329}.

\bibitem[{{Nardetto} \emph{et~al.}(2011){Nardetto}, {Fokin}, {Fouqu{\'e}},
  {Storm}, {Gieren}, {Pietrzynski}, {Mourard} and {Kervella}}]{nardetto11b}
{Nardetto}, N., {Fokin}, A., {Fouqu{\'e}}, P., {Storm}, J., {Gieren}, W.,
  {Pietrzynski}, G., {Mourard}, D.,  and {Kervella}, P. (2011). \enquote{{The
  Baade-Wesselink p-factor applicable to LMC Cepheids},} \emph{\aap}
  \textbf{534}, L16, \doi{10.1051/0004-6361/201117459}, \eprint{1109.6763}.

\bibitem[{{Nardetto} \emph{et~al.}(2006){Nardetto}, {Fokin}, {Mourard} and
  {Mathias}}]{nardetto06b}
{Nardetto}, N., {Fokin}, A., {Mourard}, D.,  and {Mathias}, P. (2006).
  \enquote{{Probing the dynamical structure of {$\delta$} Cephei atmosphere},}
  \emph{\aap} \textbf{454}, pp. 327--332, \doi{10.1051/0004-6361:20054274}.

\bibitem[{{Nardetto} \emph{et~al.}(2004){Nardetto}, {Fokin}, {Mourard},
  {Mathias}, {Kervella} and {Bersier}}]{nardetto04}
{Nardetto}, N., {Fokin}, A., {Mourard}, D., {Mathias}, P., {Kervella}, P.,  and
  {Bersier}, D. (2004). \enquote{{Self consistent modelling of the projection
  factor for interferometric distance determination},} \emph{\aap}
  \textbf{428}, pp. 131--137, \doi{10.1051/0004-6361:20041419}.

\bibitem[{{Nardetto} \emph{et~al.}(2009){Nardetto}, {Gieren}, {Kervella},
  {Fouqu{\'e}}, {Storm}, {Pietrzynski}, {Mourard} and {Queloz}}]{nardetto09}
{Nardetto}, N., {Gieren}, W., {Kervella}, P., {Fouqu{\'e}}, P., {Storm}, J.,
  {Pietrzynski}, G., {Mourard}, D.,  and {Queloz}, D. (2009).
  \enquote{{High-resolution spectroscopy for Cepheids distance determination.
  V. Impact of the cross-correlation method on the p-factor and the
  {$\gamma$}-velocities},} \emph{\aap} \textbf{502}, pp. 951--956,
  \doi{10.1051/0004-6361/200912333}, \eprint{0905.4540}.

\bibitem[{{Nardetto} \emph{et~al.}(2007){Nardetto}, {Mourard}, {Mathias},
  {Fokin} and {Gillet}}]{nardetto07}
{Nardetto}, N., {Mourard}, D., {Mathias}, P., {Fokin}, A.,  and {Gillet}, D.
  (2007). \enquote{{High-resolution spectroscopy for Cepheids distance
  determination. II. A period-projection factor relation},} \emph{\aap}
  \textbf{471}, pp. 661--669, \doi{10.1051/0004-6361:20066853},
  \eprint{0804.1331}.

\bibitem[{{Nardetto} \emph{et~al.}(2014){Nardetto}, {Poretti}, {Rainer},
  {Guiglion}, {Scardia}, {Schmid} and {Mathias}}]{nardetto14}
{Nardetto}, N., {Poretti}, E., {Rainer}, M., {Guiglion}, G., {Scardia}, M.,
  {Schmid}, V.~S.,  and {Mathias}, P. (2014). \enquote{{Understanding the
  dynamical structure of pulsating stars. HARPS spectroscopy of the {$\delta$}
  Scuti stars {$\rho$} Puppis and DX Ceti},} \emph{\aap} \textbf{561}, A151,
  \doi{10.1051/0004-6361/201322356}, \eprint{1401.2089}.

\bibitem[{{Neilson} \emph{et~al.}(2011){Neilson}, {Cantiello} and
  {Langer}}]{neilson11}
{Neilson}, H.~R., {Cantiello}, M.,  and {Langer}, N. (2011). \enquote{{The
  Cepheid mass discrepancy and pulsation-driven mass loss},} \emph{\aap}
  \textbf{529}, L9, \doi{10.1051/0004-6361/201116920}, \eprint{1104.1638}.

\bibitem[{{Nesvacil} \emph{et~al.}(2012){Nesvacil}, {L{\"u}ftinger}, {Shulyak},
  {Obbrugger}, {Weiss}, {Drake}, {Hubrig}, {Ryabchikova}, {Kochukhov},
  {Piskunov} and {Polosukhina}}]{2012A&A...537A.151N}
{Nesvacil}, N., {L{\"u}ftinger}, T., {Shulyak}, D., {Obbrugger}, M., {Weiss},
  W., {Drake}, N.~A., {Hubrig}, S., {Ryabchikova}, T., {Kochukhov}, O.,
  {Piskunov}, N.,  and {Polosukhina}, N. (2012). \enquote{{Multi-element
  Doppler imaging of the CP2 star HD 3980},} \emph{\aap} \textbf{537}, A151,
  \doi{10.1051/0004-6361/201117097}.

\bibitem[{{Norris} \emph{et~al.}(2012){Norris}, {Tuthill}, {Ireland}, {Lacour},
  {Zijlstra}, {Lykou}, {Evans}, {Stewart} and {Bedding}}]{norris2012}
{Norris}, B.~R.~M., {Tuthill}, P.~G., {Ireland}, M.~J., {Lacour}, S.,
  {Zijlstra}, A.~A., {Lykou}, F., {Evans}, T.~M., {Stewart}, P.,  and
  {Bedding}, T.~R. (2012). \enquote{{A close halo of large transparent grains
  around extreme red giant stars},} \emph{\nat} \textbf{484}, pp. 220--222,
  \doi{10.1038/nature10935}, \eprint{1204.2640}.

\bibitem[{{Nowotny} \emph{et~al.}(2005){Nowotny}, {Lebzelter}, {Hron} and
  {H{\"o}fner}}]{nowotny2005}
{Nowotny}, W., {Lebzelter}, T., {Hron}, J.,  and {H{\"o}fner}, S. (2005).
  \enquote{{Atmospheric dynamics in carbon-rich Miras. II. Models meet
  observations},} \emph{\aap} \textbf{437}, pp. 285--296,
  \doi{10.1051/0004-6361:20042572}, \eprint{astro-ph/0503653}.

\bibitem[{{Ohnaka}(2014)}]{2014A&A...568A..17O}
{Ohnaka}, K. (2014). \enquote{{Imaging the outward motions of clumpy dust
  clouds around the red supergiant Antares with VLT/VISIR},} \emph{\aap}
  \textbf{568}, A17, \doi{10.1051/0004-6361/201423893}, \eprint{1407.0715}.

\bibitem[{{Ohnaka} \emph{et~al.}(2009){Ohnaka}, {Hofmann}, {Benisty}, {Chelli},
  {Driebe}, {Millour}, {Petrov}, {Schertl}, {Stee}, {Vakili} and
  {Weigelt}}]{2009A&A...503..183O}
{Ohnaka}, K., {Hofmann}, K., {Benisty}, M., {Chelli}, A., {Driebe}, T.,
  {Millour}, F., {Petrov}, R., {Schertl}, D., {Stee}, P., {Vakili}, F.,  and
  {Weigelt}, G. (2009). \enquote{{Spatially resolving the inhomogeneous
  structure of the dynamical atmosphere of Betelgeuse with VLTI/AMBER},}
  \emph{\aap} \textbf{503}, pp. 183--195, \doi{10.1051/0004-6361/200912247},
  \eprint{0906.4792}.

\bibitem[{{Ohnaka} \emph{et~al.}(2013){Ohnaka}, {Hofmann}, {Schertl},
  {Weigelt}, {Baffa}, {Chelli}, {Petrov} and {Robbe-Dubois}}]{ohnaka2013}
{Ohnaka}, K., {Hofmann}, K.-H., {Schertl}, D., {Weigelt}, G., {Baffa}, C.,
  {Chelli}, A., {Petrov}, R.,  and {Robbe-Dubois}, S. (2013). \enquote{{High
  spectral resolution imaging of the dynamical atmosphere of the red supergiant
  Antares in the CO first overtone lines with VLTI/AMBER},} \emph{\aap}
  \textbf{555}, A24, \doi{10.1051/0004-6361/201321063}, \eprint{1304.4800}.

\bibitem[{{Ohnaka} \emph{et~al.}(2011{\natexlab{a}}){Ohnaka}, {Weigelt},
  {Millour}, {Hofmann}, {Driebe}, {Schertl}, {Chelli}, {Massi}, {Petrov} and
  {Stee}}]{ohnaka2011}
{Ohnaka}, K., {Weigelt}, G., {Millour}, F., {Hofmann}, K.-H., {Driebe}, T.,
  {Schertl}, D., {Chelli}, A., {Massi}, F., {Petrov}, R.,  and {Stee}, P.
  (2011{\natexlab{a}}). \enquote{{Imaging the dynamical atmosphere of the red
  supergiant Betelgeuse in the CO first overtone lines with VLTI/AMBER},}
  \emph{\aap} \textbf{529}, A163, \doi{10.1051/0004-6361/201016279},
  \eprint{1104.0958}.

\bibitem[{{Ohnaka} \emph{et~al.}(2011{\natexlab{b}}){Ohnaka}, {Weigelt},
  {Millour}, {Hofmann}, {Driebe}, {Schertl}, {Chelli}, {Massi}, {Petrov} and
  {Stee}}]{2011A&A...529A.163O}
{Ohnaka}, K., {Weigelt}, G., {Millour}, F., {Hofmann}, K.-H., {Driebe}, T.,
  {Schertl}, D., {Chelli}, A., {Massi}, F., {Petrov}, R.,  and {Stee}, P.
  (2011{\natexlab{b}}). \enquote{{Imaging the dynamical atmosphere of the red
  supergiant Betelgeuse in the CO first overtone lines with VLTI/AMBER},}
  \emph{\aap} \textbf{529}, p. A163, \doi{10.1051/0004-6361/201016279},
  \eprint{1104.0958}.

\bibitem[{{Oksala} \emph{et~al.}(2012){Oksala}, {Kraus}, {Arias}, {Borges
  Fernandes}, {Cidale}, {Muratore} and {Cur{\'e}}}]{Oksala}
{Oksala}, M.~E., {Kraus}, M., {Arias}, M.~L., {Borges Fernandes}, M., {Cidale},
  L., {Muratore}, M.~F.,  and {Cur{\'e}}, M. (2012). \enquote{{The sudden
  appearance of CO emission in LHA 115-S 65},} \emph{\mnras} \textbf{426}, pp.
  L56--L60, \doi{10.1111/j.1745-3933.2012.01323.x}, \eprint{1207.6294}.

\bibitem[{{Perraut} \emph{et~al.}(2013){Perraut}, {Borgniet}, {Cunha}, {Bigot},
  {Brand{\~a}o}, {Mourard}, {Nardetto}, {Chesneau}, {McAlister}, {ten
  Brummelaar}, {Sturmann}, {Sturmann}, {Turner}, {Farrington} and
  {Goldfinger}}]{2013A&A...559A..21P}
{Perraut}, K., {Borgniet}, S., {Cunha}, M., {Bigot}, L., {Brand{\~a}o}, I.,
  {Mourard}, D., {Nardetto}, N., {Chesneau}, O., {McAlister}, H., {ten
  Brummelaar}, T.~A., {Sturmann}, J., {Sturmann}, L., {Turner}, N.,
  {Farrington}, C.,  and {Goldfinger}, P.~J. (2013). \enquote{{The fundamental
  parameters of the roAp star 10 Aquilae},} \emph{\aap} \textbf{559}, A21,
  \doi{10.1051/0004-6361/201321849}, \eprint{1309.4423}.

\bibitem[{{Perraut} \emph{et~al.}(2011){Perraut}, {Brand{\~a}o}, {Mourard},
  {Cunha}, {B{\'e}rio}, {Bonneau}, {Chesneau}, {Clausse}, {Delaa}, {Marcotto},
  {Roussel}, {Spang}, {Stee}, {Tallon-Bosc}, {McAlister}, {ten Brummelaar},
  {Sturmann}, {Sturmann}, {Turner}, {Farrington} and
  {Goldfinger}}]{2011A&A...526A..89P}
{Perraut}, K., {Brand{\~a}o}, I., {Mourard}, D., {Cunha}, M., {B{\'e}rio}, P.,
  {Bonneau}, D., {Chesneau}, O., {Clausse}, J.~M., {Delaa}, O., {Marcotto}, A.,
  {Roussel}, A., {Spang}, A., {Stee}, P., {Tallon-Bosc}, I., {McAlister}, H.,
  {ten Brummelaar}, T., {Sturmann}, J., {Sturmann}, L., {Turner}, N.,
  {Farrington}, C.,  and {Goldfinger}, P.~J. (2011). \enquote{{The fundamental
  parameters of the roAp star {$\gamma$} Equulei},} \emph{\aap} \textbf{526},
  A89, \doi{10.1051/0004-6361/201015801}, \eprint{1011.2028}.

\bibitem[{{Persson} \emph{et~al.}(2004){Persson}, {Madore}, {Krzemi{\'n}ski},
  {Freedman}, {Roth} and {Murphy}}]{persson04}
{Persson}, S.~E., {Madore}, B.~F., {Krzemi{\'n}ski}, W., {Freedman}, W.~L.,
  {Roth}, M.,  and {Murphy}, D.~C. (2004). \enquote{{New Cepheid
  Period-Luminosity Relations for the Large Magellanic Cloud: 92 Near-Infrared
  Light Curves},} \emph{\aj} \textbf{128}, pp. 2239--2264,
  \doi{10.1086/424934}.

\bibitem[{{Petersen} and {Hog}(1998)}]{petersen98}
{Petersen}, J.~O. and {Hog}, E. (1998). \enquote{{HIPPARCOS parallaxes and
  period-luminosity relations of high-amplitude delta Scuti stars},}
  \emph{\aap} \textbf{331}, pp. 989--994.

\bibitem[{{Petrov}(1988)}]{petrov1988}
{Petrov}, R.~G. (1988). \enquote{{Differential Interferometry},} in D.~M.
  {Alloin} and J.-M. {Mariotti} (eds.), \emph{Diffraction-Limit.Imaging/ Very
  Large Telescopes}, p. 249.

\bibitem[{{Petrov} \emph{et~al.}(2007){Petrov}, {Malbet}, {Weigelt},
  {Antonelli}, {Beckmann}, {Bresson}, {Chelli}, {Dugu{\'e}}, {Duvert},
  {Gennari}, {Gl{\"u}ck}, {Kern}, {Lagarde}, {Le Coarer}, {Lisi}, {Millour},
  {Perraut}, {Puget}, {Rantakyr{\"o}}, {Robbe-Dubois}, {Roussel}, {Salinari},
  {Tatulli}, {Zins}, {Accardo}, {Acke}, {Agabi}, {Altariba}, {Arezki},
  {Aristidi}, {Baffa}, {Behrend}, {Bl{\"o}cker}, {Bonhomme}, {Busoni},
  {Cassaing}, {Clausse}, {Colin}, {Connot}, {Delboulb{\'e}}, {Domiciano de
  Souza}, {Driebe}, {Feautrier}, {Ferruzzi}, {Forveille}, {Fossat}, {Foy},
  {Fraix-Burnet}, {Gallardo}, {Giani}, {Gil}, {Glentzlin}, {Heiden},
  {Heininger}, {Hernandez Utrera}, {Hofmann}, {Kamm}, {Kiekebusch}, {Kraus},
  {Le Contel}, {Le Contel}, {Lesourd}, {Lopez}, {Lopez}, {Magnard}, {Marconi},
  {Mars}, {Martinot-Lagarde}, {Mathias}, {M{\`e}ge}, {Monin}, {Mouillet},
  {Mourard}, {Nussbaum}, {Ohnaka}, {Pacheco}, {Perrier}, {Rabbia}, {Rebattu},
  {Reynaud}, {Richichi}, {Robini}, {Sacchettini}, {Schertl}, {Sch{\"o}ller},
  {Solscheid}, {Spang}, {Stee}, {Stefanini}, {Tallon}, {Tallon-Bosc}, {Tasso},
  {Testi}, {Vakili}, {von der L{\"u}he}, {Valtier}, {Vannier} and
  {Ventura}}]{petrov2007}
{Petrov}, R.~G., {Malbet}, F., {Weigelt}, G., {Antonelli}, P., {Beckmann}, U.,
  {Bresson}, Y., {Chelli}, A., {Dugu{\'e}}, M., {Duvert}, G., {Gennari}, S.,
  {Gl{\"u}ck}, L., {Kern}, P., {Lagarde}, S., {Le Coarer}, E., {Lisi}, F.,
  {Millour}, F., {Perraut}, K., {Puget}, P., {Rantakyr{\"o}}, F.,
  {Robbe-Dubois}, S., {Roussel}, A., {Salinari}, P., {Tatulli}, E., {Zins}, G.,
  {Accardo}, M., {Acke}, B., {Agabi}, K., {Altariba}, E., {Arezki}, B.,
  {Aristidi}, E., {Baffa}, C., {Behrend}, J., {Bl{\"o}cker}, T., {Bonhomme},
  S., {Busoni}, S., {Cassaing}, F., {Clausse}, J.-M., {Colin}, J., {Connot},
  C., {Delboulb{\'e}}, A., {Domiciano de Souza}, A., {Driebe}, T., {Feautrier},
  P., {Ferruzzi}, D., {Forveille}, T., {Fossat}, E., {Foy}, R., {Fraix-Burnet},
  D., {Gallardo}, A., {Giani}, E., {Gil}, C., {Glentzlin}, A., {Heiden}, M.,
  {Heininger}, M., {Hernandez Utrera}, O., {Hofmann}, K.-H., {Kamm}, D.,
  {Kiekebusch}, M., {Kraus}, S., {Le Contel}, D., {Le Contel}, J.-M.,
  {Lesourd}, T., {Lopez}, B., {Lopez}, M., {Magnard}, Y., {Marconi}, A.,
  {Mars}, G., {Martinot-Lagarde}, G., {Mathias}, P., {M{\`e}ge}, P., {Monin},
  J.-L., {Mouillet}, D., {Mourard}, D., {Nussbaum}, E., {Ohnaka}, K.,
  {Pacheco}, J., {Perrier}, C., {Rabbia}, Y., {Rebattu}, S., {Reynaud}, F.,
  {Richichi}, A., {Robini}, A., {Sacchettini}, M., {Schertl}, D.,
  {Sch{\"o}ller}, M., {Solscheid}, W., {Spang}, A., {Stee}, P., {Stefanini},
  P., {Tallon}, M., {Tallon-Bosc}, I., {Tasso}, D., {Testi}, L., {Vakili}, F.,
  {von der L{\"u}he}, O., {Valtier}, J.-C., {Vannier}, M.,  and {Ventura}, N.
  (2007). \enquote{{AMBER, the near-infrared spectro-interferometric
  three-telescope VLTI instrument},} \emph{\aap} \textbf{464}, pp. 1--12,
  \doi{10.1051/0004-6361:20066496}.

\bibitem[{{Petrov} \emph{et~al.}(2012){Petrov}, {Millour}, {Lagarde},
  {Vannier}, {Rakshit}, {Marconi} and {weigelt}}]{Petrov2012}
{Petrov}, R.~G., {Millour}, F., {Lagarde}, S., {Vannier}, M., {Rakshit}, S.,
  {Marconi}, A.,  and {weigelt}, G. (2012). \enquote{{VLTI/AMBER differential
  interferometry of the broad-line region of the quasar 3C273},} in
  \emph{Society of Photo-Optical Instrumentation Engineers (SPIE) Conference
  Series}, \emph{Society of Photo-Optical Instrumentation Engineers (SPIE)
  Conference Series}, Vol. 8445, p.~0, \doi{10.1117/12.926595},
  \eprint{1410.3108}.

\bibitem[{{Pietrzy{\'n}ski} \emph{et~al.}(2013){Pietrzy{\'n}ski}, {Graczyk},
  {Gieren}, {Thompson}, {Pilecki}, {Udalski}, {Soszy{\'n}ski}, {Koz{\l}owski},
  {Konorski}, {Suchomska}, {Bono}, {Moroni}, {Villanova}, {Nardetto},
  {Bresolin}, {Kudritzki}, {Storm}, {Gallenne}, {Smolec}, {Minniti}, {Kubiak},
  {Szyma{\'n}ski}, {Poleski}, {Wyrzykowski}, {Ulaczyk}, {Pietrukowicz},
  {G{\'o}rski} and {Karczmarek}}]{gp13}
{Pietrzy{\'n}ski}, G., {Graczyk}, D., {Gieren}, W., {Thompson}, I.~B.,
  {Pilecki}, B., {Udalski}, A., {Soszy{\'n}ski}, I., {Koz{\l}owski}, S.,
  {Konorski}, P., {Suchomska}, K., {Bono}, G., {Moroni}, P.~G.~P., {Villanova},
  S., {Nardetto}, N., {Bresolin}, F., {Kudritzki}, R.~P., {Storm}, J.,
  {Gallenne}, A., {Smolec}, R., {Minniti}, D., {Kubiak}, M., {Szyma{\'n}ski},
  M.~K., {Poleski}, R., {Wyrzykowski}, {\L}., {Ulaczyk}, K., {Pietrukowicz},
  P., {G{\'o}rski}, M.,  and {Karczmarek}, P. (2013). \enquote{{An
  eclipsing-binary distance to the Large Magellanic Cloud accurate to two per
  cent},} \emph{\nat} \textbf{495}, pp. 76--79, \doi{10.1038/nature11878},
  \eprint{1303.2063}.

\bibitem[{{Pijpers} and {Hearn}(1989)}]{1989A&A...209..198P}
{Pijpers}, F.~P. and {Hearn}, A.~G. (1989). \enquote{{A model for a stellar
  wind driven by linear acoustic waves},} \emph{\aap} \textbf{209}, pp.
  198--210.

\bibitem[{{Pilecki} \emph{et~al.}(2013){Pilecki}, {Graczyk}, {Pietrzy{\'n}ski},
  {Gieren}, {Thompson}, {Freedman}, {Scowcroft}, {Madore}, {Udalski},
  {Soszy{\'n}ski}, {Konorski}, {Smolec}, {Nardetto}, {Bono}, {Prada Moroni},
  {Storm} and {Gallenne}}]{pilecki13}
{Pilecki}, B., {Graczyk}, D., {Pietrzy{\'n}ski}, G., {Gieren}, W., {Thompson},
  I.~B., {Freedman}, W.~L., {Scowcroft}, V., {Madore}, B.~F., {Udalski}, A.,
  {Soszy{\'n}ski}, I., {Konorski}, P., {Smolec}, R., {Nardetto}, N., {Bono},
  G., {Prada Moroni}, P.~G., {Storm}, J.,  and {Gallenne}, A. (2013).
  \enquote{{Physical parameters and the projection factor of the classical
  Cepheid in the binary system OGLE-LMC-CEP-0227},} \emph{\mnras} \textbf{436},
  pp. 953--967, \doi{10.1093/mnras/stt1529}, \eprint{1308.5023}.

\bibitem[{{Quirrenbach} \emph{et~al.}(1993){Quirrenbach}, {Mozurkewich},
  {Armstrong}, {Buscher} and {Hummel}}]{quirrenbach1993}
{Quirrenbach}, A., {Mozurkewich}, D., {Armstrong}, J.~T., {Buscher}, D.~F.,
  and {Hummel}, C.~A. (1993). \enquote{{Angular diameter measurements of cool
  giant stars in strong TiO bands and in the continuum},} \emph{\apj}
  \textbf{406}, pp. 215--219, \doi{10.1086/172432}.

\bibitem[{{Quirrenbach} \emph{et~al.}(1992){Quirrenbach}, {Mozurkewich},
  {Armstrong}, {Johnston}, {Colavita} and {Shao}}]{quirrenbach1992}
{Quirrenbach}, A., {Mozurkewich}, D., {Armstrong}, J.~T., {Johnston}, K.~J.,
  {Colavita}, M.~M.,  and {Shao}, M. (1992). \enquote{{Interferometric
  observations of Mira (Omicron Ceti)},} \emph{\aap} \textbf{259}, pp.
  L19--L22.

\bibitem[{{Quirrenbach} \emph{et~al.}(1994){Quirrenbach}, {Mozurkewich},
  {Hummel}, {Buscher} and {Armstrong}}]{quirrenbach1994}
{Quirrenbach}, A., {Mozurkewich}, D., {Hummel}, C.~A., {Buscher}, D.~F.,  and
  {Armstrong}, J.~T. (1994). \enquote{{Angular diameters of the carbon stars UU
  Aurigae, Y Canum Venaticorum, and TX PISCIUM from optical long-baseline
  interferometry},} \emph{\aap} \textbf{285}, pp. 541--546.

\bibitem[{{Raghavan} \emph{et~al.}(2010){Raghavan}, {McAlister}, {Henry},
  {Latham}, {Marcy}, {Mason}, {Gies}, {White} and {ten
  Brummelaar}}]{Raghavan2010}
{Raghavan}, D., {McAlister}, H.~A., {Henry}, T.~J., {Latham}, D.~W., {Marcy},
  G.~W., {Mason}, B.~D., {Gies}, D.~R., {White}, R.~J.,  and {ten Brummelaar},
  T.~A. (2010). \enquote{{A Survey of Stellar Families: Multiplicity of
  Solar-type Stars},} \emph{\apjs} \textbf{190}, pp. 1--42,
  \doi{10.1088/0067-0049/190/1/1}, \eprint{1007.0414}.

\bibitem[{{Rakshit} \emph{et~al.}(2015){Rakshit}, {Petrov}, {Meilland} and
  {H{\"o}nig}}]{Rakshit2015}
{Rakshit}, S., {Petrov}, R.~G., {Meilland}, A.,  and {H{\"o}nig}, S.~F. (2015).
  \enquote{{Differential interferometry of QSO broad-line regions - I.
  Improving the reverberation mapping model fits and black hole mass
  estimates},} \emph{\mnras} \textbf{447}, pp. 2420--2436,
  \doi{10.1093/mnras/stu2613}, \eprint{1410.4837}.

\bibitem[{{Riess} \emph{et~al.}(2011){Riess}, {Macri}, {Casertano}, {Lampeitl},
  {Ferguson}, {Filippenko}, {Jha}, {Li} and {Chornock}}]{riess11}
{Riess}, A.~G., {Macri}, L., {Casertano}, S., {Lampeitl}, H., {Ferguson},
  H.~C., {Filippenko}, A.~V., {Jha}, S.~W., {Li}, W.,  and {Chornock}, R.
  (2011). \enquote{{A 3\% Solution: Determination of the Hubble Constant with
  the Hubble Space Telescope and Wide Field Camera 3},} \emph{\apj}
  \textbf{730}, 119, \doi{10.1088/0004-637X/730/2/119}, \eprint{1103.2976}.

\bibitem[{{Riess} \emph{et~al.}(2009{\natexlab{a}}){Riess}, {Macri},
  {Casertano}, {Sosey}, {Lampeitl}, {Ferguson}, {Filippenko}, {Jha}, {Li},
  {Chornock} and {Sarkar}}]{riess09a}
{Riess}, A.~G., {Macri}, L., {Casertano}, S., {Sosey}, M., {Lampeitl}, H.,
  {Ferguson}, H.~C., {Filippenko}, A.~V., {Jha}, S.~W., {Li}, W., {Chornock},
  R.,  and {Sarkar}, D. (2009{\natexlab{a}}). \enquote{{A Redetermination of
  the Hubble Constant with the Hubble Space Telescope from a Differential
  Distance Ladder},} \emph{\apj} \textbf{699}, pp. 539--563,
  \doi{10.1088/0004-637X/699/1/539}, \eprint{0905.0695}.

\bibitem[{{Riess} \emph{et~al.}(2009{\natexlab{b}}){Riess}, {Macri}, {Li},
  {Lampeitl}, {Casertano}, {Ferguson}, {Filippenko}, {Jha}, {Chornock},
  {Greenhill}, {Mutchler}, {Ganeshalingham} and {Hicken}}]{riess09b}
{Riess}, A.~G., {Macri}, L., {Li}, W., {Lampeitl}, H., {Casertano}, S.,
  {Ferguson}, H.~C., {Filippenko}, A.~V., {Jha}, S.~W., {Chornock}, R.,
  {Greenhill}, L., {Mutchler}, M., {Ganeshalingham}, M.,  and {Hicken}, M.
  (2009{\natexlab{b}}). \enquote{{Cepheid Calibrations of Modern Type Ia
  Supernovae: Implications for the Hubble Constant},} \emph{\apjs}
  \textbf{183}, pp. 109--141, \doi{10.1088/0067-0049/183/1/109},
  \eprint{0905.0697}.

\bibitem[{{Roelofs} \emph{et~al.}(2010){Roelofs}, {Rau}, {Marsh}, {Steeghs},
  {Groot} and {Nelemans}}]{Roelofs2010}
{Roelofs}, G.~H.~A., {Rau}, A., {Marsh}, T.~R., {Steeghs}, D., {Groot}, P.~J.,
  and {Nelemans}, G. (2010). \enquote{{Spectroscopic Evidence for a 5.4 Minute
  Orbital Period in HM Cancri},} \emph{\apjl} \textbf{711}, pp. L138--L142,
  \doi{10.1088/2041-8205/711/2/L138}, \eprint{1003.0658}.

\bibitem[{{Rousselet-Perraut} \emph{et~al.}(2004){Rousselet-Perraut},
  {Stehl{\'e}}, {Lanz}, {Le Bouquin}, {Boudoyen}, {Kilbinger}, {Kochukhov} and
  {Jankov}}]{2004A&A...422..193R}
{Rousselet-Perraut}, K., {Stehl{\'e}}, C., {Lanz}, T., {Le Bouquin}, J.~B.,
  {Boudoyen}, T., {Kilbinger}, M., {Kochukhov}, O.,  and {Jankov}, S. (2004).
  \enquote{{Stellar activity and magnetism studied by optical interferometry},}
  \emph{\aap} \textbf{422}, pp. 193--203, \doi{10.1051/0004-6361:20040151}.

\bibitem[{{Ryde} \emph{et~al.}(2014){Ryde}, {Lambert}, {Farzone}, {Richter},
  {Josselin}, {Harper}, {Eriksson} and {Greathouse}}]{ryde2014}
{Ryde}, N., {Lambert}, J., {Farzone}, M., {Richter}, M.~J., {Josselin}, E.,
  {Harper}, G.~M., {Eriksson}, K.,  and {Greathouse}, T.~K. (2014).
  \enquote{{Systematic trend of water vapour absorption in red giant
  atmospheres revealed by high resolution TEXES 12 micron spectra},}
  \emph{ArXiv e-prints} \eprint{1410.3999}.

\bibitem[{{Sacuto} \emph{et~al.}(2011){Sacuto}, {Aringer}, {Hron}, {Nowotny},
  {Paladini}, {Verhoelst} and {H{\"o}fner}}]{sacuto2011}
{Sacuto}, S., {Aringer}, B., {Hron}, J., {Nowotny}, W., {Paladini}, C.,
  {Verhoelst}, T.,  and {H{\"o}fner}, S. (2011). \enquote{{Observing and
  modeling the dynamic atmosphere of the low mass-loss C-star R Sculptoris at
  high angular resolution},} \emph{\aap} \textbf{525}, A42,
  \doi{10.1051/0004-6361/200913786}, \eprint{1010.1350}.

\bibitem[{{Sacuto} \emph{et~al.}(2013){Sacuto}, {Ramstedt}, {H{\"o}fner},
  {Olofsson}, {Bladh}, {Eriksson}, {Aringer}, {Klotz} and
  {Maercker}}]{sacuto2013}
{Sacuto}, S., {Ramstedt}, S., {H{\"o}fner}, S., {Olofsson}, H., {Bladh}, S.,
  {Eriksson}, K., {Aringer}, B., {Klotz}, D.,  and {Maercker}, M. (2013).
  \enquote{{The wind of the M-type AGB star <ASTROBJ>RT Virginis</ASTROBJ>
  probed by VLTI/MIDI},} \emph{\aap} \textbf{551}, A72,
  \doi{10.1051/0004-6361/201220524}, \eprint{1301.5872}.

\bibitem[{{Sana} \emph{et~al.}(2012){Sana}, {de Mink}, {de Koter}, {Langer},
  {Evans}, {Gieles}, {Gosset}, {Izzard}, {Le Bouquin} and
  {Schneider}}]{Sana2012}
{Sana}, H., {de Mink}, S.~E., {de Koter}, A., {Langer}, N., {Evans}, C.~J.,
  {Gieles}, M., {Gosset}, E., {Izzard}, R.~G., {Le Bouquin}, J.-B.,  and
  {Schneider}, F.~R.~N. (2012). \enquote{{Binary Interaction Dominates the
  Evolution of Massive Stars},} \emph{Science} \textbf{337}, pp. 444--,
  \doi{10.1126/science.1223344}, \eprint{1207.6397}.

\bibitem[{{Shulyak} \emph{et~al.}(2014){Shulyak}, {Paladini}, {Causi},
  {Perraut} and {Kochukhov}}]{2014MNRAS.443.1629S}
{Shulyak}, D., {Paladini}, C., {Causi}, G.~L., {Perraut}, K.,  and {Kochukhov},
  O. (2014). \enquote{{Interferometry of chemically peculiar stars: theoretical
  predictions versus modern observing facilities},} \emph{\mnras} \textbf{443},
  pp. 1629--1642, \doi{10.1093/mnras/stu1259}, \eprint{1406.6093}.

\bibitem[{{Shulyak} \emph{et~al.}(2015){Shulyak}, {Sokoloff}, {Kitchatinov} and
  {Moss}}]{2015MNRAS.449.3471S}
{Shulyak}, D., {Sokoloff}, D., {Kitchatinov}, L.,  and {Moss}, D. (2015).
  \enquote{{Towards understanding dynamo action in M dwarfs},} \emph{\mnras}
  \textbf{449}, pp. 3471--3478, \doi{10.1093/mnras/stv585},
  \eprint{1503.04971}.

\bibitem[{{Soubiran} \emph{et~al.}(2010){Soubiran}, {Le Campion}, {Cayrel de
  Strobel} and {Caillo}}]{soubiran10}
{Soubiran}, C., {Le Campion}, J.-F., {Cayrel de Strobel}, G.,  and {Caillo}, A.
  (2010). \enquote{{The PASTEL catalogue of stellar parameters},} \emph{\aap}
  \textbf{515}, A111, \doi{10.1051/0004-6361/201014247}, \eprint{1004.1069}.

\bibitem[{{Stamford} and {Watson}(1981)}]{stamford81}
{Stamford}, P.~A. and {Watson}, R.~D. (1981). \enquote{{Baade-Wesselink and
  related techniques for mode discrimination in nonradial stellar pulsations},}
  \emph{\apss} \textbf{77}, pp. 131--158, \doi{10.1007/BF00648762}.

\bibitem[{{Stee} \emph{et~al.}(2012){Stee}, {Delaa}, {Monnier}, {Meilland},
  {Perraut}, {Mourard}, {Che}, {Schaefer}, {Pedretti}, {Smith}, {Lopes de
  Oliveira}, {Motch}, {Henry}, {Richardson}, {Bjorkman}, {B{\"u}cke},
  {Pollmann}, {Zorec}, {Gies}, {ten Brummelaar}, {McAlister}, {Turner},
  {Sturmann}, {Sturmann} and {Ridgway}}]{Stee2012}
{Stee}, P., {Delaa}, O., {Monnier}, J.~D., {Meilland}, A., {Perraut}, K.,
  {Mourard}, D., {Che}, X., {Schaefer}, G.~H., {Pedretti}, E., {Smith}, M.~A.,
  {Lopes de Oliveira}, R., {Motch}, C., {Henry}, G.~W., {Richardson}, N.~D.,
  {Bjorkman}, K.~S., {B{\"u}cke}, R., {Pollmann}, E., {Zorec}, J., {Gies},
  D.~R., {ten Brummelaar}, T., {McAlister}, H.~A., {Turner}, N.~H., {Sturmann},
  J., {Sturmann}, L.,  and {Ridgway}, S.~T. (2012). \enquote{{The relationship
  between {$\gamma$} Cassiopeiae's X-ray emission and its circumstellar
  environment. II. Geometry and kinematics of the disk from MIRC and VEGA
  instruments on the CHARA Array},} \emph{\aap} \textbf{545}, A59,
  \doi{10.1051/0004-6361/201219234}.

\bibitem[{{Storm} \emph{et~al.}(2011{\natexlab{a}}){Storm}, {Gieren},
  {Fouqu{\'e}}, {Barnes}, {Pietrzy{\'n}ski}, {Nardetto}, {Weber}, {Granzer} and
  {Strassmeier}}]{storm11a}
{Storm}, J., {Gieren}, W., {Fouqu{\'e}}, P., {Barnes}, T.~G.,
  {Pietrzy{\'n}ski}, G., {Nardetto}, N., {Weber}, M., {Granzer}, T.,  and
  {Strassmeier}, K.~G. (2011{\natexlab{a}}). \enquote{{Calibrating the Cepheid
  period-luminosity relation from the infrared surface brightness technique. I.
  The p-factor, the Milky Way relations, and a universal K-band relation},}
  \emph{\aap} \textbf{534}, A94, \doi{10.1051/0004-6361/201117155},
  \eprint{1109.2017}.

\bibitem[{{Storm} \emph{et~al.}(2011{\natexlab{b}}){Storm}, {Gieren},
  {Fouqu{\'e}}, {Barnes}, {Soszy{\'n}ski}, {Pietrzy{\'n}ski}, {Nardetto} and
  {Queloz}}]{storm11b}
{Storm}, J., {Gieren}, W., {Fouqu{\'e}}, P., {Barnes}, T.~G., {Soszy{\'n}ski},
  I., {Pietrzy{\'n}ski}, G., {Nardetto}, N.,  and {Queloz}, D.
  (2011{\natexlab{b}}). \enquote{{Calibrating the Cepheid period-luminosity
  relation from the infrared surface brightness technique. II. The effect of
  metallicity and the distance to the LMC},} \emph{\aap} \textbf{534}, A95,
  \doi{10.1051/0004-6361/201117154}, \eprint{1109.2016}.

\bibitem[{{Theuns} \emph{et~al.}(1996){Theuns}, {Boffin} and
  {Jorissen}}]{Theuns1996}
{Theuns}, T., {Boffin}, H.~M.~J.,  and {Jorissen}, A. (1996). \enquote{{Wind
  accretion in binary stars - II. Accretion rates},} \emph{\mnras}
  \textbf{280}, pp. 1264--1276, \eprint{astro-ph/9602089}.

\bibitem[{{Thomas} \emph{et~al.}(2013){Thomas}, {Witt}, {Aufdenberg},
  {Bjorkman}, {Dahlstrom}, {Hobbs} and {York}}]{2013MNRASThomas}
{Thomas}, J.~D., {Witt}, A.~N., {Aufdenberg}, J.~P., {Bjorkman}, J.~E.,
  {Dahlstrom}, J.~A., {Hobbs}, L.~M.,  and {York}, D.~G. (2013).
  \enquote{{Geometry and velocity structure of HD 44179's bipolar jet},}
  \emph{\mnras} , p. 585\doi{10.1093/mnras/sts693}, \eprint{1212.5735}.

\bibitem[{{Torres} \emph{et~al.}(2010){Torres}, {Andersen} and
  {Gim{\'e}nez}}]{Torres2010}
{Torres}, G., {Andersen}, J.,  and {Gim{\'e}nez}, A. (2010). \enquote{{Accurate
  masses and radii of normal stars: modern results and applications},}
  \emph{\aapr} \textbf{18}, pp. 67--126, \doi{10.1007/s00159-009-0025-1},
  \eprint{0908.2624}.

\bibitem[{{Tout} and {Eggleton}(1988)}]{Tout1988}
{Tout}, C.~A. and {Eggleton}, P.~P. (1988). \enquote{{Tidal enhancement by a
  binary companion of stellar winds from cool giants},} \emph{\mnras}
  \textbf{231}, pp. 823--831.

\bibitem[{{Tuthill} \emph{et~al.}(1997){Tuthill}, {Haniff} and
  {Baldwin}}]{1997MNRAS.285..529T}
{Tuthill}, P.~G., {Haniff}, C.~A.,  and {Baldwin}, J.~E. (1997).
  \enquote{{Hotspots on late-type supergiants},} \emph{\mnras} \textbf{285},
  pp. 529--539.

\bibitem[{{Urry} and {Padovani}(1995)}]{Urry1995}
{Urry}, C.~M. and {Padovani}, P. (1995). \enquote{{Unified Schemes for
  Radio-Loud Active Galactic Nuclei},} \emph{\pasp} \textbf{107}, p. 803,
  \doi{10.1086/133630}, \eprint{astro-ph/9506063}.

\bibitem[{{Vakili} and {Percheron}(1991)}]{vakili1991}
{Vakili}, F. and {Percheron}, I. (1991). \enquote{{On the Possibility to Detect
  NRP's by Optical Interferometry},} in D.~{Baade} (ed.), \emph{European
  Southern Observatory Conference and Workshop Proceedings}, \emph{European
  Southern Observatory Conference and Workshop Proceedings}, Vol.~36, p.~77.

\bibitem[{{van Leeuwen}(2007)}]{vanLeeuwen2007}
{van Leeuwen}, F. (2007). \enquote{{Validation of the new Hipparcos
  reduction},} \emph{\aap} \textbf{474}, pp. 653--664,
  \doi{10.1051/0004-6361:20078357}, \eprint{0708.1752}.

\bibitem[{{van Leeuwen} \emph{et~al.}(2007){van Leeuwen}, {Feast}, {Whitelock}
  and {Laney}}]{vl07}
{van Leeuwen}, F., {Feast}, M.~W., {Whitelock}, P.~A.,  and {Laney}, C.~D.
  (2007). \enquote{{Cepheid parallaxes and the Hubble constant},} \emph{\mnras}
  \textbf{379}, pp. 723--737, \doi{10.1111/j.1365-2966.2007.11972.x},
  \eprint{0705.1592}.

\bibitem[{{VandenBerg} \emph{et~al.}(2014){VandenBerg}, {Bond}, {Nelan},
  {Nissen}, {Schaefer} and {Harmer}}]{vandenberg14}
{VandenBerg}, D.~A., {Bond}, H.~E., {Nelan}, E.~P., {Nissen}, P.~E.,
  {Schaefer}, G.~H.,  and {Harmer}, D. (2014). \enquote{{Three Ancient Halo
  Subgiants: Precise Parallaxes, Compositions, Ages, and Implications for
  Globular Clusters},} \emph{\apj} \textbf{792}, 110,
  \doi{10.1088/0004-637X/792/2/110}, \eprint{1407.7591}.

\bibitem[{{von Zeipel}(1924)}]{vonzeipel1924}
{von Zeipel}, H. (1924). \enquote{{The radiative equilibrium of a rotating
  system of gaseous masses},} \emph{\mnras} \textbf{84}, pp. 665--683.

\bibitem[{{Vural} \emph{et~al.}(2014){Vural}, {Kraus}, {Kreplin}, {Weigelt},
  {Fossat}, {Massi}, {Perraut} and {Vakili}}]{Vural}
{Vural}, J., {Kraus}, S., {Kreplin}, A., {Weigelt}, G., {Fossat}, E., {Massi},
  F., {Perraut}, K.,  and {Vakili}, F. (2014). \enquote{{Study of the sub-AU
  disk of the Herbig B[e] star HD 85567 with near-infrared interferometry},}
  \emph{\aap} \textbf{569}, A25, \doi{10.1051/0004-6361/201424214},
  \eprint{1407.8190}.

\bibitem[{{Wang} \emph{et~al.}(2012){Wang}, {Weigelt}, {Kreplin}, {Hofmann},
  {Kraus}, {Miroshnichenko}, {Schertl}, {Chelli}, {Domiciano de Souza}, {Massi}
  and {Robbe-Dubois}}]{Wang}
{Wang}, Y., {Weigelt}, G., {Kreplin}, A., {Hofmann}, K.-H., {Kraus}, S.,
  {Miroshnichenko}, A.~S., {Schertl}, D., {Chelli}, A., {Domiciano de Souza},
  A., {Massi}, F.,  and {Robbe-Dubois}, S. (2012). \enquote{{AMBER/VLTI
  observations of the B[e] star MWC 300},} \emph{\aap} \textbf{545}, L10,
  \doi{10.1051/0004-6361/201219800}, \eprint{1208.5882}.

\bibitem[{{Wheelwright} \emph{et~al.}(2012{\natexlab{a}}){Wheelwright}, {de
  Wit}, {Oudmaijer} and {Vink}}]{Wheelwright2012a}
{Wheelwright}, H.~E., {de Wit}, W.~J., {Oudmaijer}, R.~D.,  and {Vink}, J.~S.
  (2012{\natexlab{a}}). \enquote{{VLTI/AMBER observations of the binary B[e]
  supergiant HD 327083},} \emph{\aap} \textbf{538}, A6,
  \doi{10.1051/0004-6361/201117766}, \eprint{1201.2866}.

\bibitem[{{Wheelwright} \emph{et~al.}(2012{\natexlab{b}}){Wheelwright}, {de
  Wit}, {Weigelt}, {Oudmaijer} and {Ilee}}]{Wheelwright2012b}
{Wheelwright}, H.~E., {de Wit}, W.~J., {Weigelt}, G., {Oudmaijer}, R.~D.,  and
  {Ilee}, J.~D. (2012{\natexlab{b}}). \enquote{{AMBER and CRIRES observations
  of the binary sgB[e] star HD 327083: evidence of a gaseous disc traced by CO
  bandhead emission},} \emph{\aap} \textbf{543}, A77,
  \doi{10.1051/0004-6361/201219325}, \eprint{1206.6252}.

\bibitem[{{Wheelwright} \emph{et~al.}(2013){Wheelwright}, {Weigelt}, {Caratti o
  Garatti} and {Garcia Lopez}}]{Wheelwright2013}
{Wheelwright}, H.~E., {Weigelt}, G., {Caratti o Garatti}, A.,  and {Garcia
  Lopez}, R. (2013). \enquote{{HD 85567: A Herbig B[e] star or an interacting
  B[e] binary?. Resolving HD 85567's circumstellar environment with the VLTI
  and AMBER},} \emph{\aap} \textbf{558}, A116,
  \doi{10.1051/0004-6361/201322128}, \eprint{1308.6000}.

\bibitem[{{White} \emph{et~al.}(2013){White}, {Huber}, {Maestro}, {Bedding},
  {Ireland}, {Baron}, {Boyajian}, {Che}, {Monnier}, {Pope}, {Roettenbacher},
  {Stello}, {Tuthill}, {Farrington}, {Goldfinger}, {McAlister}, {Schaefer},
  {Sturmann}, {Sturmann}, {ten Brummelaar} and {Turner}}]{white13}
{White}, T.~R., {Huber}, D., {Maestro}, V., {Bedding}, T.~R., {Ireland}, M.~J.,
  {Baron}, F., {Boyajian}, T.~S., {Che}, X., {Monnier}, J.~D., {Pope},
  B.~J.~S., {Roettenbacher}, R.~M., {Stello}, D., {Tuthill}, P.~G.,
  {Farrington}, C.~D., {Goldfinger}, P.~J., {McAlister}, H.~A., {Schaefer},
  G.~H., {Sturmann}, J., {Sturmann}, L., {ten Brummelaar}, T.~A.,  and
  {Turner}, N.~H. (2013). \enquote{{Interferometric radii of bright Kepler
  stars with the CHARA Array: {\thetas} Cygni and 16 Cygni A and B},}
  \emph{\mnras} \textbf{433}, pp. 1262--1270, \doi{10.1093/mnras/stt802},
  \eprint{1305.1934}.

\bibitem[{{Wilson} \emph{et~al.}(1992){Wilson}, {Baldwin}, {Buscher} and
  {Warner}}]{1992MNRAS.257..369W}
{Wilson}, R.~W., {Baldwin}, J.~E., {Buscher}, D.~F.,  and {Warner}, P.~J.
  (1992). \enquote{{High-resolution imaging of Betelgeuse and Mira},}
  \emph{\mnras} \textbf{257}, pp. 369--376.

\bibitem[{{Wilson} \emph{et~al.}(1997){Wilson}, {Dhillon} and
  {Haniff}}]{1997MNRAS.291..819W}
{Wilson}, R.~W., {Dhillon}, V.~S.,  and {Haniff}, C.~A. (1997). \enquote{{The
  changing face of Betelgeuse},} \emph{\mnras} \textbf{291}, p. 819.

\bibitem[{{Wilson} \emph{et~al.}(1993){Wilson}, {Milone} and {Fry}}]{wilson93}
{Wilson}, W.~J.~F., {Milone}, E.~F.,  and {Fry}, D.~J.~I. (1993).
  \enquote{{Studies of large-amplitude Delta Scuti variables. I - A case study
  of EH Librae},} \emph{\pasp} \textbf{105}, pp. 809--820,
  \doi{10.1086/133237}.

\bibitem[{{Wilson} \emph{et~al.}(1998){Wilson}, {Milone}, {Fry} and {van
  Leeuwen}}]{wilson98}
{Wilson}, W.~J.~F., {Milone}, E.~F., {Fry}, D.~J.~I.,  and {van Leeuwen}, J.
  (1998). \enquote{{Studies of Large-Amplitude delta Scuti Variables. III. DY
  Pegasi},} \emph{\pasp} \textbf{110}, pp. 433--450, \doi{10.1086/316148}.

\bibitem[{{Wittkowski} \emph{et~al.}(2002){Wittkowski}, {Sch{\"o}ller},
  {Hubrig}, {Posselt} and {von der L{\"u}he}}]{2002AN....323..241W}
{Wittkowski}, M., {Sch{\"o}ller}, M., {Hubrig}, S., {Posselt}, B.,  and {von
  der L{\"u}he}, O. (2002). \enquote{{Measuring starspots on magnetically
  active stars with the VLTI},} \emph{Astronomische Nachrichten} \textbf{323},
  pp. 241--250,
  \doi{10.1002/1521-3994(200208)323:3/4<241::AID-ASNA241>3.0.CO;2-D},
  \eprint{astro-ph/0206090}.

\bibitem[{{Yadav} \emph{et~al.}(2015){Yadav}, {Gastine}, {Christensen} and
  {Reiners}}]{2015A&A...573A..68Y}
{Yadav}, R.~K., {Gastine}, T., {Christensen}, U.~R.,  and {Reiners}, A. (2015).
  \enquote{{Formation of starspots in self-consistent global dynamo models:
  Polar spots on cool stars},} \emph{\aap} \textbf{573}, A68,
  \doi{10.1051/0004-6361/201424589}, \eprint{1407.3187}.

\bibitem[{{Young} \emph{et~al.}(2000){Young}, {Baldwin}, {Boysen}, {Haniff},
  {Lawson}, {Mackay}, {Pearson}, {Rogers}, {St.-Jacques}, {Warner}, {Wilson}
  and {Wilson}}]{2000MNRAS.315..635Y}
{Young}, J.~S., {Baldwin}, J.~E., {Boysen}, R.~C., {Haniff}, C.~A., {Lawson},
  P.~R., {Mackay}, C.~D., {Pearson}, D., {Rogers}, J., {St.-Jacques}, D.,
  {Warner}, P.~J., {Wilson}, D.~M.~A.,  and {Wilson}, R.~W. (2000).
  \enquote{{New views of Betelgeuse: multi-wavelength surface imaging and
  implications for models of hotspot generation},} \emph{\mnras} \textbf{315},
  pp. 635--645.

\bibitem[{{Zhao} \emph{et~al.}(2009){Zhao}, {Monnier}, {Pedretti}, {Thureau},
  {M{\'e}rand}, {ten Brummelaar}, {McAlister}, {Ridgway}, {Turner}, {Sturmann},
  {Sturmann}, {Goldfinger} and {Farrington}}]{zhao2009}
{Zhao}, M., {Monnier}, J.~D., {Pedretti}, E., {Thureau}, N., {M{\'e}rand}, A.,
  {ten Brummelaar}, T., {McAlister}, H., {Ridgway}, S.~T., {Turner}, N.,
  {Sturmann}, J., {Sturmann}, L., {Goldfinger}, P.~J.,  and {Farrington}, C.
  (2009). \enquote{{Imaging and Modeling Rapidly Rotating Stars: {$\alpha$}
  Cephei and {$\alpha$} Ophiuchi},} \emph{\apj} \textbf{701}, pp. 209--224,
  \doi{10.1088/0004-637X/701/1/209}, \eprint{0906.2241}.

\bibitem[{{Zickgraf} \emph{et~al.}(1985){Zickgraf}, {Wolf}, {Stahl},
  {Leitherer} and {Klare}}]{Zickgraf}
{Zickgraf}, F.-J., {Wolf}, B., {Stahl}, O., {Leitherer}, C.,  and {Klare}, G.
  (1985). \enquote{{The hybrid spectrum of the LMC hypergiant R126},}
  \emph{\aap} \textbf{143}, pp. 421--430.

\end{thebibliography}


\begin{thebibliography}{23}
\newcommand{\enquote}[1]{#1}
\providecommand{\natexlab}[1]{#1}
\providecommand{\url}[1]{\texttt{#1}}
\providecommand{\urlprefix}{URL }
\providecommand{\eprint}{eprint }
\expandafter\ifx\csname urlstyle\endcsname\relax
  \providecommand{\doi}[1]{doi:\discretionary{}{}{}#1}\else
  \providecommand{\doi}{doi:\discretionary{}{}{}\begingroup
  \urlstyle{rm}\Url}\fi

\bibitem[{{Baines} \emph{et~al.}(2014){Baines}, {Armstrong}, {Schmitt},
  {Benson}, {Zavala} and {van Belle}}]{Baines2014}
{Baines}, E.~K., {Armstrong}, J.~T., {Schmitt}, H.~R., {Benson}, J.~A.,
  {Zavala}, R.~T.,  and {van Belle}, G.~T. (2014). \enquote{{Navy Precision
  Optical Interferometer Measurements of 10 Stellar Oscillators},} \emph{\apj}
  \textbf{781}, 90, \doi{10.1088/0004-637X/781/2/90}.

\bibitem[{{Boyajian} \emph{et~al.}(2013){Boyajian}, {von Braun}, {van Belle},
  {Farrington}, {Schaefer}, {Jones}, {White}, {McAlister}, {ten Brummelaar},
  {Ridgway}, {Gies}, {Sturmann}, {Sturmann}, {Turner}, {Goldfinger} and
  {Vargas}}]{2013ApJ...771...40B}
{Boyajian}, T.~S., {von Braun}, K., {van Belle}, G., {Farrington}, C.,
  {Schaefer}, G., {Jones}, J., {White}, R., {McAlister}, H.~A., {ten
  Brummelaar}, T.~A., {Ridgway}, S., {Gies}, D., {Sturmann}, L., {Sturmann},
  J., {Turner}, N.~H., {Goldfinger}, P.~J.,  and {Vargas}, N. (2013).
  \enquote{{Stellar Diameters and Temperatures. III. Main-sequence A, F, G, and
  K Stars: Additional High-precision Measurements and Empirical Relations},}
  \emph{\apj} \textbf{771}, 40, \doi{10.1088/0004-637X/771/1/40},
  \eprint{1306.2974}.

\bibitem[{{Bruntt} \emph{et~al.}(2010){Bruntt}, {Kervella}, {M{\'e}rand},
  {Brand{\~a}o}, {Bedding}, {ten Brummelaar}, {Coud{\'e} du Foresto}, {Cunha},
  {Farrington}, {Goldfinger}, {Kiss}, {McAlister}, {Ridgway}, {Sturmann},
  {Sturmann}, {Turner} and {Tuthill}}]{2010A&A...512A..55B}
{Bruntt}, H., {Kervella}, P., {M{\'e}rand}, A., {Brand{\~a}o}, I.~M.,
  {Bedding}, T.~R., {ten Brummelaar}, T.~A., {Coud{\'e} du Foresto}, V.,
  {Cunha}, M.~S., {Farrington}, C., {Goldfinger}, P.~J., {Kiss}, L.~L.,
  {McAlister}, H.~A., {Ridgway}, S.~T., {Sturmann}, J., {Sturmann}, L.,
  {Turner}, N.,  and {Tuthill}, P.~G. (2010). \enquote{{The radius and
  effective temperature of the binary Ap star {$\beta$} CrB from CHARA/FLUOR
  and VLT/NACO observations},} \emph{\aap} \textbf{512}, A55,
  \doi{10.1051/0004-6361/200913405}, \eprint{0912.3215}.

\bibitem[{{Bruntt} \emph{et~al.}(2008){Bruntt}, {North}, {Cunha},
  {Brand{\~a}o}, {Elkin}, {Kurtz}, {Davis}, {Bedding}, {Jacob}, {Owens},
  {Robertson}, {Tango}, {Gameiro}, {Ireland} and
  {Tuthill}}]{2008MNRAS.386.2039B}
{Bruntt}, H., {North}, J.~R., {Cunha}, M., {Brand{\~a}o}, I.~M., {Elkin},
  V.~G., {Kurtz}, D.~W., {Davis}, J., {Bedding}, T.~R., {Jacob}, A.~P.,
  {Owens}, S.~M., {Robertson}, J.~G., {Tango}, W.~J., {Gameiro}, J.~F.,
  {Ireland}, M.~J.,  and {Tuthill}, P.~G. (2008). \enquote{{The fundamental
  parameters of the roAp star {$\alpha$} Circini},} \emph{\mnras} \textbf{386},
  pp. 2039--2046, \doi{10.1111/j.1365-2966.2008.13167.x}, \eprint{0803.1518}.

\bibitem[{{Chiavassa} \emph{et~al.}(2012){Chiavassa}, {Bigot}, {Kervella},
  {Matter}, {Lopez}, {Collet}, {Magic} and {Asplund}}]{2012A&A...540A...5C}
{Chiavassa}, A., {Bigot}, L., {Kervella}, P., {Matter}, A., {Lopez}, B.,
  {Collet}, R., {Magic}, Z.,  and {Asplund}, M. (2012).
  \enquote{{Three-dimensional interferometric, spectrometric, and planetary
  views of Procyon},} \emph{\aap} \textbf{540}, A5,
  \doi{10.1051/0004-6361/201118652}, \eprint{1201.3264}.

\bibitem[{{Chiavassa} \emph{et~al.}(2010){Chiavassa}, {Collet}, {Casagrande}
  and {Asplund}}]{2010A&A...524A..93C}
{Chiavassa}, A., {Collet}, R., {Casagrande}, L.,  and {Asplund}, M. (2010).
  \enquote{{Three-dimensional hydrodynamical simulations of red giant stars:
  semi-global models for interpreting interferometric observations},}
  \emph{\aap} \textbf{524}, A93, \doi{10.1051/0004-6361/201015507},
  \eprint{1009.1745}.

\bibitem[{{Chiavassa} \emph{et~al.}(2014){Chiavassa}, {Ligi}, {Magic},
  {Collet}, {Asplund} and {Mourard}}]{2014A&A...567A.115C}
{Chiavassa}, A., {Ligi}, R., {Magic}, Z., {Collet}, R., {Asplund}, M.,  and
  {Mourard}, D. (2014). \enquote{{Planet transit and stellar granulation
  detection with interferometry. Using the three-dimensional stellar atmosphere
  Stagger-grid simulations},} \emph{\aap} \textbf{567}, A115,
  \doi{10.1051/0004-6361/201323207}, \eprint{1404.7049}.

\bibitem[{{Creevey} \emph{et~al.}(2014){Creevey}, {Th{\'e}venin}, {Berio},
  {Heiter}, {von Braun}, {Mourard}, {Bigot}, {Boyajian}, {Kervella}, {Morel},
  {Pichon}, {Chiavassa}, {Nardetto}, {Perraut}, {Meilland}, {Mc Alister}, {ten
  Brummelaar}, {Sturmann}, {Sturmann} and {Turner}}]{creevey14}
{Creevey}, O., {Th{\'e}venin}, F., {Berio}, P., {Heiter}, U., {von Braun}, K.,
  {Mourard}, D., {Bigot}, L., {Boyajian}, T.~S., {Kervella}, P., {Morel}, P.,
  {Pichon}, B., {Chiavassa}, A., {Nardetto}, N., {Perraut}, K., {Meilland}, A.,
  {Mc Alister}, H.~A., {ten Brummelaar}, T.~A., {Sturmann}, C.~F.~J.,
  {Sturmann}, L.,  and {Turner}, N. (2014). \enquote{{Benchmark stars for Gaia:
  fundamental properties of the Population II star HD140283 from
  interferometric, spectroscopic and photometric data},} \emph{ArXiv e-prints}
  \eprint{1410.4780}.

\bibitem[{{Creevey} \emph{et~al.}(2013){Creevey}, {Th{\'e}venin}, {Basu},
  {Chaplin}, {Bigot}, {Elsworth}, {Huber}, {Monteiro} and
  {Serenelli}}]{creevey13}
{Creevey}, O.~L., {Th{\'e}venin}, F., {Basu}, S., {Chaplin}, W.~J., {Bigot},
  L., {Elsworth}, Y., {Huber}, D., {Monteiro}, M.~J.~P.~F.~G.,  and
  {Serenelli}, A. (2013). \enquote{{A large sample of calibration stars for
  Gaia: log g from Kepler and CoRoT fields},} \emph{\mnras} \textbf{431}, pp.
  2419--2432, \doi{10.1093/mnras/stt336}, \eprint{1302.7158}.

\bibitem[{{Creevey} \emph{et~al.}(2012){Creevey}, {Th{\'e}venin}, {Boyajian},
  {Kervella}, {Chiavassa}, {Bigot}, {M{\'e}rand}, {Heiter}, {Morel}, {Pichon},
  {Mc Alister}, {ten Brummelaar}, {Collet}, {van Belle}, {Coud{\'e} du
  Foresto}, {Farrington}, {Goldfinger}, {Sturmann}, {Sturmann} and
  {Turner}}]{creevey12}
{Creevey}, O.~L., {Th{\'e}venin}, F., {Boyajian}, T.~S., {Kervella}, P.,
  {Chiavassa}, A., {Bigot}, L., {M{\'e}rand}, A., {Heiter}, U., {Morel}, P.,
  {Pichon}, B., {Mc Alister}, H.~A., {ten Brummelaar}, T.~A., {Collet}, R.,
  {van Belle}, G.~T., {Coud{\'e} du Foresto}, V., {Farrington}, C.,
  {Goldfinger}, P.~J., {Sturmann}, J., {Sturmann}, L.,  and {Turner}, N.
  (2012). \enquote{{Fundamental properties of the Population II fiducial stars
  <ASTROBJ>HD 122563</ASTROBJ> and <ASTROBJ>Gmb 1830</ASTROBJ> from CHARA
  interferometric observations},} \emph{\aap} \textbf{545}, A17,
  \doi{10.1051/0004-6361/201219651}, \eprint{1207.5954}.

\bibitem[{{Ireland} \emph{et~al.}(2008){Ireland}, {M{\'e}rand}, {ten
  Brummelaar}, {Tuthill}, {Schaefer}, {Turner}, {Sturmann}, {Sturmann} and
  {McAlister}}]{Ireland2008}
{Ireland}, M.~J., {M{\'e}rand}, A., {ten Brummelaar}, T.~A., {Tuthill}, P.~G.,
  {Schaefer}, G.~H., {Turner}, N.~H., {Sturmann}, J., {Sturmann}, L.,  and
  {McAlister}, H.~A. (2008). \enquote{{Sensitive visible interferometry with
  PAVO},} in \emph{Society of Photo-Optical Instrumentation Engineers (SPIE)
  Conference Series}, Vol. 7013, \doi{10.1117/12.788386}.

\bibitem[{{Jankov} \emph{et~al.}(2003){Jankov}, {Domiciano de Souza}, {Stehle},
  {Vakili}, {Perraut-Rousselet} and {Chesneau}}]{2003SPIE.4838..587J}
{Jankov}, S., {Domiciano de Souza}, A., Jr., {Stehle}, C., {Vakili}, F.,
  {Perraut-Rousselet}, K.,  and {Chesneau}, O. (2003).
  \enquote{{Interferometric-Doppler imaging of stellar surface abundances},} in
  W.~A. {Traub} (ed.), \emph{Interferometry for Optical Astronomy II},
  \emph{Society of Photo-Optical Instrumentation Engineers (SPIE) Conference
  Series}, Vol. 4838, pp. 587--593.

\bibitem[{{Ligi} \emph{et~al.}(2014){Ligi}, {Mourard}, {Lagrange}, {Perraut}
  and {Chiavassa}}]{Ligi2014}
{Ligi}, R., {Mourard}, D., {Lagrange}, A.-M., {Perraut}, K.,  and {Chiavassa},
  A. (2014). \enquote{{On the characterization of transiting exoplanets and
  magnetic spots with optical interferometry},} \emph{ArXiv e-prints}
  \eprint{1410.5333}.

\bibitem[{Ligi \emph{et~al.}(2013)Ligi, Mourard, Nardetto and
  Clausse}]{Ligi2013}
Ligi, R., Mourard, D., Nardetto, N.,  and Clausse, J.-M. (2013). \enquote{The
  operation of vega/chara: From the scientific idea to the final products,}
  \emph{Journal of Astronomical Instrumentation} \textbf{02}, 02, p. 1340003,
  \doi{10.1142/S2251171713400035},
  \eprint{http://www.worldscientific.com/doi/pdf/10.1142/S2251171713400035}.

\bibitem[{{Maestro} \emph{et~al.}(2013){Maestro}, {Che}, {Huber}, {Ireland},
  {Monnier}, {White}, {Kok}, {Robertson}, {Schaefer}, {ten Brummelaar} and
  {Tuthill}}]{2013MNRAS.434.1321M}
{Maestro}, V., {Che}, X., {Huber}, D., {Ireland}, M.~J., {Monnier}, J.~D.,
  {White}, T.~R., {Kok}, Y., {Robertson}, J.~G., {Schaefer}, G.~H., {ten
  Brummelaar}, T.~A.,  and {Tuthill}, P.~G. (2013). \enquote{{Optical
  interferometry of early-type stars with PAVO@CHARA - I. Fundamental stellar
  properties},} \emph{\mnras} \textbf{434}, pp. 1321--1331,
  \doi{10.1093/mnras/stt1092}, \eprint{1306.5937}.

\bibitem[{{McAlister} \emph{et~al.}(2012){McAlister}, {ten Brummelaar},
  {Ridgway}, {Gies}, {Sturmann}, {Sturmann}, {Turner}, {Schaefer}, {Boyajian},
  {Farrington}, {Goldfinger} and {Webster}}]{McAlister2012}
{McAlister}, H.~A., {ten Brummelaar}, T.~A., {Ridgway}, S.~T., {Gies}, D.~R.,
  {Sturmann}, J., {Sturmann}, L., {Turner}, N.~H., {Schaefer}, G.~H.,
  {Boyajian}, T.~S., {Farrington}, C.~D., {Goldfinger}, P.~J.,  and {Webster},
  L. (2012). \enquote{{Recent technical and scientific highlights from the
  CHARA Array},} in \emph{Society of Photo-Optical Instrumentation Engineers
  (SPIE) Conference Series}, Vol. 8445, \doi{10.1117/12.926452}.

\bibitem[{{Mourard} \emph{et~al.}(2009){Mourard}, {Clausse}, {Marcotto},
  {Perraut}, {Tallon-Bosc}, {B{\'e}rio}, {Blazit}, {Bonneau}, {Bosio},
  {Bresson}, {Chesneau}, {Delaa}, {H{\'e}nault}, {Hughes}, {Lagarde}, {Merlin},
  {Roussel}, {Spang}, {Stee}, {Tallon}, {Antonelli}, {Foy}, {Kervella},
  {Petrov}, {Thiebaut}, {Vakili}, {McAlister}, {ten Brummelaar}, {Sturmann},
  {Sturmann}, {Turner}, {Farrington} and {Goldfinger}}]{Mourard2009}
{Mourard}, D., {Clausse}, J.~M., {Marcotto}, A., {Perraut}, K., {Tallon-Bosc},
  I., {B{\'e}rio}, P., {Blazit}, A., {Bonneau}, D., {Bosio}, S., {Bresson}, Y.,
  {Chesneau}, O., {Delaa}, O., {H{\'e}nault}, F., {Hughes}, Y., {Lagarde}, S.,
  {Merlin}, G., {Roussel}, A., {Spang}, A., {Stee}, P., {Tallon}, M.,
  {Antonelli}, P., {Foy}, R., {Kervella}, P., {Petrov}, R., {Thiebaut}, E.,
  {Vakili}, F., {McAlister}, H., {ten Brummelaar}, T., {Sturmann}, J.,
  {Sturmann}, L., {Turner}, N., {Farrington}, C.,  and {Goldfinger}, P.~J.
  (2009). \enquote{{VEGA: Visible spEctroGraph and polArimeter for the CHARA
  array: principle and performance},} \emph{\aap} \textbf{508}, pp. 1073--1083,
  \doi{10.1051/0004-6361/200913016}.

\bibitem[{{Perraut} \emph{et~al.}(2013){Perraut}, {Borgniet}, {Cunha}, {Bigot},
  {Brand{\~a}o}, {Mourard}, {Nardetto}, {Chesneau}, {McAlister}, {ten
  Brummelaar}, {Sturmann}, {Sturmann}, {Turner}, {Farrington} and
  {Goldfinger}}]{2013A&A...559A..21P}
{Perraut}, K., {Borgniet}, S., {Cunha}, M., {Bigot}, L., {Brand{\~a}o}, I.,
  {Mourard}, D., {Nardetto}, N., {Chesneau}, O., {McAlister}, H., {ten
  Brummelaar}, T.~A., {Sturmann}, J., {Sturmann}, L., {Turner}, N.,
  {Farrington}, C.,  and {Goldfinger}, P.~J. (2013). \enquote{{The fundamental
  parameters of the roAp star 10 Aquilae},} \emph{\aap} \textbf{559}, A21,
  \doi{10.1051/0004-6361/201321849}, \eprint{1309.4423}.

\bibitem[{{Perraut} \emph{et~al.}(2011){Perraut}, {Brand{\~a}o}, {Mourard},
  {Cunha}, {B{\'e}rio}, {Bonneau}, {Chesneau}, {Clausse}, {Delaa}, {Marcotto},
  {Roussel}, {Spang}, {Stee}, {Tallon-Bosc}, {McAlister}, {ten Brummelaar},
  {Sturmann}, {Sturmann}, {Turner}, {Farrington} and
  {Goldfinger}}]{2011A&A...526A..89P}
{Perraut}, K., {Brand{\~a}o}, I., {Mourard}, D., {Cunha}, M., {B{\'e}rio}, P.,
  {Bonneau}, D., {Chesneau}, O., {Clausse}, J.~M., {Delaa}, O., {Marcotto}, A.,
  {Roussel}, A., {Spang}, A., {Stee}, P., {Tallon-Bosc}, I., {McAlister}, H.,
  {ten Brummelaar}, T., {Sturmann}, J., {Sturmann}, L., {Turner}, N.,
  {Farrington}, C.,  and {Goldfinger}, P.~J. (2011). \enquote{{The fundamental
  parameters of the roAp star {$\gamma$} Equulei},} \emph{\aap} \textbf{526},
  A89, \doi{10.1051/0004-6361/201015801}, \eprint{1011.2028}.

\bibitem[{{Rousselet-Perraut} \emph{et~al.}(2004){Rousselet-Perraut},
  {Stehl{\'e}}, {Lanz}, {Le Bouquin}, {Boudoyen}, {Kilbinger}, {Kochukhov} and
  {Jankov}}]{2004A&A...422..193R}
{Rousselet-Perraut}, K., {Stehl{\'e}}, C., {Lanz}, T., {Le Bouquin}, J.~B.,
  {Boudoyen}, T., {Kilbinger}, M., {Kochukhov}, O.,  and {Jankov}, S. (2004).
  \enquote{{Stellar activity and magnetism studied by optical interferometry},}
  \emph{\aap} \textbf{422}, pp. 193--203, \doi{10.1051/0004-6361:20040151}.

\bibitem[{{Shulyak} \emph{et~al.}(2014){Shulyak}, {Paladini}, {Causi},
  {Perraut} and {Kochukhov}}]{2014MNRAS.443.1629S}
{Shulyak}, D., {Paladini}, C., {Causi}, G.~L., {Perraut}, K.,  and {Kochukhov},
  O. (2014). \enquote{{Interferometry of chemically peculiar stars: theoretical
  predictions versus modern observing facilities},} \emph{\mnras} \textbf{443},
  pp. 1629--1642, \doi{10.1093/mnras/stu1259}, \eprint{1406.6093}.

\bibitem[{{Soubiran} \emph{et~al.}(2010){Soubiran}, {Le Campion}, {Cayrel de
  Strobel} and {Caillo}}]{soubiran10}
{Soubiran}, C., {Le Campion}, J.-F., {Cayrel de Strobel}, G.,  and {Caillo}, A.
  (2010). \enquote{{The PASTEL catalogue of stellar parameters},} \emph{\aap}
  \textbf{515}, A111, \doi{10.1051/0004-6361/201014247}, \eprint{1004.1069}.

\bibitem[{{Wittkowski} \emph{et~al.}(2002){Wittkowski}, {Sch{\"o}ller},
  {Hubrig}, {Posselt} and {von der L{\"u}he}}]{2002AN....323..241W}
{Wittkowski}, M., {Sch{\"o}ller}, M., {Hubrig}, S., {Posselt}, B.,  and {von
  der L{\"u}he}, O. (2002). \enquote{{Measuring starspots on magnetically
  active stars with the VLTI},} \emph{Astronomische Nachrichten} \textbf{323},
  pp. 241--250,
  \doi{10.1002/1521-3994(200208)323:3/4<241::AID-ASNA241>3.0.CO;2-D},
  \eprint{astro-ph/0206090}.

\end{thebibliography}

%\clearpage                        % when previous chapter ends on even page
%\blankpage                         % when previous chapter ends on odd page
%\title{Index}                      % to set `Index` as even page running title
%\printindex
\end{document}